\documentclass[12pt,final]{uvathesis}         

\usepackage{graphicx}
\usepackage{amssymb}

\graphicspath{{eps/}}

\shorttitle{Resonant Processes in a Frozen Gas}
\title{Resonant Processes in a Frozen Gas}
\name{Jeremy Shane Frasier}
\home{Charlottesville, Virginia}
\degrees{B.S.\ University of Virginia, 1996}
\department{Physics}
\monthyear{May, 2001}
\numberofmembers{4}

\newlength{\figwidth}

\begin{document}

\maketitle
\copyrightpage
\begin{frontmatter}

\begin{acknowledgements}
First I would like to thank my advisor, Vittorio Celli, for
agreeing to take me on as a student and for finding an interesting
and novel problem on which to work.

\medskip

Thanks to Michael Fowler, we were lucky enough to have Tom Blum
work with us during the summer of 1998. We made a great deal of
progress and submitted our first publication on this work that
summer, in large part because of Tom's diligent efforts. He also
kept me from becoming lazy while my advisor was away.

\medskip

Tom Gallagher first brought the results of his experiments to our
attention, and he was always willing to discuss this work with us.
He also provided me with the opportunity to travel to the
University of Paris and work with Vladimir Akulin during the
spring of 1999.

\medskip

Ziad Maassarani, Tim Newman, and Mike Robinson provided many
useful discussions. Bryan Wright was always willing to lend his
considerable computer expertise. Bobby Anderson and Jay Lowell
kindly provided the raw data for the figures that appear in their
theses. Jay also donated a postscript file he created of the
energy levels of rubidium, which saved me the enormous trouble of
recreating that figure from scratch.

\medskip

Acknowledgement is made to the Thomas F. and Kate Miller Jeffress
Memorial Trust for the support of this research. I was supported
by a Dissertation Year Fellowship from the Graduate School of Arts
and Sciences during the fall and spring semesters of the
1999--2000 school year. During the 1996--1999 school years I was
supported by a Presidential Fellowship and a Hugh P. Kelly
Fellowship from the Department of Physics.
\end{acknowledgements}

\begin{abstractpage}
In this thesis are presented analytical results and numerical
simulations concerning resonant processes in a frozen gas. This
work was begun with the intent of modeling experiments involving
energy transfer due to dipole-dipole interactions between
ultracold Rydberg atoms, and the subsequent propagation of dipolar
excitons. However, the results and the techniques developed to
obtain them are more generally applicable to a wide variety of
problems.

\medskip

We start by studying a simplified model that considers only the
energy transfer process in an infinite medium of randomly
distributed atoms. This model is very instructive, because it
gives a simplified arena in which to develop the techniques that
are used again and again throughout the thesis. Exact analytical
results are obtained for the time evolution of energy transfer,
its dependence on the detuning from resonance, and the
localization distance over which the transfer process is
effective. The result for the initial rate of energy transfer
turns out to have general validity. This soluble model also
provides an excellent means for testing the numerical simulations
that are carried out for a finite sample.

\medskip

When we allow for the propagation of dipolar excitons, it is no
longer possible to obtain exact analytical solutions, but an
approximation has been found that agrees quite well with the
numerical simulations for the amplitude of the overall process.
Simulations are necessary to find the sample-averaged probability
of the overall process, and also to take into account the
dependence of the results on the sample shape.

\medskip

In the experiments we wish to model, the atomic states are
degenerate with respect to the sign of the projection of total
angular momentum in the direction of the applied electric field.
We show that, in one case of interest, this degeneracy simply
leads to a rescaling of the dipole-dipole interaction in the final
formulas for the simple model.
\end{abstractpage}

\tableofcontents
\listoffigures
\listoftables
\clearpage
\end{frontmatter}


\newcommand{\mc}[1]{\ensuremath{{\mathcal{#1}}}}

\newcommand{\mbf}[1]{\ensuremath{{\mathbf{#1}}}}

\newcommand{\mrm}[1]{\ensuremath{{\mathrm{#1}}}}

\newcommand{\erfc}{\mrm{erfc}}

\newcommand{\G}[1]{\Gamma\left(#1\right)}

\newcommand{\Tr}{\mrm{Tr}}

\chapter{Introduction}
\markright{Chapter \arabic{chapter}: Introduction}
\label{introduction}

Frozen gases are a new and in many ways ideal laboratory to test
our understanding of quantum theory in a complex system. With
present technology one can manipulate and detect electronic
processes with an extraordinary selectivity and precision.
Furthermore, the translational temperature of the gas can be
lowered to the point where it can be ignored when discussing
electronic processes. Frozen Rydberg gases have the added
advantages that their states are well understood and many
processes occur on a microsecond time scale, easily allowing for
time-resolved spectroscopy.

Pioneering experiments on resonant processes in these gases have
been carried out by Anderson \emph{et
al.\/}~\cite{Anderson1996a,Anderson1998a} and Mourachko \emph{et
al.\/}~\cite{Mourachko1998a,Mourachko1999a}. The present work was
motivated by the desire to understand those experiments and the
subsequent work of Lowell \emph{et al.\/}~\cite{Lowell1998a}
highlighting the dynamic aspects of these resonant, many particle
systems. The measurements of Anderson \emph{et al.\/} and Lowell
\emph{et al.\/} are performed on mixtures of cold (about $150$
$\mu$K) ${}^{85}Rb$ atoms initially prepared in the $25s_{1/2}$
and $33s_{1/2}$ states, henceforth to be called the $s$ and
$s^{\prime}$ states respectively. There are initially $N$ atoms in
the $s$ state and $N^{\prime}$ atoms in the $s^{\prime}$ state. A
dipole matrix element $\mu$ connects the state $s$ to the state
$p$, and another dipole matrix element $\mu^{\prime}$ connects the
state $s^{\prime}$ to the state $p^{\prime}$, where $p$ refers to
the $24p_{1/2}$ state and $p^{\prime}$ refers to the $34p_{3/2}$
state. Dipole-dipole interactions cause $ss^{\prime}\rightarrow
pp^{\prime}$ interactions to occur, and then $sp\rightarrow ps$
and $s^{\prime}p^{\prime}\rightarrow  p^{\prime}s^{\prime}$
flippings become possible.\footnote{Although the
$s^{\prime}p^{\prime}\rightarrow p^{\prime}s^{\prime}$ process is
present, it is smaller than the $sp\rightarrow ps$ process by a
factor of $16$ in this case and can therefore be neglected to a
first approximation, except when $N \ll N^{\prime}$.} The number
of $p^{\prime}$ states is measured after a given time has elapsed,
up to about 5 $\mu$s. As a function of time, the number of
$p^{\prime}$ states rises rapidly at first and then slowly
saturates. The energy of a $pp^{\prime}$ pair can be swept through
resonance with an $ss^{\prime}$ pair, where $\epsilon_{s^{\prime}}
+ \epsilon_{s} = \epsilon_{p^{\prime}} + \epsilon_{p}$, via the
Stark effect by the application of an electric field.\footnote{For
a discussion of the Stark effect in hydrogenic atoms, see pages
228--241 of Ref.~\cite{Bethe1977a} or pages 164--171 of
Ref.~\cite{Friedrich1991a}.} At any one time the resonance
lineshape can be obtained as a function of the detuning $\Delta =
\epsilon_{p^{\prime}}+\epsilon_{p} -
\epsilon_{s^{\prime}}-\epsilon_{s}$, which is a known function of
the applied electric field.

\begin{figure}
\resizebox{\textwidth}{!}{\includegraphics[0in,0in][8in,10in]{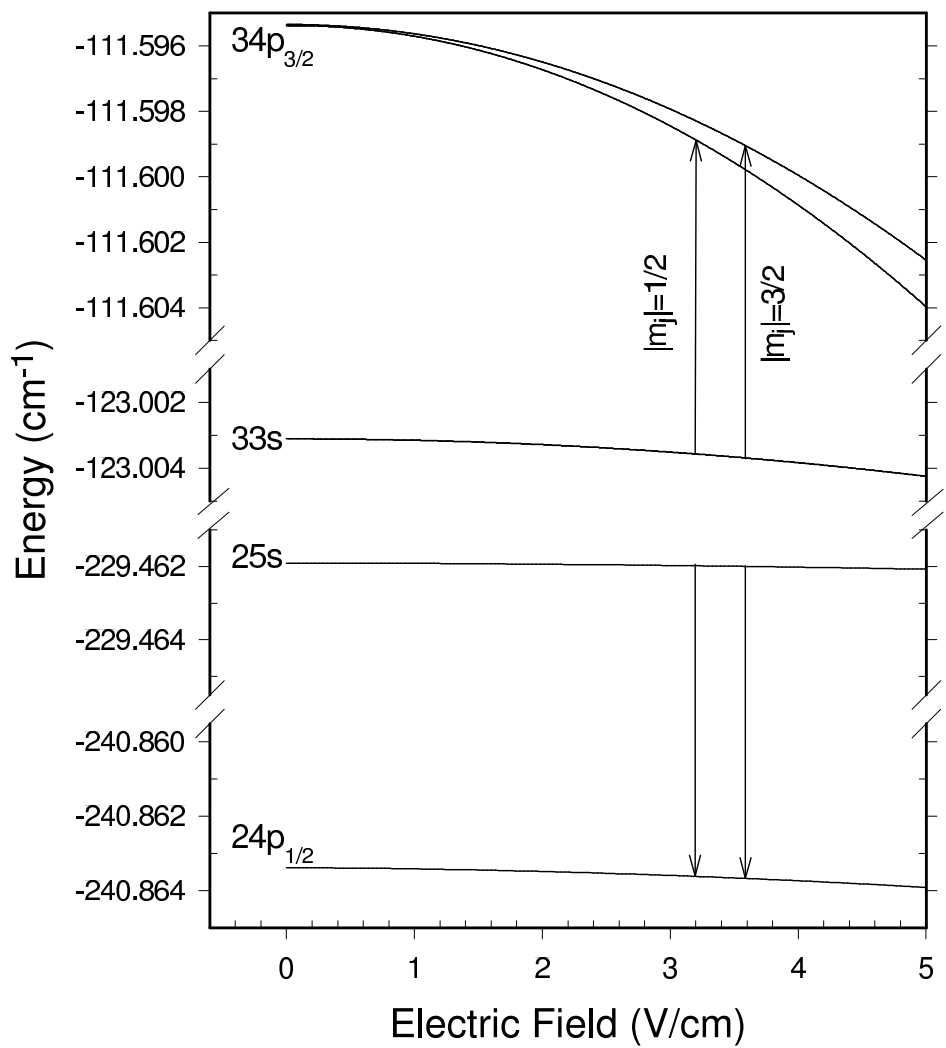}}
\caption[The $s$, $s^{\prime}$, $p$, and $p^{\prime}$ energy
levels of Rb as a function of electric field]{The $s$,
$s^{\prime}$, $p$, and $p^{\prime}$ energy levels of Rb as a
function of electric field.} \label{introduction_states}
\end{figure}

Figure~\ref{introduction_states} shows the $s$, $s^{\prime}$, $p$,
and $p^{\prime}$ energies as a function of the applied electric
field. This figure is a slightly modified version of Fig.~3.1 on
page~54 of Ref.~\cite{Lowell1998a}. Because the $p^{\prime}$ state
is in the $j = 3/2$ state, there are actually two resonances. One
corresponds to the two $\left| m_{j}\right| = 1/2$ states, and the
other corresponds to the two $\left| m_{j}\right| = 3/2$ states.
To a first approximation the two resonances can be treated
separately.

\begin{figure}
\resizebox{0.5\textwidth}{!}{\includegraphics[0in,0in][8in,10in]{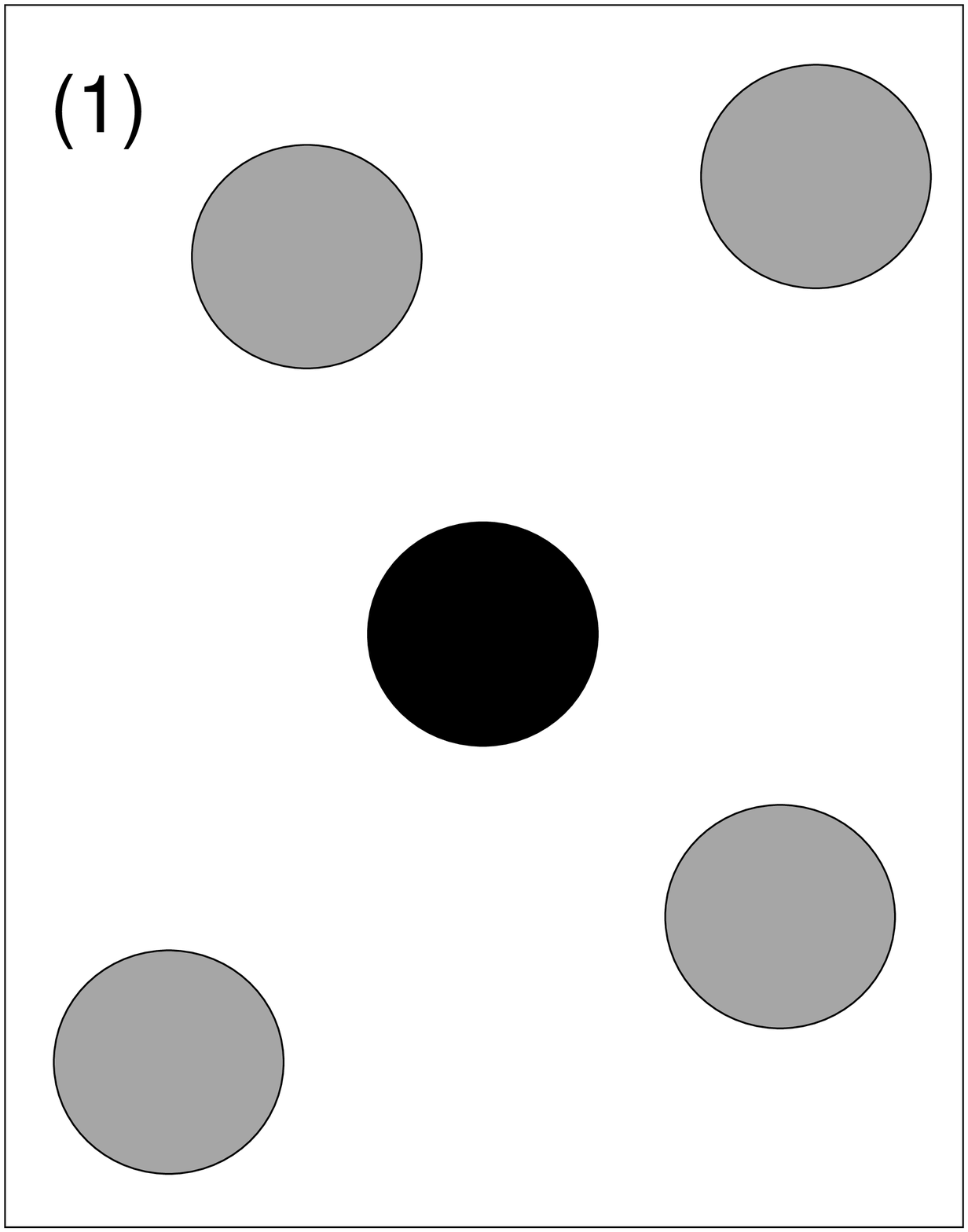}}
\resizebox{0.5\textwidth}{!}{\includegraphics[0in,0in][8in,10in]{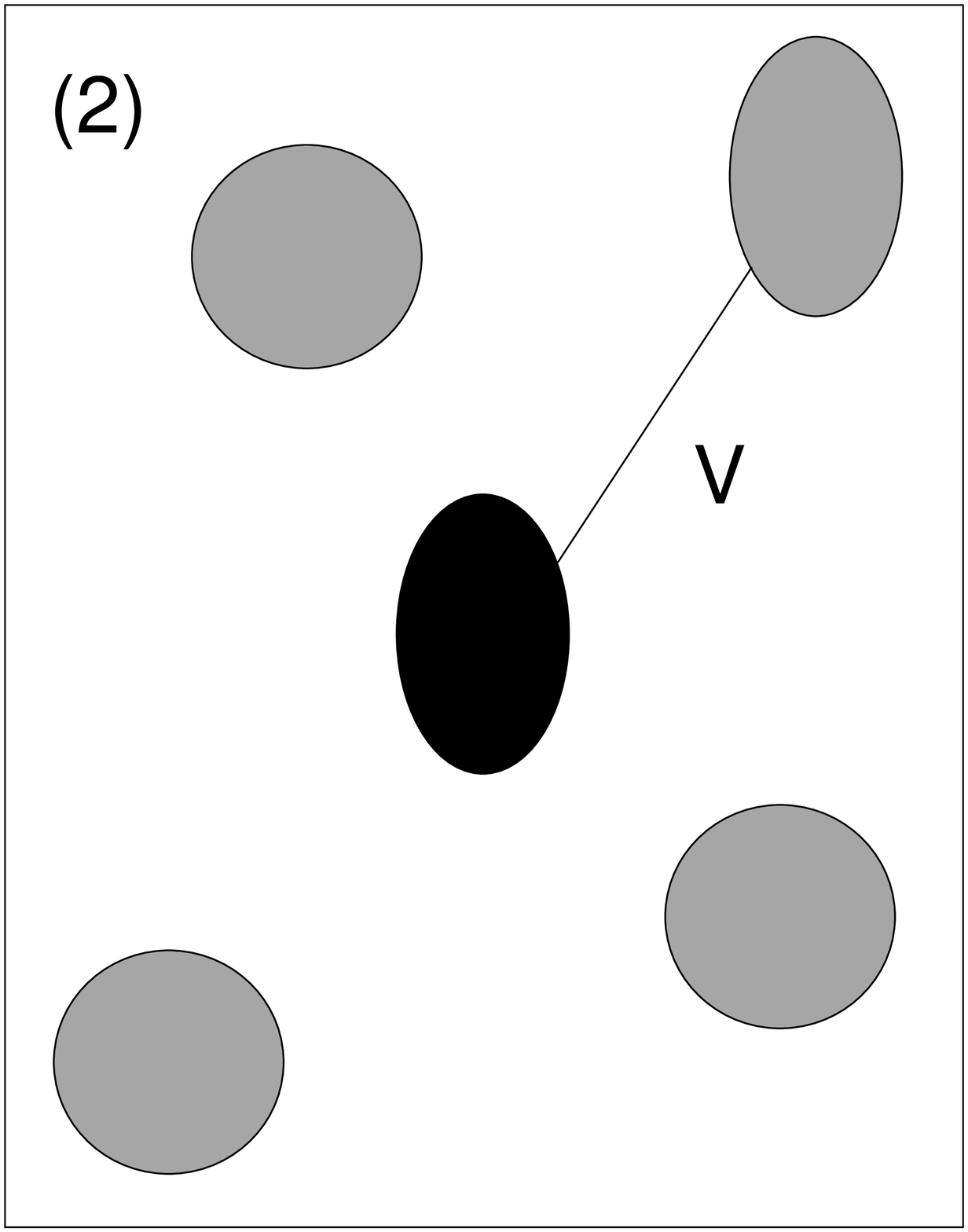}}
\\
\resizebox{0.5\textwidth}{!}{\includegraphics[0in,0in][8in,10in]{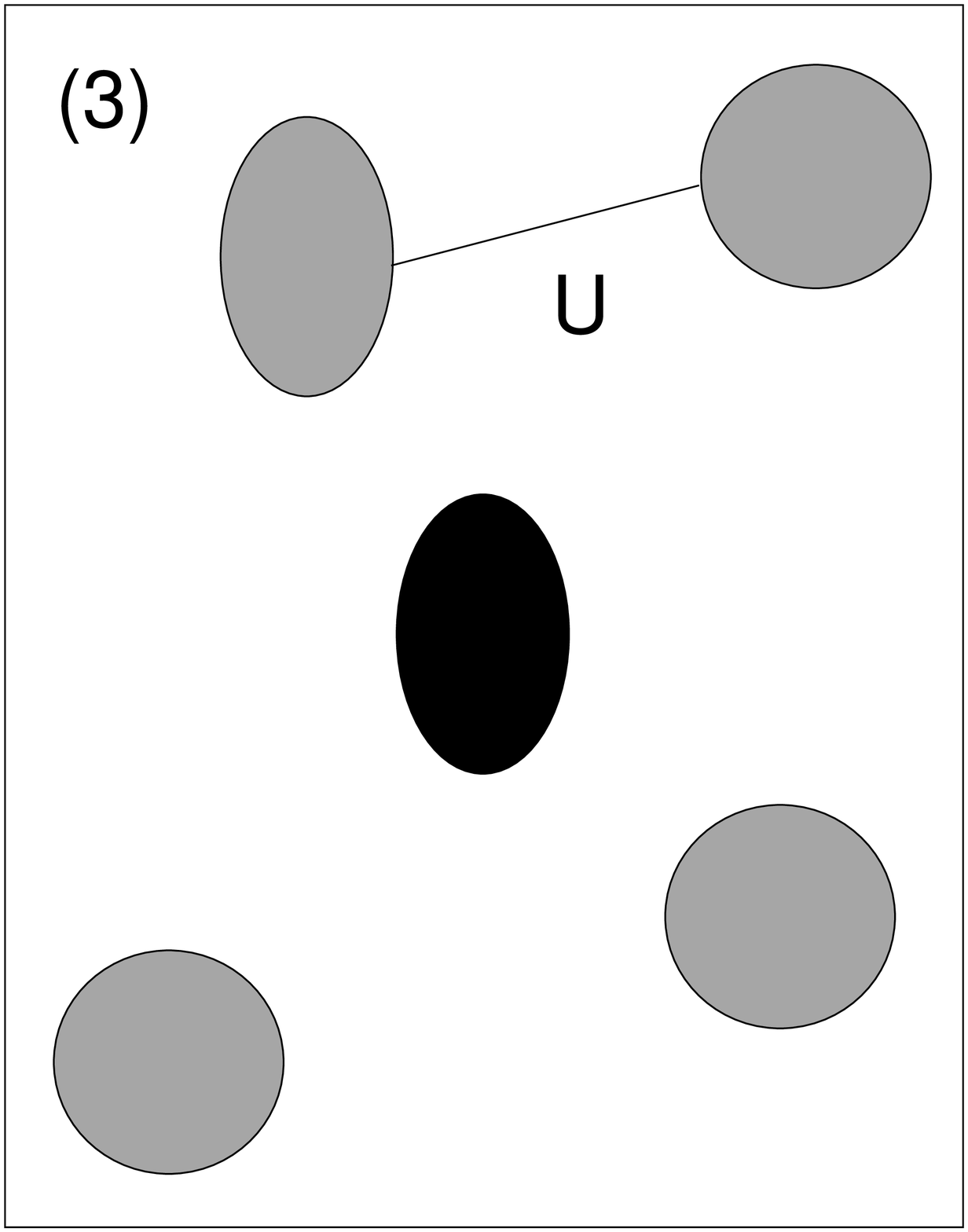}}
\resizebox{0.5\textwidth}{!}{\includegraphics[0in,0in][8in,10in]{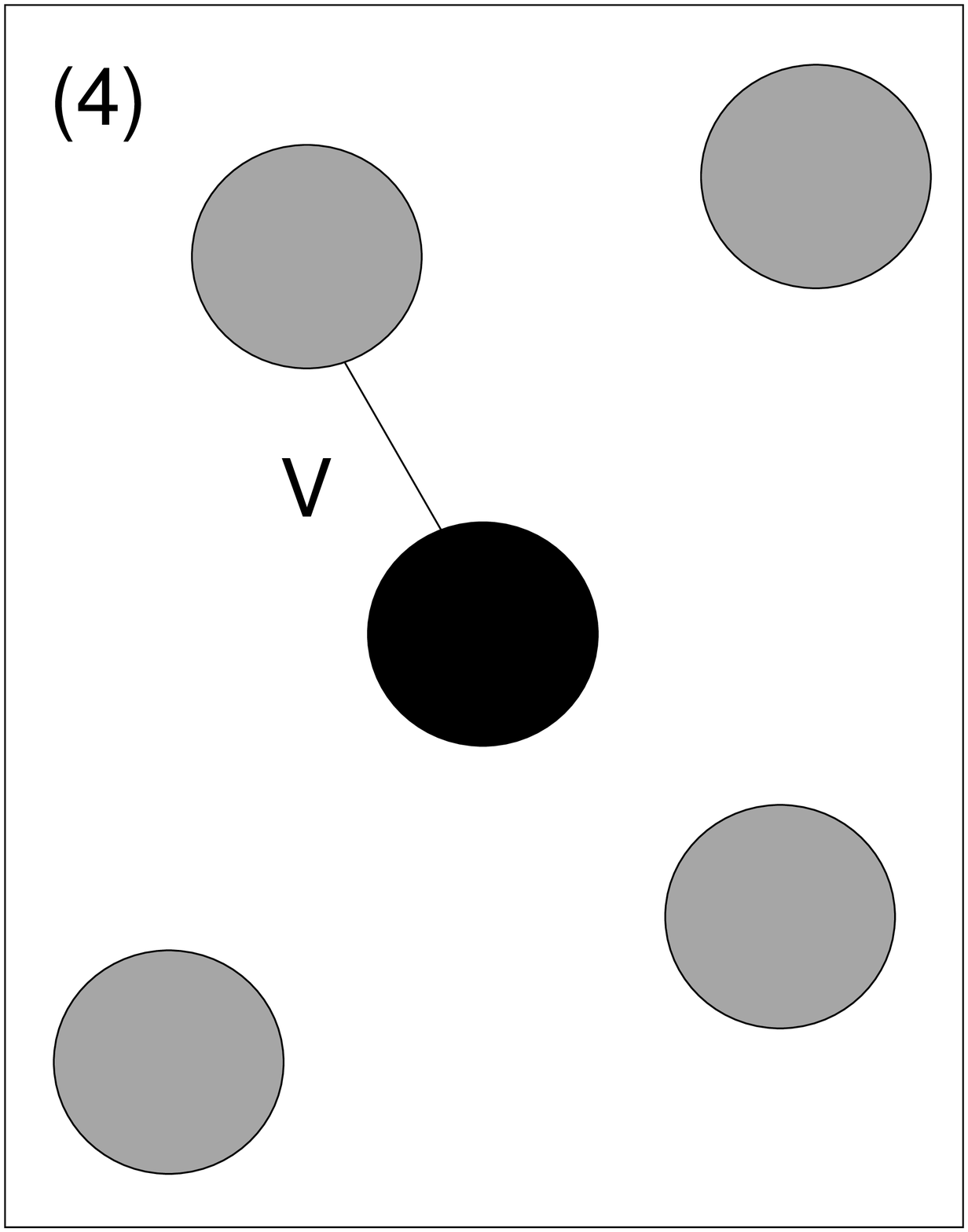}}
\caption[Illustration of the $ss^{\prime}\rightarrow pp^{\prime}$
and $sp\rightarrow ps$ processes]{The circles represent atoms in
an $s$ state and the ovals represent atoms in a $p$ state. The
black shapes are in primed states, while the grey shapes are in
unprimed states. The first panel shows a single atom in the
$s^{\prime}$ state surrounded by atoms in the $s$ state. In the
second panel an $ss^{\prime}$ pair has made the
$ss^{\prime}\rightarrow pp^{\prime}$ transition via an interaction
potential $V$. In the third panel an $sp$ pair has made the
$sp\rightarrow ps$ transition via an interaction potential $U$. In
the final panel the $pp^{\prime}$ pair has made the
$ss^{\prime}\rightarrow pp^{\prime}$ transition, and the system is
again in its initial state.} \label{introduction_processes}
\end{figure}

Figure~\ref{introduction_processes} is a schematic of the
$ss^{\prime}\rightarrow pp^{\prime}$ and $sp\rightarrow ps$
processes. In this illustration the circles represent atoms in an
$s$ state and the ovals represent atoms in a $p$ state. In
addition, the black shapes represent primed states, while the grey
shapes represent unprimed states. The first panel shows a single
atom in the $s^{\prime}$ state surrounded by atoms in the $s$
state. In going from the first to the second panel an
$ss^{\prime}$ pair has made the $ss^{\prime}\rightarrow
pp^{\prime}$ transition via an interaction potential $V$. In the
third panel an $sp$ pair has made the $sp\rightarrow ps$
transition via an interaction potential $U$. In the final panel
the $pp^{\prime}$ pair has made the $ss^{\prime}\rightarrow
pp^{\prime}$ transition, and the system has returned to the
initial state of the first panel.

\begin{figure}
\resizebox{\textwidth}{!}{\includegraphics[0in,0in][8in,10in]{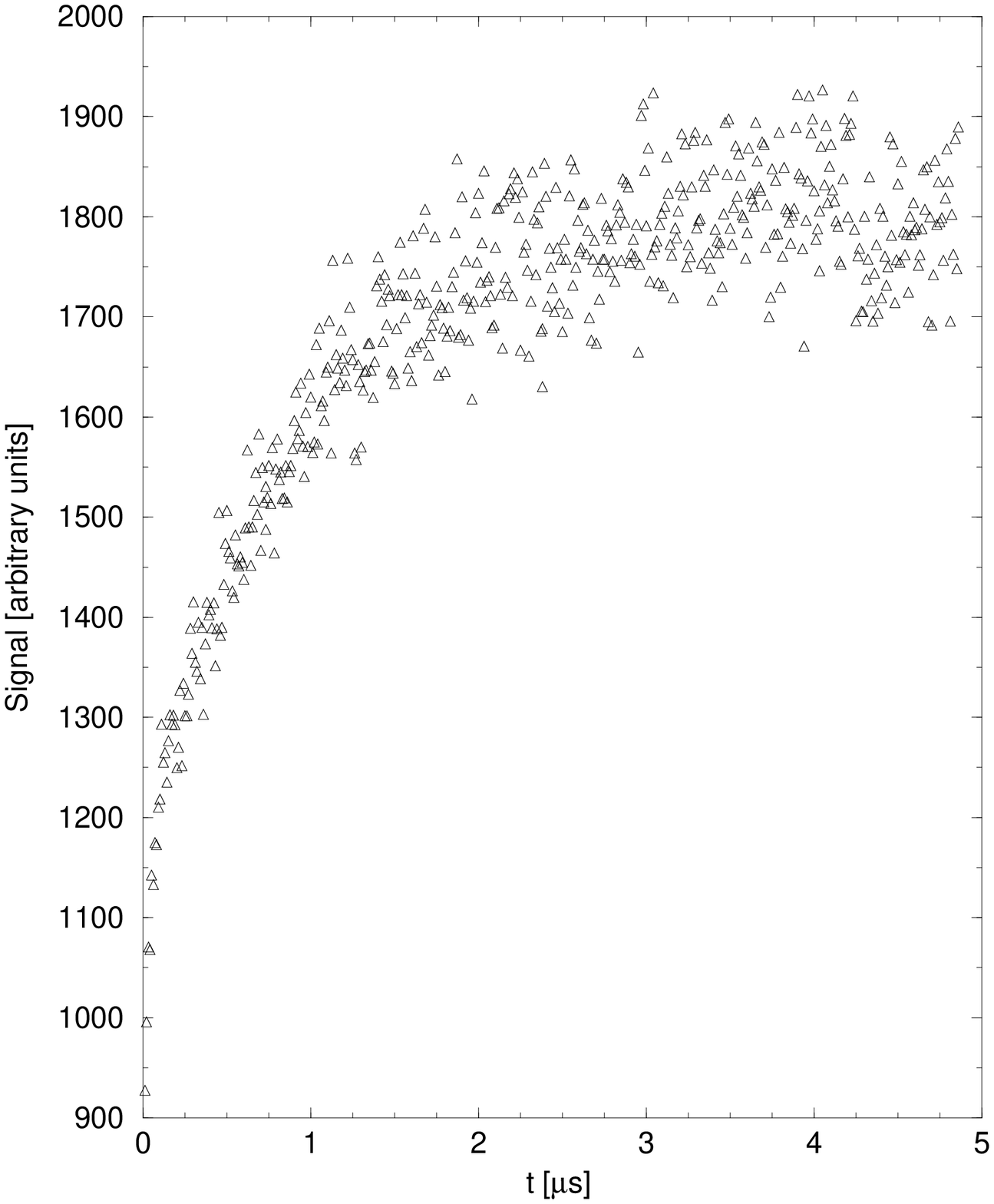}}
\caption[Experimental signal as a function of time]{A
representative plot of the experimental signal as a function of
time. The data are taken from Fig.~3.8 on page~71 of
Ref.~\cite{Lowell1998a} with all the data points shown.}
\label{introduction_expsig}
\end{figure}

Several features of the experimental data present a challenge to
theorists and are investigated in this thesis. The first feature
we wish to understand is the rapid rise followed by slow approach
to saturation of the signal. This behavior can be seen in
Fig.~\ref{introduction_expsig}, which is a plot of the
experimental data taken from Fig.~3.8 on page~71 of
Ref.~\cite{Lowell1998a} with all the data points shown. Each
individual interaction between pairs of atoms leads to a coherent
oscillatory behavior, but as we will see in
Chapter~\ref{sparse_no_u}, averaging the $ss^{\prime}\rightarrow
pp^{\prime}$ interaction over the random positions of the atoms
greatly smooths out the signal. The effective incoherence brought
about by the $sp\rightarrow ps$ process (the ``walking away''
discussed by Mourachko \emph{et
al.\/}~\cite{Mourachko1998a,Mourachko1999a}) completes the
smoothing out of the on-resonance signal but has less effect on
the off-resonance signal.

\begin{figure}
\resizebox{\textwidth}{!}{\includegraphics[0in,0in][8in,10in]{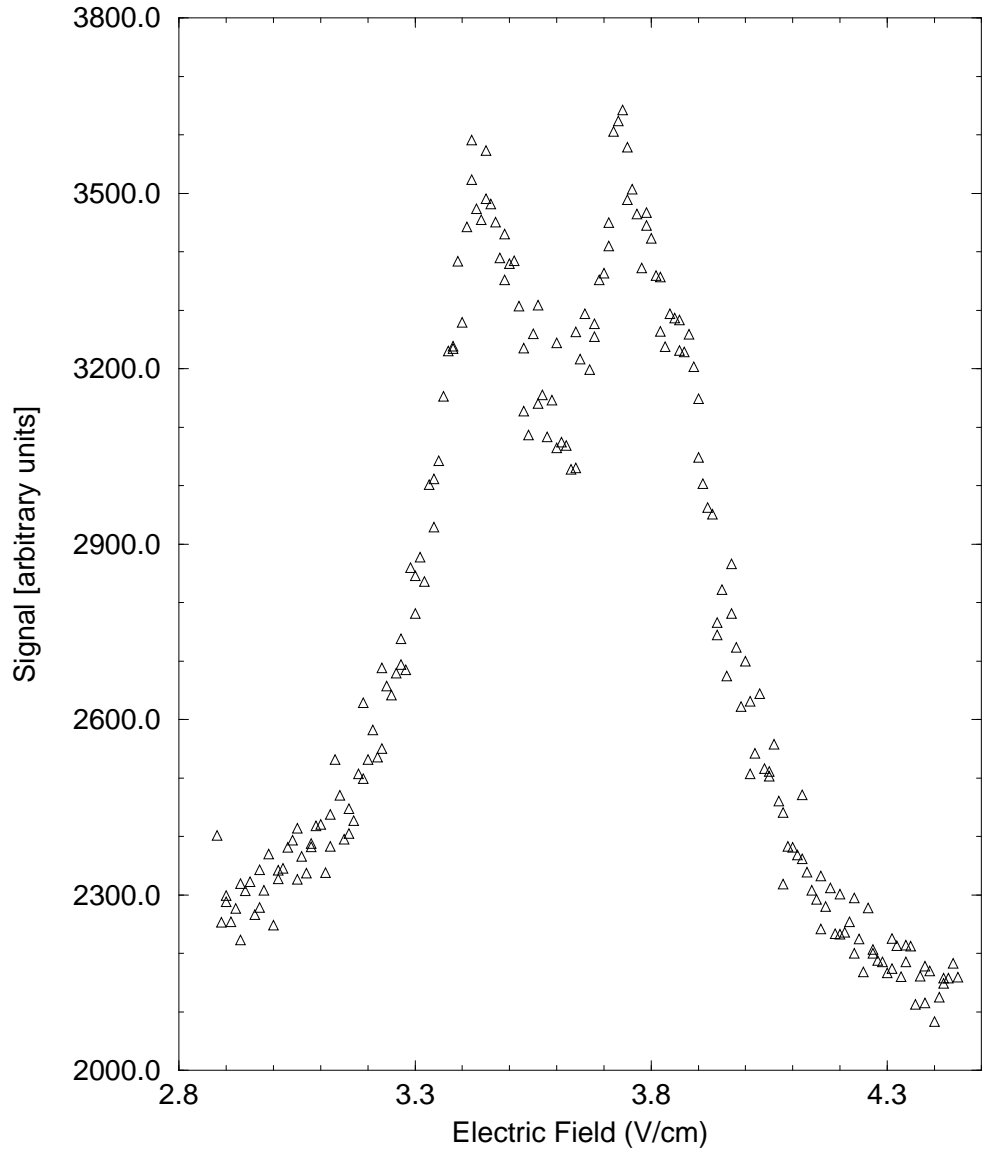}}
\caption[Experimental lineshape as a function of detuning]{A
representative plot of the experimental lineshape as a function of
applied electric field. The detuning $\Delta$ varies linearly with
the field. The data are taken from Fig.~3.9 on page~74 of
Ref.~\cite{Lowell1998a}.} \label{introduction_explineshape}
\end{figure}

We also want to understand the width of the lineshape in the
detuning $\Delta$ as a function of time.
Fig.~\ref{introduction_explineshape} shows a plot of the
experimental data for the lineshape as a function of detuning for
an interaction time of $1.78$ $\mu$s. These data are taken from
Fig.~3.9 on page~74 of Ref.~\cite{Lowell1998a}. We note the
presence of two resonances, which as was stated before correspond
to the $\left| m_{j}\right| = 1/2$ and $\left| m_{j}\right| = 3/2$
states of the $p^{\prime}$ atom. The behavior of the collisional
resonance width as a function of temperature for hot (i.e.,
roughly room temperature) Rydberg gases is known to be
\begin{equation}
\Delta\nu \propto T^{3/4},
\end{equation}
as can be found in Ref.~\cite{Gallagher1992a} or on page~293 of
Ref.~\cite{Gallagher1994a}. This result agrees quite well with
experiments on such gases,\footnote{See, for example,
pages~309--310 of Ref.~\cite{Gallagher1994a}.} and is obtained by
considering only two-body collisions. Simply plugging the
temperature $150$ $\mu$K into this result gives a width that is
smaller than the room temperature width by a factor of about
$2\times 10^{4}$. The experimental results, however, show a
reduction in the width of only a factor of ten. Thus we can
conclude that the resonance effects of the frozen gas are not
determined by two-body interactions alone.

It was surmised by Anderson \emph{et al.\/}~\cite{Anderson1998a}
that the linewidth must be of the order of the average interaction
energy. This interaction splits the $ss^{\prime}$--$pp^{\prime}$
degeneracy and the subsequent migration of the $p$ state to other
atoms broadens the energy of this ``elementary excitation'' into a
band, as appropriate to an amorphous solid. Cluster calculations
were performed to illustrate this band
formation~\cite{Lowell1998a}, but averaging over atom positions
was done by order-of-magnitude arguments only. Mourachko \emph{et
al.\/}~\cite{Mourachko1998a,Mourachko1999a} pointed out, in
connection with their experiments on a different system, that the
``walking away'' of the $p$ excitation from its original location
should be regarded as a diffusion process, and dynamical equations
for the resonance transition in the presence of this diffusion
were written down~\cite{Akulin1999a,Mourachko1999a}.

The present work builds on these earlier insights, but goes
considerably further in the process of accurately averaging over
atomic positions. It turns out that this averaging (effectively, a
phase averaging) itself produces an
$\exp\left(-\sqrt{\gamma_{eq}t}\right)$ dependence at large times,
characteristic of a diffusive process~\cite{Frasier1999a}. We
discuss in detail in Chapter~\ref{sparse_no_u} a particular case
that cleanly shows the effects of randomness and phase averaging.
In this simple case, the averaging process can be carried out
exactly. For the practically important $r^{-3}$ interaction
potential, one can proceed in a mathematically elegant way and
obtain closed formulas that can be evaluated by a computer package
such as \emph{Mathematica\/}, or in some cases by hand. It appears
to be a fortunate coincidence that the $r^{-3}$ potential
accurately describes the dipole-dipole interactions that are so
prevalent in nature. Some aspects of this approach are easily
extended to more general interactions; however, the $r^{-3}$
dependence has special properties which, when coupled with the
assumption of a random distribution of atoms, lead to relatively
simple results.

Although this work was stimulated by experiments on Rydberg atoms,
it is generally applicable to resonances induced by dipole-dipole
interactions. These may be common in highly excited gases and in
molecular systems. Frozen, resonant gases may even exist in
interstellar clouds.

First, in Chapter~\ref{sparse_no_u}, we discuss a single
$s^{\prime}$ atom interacting resonantly with a surrounding gas of
$s$ atoms, without any possibility for the $s$ atom to interact
with $p$ atoms through an $sp\rightarrow ps$ process. When the
result is averaged over the positions of the $s$ atoms, it
corresponds to a ``sparse'' system of $s^{\prime}$ atoms, i.e.\ a
system where $N^{\prime}\ll N$. We do not know of an existing
experiment to which this treatment applies, so this section can be
viewed as a theoretical prediction of the outcome of a possible
experiment. Our purpose here is also one of exposition, since this
example allows us to introduce in the simplest context some of the
mathematical techniques used throughout the rest of this thesis.
Some plots of the results are presented, and in the final sections
of this chapter we consider the effect of interaction potentials
more complicated than just a simple $1/r^{3}$ potential.

In Chapter~\ref{localization} we continue with our discussion of
the simple model of Chapter~\ref{sparse_no_u} by examining the
issue of localization. We study localization in this system
because we need to have some idea of the relative importance of
close pairs of atoms, as well as some idea of the effective range
of the $1/r^{3}$ interaction. Naively, it seems that the close
pairs are dominant and that the range of the $1/r^{3}$ potential
is infinite because $1/r^{3}$ diverges at $r = 0$ and yet the
integral $\int d^{3}r/r^{3}$ over all space diverges at both
limits.

In Chapter~\ref{sparse_with_u} we come one step closer to the
actual system used in the experiments by allowing the
$sp\rightarrow ps$ process to take place. This renders the problem
analytically insoluble but we consider several approximations,
each of which allows us to make some progress. When compared with
the results of numerical simulations, we see that the coherent
potential approximation compares most favorably.

In Chapter~\ref{cesiumappendix}, we introduce an effective spin
Hamiltonian which fully models the Rb system of Anderson \emph{et
al.\/}~\cite{Anderson1996a,Anderson1998a} and Lowell \emph{et
al.\/}~\cite{Lowell1998a} for all values of $N$ and $N^{\prime}$,
as well as for any strength of the resonant process $ss^{\prime}
\rightarrow pp^{\prime}$ and of the mixing processes $sp
\rightarrow ps$ and $s^{\prime}p^{\prime} \rightarrow
p^{\prime}s^{\prime}$. This Hamiltonian corresponds to a system of
two interpenetrating spin glasses. We show its equivalence to the
results of Chapter~\ref{sparse_with_u} in the sparse limit $\left(
N^{\prime}\ll N \right)$. We then derive an entirely equivalent
but seemingly more cumbersome Hamiltonian to describe this system.
This second derivation embodies a more general technique, which we
demonstrate by constructing a spin Hamiltonian that describes the
cesium system studied by Mourachko \emph{et
al.\/}~\cite{Mourachko1998a,Mourachko1999a}.

In Chapter~\ref{spin} we take into account fully the complication
introduced by the consideration of spin, and we determine exactly
what the spin-dependent interatomic potentials are for the
$sp\rightarrow ps$ and $ss^{\prime}\rightarrow pp^{\prime}$
processes in the experiment we are
modeling~\cite{Anderson1998a,Lowell1998a}.

In Chapter~\ref{simulations} we discuss numerical simulations of
the system we are considering. We start by describing in detail
how the simulations are carried out, then we present the results,
including comparisons of the simulations with the experimental
data.

The limitations of the present theory and some of the many
possible extensions are discussed in Chapter~\ref{finalthoughts}.

\chapter{Sparse $ss^{\prime}\rightarrow pp^{\prime}$ Processes --- The Simple Model}
\markright{Chapter \arabic{chapter}: Sparse
$ss^{\prime}\rightarrow pp^{\prime}$ Processes --- The Simple
Model}
\label{sparse_no_u}

\section{Introduction}

In this chapter we discuss a single $s^{\prime}$ atom interacting
resonantly with a surrounding gas of $s$ atoms, without any
possibility for the $s$ atom to interact with $p$ atoms through an
$sp\rightarrow ps$ process. In Section~\ref{basicequations} we
write down and solve the Schr\"{o}dinger equation for a given
spatial configuration of the $s$ atoms. Then in
Section~\ref{averaging} this result is averaged over the positions
of the $s$ atoms, so that it corresponds to a sparse system of
$s^{\prime}$ atoms where $N^{\prime}\ll N$. In
Section~\ref{averagedsignal} we compute the averaged signal and
the lineshape. Specifically, in Sections~\ref{laplacefourier} and
\ref{sparse_no_u_inverting} we start from the Laplace transform of
the signal and develop series expansions for the signal as a
function of time. We also discuss the use of the fast Fourier
transform to obtain the signal as a function of time. In
Section~\ref{pathmethod} we develop a numerical method to compute
the averaged signal and lineshape that works exceedingly well, and
in Section~\ref{compmeth} we discuss the good and bad points of
the three methods. Next, in Section~\ref{no_u_plots} we present
plots of the averaged signal and lineshape. In
Section~\ref{angular} we consider the effect of interaction
potentials more complicated than just a simple $1/r^{3}$
potential. Finally, we summarize the results of this chapter in
Section~\ref{sparse_no_u_conclusions}.

\section{Basic equations}
\label{basicequations}

We consider one atom at the origin, initially in the state
$s^{\prime}$, in interaction with a gas of $N$ randomly
distributed $s$ atoms through an $ss^{\prime}\rightarrow
pp^{\prime}$ process. Like Mourachko \emph{et
al.\/}~\cite{Mourachko1998a}, we describe the system by the
equations
\newcounter{tlet}
\setcounter{tlet}{1}
\renewcommand{\theequation}{\arabic{chapter}.\arabic{equation}\alph{tlet}}
\begin{eqnarray}
\label{dilute-eqns}
 i\dot{a}_{0} &=& \Delta a_{0}+\sum_{k=1}^{N}V_{k}c_{k}, \label{a0dot} \\
\stepcounter{tlet}\addtocounter{equation}{-1}
 i\dot{c}_{k} &=& V_{k}a_{0}.
\label{ckdot}
\end{eqnarray}
\renewcommand{\theequation}{\arabic{chapter}.\arabic{equation}}
Here $a_{0}\left(\Delta,t\right)$ is the amplitude of the state in
which the atom at the origin is in state $s^{\prime}$ and all
other atoms are in state $s$, while $c_{k}\left(\Delta,t\right)$
(with $k$ running from $1$ to $N$) is the amplitude of the state
in which the atom at the origin is in state $p^{\prime}$ and the
atom at ${\bf r}_{k}$ is in state $p$, while all the others remain
in state $s$. The quantity $V_{k}$ is the interaction potential
and
$\Delta=\epsilon_{p}^{\prime}+\epsilon_{p}-\epsilon_{s}^{\prime}-\epsilon_{s}$
is the detuning from resonance. For simplicity, we will assume for
now that $V_{k}$ is of the dipole-dipole form
\begin{equation}
V_{k}=\frac{\mu\mu^{\prime}}{r_{k}^{3}}, \label{Vk}
\end{equation}
and we will discuss later in Section~\ref{angular} what happens
when the interaction potential has a different dependence on $r$
or contains an angular dependence. Furthermore, the atoms are
sufficiently cold that during the time scale of interest they move
only a very small fraction of their separation, and therefore
$V_{k}$ can be taken to be independent of time. As the temperature
increases one approaches the opposite limit, where binary
collisions control the resonant process \cite{Stoneman1987a}.

The set of differential equations (\ref{a0dot}) and (\ref{ckdot})
can easily be solved with the initial condition
$a_{0}\left(\Delta,t=0\right)=1$ and
$c_{k}\left(\Delta,t=0\right)=0$ for all $k$, yielding
\begin{equation}
c_{k}\left( \Delta,t\right) = -\frac{2iV_{k}e^{-i\Delta t/2}\sin
\left( \sqrt{\Delta^{2}+4\mc{V}^{2}}t/2\right)}{\sqrt{\Delta
^{2}+4\mc{V}^{2}}}, \label{ck}
\end{equation}
with
\begin{equation}
\mc{V}^{2}=\sum_{k=1}^{N}V_{k}^{2}.
\end{equation}
Eq.~(\ref{ckdot}) then gives
\begin{equation}
a_{0}\left(\Delta,t\right) = e^{-i\Delta
t/2}\left[\cos\left(\sqrt{\Delta^{2}+4\mc{V}^{2}}t/2\right)
-
i\frac{\Delta}{\sqrt{\Delta^{2}+4\mc{V}^{2}}}
\sin\left(\sqrt{\Delta^{2}+4\mc{V}^{2}}t/2\right) \right].
\label{a0tsimple}
\end{equation}
The experiment detects transitions to any one of the $p^{\prime}$
states (or, equivalently, to any one of the $p$ states), and is
thus a measure of
\begin{eqnarray}
S\left(\Delta,t\right) &=& 1-\left| a_{0}\left( \Delta,t\right)
\right| ^{2}=\sum_{k}\left| c_{k}\left( \Delta,t\right) \right|
^{2} \nonumber \\ &=& \frac{4\mc{V}^{2}\sin^{2} \left(
\sqrt{\Delta^{2}+4\mc{V}^{2}}\,t/2 \right)}
{\Delta^{2}+4\mc{V}^{2}}. \label{S}
\end{eqnarray}
We see that the signal $S\left( \Delta,t \right)$ exhibits Rabi
oscillations.

We apply this result to a system of $N^{\prime}$ atoms, initially
in state $s^{\prime}$, randomly dispersed among a much larger
number, $N$, of $s$ atoms. In this sparse limit, the experimental
signal is proportional to $N^{\prime}$ times the sample average of
$S\left(\Delta,t\right)$, and the sample average is equivalent to
an ensemble average over the atomic positions ${\bf r}_{k}$ that
are hidden in $\mc{V}^{2}$ (see Eq.~(\ref{avexp}) below). In
general, averaging is more easily done on the Laplace transform of
the signal, which in this case is
\begin{eqnarray}
\tilde{S}\left(\Delta,\alpha \right) &=&
\int_{0}^{\infty}e^{-\alpha t}\,S\left(\Delta,t\right)\,dt
\nonumber
\\ &=& \frac{1}{\alpha} \frac{2\mc{V}^{2}}{\alpha^{2} + \Delta^{2}
+ 4\mc{V}^{2}} \nonumber \\ &=&\frac{1}{2\alpha }\left(
1-\frac{\alpha ^{2}+\Delta ^{2}}{\alpha ^{2}+\Delta
^{2}+4\mc{V}^{2}}\right). \label{Stild}
\end{eqnarray}
This Laplace transform can also be found as Eq.~(5.9) on page~28
of Ref.~\cite{Roberts1966a}.

\section{Averaging over atom positions}
\label{averaging}

To compute ensemble averages we use the following result, valid in
the limit $N\gg 1$, for a random distribution of the variables
${\bf r}_{k}$:\footnote{The procedure here is, in effect, that
used in the derivation of the second virial coefficient. See, for
example, page~421 of Ref.~\cite{Reif1965a}.}
\begin{eqnarray}
\left\langle e^{-\beta \mc{V}^{2}}\right\rangle &=&
\frac{1}{\Omega ^{N}}\int d^{3}r_{1}\ldots d^{3}r_{N}\,\exp \left[
-\beta \sum_{k=1}^{N}V^{2}(r_{k})\right] \nonumber \\ &=& \left\{
\frac{1}{\Omega }\int d^{3}r\exp \left[ -\beta V^{2}(r)\right]
\right\} ^{N}  \nonumber \\ &=& \left\{ 1-\frac{1}{\Omega }\int
d^{3}r\left[ 1-e^{-\beta V^{2}(r)}\right] \right\} ^{N}  \nonumber
\\ &\longrightarrow &\exp \left\{ -\frac{N}{\Omega }\int
d^{3}r\,\left[ 1-e^{-\beta V^{2}(r)}\right] \right\} ,
\label{avexp}
\end{eqnarray}
where $\Omega$ is the volume of the gas in the trap and
\begin{equation}
\left\langle X\right\rangle = \frac{1}{\Omega ^{N}}\int
d^{3}r_{1}\ldots d^{3}r_{N}\, X.
\end{equation}
In particular, for the dipolar interaction $V(r)=\mu
\mu^{\prime}/r^{3}$, we have
\begin{equation}
\int d^{3}r\left[ 1-e^{-\beta \left( \mu \mu ^{\prime }\right)
^{2}/r^{6}}\right] =\frac{4\pi ^{3/2}}{3}~\mu \mu ^{\prime
}~\sqrt{\beta }, \label{aveintsimple}
\end{equation}
leading to
\begin{equation}
\left\langle e^{-\beta \mc{V}^{2}}\right\rangle =
e^{-v\sqrt{\beta}}, \label{ave}
\end{equation}
where
\begin{equation}
v=\frac{4\pi^{3/2}}{3}\frac{N}{\Omega }\mu\mu^{\prime}. \label{v}
\end{equation}
For the typical densities in the experiments of Anderson \emph{et
al.\/}~\cite{Anderson1998a} and Lowell \emph{et
al.\/}~\cite{Lowell1998a}, $v$ is on the order of MHz. Since $v$
has the units of an energy, it must be of order
$\mu\mu^{\prime}N/\Omega$ on dimensional grounds. It is still
remarkable, however, that Eq.~(\ref{v}) gives simply and exactly
the quantity $v$ that will enter in all the averaged quantities in
this section.\footnote{In this section we perform an average over
many configurations of randomly positioned atoms. The
corresponding dipolar sums for regular lattices are discussed in
Refs.~\cite{Cohen1955a} and \cite{Fujiki1987a}.}

Using Eq.~(\ref{ave}), we can evaluate the average of any function
$F\left(\mc{V}^{2}\right)$. We simply need to determine the
probability distribution of the variable $\mc{V}^{2}$, which we
will denote as $P\left( \mc{V}^{2}\right)$. By definition, we have
\begin{equation}
P\left( \mc{V}^{2}\right) = \left\langle\delta\left(\mc{V}^{2} -
\sum_{k=1}^{N}\left( \frac{\mu\mu^{\prime}}{r_{k}^{3}}\right)^{2}
\right)\right\rangle.
\end{equation}
We proceed by writing the delta function as a Fourier transform
over the dummy variable $q$. This leads to
\begin{equation}
P\left( \mc{V}^{2}\right) = \int_{-\infty}^{\infty}
\frac{dq}{2\pi}\,e^{iq\mc{V}^{2}}\left\langle \exp\left[
-iq\sum_{k=1}^{N}\left(
\frac{\mu\mu^{\prime}}{r_{k}^{3}}\right)^{2} \right]\right\rangle.
\end{equation}
The average can now be performed using Eq.~(\ref{ave}), and we
have
\begin{equation}
P\left( \mc{V}^{2}\right) = \int_{-\infty}^{\infty}
\frac{dq}{2\pi}\,\exp\left(iq\mc{V}^{2} - \sqrt{iq}v\right).
\label{crazyPintegral}
\end{equation}

The remaining integral requires a little finesse to evaluate.
First, we split the integral into two parts, so that
\begin{equation}
P_{1}\left( \mc{V}^{2}\right) = \int_{0}^{\infty}
\frac{dq}{2\pi}\,\exp\left(iq\mc{V}^{2} - \sqrt{iq}v\right),
\end{equation}
and
\begin{eqnarray}
P_{2}\left( \mc{V}^{2}\right) &=& \int_{-\infty}^{0}
\frac{dq}{2\pi}\,\exp\left(iq\mc{V}^{2} - \sqrt{iq}v\right)
\nonumber \\ &=& \int_{0}^{\infty}
\frac{dq}{2\pi}\,\exp\left(-iq\mc{V}^{2} - \sqrt{-iq}v\right)
\nonumber \\ &=& \left[ P_{1}\left( \mc{V}^{2}\right)\right]^{*}.
\end{eqnarray}
Now we note that
\begin{eqnarray}
P\left( \mc{V}^{2}\right) &=& P_{1}\left( \mc{V}^{2}\right) +
P_{2}\left( \mc{V}^{2}\right) \nonumber \\ &=& 2\,
\mrm{Re}\,\int_{0}^{\infty}
\frac{dq}{2\pi}\,\exp\left(iq\mc{V}^{2} - \sqrt{iq}v\right).
\end{eqnarray}
Changing variables to $\sigma = \sqrt{-iq\mc{V}^{2}} =
e^{-i\pi/4}\sqrt{qx}$, we see that $q = i\sigma^{2}/\mc{V}^{2}$
and hence
\begin{eqnarray}
P\left( \mc{V}^{2}\right) &=&
\frac{2}{\pi}\frac{1}{\mc{V}^{2}}\,\mrm{Re}\left[i\int_{0}^{\infty}
d\sigma\,\sigma\exp\left(-\sigma^{2} -
\frac{iv\sigma}{\sqrt{\mc{V}^{2}}}\right)\right] \nonumber \\ &=&
\frac{2}{\pi}\frac{1}{\sqrt{\mc{V}^{2}}}\,
\mrm{Re}\left[\left(-\frac{\partial}{\partial v}\right)
\int_{0}^{\infty} d\sigma\,\exp\left(-\sigma^{2} -
\frac{iv\sigma}{\sqrt{\mc{V}^{2}}}\right)\right].
\end{eqnarray}
This can be rewritten as
\begin{equation}
P\left( \mc{V}^{2}\right) =
\frac{2}{\pi}\frac{1}{\sqrt{\mc{V}^{2}}}\left(-\frac{\partial}{\partial
v}\right) \int_{0}^{\infty} d\sigma\,\exp\left(-\sigma^{2}\right)
\cos\left(\frac{v\sigma}{\sqrt{\mc{V}^{2}}}\right),
\end{equation}
and the remaining integral is easily looked up in tables, such as
Ref.~\cite{Gradshteyn1994a}. One finds that
\begin{equation}
\int_{0}^{\infty}dx\,\exp\left(-\beta x^{2}\right)
\cos\left(bx\right) =
\frac{1}{2}\sqrt{\frac{\pi}{\beta}}\exp\left(-\frac{b^{2}}{4\beta}\right),
\end{equation}
and hence we have
\begin{eqnarray}
P\left( \mc{V}^{2}\right) =
\frac{1}{\sqrt{\pi}}\frac{1}{\sqrt{\mc{V}^{2}}}\left(-\frac{\partial}{\partial
v}\right) \exp\left(-\frac{v^{2}}{4\mc{V}^{2}}\right).
\end{eqnarray}
When the derivative is evaluated we are left with
\begin{equation}
P\left( \mc{V}^{2}\right) =
\frac{1}{2\sqrt{\pi}}\frac{v}{\mc{V}^{3}}\exp\left(
-\frac{v^{2}}{4\mc{V}^{2}}\right). \label{Pv}
\end{equation}
It is interesting to note that $P\left( \mc{V}^{2}\right)$ is not
Gaussian, and the central limit theorem does not apply, because
$P\left(V\right) \propto V^{-2}$ does not have a root mean square
value. In fact, $P\left(V\right)$ is not even normalizable.

Therefore, using Eq.~(\ref{Pv}) we can compute the average of any
function $F\left( \mc{V}^{2} \right)$ as
\begin{equation}
\left\langle F\left( \mc{V}^{2}\right)\right\rangle =
\int_{0}^{\infty}d\left(\mc{V}^{2}\right)\,
P\left(\mc{V}^{2}\right) F\left(\mc{V}^{2}\right). \label{trik}
\end{equation}
We can further define $y^{2} = v^{2}/{\mc{V}}^{2}$, and then by a
simple change of variable the average of $F$ can be written in the
form
\begin{equation}
\left\langle F\left( \mc{V}^{2}\right)\right\rangle =
\int_{0}^{\infty}dy\,
\frac{1}{\sqrt{\pi}}\exp\left(-\frac{y^{2}}{4}\right)
F\left(\frac{v^{2}}{y^{2}}\right), \label{trik2}
\end{equation}
which agrees with the result derived by an alternative method in
Ref.~\cite{Frasier1999a}.

This last set of relations implies that we can average a function
$F(\mc{V}^{2})$ over the positions of the interacting dipoles
(atoms in our case) by replacing $\mc{V}^{2}$ with $v^{2}/y^{2}$
and integrating over the kernel $\exp \left(
-y^{2}/4\right)/\sqrt{\pi}$. This trick is not always useful as
such, because it can lead to highly oscillatory integrals.
However, if the function $F$ has suitable analytic properties,
then the integration path of Eq.~(\ref{trik2}) can be shifted in
the complex $y$ plane in such a way that the oscillations are
damped out and the integral is rather easy to evaluate
numerically. An example of this procedure is discussed in detail
in Section~\ref{pathmethod}.

\section{The averaged signal and lineshape}
\label{averagedsignal}

The average over atomic positions can be obtained in closed form
for the Laplace transform of the signal, $\left\langle
\tilde{S}\left(\Delta,\alpha \right)\right\rangle$. Starting from
Eq.~(\ref{Stild}) in the form
\begin{equation}
\tilde{S}\left(\Delta,\alpha\right)=\frac{1}{2\alpha
}-\frac{\alpha ^{2}+\Delta ^{2}}{2\alpha
}\int_{0}^{\infty}d\beta\, e^{-\beta (\alpha ^{2}+\Delta
^{2}+4\mc{V}^{2})}, \label{unaveragedS}
\end{equation}
and using Eq.~(\ref{ave}), we obtain
\begin{eqnarray}
\left\langle\tilde{S}\left(\Delta,\alpha\right)\right\rangle
&=&\frac{1}{2\alpha}-\frac{A^{2}}{2\alpha }\int_{0}^{\infty
}d\beta ~e^{-\beta A^{2}-2v\sqrt{\beta }} \nonumber \\ &=&
\frac{\sqrt{\pi}}{2}\frac{v}{\alpha A}\exp \left(
\frac{v^{2}}{A^{2}}\right) \erfc\left( \frac{v}{A}\right),
\label{Stld2}
\end{eqnarray}
where $A^{2}=\alpha ^{2}+\Delta ^{2}$. For the sake of
completeness, and because it will be used later on, we note that a
similar result can be obtained for
$\left\langle\tilde{a}_{0}\left(\Delta,\alpha\right)\right\rangle$
by applying the same method to Eq.~(\ref{a0tsimple}). One finds
\begin{equation}
\left\langle\tilde{a}_{0}\left(\Delta,\alpha\right)\right\rangle =
\frac{1}{\alpha+i\Delta} -
\frac{\sqrt{\pi}}{2}\frac{v}{\left(\alpha+i\Delta\right)^{3/2}
\sqrt{\alpha}}
\exp\left[\frac{v^{2}}{4\alpha\left(\alpha+i\Delta\right)}\right]
{\mathrm{erfc}}\left[\frac{v}{2\sqrt{\alpha\left(\alpha +
i\Delta\right)}}\right]. \label{a0exact}
\end{equation}

The problem is then to invert the Laplace transform of
Eq.~(\ref{Stld2}). As is shown in Section~\ref{laplacefourier}, it
is possible in this case to convert the Laplace transform to a
Fourier transform, which can then be inverted numerically by a
fast Fourier transform algorithm. A more convenient inversion
method was presented in Ref.~\cite{Frasier1999a} and is described
in more detail in Section~\ref{sparse_no_u_inverting}. It leads to
an expression for $\left\langle
S\left(\Delta,t\right)\right\rangle$ in terms of generalized
hypergeometric functions, which can be conveniently handled by
\emph{Mathematica\/}.

A third method to obtain $\left\langle
S\left(\Delta,t\right)\right\rangle$ is to apply Eq.~(\ref{trik2})
directly to $\left\langle
\tilde{S}\left(\Delta,\alpha\right)\right\rangle$ as given by
Eq.~(\ref{S}). The remaining integral is difficult to perform
numerically, as we already noted, but this difficulty can be
circumvented as is shown in Section~\ref{pathmethod}. The
advantages and limitations of the three numerical methods for
evaluating $\left\langle S\left(\Delta,t\right)\right\rangle$ are
compared in Section~\ref{compmeth}.

Plots of $\left\langle S\left(\Delta,t\right)\right\rangle$ are
presented and discussed in Section~\ref{no_u_plots}. Readers who
are not interested in mathematical methods can proceed directly to
Section~\ref{no_u_plots}.

\subsection{The Laplace and Fourier transforms}
\label{laplacefourier}

We see from Eq.~(7.1.6) on page~297 of Ref.~\cite{Abramowitz1972a}
that
\begin{eqnarray}
e^{z^{2}} \erfc\left( z\right) &=& e^{z^{2}} -
\frac{2}{\sqrt{\pi}}\sum_{n=0}^{\infty}\frac{2^{n}}{\left(
2n+1\right)!!}z^{2n+1} \nonumber \\ &=&
\sum_{n=0}^{\infty}\frac{z^{2n}}{\G{n+1}} -
\frac{2}{\sqrt{\pi}}\sum_{n=0}^{\infty}\frac{4^{n}\G{n+1}}{\G{2n+2}}
z^{2n+1}.
\end{eqnarray}
Now we use the fact that
\begin{equation}
\G{2z} = \frac{1}{2\sqrt{\pi}}2^{2z}\G{z}\G{z+\frac{1}{2}},
\label{gamma2z}
\end{equation}
and we find
\begin{eqnarray}
e^{z^{2}} \erfc\left( z\right) &=&
\sum_{n=0}^{\infty}\frac{z^{2n}}{\G{n+1}} -
\sum_{n=0}^{\infty}\frac{z^{2n+1}}{\G{n+\frac{3}{2}}} \nonumber \\
&=&
\sum_{n=0}^{\infty}\frac{\left(-z\right)^{n}}{\G{\frac{n}{2}+1}}.
\label{expzerfcz}
\end{eqnarray}
Therefore Eq.~(\ref{Stld2}) gives
\begin{equation}
\left\langle\tilde{S}\left(\Delta,\alpha\right)\right\rangle =
\frac{\sqrt{\pi}}{2}
\sum_{n=0}^{\infty}\frac{1}{\G{\frac{n+2}{2}}} \left(
-1\right)^{n} \frac{v^{n+1}}{\alpha A^{n+1}}.
\end{equation}

This equation shows that $\left\langle\tilde{S}\left(\Delta,
\alpha\right)\right\rangle$ can be analytically continued to
complex $\alpha$ with $\mrm{Re}\,\alpha > 0$, although it has a
branch cut ending at essential singularities located at $\alpha =
i\Delta$ and $\alpha = -i\Delta$. In particular, the Fourier
transform is obtained by putting $\alpha = \omega/i$. We have
obtained $\left\langle S\left(\Delta,t\right)\right\rangle$ by
inverting this Fourier transform numerically using a fast Fourier
transform routine. However, the following method of inverting the
Laplace transform is much more convenient, for reasons that are
explained in Section~\ref{compmeth}.

\subsection{Inverting the Laplace transform}
\label{sparse_no_u_inverting}

According to Eq.~(2.66) of Ref.~\cite{Roberts1966a}, the inverse
Laplace transform of
\begin{equation}
\frac{1}{\alpha^{2\mu}\left( \alpha^{2} + a^{2}\right)^{\nu}}
\label{before}
\end{equation}
is
\begin{equation}
\frac{t^{2\mu + 2\nu - 1}}{\G{2\mu + 2\nu}} {}_{1}F_{2}\left(
\nu;\mu+\nu,\mu+\nu+\frac{1}{2};- \frac{a^{2}t^{2}}{4}\right),
\label{after}
\end{equation}
where the function ${}_{p}F_{q}\left(a_{1},a_{2},\ldots,a_{p};
b_{1},b_{2},\ldots,b_{q}; z\right)$ is a generalized
hypergeometric function, and is defined on pages 750--751 of
Ref.~\cite{Wolfram1996a} to be
\begin{equation}
{}_{p}F_{q}\left(a_{1},a_{2},\ldots,a_{p};
b_{1},b_{2},\ldots,b_{q}; z\right) =
\frac{\prod_{i=1}^{q}\G{b_{i}}}{\prod_{j=1}^{p}\G{a_{j}}}
\sum_{k=0}^{\infty}\frac{\prod_{l=1}^{p}\G{a_{l}+k}}{\prod_{m=1}^{q}
\G{b_{m}+k}}\frac{z^{k}}{\G{k+1}}. \label{hypergeodef}
\end{equation}
Hence we find that the inverse Laplace transform of
$\left\langle\tilde{S}\left(\Delta,\alpha\right)\right\rangle$ is
\begin{equation}
\left\langle S\left( \Delta,t\right)\right\rangle =
\frac{\sqrt{\pi}}{2} \sum_{n=0}^{\infty}\left( -1\right)^{n}
\frac{\left(vt\right)^{n+1}}{\G{\frac{n+2}{2}} \G{n+2}}
{}_{1}F_{2}\left(
\frac{n+1}{2};\frac{n+2}{2},\frac{n+3}{2};-\frac{\Delta^{2}t^{2}}{4}\right).
\label{hypergeo}
\end{equation}
Expanding ${}_{1}F_{2}$ using Eq.~(\ref{hypergeodef}) yields
\begin{eqnarray}
\left\langle S\left(\Delta,t\right)\right\rangle &=&
\frac{\sqrt{\pi}}{2}vt \sum_{n=0}^{\infty}
\frac{\left(-vt\right)^{n}}{\G{\frac{n+2}{2}}
\G{n+2}}\frac{\G{\frac{n+2}{2}}
\G{\frac{n+3}{2}}}{\G{\frac{n+1}{2}}}\times \nonumber
\\ && \sum_{m=0}^{\infty}
\frac{\G{\frac{n+2m+1}{2}}}{\G{\frac{n+2m+2}{2}}
\G{\frac{n+2m+3}{2}} \G{m+1}}\left(-\frac{\Delta^{2}t^{2}}{4}
\right)^{m} \nonumber
\\ &=& \frac{\sqrt{\pi}}{2}vt \sum_{n=0}^{\infty}\sum_{m=0}^{\infty}
\left(-vt\right)^{n}\frac{\G{\frac{n+3}{2}}}{\G{n+2}\G{\frac{n+1}{2}}}
\times \nonumber \\ &&
\frac{\G{\frac{n+2m+1}{2}}}{\G{\frac{n+2m+2}{2}}
\G{\frac{n+2m+3}{2}} \G{m+1}}\left(-\frac{\Delta^{2}t^{2}}{4}
\right)^{m} \nonumber
\\ &=& \frac{\sqrt{\pi}}{2}vt
\sum_{n=0}^{\infty}\sum_{m=0}^{\infty}
\left(-vt\right)^{n}\frac{n+1}{2}\frac{1}{\G{n+2}} \times
\nonumber \\ && \frac{1}{\G{\frac{n+2m+2}{2}}
\left(\frac{n+2m+1}{2}\right)
\G{m+1}}\left(-\frac{\Delta^{2}t^{2}}{4}\right)^{m} \nonumber
\\ &=& \frac{\sqrt{\pi
}}{2}vt\sum_{n=0}^{\infty}\sum_{m=0}^{\infty
}\frac{(-1)^{n+m}v^{n}\Delta^{2m}
t^{n+2m}}{4^{m}\G{n+1}\G{m+1}\G{\frac{n+2m+2}{2}}
\left(n+2m+1\right)}, \label{needsanumber}
\end{eqnarray}
which can also be obtained directly by expanding Eq.~(\ref{Stld2})
in $1/\alpha $ and then taking the inverse Laplace transform term
by term.

For some purposes, it is convenient to rearrange
Eq.~(\ref{needsanumber}) in the form of a series in powers of the
detuning squared. To achieve this, we use the second line of this
equation and find
\begin{equation}
\left\langle S\left(\Delta,t\right)\right\rangle =
\sum_{m=0}^{\infty}\left( -\frac{\Delta^{2}t^{2}}{4}
\right)^{m}\frac{1}{\G{m+1}} Z_{m},
\end{equation}
where
\begin{eqnarray}
Z_{m} &=& \frac{\sqrt{\pi}}{2}vt\sum_{n=0}^{\infty}
\left(-vt\right)^{n}
\frac{\G{\frac{n+3}{2}}\G{\frac{n+2m+1}{2}}}{\G{n+2}
\G{\frac{n+1}{2}}\G{\frac{n+2m+2}{2}} \G{\frac{n+2m+3}{2}}}
\nonumber \\ &=& \frac{\sqrt{\pi}}{2}vt\sum_{n=0}^{\infty}
\left(-vt\right)^{2n}\frac{\G{n+\frac{3}{2}}\G{n+m+
\frac{1}{2}}}{\G{2n+2}\G{n+\frac{1}{2}}\G{n+m+1}
\G{n+m+\frac{3}{2}}} + \nonumber \\ &&
\frac{\sqrt{\pi}}{2}vt\sum_{n=0}^{\infty}
\left(-vt\right)^{2n+1}\frac{\G{n+2}\G{n+m+1}}{\G{2n+3}\G{n+1}
\G{n+m+\frac{3}{2}} \G{n+m+2}}.
\end{eqnarray}
Again using Eq.~(\ref{gamma2z}), we see that
\begin{eqnarray}
Z_{m} &=& \frac{\sqrt{\pi}}{2}vt\sum_{n=0}^{\infty}
\frac{2\sqrt{\pi}}{4^{n+1}}\left(
v^{2}t^{2}\right)^{n}\frac{\G{n+\frac{3}{2}}
\G{n+m+\frac{1}{2}}}{\G{n+1}\G{n+\frac{3}{2}}\G{n+\frac{1}{2}}
\G{n+m+1} \G{n+m+\frac{3}{2}}} - \nonumber
\\ && \frac{\sqrt{\pi}}{2}\sum_{n=0}^{\infty}
\frac{2\sqrt{\pi}}{4^{n+3/2}}
\left(v^{2}t^{2}\right)^{n+1}\frac{\G{n+2}\G{n+m+1}}{\G{n+\frac{3}{2}}
\G{n+2} \G{n+1}\G{n+m+\frac{3}{2}} \G{n+m+2}} \nonumber
\\ &=& \frac{\pi}{4}vt\sum_{n=0}^{\infty} \left(
\frac{v^{2}t^{2}}{4}\right)^{n}\frac{\G{n+m+\frac{1}{2}}}{\G{n+1}
\G{n+\frac{1}{2}} \G{n+m+1} \G{n+m+\frac{3}{2}}} - \nonumber
\\ && \frac{\pi}{2}\frac{v^{2}t^{2}}{4}\sum_{n=0}^{\infty} \left(
\frac{ v^{2}t^{2}}{4}\right)^{n}\frac{\G{n+m+1}}{\G{n+\frac{3}{2}}
\G{n+1} \G{n+m+\frac{3}{2}} \G{n+m+2}} \nonumber
\\ &=& \frac{\pi}{4}vt\frac{\G{m+\frac{1}{2}}}{\G{\frac{1}{2}}
\G{m+1}\G{m+\frac{3}{2}}}\,{}_{1}F_{3}\left(m+\frac{1}{2};
\frac{1}{2},m+1,m+\frac{3}{2}; \frac{v^{2}t^{2}}{4} \right) -
\nonumber \\ &&
\frac{\pi}{8}v^{2}t^{2}\frac{\G{m+1}}{\G{\frac{3}{2}}
\G{m+\frac{3}{2}}\G{m+2}}
{}_{1}F_{3}\left(m+1;\frac{3}{2},m+\frac{3}{2},m+2;\frac{v^{2}t^{2}}{4}\right)
\nonumber
\\ &=& \frac{\sqrt{\pi}}{4}vt\frac{1}{\left(m+\frac{1}{2}\right)\G{m+1}}\,
{}_{1}F_{3}\left(m+\frac{1}{2};\frac{1}{2},m+1,m+\frac{3}{2};
\frac{v^{2}t^{2}}{4} \right) - \nonumber \\ &&
\frac{\sqrt{\pi}}{4}v^{2}t^{2}
\frac{1}{\left(m+1\right)\G{m+\frac{3}{2}}}
{}_{1}F_{3}\left(m+1;\frac{3}{2},m+\frac{3}{2},m+2;\frac{v^{2}t^{2}}{4}\right).
\end{eqnarray}
Thus the expression for the signal becomes
\begin{eqnarray}
\left\langle S\left(\Delta,t\right)\right\rangle &=&
\sum_{m=0}^{\infty}\left( -\frac{\Delta^{2}t^{2}}{4}
\right)^{m}\frac{1}{\G{m+1}} \times \nonumber \\ &&
\left[\frac{\sqrt{\pi}}{4}vt\frac{1}{\left(m+\frac{1}{2}\right)
\G{m+1}}\,
{}_{1}F_{3}\left(m+\frac{1}{2};\frac{1}{2},m+1,m+\frac{3}{2};
\frac{v^{2}t^{2}}{4} \right) - \right. \nonumber \\ && \left.
\frac{\sqrt{\pi}}{4}v^{2}t^{2}
\frac{1}{\left(m+1\right)\G{m+\frac{3}{2}}}
{}_{1}F_{3}\left(m+1;\frac{3}{2},m+\frac{3}{2},m+2;
\frac{v^{2}t^{2}}{4}\right)\right] \nonumber \\ &=&
\sum_{m=0}^{\infty}\left( -\frac{\Delta^{2}t^{2}}{4} \right)^{m}
\times \nonumber \\ &&
\left\{\frac{\sqrt{\pi}}{4}vt\frac{1}{\left(m+\frac{1}{2}\right)
\left[\G{m+1}\right]^{2}}\,
{}_{1}F_{3}\left(m+\frac{1}{2};\frac{1}{2},m+1,m+\frac{3}{2};
\frac{v^{2}t^{2}}{4} \right) - \right. \nonumber \\ && \left.
\frac{\sqrt{\pi}}{4}v^{2}t^{2} \frac{1}{\G{m+\frac{3}{2}}\G{m+2}}
{}_{1}F_{3}\left(m+1;\frac{3}{2},m+\frac{3}{2},m+2;
\frac{v^{2}t^{2}}{4}\right)\right\}.
\label{deltaexpansion}
\end{eqnarray}

If we consider the signal at resonance, then the expression in
Eq.~(\ref{deltaexpansion}) simplifies considerably. We note that
only the $m=0$ term contributes when $\Delta = 0$. Thus we have
\begin{eqnarray}
\left\langle S\left(\Delta = 0,t\right)\right\rangle &=&
\frac{\sqrt{\pi}}{4}vt\frac{1}{\frac{1}{2}\left[\G{1}\right]^{2}}\,
{}_{1}F_{3}\left(\frac{1}{2};\frac{1}{2},1,\frac{3}{2};
\frac{v^{2}t^{2}}{4} \right) - \nonumber \\ &&
\frac{\sqrt{\pi}}{4}v^{2}t^{2} \frac{1}{\G{\frac{3}{2}}\G{2}}
{}_{1}F_{3}\left(1;\frac{3}{2},\frac{3}{2},2;\frac{v^{2}t^{2}}{4}\right)
\nonumber \\ &=& \frac{\sqrt{\pi}}{2}vt\,
{}_{0}F_{2}\left(;1,\frac{3}{2}; \frac{v^{2}t^{2}}{4} \right) -
\frac{1}{2}v^{2}t^{2}\,
{}_{1}F_{3}\left(1;\frac{3}{2},\frac{3}{2},2;\frac{v^{2}t^{2}}{4}\right).
\label{Sres}
\end{eqnarray}

We further note that
\begin{eqnarray}
\frac{1}{2}v^{2}t^{2}
{}_{1}F_{3}\left(1;\frac{3}{2},\frac{3}{2},2;\frac{v^{2}t^{2}}{4}\right)
&=& \frac{1}{2}v^{2}t^{2} \G{\frac{3}{2}}\G{\frac{3}{2}}\G{2}
\times \nonumber
\\ && \sum_{n=0}^{\infty}
\frac{\G{n+1}}{\G{n+1}\G{n+\frac{3}{2}} \G{n+\frac{3}{2}}\G{n+2}}
\left(\frac{v^{2}t^{2}}{4}\right)^{n} \nonumber \\ &=&
\frac{\pi}{2} \sum_{n=0}^{\infty} \frac{1}{\G{n+\frac{3}{2}}
\G{n+\frac{3}{2}}\G{n+2}} \left(\frac{v^{2}t^{2}}{4}\right)^{n+1}
\nonumber \\ &=& \frac{\pi}{2} \left[\sum_{n=0}^{\infty}
\frac{1}{\G{n+\frac{1}{2}} \G{n+\frac{1}{2}}\G{n+1}}
\left(\frac{v^{2}t^{2}}{4}\right)^{n} - \frac{1}{\pi}\right]
\nonumber \\ &=& \frac{\pi}{2} \frac{1}{\G{\frac{1}{2}}
\G{\frac{1}{2}}}\,{}_{0}F_{2}\left(
;\frac{1}{2},\frac{1}{2};\frac{v^{2}t^{2}}{4}\right) - \frac{1}{2}
\nonumber \\ &=& \frac{1}{2}\,{}_{0}F_{2}\left(
;\frac{1}{2},\frac{1}{2};\frac{v^{2}t^{2}}{4}\right) -
\frac{1}{2}.
\end{eqnarray}
Thus the signal at resonance can be written as
\begin{equation}
\left\langle S\left(\Delta = 0,t\right)\right\rangle =
\frac{\sqrt{\pi}}{2}vt\, {}_{0}F_{2}\left(;1,\frac{3}{2};
\frac{v^{2}t^{2}}{4} \right) + \frac{1}{2} -
\frac{1}{2}\,{}_{0}F_{2}\left(
;\frac{1}{2},\frac{1}{2};\frac{v^{2}t^{2}}{4}\right),
\end{equation}
which agrees perfectly with Eq.~(19) of Ref.~\cite{Frasier1999a}.

\subsection{A numerical method using Eq.~(\ref{trik2})}
\label{pathmethod}

We have from Eq.~(\ref{trik2}) that the averaged signal can be
written as
\begin{equation}
\left\langle S\left( \Delta,t\right)\right\rangle =
\int_{0}^{\infty}dy\,
\frac{1}{\sqrt{\pi}}\exp\left(-\frac{y^{2}}{4}\right)
S\left(\Delta,t\right)_{\mc{V}^{2}\rightarrow v^{2}/y^{2}}.
\end{equation}
Then, using Eq.~(\ref{S}), we have
\begin{equation}
\left\langle S\left( \Delta,t\right)\right\rangle =
\frac{4}{\sqrt{\pi}}\int_{0}^{\infty}dy\,
\exp\left(-\frac{y^{2}}{4}\right) \frac{v^{2}}{y^{2}}
\frac{\sin^{2}\left(
\frac{1}{2}\sqrt{\Delta^{2}+4\frac{v^{2}}{y^{2}}}\,t \right)}
{\Delta^{2}+4\frac{v^{2}}{y^{2}}}.
\end{equation}
We can rewrite this as
\begin{eqnarray}
\left\langle S\left( \Delta,t\right)\right\rangle &=&
\frac{2}{\sqrt{\pi}} \int_{0}^{\infty}dy\,
\exp\left(-\frac{y^{2}}{4}\right)
\frac{\frac{v^{2}}{y^{2}}}{\Delta^{2}+4\frac{v^{2}}{y^{2}}} \left[
1 - \cos\left(
\sqrt{\Delta^{2}+4\frac{v^{2}}{y^{2}}}\,t\right)\right] \nonumber
\\ &=& \frac{2}{\sqrt{\pi}}\int_{0}^{\infty}dy\,
\exp\left(-\frac{y^{2}}{4}\right)
\frac{\frac{v^{2}}{y^{2}}}{\Delta^{2}+4\frac{v^{2}}{y^{2}}} -
\nonumber \\ && \frac{2}{\sqrt{\pi}}\,
\mrm{Re}\int_{0}^{\infty}dy\,
\frac{\frac{v^{2}}{y^{2}}}{\Delta^{2}+4\frac{v^{2}}{y^{2}}}
\exp\left(-\frac{y^{2}}{4} +
it\sqrt{\Delta^{2}+4\frac{v^{2}}{y^{2}}} \right).
\label{S_with_TT}
\end{eqnarray}

If $\Delta = 0$, then we have
\begin{equation}
\left\langle S\left( \Delta = 0,t\right)\right\rangle =
\frac{1}{2} - I\left(\Delta = 0,t\right), \label{sdD0}
\end{equation}
where
\begin{equation}
I\left(\Delta = 0,t\right) = \frac{1}{2\sqrt{\pi}}\,
\mrm{Re}\int_{0}^{\infty}dy\, \exp\left(-\frac{y^{2}}{4} +
2it\frac{v}{y}\right). \label{sdID0}
\end{equation}
We can change the path of integration from the positive real axis
to a path where we start from the origin and proceed along the
negative imaginary axis to the point $y = -i$, then proceed to the
point $y = \infty - i$ along the line $\mrm{Im}\, y = -i$. Then we
have
\begin{eqnarray}
I\left(\Delta = 0,t\right) &=& \frac{1}{2\sqrt{\pi}}\,
\mrm{Re}\left[ -\frac{1}{i}\int_{0}^{1}dz\,
\exp\left(\frac{z^{2}}{4} - 2t\frac{v}{z}\right)\right] +
\nonumber
\\ && \frac{1}{2\sqrt{\pi}}\, \mrm{Re}\int_{0}^{\infty}dw\,
\exp\left[-\frac{\left(w-i\right)^{2}}{4} +
2it\frac{v}{w-i}\right].
\end{eqnarray}
Clearly the integral over $z$ is purely real, and so
\begin{equation}
\left\langle S\left( \Delta = 0,t\right)\right\rangle =
\frac{1}{2} - \frac{1}{2\sqrt{\pi}}\,
\mrm{Re}\int_{0}^{\infty}dw\,
\exp\left[-\frac{\left(w-i\right)^{2}}{4} +
\frac{2ivt}{w-i}\right]. \label{d0path}
\end{equation}
\emph{Mathematica\/} is quite adept at quickly evaluating this
integral for any value of $vt$.

The case where $\Delta\neq 0$ is only slightly more complicated.
We have from Eq.~(\ref{S_with_TT}) that
\begin{equation}
\left\langle S\left( \Delta,t\right)\right\rangle =
\frac{\sqrt{\pi}}{2} \frac{v}{\Delta}
\exp\left(\frac{v^{2}}{\Delta^{2}}\right)
\erfc\left(\frac{v}{\Delta}\right) - I\left(\Delta,t\right),
\end{equation}
where
\begin{equation}
I\left(\Delta,t\right) = \frac{2}{\sqrt{\pi}}\,
\mrm{Re}\int_{0}^{\infty}dy\,
\frac{1}{y^{2}}\frac{1}{\frac{\Delta^{2}}{v^{2}}+\frac{4}{y^{2}}}
\exp\left(-\frac{y^{2}}{4} +
ivt\sqrt{\frac{\Delta^{2}}{v^{2}}+\frac{4}{y^{2}}} \right).
\end{equation}
The path we used in the resonant case will still work well for
small $\Delta/v$, but will fail for $\Delta \gtrapprox 2v$. This
is clear because the radicand
$\frac{\Delta^{2}}{v^{2}}+\frac{4}{y^{2}}$ will change sign as $y$
goes from zero to $-i$. By analogy with the resonant case and
through a bit of trial and error, however, one finds that for
$\Delta \gtrapprox 2v$ the path where we start from the origin and
proceed along the positive imaginary axis to the point $y =
-\frac{iv}{2\Delta}$, then proceed to the point $y = \infty -
\frac{iv}{2\Delta}$ along the line $\mrm{Im}\, y =
-\frac{iv}{2\Delta}$ works well. Using this path we have
\begin{eqnarray}
I\left(\Delta,t\right) &=& -\frac{2}{\sqrt{\pi}}\,
\mrm{Re}\left[-\frac{1}{i}\int_{0}^{\frac{v}{2\Delta}}dz\,
\frac{1}{z^{2}}\frac{1}{\frac{\Delta^{2}}{v^{2}}-\frac{4}{z^{2}}}
\exp\left(\frac{z^{2}}{4} +
ivt\sqrt{\frac{\Delta^{2}}{v^{2}}-\frac{4}{z^{2}}}\right)\right] +
\nonumber \\ && \frac{2}{\sqrt{\pi}}\,
\mrm{Re}\int_{0}^{\infty}dw\,
\frac{1}{\left(w-\frac{iv}{2\Delta}\right)^{2}} \left(
\frac{\Delta^{2}}{v^{2}} +
\frac{4}{\left(w-\frac{iv}{2\Delta}\right)^{2}}\right)^{-1} \times
\nonumber \\ &&
\exp\left[-\frac{\left(w-\frac{iv}{2\Delta}\right)^{2}}{4} +
ivt\sqrt{\frac{\Delta^{2}}{v^{2}} +
\frac{4}{\left(w-\frac{iv}{2\Delta}\right)^{2}}} \right].
\end{eqnarray}
It is again the case that the $z$ integral gives zero, and we have
that
\begin{eqnarray}
\left\langle S\left( \Delta,t\right)\right\rangle &=&
\frac{\sqrt{\pi}}{2} \frac{v}{\Delta}
\exp\left(\frac{v^{2}}{\Delta^{2}}\right)
\erfc\left(\frac{v}{\Delta}\right) - \nonumber \\ &&
\frac{2}{\sqrt{\pi}}\, \mrm{Re}\int_{0}^{\infty}dw\,
\frac{1}{\left(w-\frac{iv}{2\Delta}\right)^{2}} \left(
\frac{\Delta^{2}}{v^{2}} +
\frac{4}{\left(w-\frac{iv}{2\Delta}\right)^{2}}\right)^{-1} \times
\nonumber \\ &&
\exp\left[-\frac{\left(w-\frac{iv}{2\Delta}\right)^{2}}{4} +
ivt\sqrt{\frac{\Delta^{2}}{v^{2}} +
\frac{4}{\left(w-\frac{iv}{2\Delta}\right)^{2}}} \right].
\label{dnot0path}
\end{eqnarray}
\emph{Mathematica\/} is again fairly adept at evaluating this
expression numerically for any values of $vt$ and $\Delta/v$,
although the computation time increases with increasing
$\Delta/v$. In practice, we have used Eq.~(\ref{d0path}) for
$\Delta < v$ and Eq.~(\ref{dnot0path}) for $\Delta > v$.

Note that, using this method, it is possible to obtain a plot of
the resonance width versus time by solving numerically for a value
of $\Delta$ such that
\begin{equation}
\left\langle S\left( \Delta,t\right)\right\rangle = \frac{1}{2}
\left\langle S\left( 0,t\right)\right\rangle,
\label{widthpathmethod}
\end{equation}
at many values of $t$.

We can also use this method to obtain asymptotic expressions for
the signal $\left\langle S\left(\Delta,t\right)\right\rangle$.
Consider the resonant case. We see from Eqs.~(\ref{sdD0}) and
(\ref{sdID0}) that we want to obtain an asymptotic expression for
the integral
\begin{equation}
I\left(\Delta = 0,t\right) =
\frac{1}{2\sqrt{\pi}}\mrm{Re}\int_{0}^{\infty}dy\,
\exp\left[\phi\left(y\right)\right],
\end{equation}
where
\begin{equation}
\phi\left(y\right) = -\frac{y^{2}}{4} + \frac{2ivt}{y}.
\end{equation}
We can simply apply the method of steepest descent to this
integral, as discussed on pages~82--90 of
Ref.~\cite{Matthews1970a}. We see that
\begin{equation}
\phi^{\prime}\left(y\right) = -\frac{y}{2} - \frac{2ivt}{y^{2}},
\end{equation}
and
\begin{equation}
\phi^{\prime\prime}\left(y\right) = -\frac{1}{2} +
\frac{4it}{y^{3}}.
\end{equation}
Thus we find that $\phi^{\prime}\left(y\right)$ vanishes when
$y^{3} = -4ivt$. To ensure convergence, we want to pick the root
closest to the original integration path in the lower half plane,
and so we choose the root
\begin{equation}
y_{0} = \exp\left(-\frac{i\pi}{6}\right)\left(4vt\right)^{1/3}.
\end{equation}
We then have
\begin{equation}
\phi\left(y\right) \approx \phi\left(y_{0}\right) +
\frac{1}{2}\phi^{\prime\prime}\left(y_{0}\right) \left(y -
y_{0}\right).
\end{equation}
Proceeding carefully we find that
\begin{equation}
\phi\left(y_{0}\right) = \frac{3}{2}\left(-1 + i\sqrt{3}\right)
\left(vt\right)^{2/3},
\end{equation}
and
\begin{equation}
\phi^{\prime\prime}\left(y_{0}\right) = -6.
\end{equation}
Finally there is only a Gaussian integral to do, and one easily
finds that
\begin{equation}
\left\langle S\left(\Delta=0 ,t\right)\right\rangle \approx
\frac{1}{2} - \frac{1}{\sqrt{3}}\exp\left[
-\frac{3}{2}\left(vt\right)^{2/3}\right] \cos\left[
\frac{3\sqrt{3}}{2}\left(vt\right)^{2/3}\right], \label{asympSD0}
\end{equation}
for large $t$.

The same procedure can be applied when $\Delta\neq 0$, although
this situation is somewhat more complicated because in order to
determine the position of the saddle point one must first solve a
cubic equation. This is best done numerically, and one finds that
the result is qualitatively the same as Eq.~(\ref{asympSD0}).

\subsection{Comparison of the three methods}
\label{compmeth}

Each of the three methods we have discussed has its drawbacks.

In order for the fast Fourier transform method of
Section~\ref{laplacefourier} to be practical, we must first
subtract off certain parts of the Fourier transform that are not
well behaved.  For instance, we must subtract off the part of the
Fourier transform that goes as $1/\omega$ at small $\omega$.
Because we must sample points over a wide range of values of
$\omega$ in Fourier space in order to obtain an accurate picture
of the signal in $t$ space, however, we must also ensure that the
Fourier transform falls off suitably fast as $\omega$ increases.
We actually subtract off the quantity
\begin{equation}
-i\frac{\sqrt{\pi}}{2} \frac{\exp\left(\frac{1}{\Delta^{2}}\right)
\erfc\left(\frac{1}{\Delta}\right)}{\Delta\omega +
i\exp\left(\frac{1}{\Delta^{2}}\right)
\erfc\left(\frac{1}{\Delta}\right)\omega^{2}},
\end{equation}
so that the coefficient of $1/\omega^{2}$ at large $\omega$ also
vanishes. The parts that are subtracted off must have a known
inverse transform, which at the end is added to the remaining part
that is transformed by a fast Fourier transform algorithm. Finding
a suitable subtraction is a rather cumbersome process. In
addition, because the fast Fourier transform method is a numerical
method (as opposed to an analytical method), we do not obtain an
explicit form for $\left\langle
S\left(\Delta,t\right)\right\rangle$.

The series obtained using the method of
Section~\ref{sparse_no_u_inverting}, on the other hand, do give an
explicit form for $\left\langle
S\left(\Delta,t\right)\right\rangle$. Unfortunately, however, they
have the very serious limitation that they do not always converge
for larger values of $t$ unless $\Delta$ is small.

The method discussed in Section~\ref{pathmethod} shares a drawback
with the fast Fourier transform method, namely that unless the
integral can be done analytically, an explicit form for
$\left\langle S\left(\Delta,t\right)\right\rangle$ cannot be
obtained.  This method is, however, superior to the fast Fourier
transform method because it does not require the cumbersome
subtractions.

As a general rule, the series method of
Section~\ref{sparse_no_u_inverting} is best when it converges, and
the integral method of Section~\ref{pathmethod} is the best
otherwise.

\subsection{Plots and discussion}
\label{no_u_plots}

\begin{figure}
\resizebox{\textwidth}{!}{\includegraphics[0in,0in][8in,10in]{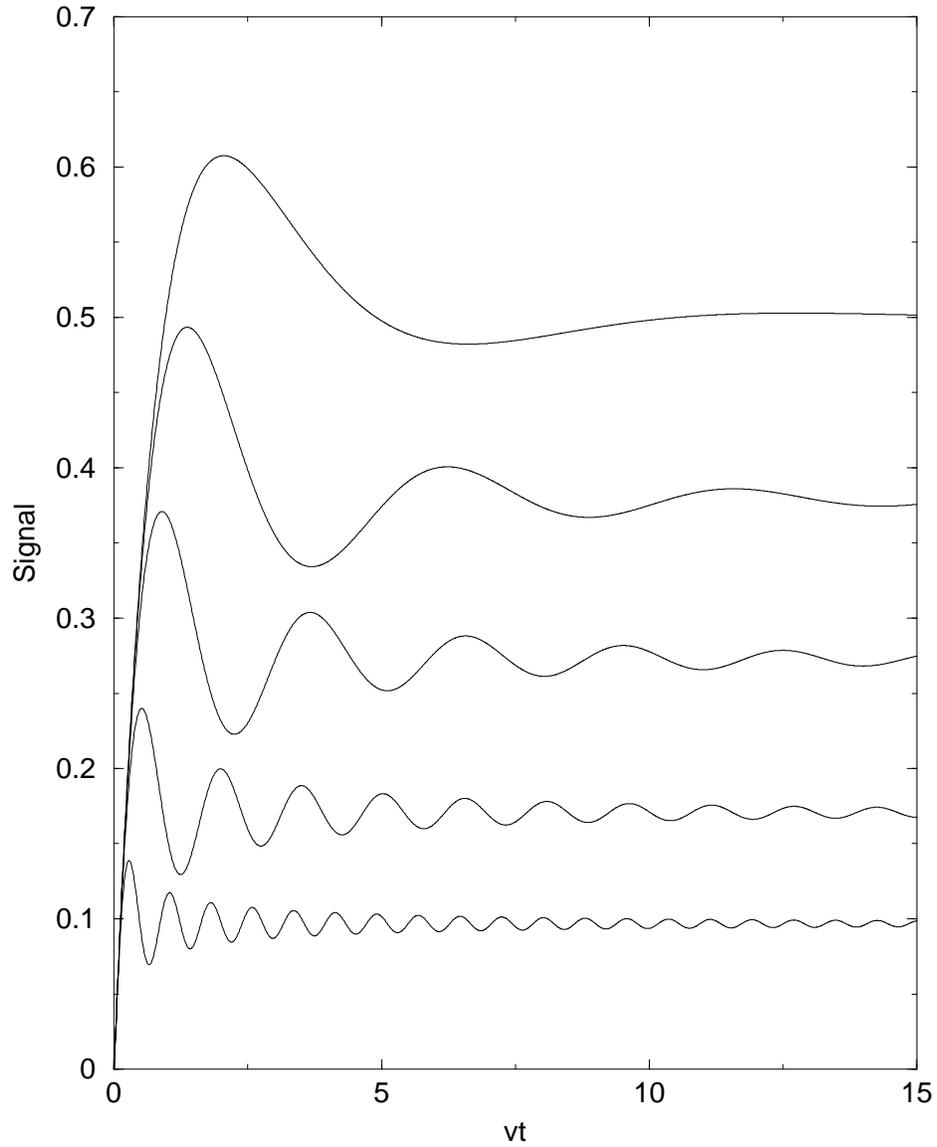}}
\caption[Signal as a function of time]{Averaged signal
$S\left(\Delta,t\right)$ for (from top to bottom) $\Delta/v = 0$,
$1$, $2$, $4$, and $8$, calculated using the method of
Section~\ref{pathmethod} for the sparse limit. The
$ss^{\prime}\rightarrow pp^{\prime}$ process is averaged, and the
$sp\rightarrow ps$ process is neglected. The Rabi oscillations are
not completely washed out by the averaging process and are more
pronounced off resonance.} \label{sparse_no_u_fig_1}
\end{figure}

Plots of $\left\langle S\left(\Delta,t\right)\right\rangle$ for
several values of $\Delta$ are shown in
Fig.~\ref{sparse_no_u_fig_1}. Notice that the initial slope of the
signal is independent of $\Delta$. This feature translates into
resonance widths that vary as $1/t$ for small $t$ -- the so-called
transform broadening discussed by Thomson \emph{et
al.\/}~\cite{Thomson1990a,Thomson1990b}. With $v$ on the order of
MHz, this initial rise occurs in a fraction of a $\mu$s. We will
see later that when the $sp\rightarrow ps$ process is included,
the initial rise is independent of both the detuning $\Delta$ and
the strength of the $sp\rightarrow ps$ interaction. Another point
of interest is that the averaging is much more effective at
smoothing out the oscillations for $\Delta =0$ than it is for
$\Delta \neq 0$.

\begin{figure}
\resizebox{\textwidth}{!}{\includegraphics[0in,0in][8in,10in]{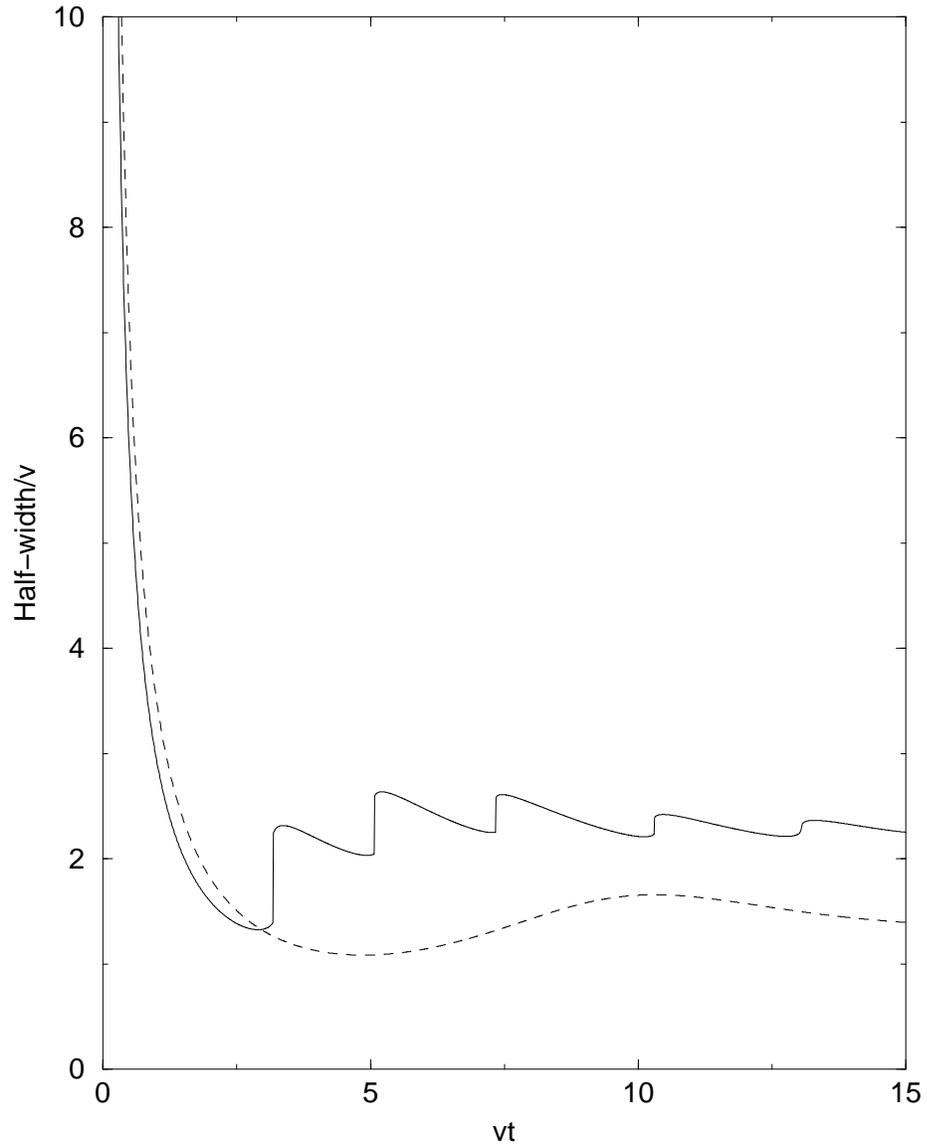}}
\caption[Resonance width as a function of time]{The dashed line is
the width $w$, as given by Eq.~(\ref{width}), versus time. The
solid line is the width as determined using
Eq.~(\ref{widthpathmethod}) of Section~\ref{pathmethod}. The
``dip'' seen in both curves is associated with the stronger and
higher frequency oscillations seen at larger values of $\Delta$,
as shown in Fig.~\ref{sparse_no_u_fig_1}.}
\label{sparse_no_u_fig_2}
\end{figure}

We can construct a measure of the width from the first two terms
in the $\Delta$-expansion, Eq.~(\ref{deltaexpansion}). If
\begin{equation}
\left\langle S\left(\Delta,t\right)\right\rangle = S_{0}\left(
t\right) +\Delta ^{2}S_{1}\left( t\right) +\mc{O}\left( \Delta
^{4}\right),
\end{equation}
and we define
\begin{equation}
w = \sqrt{-\frac{S_{0}\left( t \right)}{S_{1}\left( t \right)}},
\label{width}
\end{equation}
then $2w$ would be the FWHM if the lineshape were Lorentzian. We
note that the functions $S_{0}\left(t\right)$ and
$S_{1}\left(t\right)$ are easily obtained in closed form, since
they are essentially just the $m=0$ and $m=1$ terms in the sum of
Eq.~(\ref{deltaexpansion}), respectively. The quantity $w/v$ is
plotted in Fig.~\ref{sparse_no_u_fig_2} as the dashed curve. For
small $vt$, we have explicitly $w = \sqrt{12}/t$. While the $1/t$
behavior follows from the transform broadening argument, or even
simply from dimensional analysis, the coefficient $\sqrt{12}$ is a
prediction of the detailed theory.

The solid curve in Fig.~\ref{sparse_no_u_fig_2} is the exact half
width, as determined using Eq.~(\ref{widthpathmethod}) of
Section~\ref{pathmethod}. We see that the width of
Eq.~(\ref{width}) is a good approximation to the exact result only
at small times.

\begin{figure}
\resizebox{\textwidth}{!}{\includegraphics[0in,0in][8in,10in]{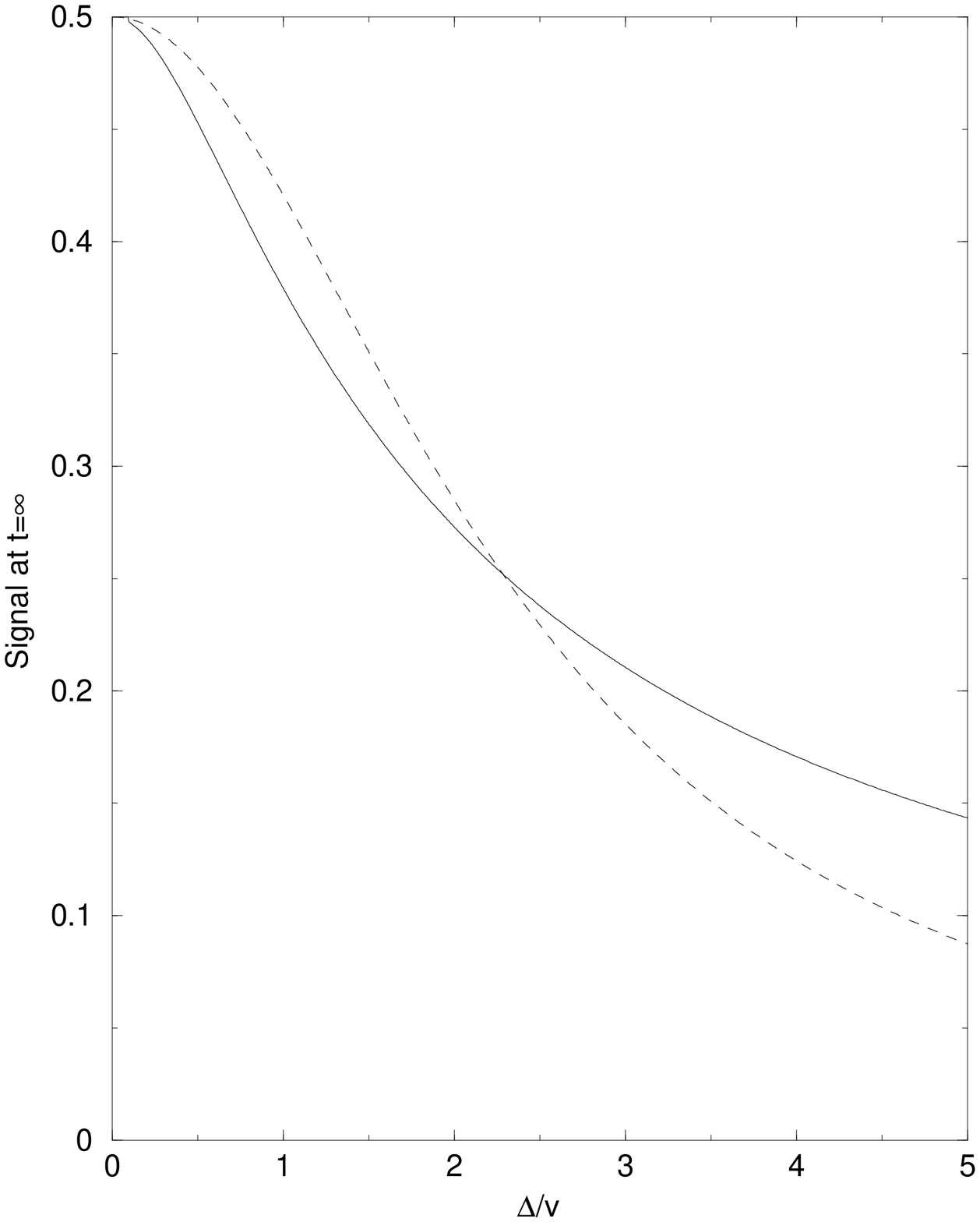}}
\caption[Saturation lineshape]{The solid curve is the saturation
$\left(t\rightarrow\infty\right)$ lineshape of Eq.~(\ref{Sinf}).
It is not a Lorentzian but a Lorentzian averaged as in
Eq.~(\ref{trik2}). The dashed curve is a Lorentzian with the same
height and FWHM, drawn for comparison.} \label{sparse_no_u_fig_3}
\end{figure}

The lineshape at saturation ($t\rightarrow \infty $) can be
obtained directly from the first line of Eq.~(\ref{S_with_TT}).
For large times the cosine term averages to zero, giving
\begin{equation}
\left\langle S\left(\Delta,t\rightarrow \infty\right)\right\rangle
= \frac{\sqrt{\pi }}{2}\frac{v}{\Delta}\exp \left(
\frac{v^{2}}{\Delta ^{2}}\right) \erfc\left( \frac{v}{\Delta
}\right), \label{Sinf}
\end{equation}
which can also be extracted from the small $\alpha $ behavior of
Eq.~(\ref{Stld2}). This lineshape is plotted in
Fig.~\ref{sparse_no_u_fig_3}. The FWHM is approximately $4.6v$.
Also plotted is a Lorentzian lineshape with the same height and
FWHM. Notice that this lineshape is sharper than the Lorentzian
for small $\Delta $ and falls off more slowly for large $\Delta$.

\section{Averaging more complicated interaction potentials}
\label{angular}

In Section~\ref{averaging} we averaged the quantity $e^{-\beta
\mc{V}^{2}}$, where $\mc{V}^{2}=\sum_{k=1}^{N}V_{k}^{2}$ and
$V_{k}=\mu \mu ^{\prime}/r_{k}^{3}$. In this section we consider
the problem of averaging a more complicated interaction potential.
Here we perform the averaging procedure for a potential that
varies as
\begin{equation}
V_{k}\left( \mbf{r}_{k}\right) = \frac{C}{r_{k}^{p}}
f\left(\theta_{k},\phi_{k} \right),
\end{equation}
where $C$ is a constant and $r_{k}$, $\theta_{k}$, and $\phi_{k}$
are the spherical coordinates corresponding to $\mbf{r}_{k}$. We
have seen that we can evaluate the average of the signal if we can
compute the average of the quantity $\exp\left( -\beta\mc{V}^{2}
\right)$. A careful look at Eq.~(\ref{avexp}) shows that the steps
do not make any assumptions on how $V\left(\mbf{r}_{k}\right)$
depends on $\mbf{r}_{k}$, so this equation still holds and we have
\begin{equation}
\left\langle e^{-\beta \mc{V}^{2}}\right\rangle = \exp \left\{
-\frac{N}{\Omega}\int d^{3}r\,\left[ 1-e^{-\beta
V^{2}\left(\mbf{r}\right)}\right] \right\}. \label{avgagain}
\end{equation}
Thus we are left with the task of evaluating the integral
\begin{eqnarray}
I &=& \int d^{3}r\,\left[ 1-e^{-\beta
V^{2}\left(\mbf{r}\right)}\right] \\ \nonumber &=&
\int_{0}^{\infty} dr\,r^{2}\int_{0}^{2\pi}
d\phi\int_{0}^{\pi}d\theta\,\sin\theta \left[ 1-\exp\left(-\beta
\frac{C^{2}}{r^{2p}}f^{2}\left(\theta,\phi\right)\right)\right].
\end{eqnarray}
We can define the variable $z =
r^{3}\left[C^{2}f^{2}\left(\theta,\phi\right)\right]^{-3/2p}$, and
then
\begin{eqnarray}
I &=& \frac{1}{3}\int_{0}^{\infty} dz\int_{0}^{2\pi}
d\phi\int_{0}^{\pi}d\theta\,\sin\theta
\left[C^{2}f^{2}\left(\theta,\phi\right)\beta\right]^{3/2p}\left[
1-\exp\left(-\frac{1}{z^{2p/3}}\right)\right] \nonumber \\ &=&
\frac{1}{3}\left\vert C \right\vert^{3/p}\beta^{3/2p} \nonumber \\
&& \left\{\int_{0}^{\infty} dz\,\left[
1-\exp\left(-\frac{1}{z^{2p/3}}\right)\right]\right\}\left\{\int_{0}^{2\pi}
d\phi\int_{0}^{\pi}d\theta\,\sin\theta \left\vert
f\left(\theta,\phi\right)\right\vert^{3/p}\right\}. \label{fancyv}
\end{eqnarray}
The remaining integrals over $z$ and the angular variables can now
be looked up in tables or done numerically.

It is very important to note that the introduction of an angular
dependence into the potential does not affect the averaging
process except to change the definition of the scaling parameter
$v$ by a constant factor. This means that we can, for simplicity,
use a simple potential that contains no angular dependence to get
an answer, and then take into account any angular dependence by
simply changing the definition of $v$ in the formulas.

\subsection{Interaction potential of aligned dipoles}
\label{aligneddipoles}

The induced dipoles in the magneto-optical trap become aligned
when the applied electric field is sufficiently large that the
Stark effect has passed from the quadratic to the linear regime.
The aligned dipole case is the one considered in
Refs.~\cite{Jaksch2000a} and \cite{Santos2000a}, for example. It
is also the potential we use in performing the numerical
simulations discussed in Chapter~\ref{simulations}, for reasons
that are explained fully in Section~\ref{theoryreview}.
Specifically, we have
\begin{equation}
V\left(\mbf{r}\right) = -\frac{\mu_{1}\mu_{2}}{r^{3}}\left(
3\cos^{2}\theta -1\right).
\end{equation}
Thus we need to evaluate the integral
\begin{eqnarray}
&&\frac{1}{3}\mu_{1}\mu_{2}\sqrt{\beta}\left\{\int_{0}^{\infty}
dz\,\left[
1-\exp\left(-\frac{1}{z^{2}}\right)\right]\right\}\left\{\int_{0}^{2\pi}
d\phi\int_{0}^{\pi}d\theta\,\sin\theta \left\vert 3\cos^{2}\theta
-1\right\vert\right\} \nonumber \\ &=&
\frac{1}{3}\mu_{1}\mu_{2}\sqrt{\beta}\left\{\sqrt{\pi}\right\}\left\{2\pi
\frac{8}{3\sqrt{3}}\right\} \nonumber \\ &=& \frac{16\pi
^{3/2}}{9\sqrt{3}}\mu_{1}\mu_{2}\sqrt{\beta}. \label{valigned}
\end{eqnarray}

We see that the additional angular dependence changes $v$ by a
factor of $4/3\sqrt{3}\approx 0.77$.

\subsection{Effect of a different radial dependence on the initial rise}

In this section, we want to determine the effect of a different
radial dependence in the interaction potential on the initial
rise. We saw that if the potential goes as $1/r^{3}$, then the
signal starts linearly with $t$, with a slope that is independent
of the detuning $\Delta$. We will see that if the interaction
potential goes as $1/r^{p}$, then the initial rise will go as
$t^{3/p}$, with a coefficient that is again independent of the
detuning.

Consider again Eq.~(\ref{unaveragedS}) in the form
\begin{equation}
\tilde{S}\left(\Delta,\alpha\right) = \frac{1}{2\alpha} -
\frac{\alpha^{2} + \Delta^{2}}{2\alpha}\int_{0}^{\infty}d\beta\,
e^{-\beta\left(\alpha^{2} + \Delta^{2} + 4\mc{V}^{2}\right)}.
\end{equation}
We noted previously that any angular dependence in the interaction
potential will only change the scale factor $v$, and will not
change the functional form of the signal. Therefore we can simply
consider a potential of the form
\begin{equation}
V\left(r_{k}\right) = \frac{C}{r_{k}^{p}}.
\end{equation}
Of course, we can incorporate the effect of an angular
contribution by simply rescaling $C$.

It follows, then, from Eqs.~(\ref{avgagain}) and (\ref{fancyv})
that
\begin{equation}
\left\langle\tilde{S}\left(\Delta,\alpha\right)\right\rangle =
\frac{1}{2\alpha} - \frac{\alpha^{2} +
\Delta^{2}}{2\alpha}\int_{0}^{\infty}d\beta\,
e^{-\beta\left(\alpha^{2} + \Delta^{2}\right) - \nu\beta^{3/2p}},
\end{equation}
where
\begin{equation}
\nu = 4\pi\frac{4^{3/2p}}{3}\left\vert
C\right\vert^{3/p}\int_{0}^{\infty}dz\,\left[1 -
\exp\left(-\frac{1}{z^{2p/3}}\right)\right].
\end{equation}
If we define $\xi = \beta\left(\alpha^{2} + \Delta^{2}\right)$,
then
\begin{equation}
\left\langle\tilde{S}\left(\Delta,\alpha\right)\right\rangle =
\frac{1}{2\alpha} - \frac{1}{2\alpha}\int_{0}^{\infty}d\xi\,
e^{-\xi}\exp\left(-\xi^{3/2p}\frac{\nu}{\left(\alpha^{2} +
\Delta^{2}\right)^{3/2p}}\right).
\end{equation}
We want the behavior at small times, which corresponds to large
$\alpha$. Therefore we can replace $\alpha^{2} + \Delta^{2}$ with
$\alpha^{2}$ and expand the second exponential, finding
\begin{eqnarray}
\left\langle\tilde{S}\left(\Delta,\alpha\right)\right\rangle &=&
\frac{1}{2\alpha} - \frac{1}{2\alpha}\int_{0}^{\infty}d\xi\,
e^{-\xi}\left[1 - \xi^{3/2p}\frac{\nu}{\alpha^{3/p}} +
\cdots\right] \nonumber \\ &=& \G{1 +
\frac{3}{2p}}\frac{\nu}{2\alpha^{1 + 3/p}} - \cdots.
\end{eqnarray}
It immediately follows that
\begin{equation}
\left\langle S\left(\Delta,t\right)\right\rangle = \frac{\G{1 +
\frac{3}{2p}}}{2\G{1+\frac{3}{p}}}\nu t^{3/p} - \cdots.
\end{equation}

Thus we see that if the interaction potential goes as $1/r^{p}$,
then the initial rise is independent of the detuning, $\Delta$,
and is proportional to $t^{3/p}$. As a check, we note that if
$p=3$, then we have
\begin{eqnarray}
\left\langle S\left(\Delta,t\right)\right\rangle &=&
\frac{\G{\frac{3}{2}}}{2\G{2}}\nu t - \cdots \nonumber
\\ &=& \frac{\sqrt{\pi}}{4}\nu t - \cdots,
\end{eqnarray}
where
\begin{eqnarray}
\nu &=& 4\pi\frac{2}{3}\left\vert
C\right\vert\int_{0}^{\infty}dz\,\left[1 -
\exp\left(-\frac{1}{z^{2}}\right)\right] \nonumber \\ &=&
\frac{8\pi^{3/2}}{3} \left\vert C\right\vert,
\end{eqnarray}
and so
\begin{equation}
\left\langle S\left(\Delta,t\right)\right\rangle =
\frac{\sqrt{\pi}}{2}vt - \cdots, \label{needsanumber2}
\end{equation}
which agrees perfectly with the $m=0$, $n=0$ term of
Eq.~(\ref{needsanumber}).

\section{Summary}
\label{sparse_no_u_conclusions}

In this chapter we discussed the case of a single $s^{\prime}$
atom interacting resonantly with a surrounding gas of $s$ atoms,
without the possibility of any $sp$--$ps$ interaction. The
results, when averaged over the positions of the $s$ atoms, are
applicable to a sparse distribution of $s^{\prime}$ atoms among
the $s$ atoms (or vice versa, with the appropriate changes). This
case is important to study for several reasons. First, exact
analytical results can be obtained. This not only provides an
excellent means for testing the validity of the numerical
simulations discussed in Chapter~\ref{simulations}, but it also
provides a simpler arena for the development of the techniques
that are used again and again in the more complicated models of
the later chapters. Two key examples of the latter are the
averaging technique of Sections~\ref{averaging} and \ref{angular},
and the writing of sums as generalized hypergeometrics, as was
done in Section~\ref{sparse_no_u_inverting}.

The averaging is non-trivial because the interaction potential
itself, being proportional to $1/r^{3}$, does not have an average
value. Nevertheless, an effective average potential $v$ given by
Eq.~(\ref{v}) appears naturally in the theory. The basic averaging
equation is Eq.~(\ref{ave}), which enables the computation of
averages of functions that depend on the sum of the squares of the
interaction potentials in terms of $v$. This is equivalent to the
distribution of Eq.~(\ref{Pv}) for the quantity $\mc{V}^{2} =
\sum_{k} V_{k}^{2}$, where $V_{k}$ is the dipole-dipole
interaction potential defined by Eq.~(\ref{Vk}). The initial rate
of conversion from $ss^{\prime}$ to $pp^{\prime}$ is given by
Eq.~(\ref{needsanumber}) or Eq.~(\ref{needsanumber2}) to be
$\sqrt{\pi}v/2$, and is independent of the detuning $\Delta$. As
we will see later, this initial rate remains the same even when
the $sp$--$ps$ interaction is switched on.

\chapter{Localization in the Simple Model}
\markright{Chapter \arabic{chapter}: Localization in the Simple
Model}
\label{localization}

\section{Introduction}

In this chapter we continue with our discussion of the simple
model of Chapter~\ref{sparse_no_u} by examining the issue of
localization. We want to determine the effective range, or
localization length, of the disturbance caused by the presence of
an $s^{\prime}$ atom in the frozen gas of $s$ atoms. Since the
$1/r^{3}$ interaction does not have a length scale, it follows
that the localization length must be of the form
$\left(\Omega/N\right)^{1/3} g\left(vt,\Delta/v\right)$, where
$N/\Omega$ is the density of $s$ atoms, $t$ is the time, $\Delta$
is the detuning from resonance, and $v$ is defined in
Eq.~(\ref{v}), or more generally in Eq.~(\ref{fancyv}). The task,
then, is to give a precise definition of the localization length
and to compute the function $g\left(vt,\Delta/v\right)$.

First, in Section~\ref{calSDeltart} we compute a distribution
function $\left\langle\mc{S}\left(
\Delta,\mbf{r},t\right)\right\rangle$, which describes the
contribution to the $ss^{\prime}\rightarrow pp^{\prime}$ process
due to $s$ atoms at position $\mbf{r}$ when the $s^{\prime}$ atom
is at the origin. We begin by computing the Laplace transform of
this function in Section~\ref{LTSDeltart}, then we invert this
Laplace transform in Section~\ref{invertLTSDeltart}. In
Section~\ref{check} we perform a check on the results of
Section~\ref{LTSDeltart} and \ref{invertLTSDeltart} by using them
to rederive the result for the signal that was found in
Chapter~\ref{sparse_no_u}. In Section~\ref{distribution_plots} we
present plots of the distribution $\left\langle\mc{S}\left(
\Delta,\mbf{r},t\right)\right\rangle$.

In Section~\ref{moments} we compute the moments of the
distribution $\left\langle\mc{S}\left(
\Delta,\mbf{r},t\right)\right\rangle$. We again start by computing
the Laplace transform of the moments in Section~\ref{LTmoments},
then we invert this Laplace transform in
Section~\ref{invertLTmoments}. The result of
Section~\ref{invertLTmoments} is rearranged in
Section~\ref{moreconvenient} to obtain a more convenient form, and
in this section we also define localization lengths in terms of
the moments in the usual way. The end result for the moments is
checked in Section~\ref{checkint} by seeing that the zeroth moment
agrees with the result for the signal as a function of time that
was found in Chapter~\ref{sparse_no_u}. In
Section~\ref{localization_plots} we present plots of the results.

The summary and conclusions are given in
Section~\ref{localization_conclusions}. It is possible to go to
this section directly and refer back to final formulas and the
plots of Section~\ref{localization_plots}, skipping the
intervening mathematics.

\section{The function
$\left\langle\mc{S}\left( \Delta,\mbf{r},t\right)\right\rangle$}
\label{calSDeltart}

In order to study localization in the simple model of
Chapter~\ref{sparse_no_u}, we consider the quantity
\begin{equation}
\left\langle\mc{S}\left( \Delta,\mbf{r},t\right)\right\rangle =
\left\langle \sum_{k=1}^{N}\left| c_{k}\left( \Delta,t\right)
\right| ^{2}\delta \left( \mbf{r}-\mbf{r}_{k}\right)
\right\rangle. \label{Srt}
\end{equation}
As is illustrated explicitly below, the computation of the average
is very similar to the computation of the average signal,
$\left\langle\tilde{S}\left(\Delta,\alpha\right)\right\rangle$, of
Chapter~\ref{sparse_no_u}.

\subsection{Computation of the Laplace transform of
$\left\langle\mc{S}\left(\Delta,\mbf{r},t\right)\right\rangle$}
\label{LTSDeltart}

From Eq.~(\ref{ck}) of this thesis or Eq.~(3) of
Ref.~\cite{Frasier1999a} we have that
\begin{equation}
\int_{0}^{\infty}dt\,e^{-\alpha t}\left|
c_{k}\left(t\right)\right|^{2} = \frac{1}{\alpha}
\frac{2V_{k}^{2}}{\alpha^{2} + \Delta^{2} + 4\mc{V}^{2}},
\end{equation}
where $\mc{V}^{2} = \sum_{k} V_{k}^{2}$. Therefore we define
\begin{equation}
\left\langle\tilde{\mc{S}}\left( \Delta,\mbf{r},\alpha
\right)\right\rangle = \frac{1}{\Omega^{N}}\int d^{3}r_{1}\cdots
d^{3}r_{N}\frac{2}{\alpha }\sum_{k=1}^{N}\frac{V_{k}^{2}\delta
\left( \mbf{r}-\mbf{r}_{k}\right) }{\alpha ^{2}+\Delta
^{2}+4{\mathcal{V}}^{2}}. \label{Sralphaintform}
\end{equation}

Consider for now just the $k=1$ term in the sum. For convenience,
we will call this term $\mc{A}$. We have
\begin{equation}
\mc{A} = \frac{2}{\alpha}\frac{1}{\Omega^{N}}\int d^{3}r_{1}\cdots
d^{3}r_{N} \frac{V^{2}\left(\mbf{r}_{1}\right)
\delta\left(\mbf{r}-\mbf{r}_{1}\right)}{\alpha^{2} + \Delta^{2} +
4V^{2}\left(\mbf{r}_{1}\right) + 4\sum_{l=2}^{N} V_{l}^{2}},
\end{equation}
where I have written $V_{1}$ as $V\left(\mbf{r}_{1}\right)$ to be
explicit. The integral over $\mbf{r}_{1}$ is trivial, and we find
\begin{equation}
\mc{A} =
\frac{2}{\alpha}V^{2}\left(\mbf{r}\right)\frac{1}{\Omega^{N}}\int
d^{3}r_{2}\cdots d^{3}r_{N} \frac{1}{\alpha^{2} + \Delta^{2} +
4V^{2}\left(\mbf{r}\right) + 4\sum_{l=2}^{N} V_{l}^{2}}.
\end{equation}
We observe now that we can write the denominator as an integral
over an exponential and use the same averaging technique as in
Chapter~\ref{sparse_no_u}. The fact that the sum over $l$ starts
from $l=2$ does not affect the validity of Eq.~(\ref{ave}), as
long as $N$ is large. Thus we have
\begin{eqnarray}
\mc{A} &=&
\frac{2}{\alpha}V^{2}\left(\mbf{r}\right)\frac{1}{\Omega^{N}}\int_{0}^{\infty}
d\Lambda \int d^{3}r_{2}\cdots d^{3}r_{N}
\exp\left\{-\Lambda\left[\alpha^{2} + \Delta^{2} +
4V^{2}\left(\mbf{r}\right) + 4\sum_{l=2}^{N}
V_{l}^{2}\right]\right\} \nonumber \\ &=&
\frac{2}{\alpha}V^{2}\left(\mbf{r}\right)\frac{1}{\Omega}\int_{0}^{\infty}
d\Lambda \exp\left\{-\Lambda\left[\alpha^{2} + \Delta^{2} +
4V^{2}\left(\mbf{r}\right)\right] - 2v\sqrt{\Lambda}\right\}.
\label{A1}
\end{eqnarray}

It is clear that it did not matter that we chose the $k=1$ term
from the sum in Eq.~(\ref{Sralphaintform}). We would have gotten
the result of Eq.~(\ref{A1}) no matter which term we picked. It
follows that all the terms are equal, and so
\begin{equation}
\left\langle
\tilde{\mc{S}}\left(\Delta,\mbf{r},\alpha\right)\right\rangle =
\frac{2}{\alpha} \frac{N}{\Omega} V^{2}\left(\mbf{r}\right)
\int_{0}^{\infty}d\Lambda\, \exp\left\{-\Lambda\left[\alpha^{2} +
\Delta^{2} + 4V^{2}\left(\mbf{r}\right)\right] -
2v\sqrt{\Lambda}\right\}. \label{avSintform}
\end{equation}
The remaining integral over $\Lambda$ is familiar from
Chapter~\ref{sparse_no_u} and can be looked up in tables (see, for
example, Ref.~\cite{Gradshteyn1994a}) or done by
\emph{Mathematica\/}. The result is
\begin{eqnarray}
\left\langle\tilde{\mc{S}}\left( \Delta,\mbf{r},\alpha
\right)\right\rangle &=&
\frac{N}{\Omega}\frac{2V^{2}\left(\mbf{r}\right)}{\alpha\left[\alpha^{2}
+ \Delta^{2} + 4V^{2}\left(\mbf{r}\right)\right]} - \nonumber
\\ && \frac{N}{\Omega}\frac{2v\sqrt{\pi}V^{2}\left(\mbf{r}\right)}{\alpha\left[\alpha
^{2}+\Delta ^{2}+4V^{2}\left( \mbf{r}\right)\right]^{3/2}}\times
\nonumber \\ && \exp \left[ \frac{v^{2}}{\alpha ^{2}+\Delta
^{2}+4V^{2}\left( \mbf{r}\right) }\right] \erfc \left[
\frac{v}{\sqrt{\alpha ^{2}+\Delta ^{2}+4V^{2}\left(
\mbf{r}\right)}}\right]. \label{calSalpha}
\end{eqnarray}

\subsection{Inverting the Laplace transform}
\label{invertLTSDeltart}

Now that we have $\left\langle\tilde{\mc{S}}\left(
\Delta,\mbf{r},\alpha \right)\right\rangle$, the next step is to
invert the Laplace transform. Because Eq.~(\ref{calSalpha}) is of
a form that is similar to Eq.~(\ref{Stld2}), we can again use
Eqs.~(\ref{expzerfcz}), (\ref{before}), and (\ref{after}) to
achieve this.

Using Eq.~(\ref{expzerfcz}), we have
\begin{eqnarray}
\left\langle\tilde{\mc{S}}\left( \Delta,\mbf{r},\alpha
\right)\right\rangle &=&
\frac{N}{\Omega}\frac{2V^{2}\left(\mbf{r}\right)}{\alpha\left[\alpha^{2}
+ \Delta^{2} + 4V^{2}\left(\mbf{r}\right)\right]} - \nonumber
\\ && \frac{N}{\Omega}\frac{2v\sqrt{\pi}V^{2}\left(\mbf{r}\right)}{\alpha\left[\alpha
^{2}+\Delta ^{2}+4V^{2}\left(
\mbf{r}\right)\right]^{3/2}}\sum_{n=0}^{\infty}
\frac{1}{\G{\frac{n}{2}+1}} \left[-\frac{v}{\sqrt{\alpha
^{2}+\Delta ^{2}+4V^{2}\left( \mbf{r}\right)}}\right]^{n}
\nonumber \\ &=&
\frac{N}{\Omega}\frac{2V^{2}\left(\mbf{r}\right)}{\alpha\left[\alpha^{2}
+ \Delta^{2} + 4V^{2}\left(\mbf{r}\right)\right]} - \nonumber
\\ && \frac{N}{\Omega}2v\sqrt{\pi}V^{2}\left(\mbf{r}\right)\sum_{n=0}^{\infty}
\frac{1}{\G{\frac{n}{2}+1}} \frac{\left(
-v\right)^{n}}{\alpha\left[\alpha^{2}+\Delta ^{2}+4V^{2}\left(
\mbf{r}\right)\right]^{\frac{n+3}{2}}}.
\end{eqnarray}
Next, using the fact (from Eq.~(2.66) of Ref.~\cite{Roberts1966a})
that the inverse Laplace transform of Eq.~(\ref{before}) is
Eq.~(\ref{after}), we see that
\begin{eqnarray}
\left\langle\mc{S}\left( \Delta,\mbf{r},t \right)\right\rangle &=&
\frac{N}{\Omega}2V^{2}\left(\mbf{r}\right)\frac{t^{2}}{\G{3}}\,
{}_{1}F_{2}\left\{ 1;\frac{3}{2},2;-\frac{1}{4}\left[\Delta^{2} +
4V^{2}\left(\mbf{r}\right)\right]t^{2}\right\} - \nonumber
\\ && \frac{N}{\Omega}2v\sqrt{\pi}V^{2}\left(\mbf{r}\right)\sum_{n=0}^{\infty}
\frac{\left(
-v\right)^{n}t^{n+3}}{\G{\frac{n}{2}+1}\G{n+4}}\,\times \nonumber
\\ && {}_{1}F_{2}\left\{
\frac{n+3}{2};\frac{n+4}{2},\frac{n+5}{2};-\frac{1}{4}\left[\Delta^{2}
+ 4V^{2}\left(\mbf{r}\right)\right]t^{2}\right\} \nonumber \\ &=&
\frac{N}{\Omega}V^{2}\left(\mbf{r}\right)t^{2}\,
{}_{1}F_{2}\left\{ 1;\frac{3}{2},2;-\frac{1}{4}\left[\Delta^{2} +
4V^{2}\left(\mbf{r}\right)\right]t^{2}\right\} - \nonumber
\\ && \frac{N}{\Omega}2\sqrt{\pi}vV^{2}\left(\mbf{r}\right)t^{3}\sum_{n=0}^{\infty}
\frac{\left( -vt\right)^{n}}{\G{\frac{n}{2}+1}\G{n+4}}\,\nonumber
\times
\\ && {}_{1}F_{2}\left\{
\frac{n+3}{2};\frac{n+4}{2},\frac{n+5}{2};-\frac{1}{4}\left[\Delta^{2}
+ 4V^{2}\left(\mbf{r}\right)\right]t^{2}\right\}.
\label{calSdeltat}
\end{eqnarray}

\subsection{A check on the result for
$\left\langle\tilde{\mc{S}}\left(\Delta, \mbf{r},\alpha
\right)\right\rangle$} \label{check}

It is clear from Eq.~(\ref{Srt}) that
\begin{equation}
\int d^{3}r\,\left\langle\mc{S}\left( \Delta,\mbf{r},t
\right)\right\rangle = \left\langle S\left(
\Delta,t\right)\right\rangle, \label{M_0result}
\end{equation}
and since taking the Laplace transform with respect to time and
integrating over the variable $\mbf{r}$ clearly commute, it must
also be the case that
\begin{equation}
\int d^{3}r\,\left\langle\tilde{\mc{S}}\left(
\Delta,\mbf{r},\alpha \right)\right\rangle =
\left\langle\tilde{S}\left( \Delta,\alpha\right)\right\rangle.
\end{equation}
For the sake of completeness, we now check that this is true. We
will also find this exercise useful in the next section. From
Eq.~(\ref{avSintform}) we have
\begin{eqnarray}
\int d^{3}r\left\langle\tilde{\mc{S}}\left(\Delta,
\mbf{r},\alpha\right)\right\rangle &=& \int d^{3}r\frac{N}{\Omega
}\frac{2V^{2}\left( \mbf{r}\right)}{\alpha}\int_{0}^{\infty
}d\Lambda\,\exp\left\{ -\Lambda \left[ \alpha ^{2}+\Delta
^{2}+4V^{2}\left( \mbf{r}\right)\right] - 2v\sqrt{\Lambda}\right\}
\nonumber \\ &=&
\frac{2}{\alpha}\int_{0}^{\infty}d\Lambda\,\exp\left[ -\Lambda
\left( \alpha ^{2}+\Delta ^{2}\right) - 2v\sqrt{\Lambda}\right]
\frac{N}{\Omega}\int d^{3}r\,V^{2}\left( \mbf{r}\right) \exp
\left[ -4\Lambda V^{2}\left( \mbf{r}\right) \right] \nonumber
\\ &=& \frac{2}{\alpha }\int_{0}^{\infty }d\Lambda \exp \left[
-\Lambda \left( \alpha ^{2}+\Delta ^{2}\right) -2v\sqrt{\Lambda
}\right] \times \nonumber \\ && \left( -\frac{1}{4} \right)
\frac{\partial }{\partial \Lambda }\left[ \frac{N}{\Omega }\int
d^{3}r\,\exp \left( -4\Lambda V^{2}\left( \mbf{r}\right) \right)
\right].
\end{eqnarray}

We now recall from Eqs.~(\ref{aveintsimple}) and (\ref{v}), or
more generally from Section \ref{angular} of this thesis, that
\begin{equation}
\frac{N}{\Omega}\int d^{3}r\left[ 1-e^{-\beta
V^{2}\left(\mbf{r}\right)}\right] = v\sqrt{\beta}.
\end{equation}
It follows from this that
\begin{equation}
\frac{N}{\Omega }\int d^{3}r\,\exp \left[ -4\Lambda V^{2}\left(
\mbf{r}\right) \right] = N - 2v\sqrt{\Lambda},
\end{equation}
and so
\begin{eqnarray}
\int d^{3}r\left\langle\tilde{\mc{S}}\left(\Delta,
\mbf{r},\alpha\right)\right\rangle &=&
\frac{1}{2\alpha}\int_{0}^{\infty}d\Lambda\,\exp\left[ -\Lambda
\left( \alpha ^{2}+\Delta ^{2}\right) -2v\sqrt{\Lambda}\right]
\left(-\frac{\partial}{\partial\Lambda}\right)\left( N -
2v\sqrt{\Lambda}\right) \nonumber \\ &=&
\frac{v}{2\alpha}\int_{0}^{\infty}\frac{d\Lambda}{\sqrt{\Lambda}}\,\exp\left[
-\Lambda \left( \alpha ^{2}+\Delta ^{2}\right)
-2v\sqrt{\Lambda}\right].
\end{eqnarray}
The integral over $\Lambda$ can again be found in tables or done
by \emph{Mathematica\/}. The result is
\begin{eqnarray}
\int d^{3}r\left\langle\tilde{\mc{S}}\left(\Delta,
\mbf{r},\alpha\right)\right\rangle &=&
\frac{\sqrt{\pi}}{2}\frac{v}{\alpha\sqrt{\alpha^{2} +
\Delta^{2}}}\exp\left(\frac{v^{2}}{\alpha^{2} +
\Delta^{2}}\right)\erfc\left(\frac{v}{\sqrt{\alpha^{2} +
\Delta^{2}}}\right) \nonumber \\ &=&
\left\langle\tilde{S}\left(\Delta, \alpha\right)\right\rangle,
\end{eqnarray}
where we have used Eq.~(\ref{Stld2}).

A similar check for $\left\langle
\mc{S}\left(\Delta,\mbf{r},t\right)\right\rangle$ is carried out
in Section~\ref{checkint}.

\subsection{Plots}
\label{distribution_plots}

\begin{figure}
\resizebox{\textwidth}{!}{\includegraphics[0in,0in][8in,10in]{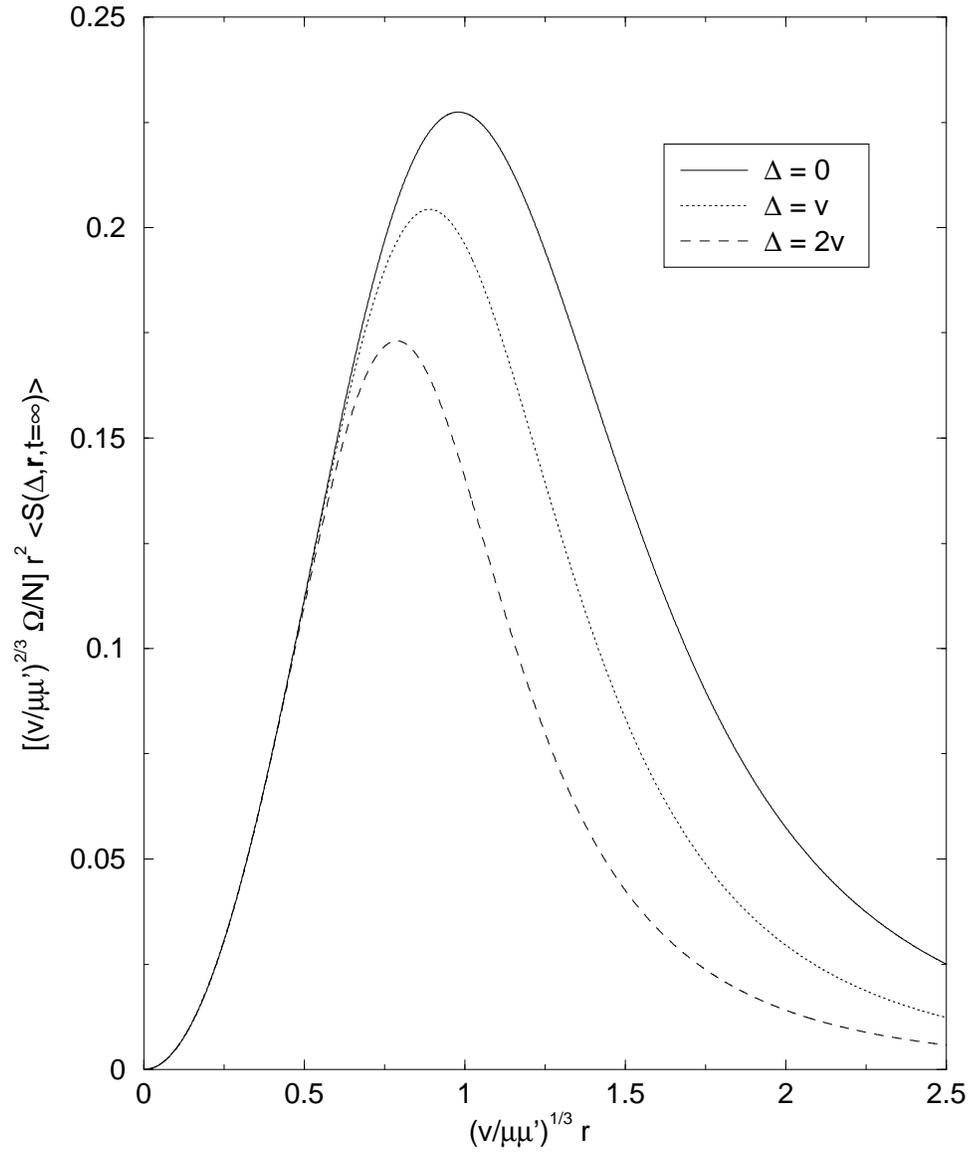}}
\caption[The quantity $r^{2} \left\langle\mc{S}\left(
\Delta,\mbf{r},t=\infty \right)\right\rangle$ as a function of $r$
for several values of $\Delta$]{Plots of the quantity $r^{2}
\left\langle\mc{S}\left( \Delta,\mbf{r},t=\infty
\right)\right\rangle$ as a function of $r$ for $\Delta$ equal to
$0$, $v$, and $2v$.} \label{sparse_no_u_S}
\end{figure}

In Fig.~\ref{sparse_no_u_S} we see plots of the dimensionless
quantity
\begin{equation}
\left[ \left(\frac{v}{\mu\mu^{\prime}}\right)^{2/3}
\frac{\Omega}{N}\right] r^{2} \left\langle\mc{S}\left(
\Delta,\mbf{r},t=\infty \right)\right\rangle,
\end{equation}
as a function of $\left(v/\mu\mu^{\prime}\right)^{1/3} r$, for
$\Delta = 0$, $v$, and $2v$. The quantity
$\left\langle\mc{S}\left( \Delta,\mbf{r},t=\infty
\right)\right\rangle$ is most easily computed by multiplying
Eq.~(\ref{calSalpha}) by $\alpha$ and then taking the limit
$\alpha\rightarrow 0$. We see that the peak of the distribution
shifts to the left with increasing $\Delta$.

\section{The moments of $\left\langle\mc{S}\left(\Delta,
\mbf{r},t \right)\right\rangle$}
\label{moments}

In order to compute the moments of
$\left\langle\mc{S}\left(\Delta,\mbf{r},t \right)\right\rangle$,
we first define
\begin{equation}
\tilde{\mc{M}}_{p}\left( \Delta,\alpha \right) = \int
d^{3}r\,r^{p}\left\langle\tilde{\mc{S}}\left(
\Delta,\mbf{r},\alpha \right)\right\rangle,
\end{equation}
so that $\tilde{\mc{M}}_{p}\left( \Delta,\alpha \right)$ is the
$p$th moment of $\left\langle\tilde{\mc{S}}\left(
\Delta,\mbf{r},\alpha \right)\right\rangle$. We will see that we
can then determine $\mc{M}_{p}\left(\Delta, t \right)$ by applying
the inverse Laplace transform methods we have already refined.

\subsection{Computation of $\tilde{\mc{M}}_{p}\left(\Delta,
\alpha \right)$}
\label{LTmoments}

Much of this computation is very similar to that done in Section
\ref{check}. Using Eq.~(\ref{avSintform}), we have
\begin{eqnarray}
\tilde{\mc{M}}_{p}\left( \Delta,\alpha \right) &=& \int
d^{3}r\,r^{p}\frac{N}{\Omega }\frac{2V^{2}\left(
\mbf{r}\right)}{\alpha}\int_{0}^{\infty }d\Lambda\,\exp \left[
-\Lambda \left( \alpha ^{2}+\Delta ^{2}+4V^{2}\left(
\mbf{r}\right) \right) - 2v\sqrt{\Lambda }\right] \nonumber
\\ &=& \frac{2}{\alpha}\int_{0}^{\infty}d\Lambda\,\exp \left[
-\Lambda \left( \alpha ^{2}+\Delta ^{2}\right) -2v\sqrt{\Lambda
}\right] \frac{N}{\Omega} \int d^{3}r\,r^{p}V^{2}\left(
\mbf{r}\right) \exp\left[ -4\Lambda V^{2}\left( \mbf{r}\right)
\right] \nonumber \\ &=& \frac{2}{\alpha }\int_{0}^{\infty
}d\Lambda\,\exp\left[ -\Lambda \left( \alpha ^{2}+\Delta
^{2}\right) - 2v\sqrt{\Lambda}\right] \times \nonumber
\\ && \left( -\frac{1}{4} \frac{\partial }{\partial \Lambda }\right)
\left\{ \frac{N}{\Omega }\int d^{3}r\,r^{p}\exp \left[ -4\Lambda
V^{2}\left( \mbf{r}\right) \right] \right\}.
\end{eqnarray}

We now note that
\begin{equation}
\int d^{3}r\,r^{p}\exp\left[-4\Lambda
V^{2}\left(\mbf{r}\right)\right] = \int d^{3}r\,r^{p} - \int
d^{3}r\,r^{p}\left\{1 - \exp\left[-4\Lambda
V^{2}\left(\mbf{r}\right)\right]\right\}.
\end{equation}
Although the first term diverges, it will disappear when we take
the derivative over $\Lambda$ because it does not depend on that
variable.

We now need to compute the remaining integral over $\mbf{r}$. I
will compute it here for the case where $V\left(\mbf{r}\right) =
\mu\mu^{\prime}/r^{3}$. If there is an angular dependence in the
potential, then the computation is somewhat more complicated
because the extra factors of $r$ in the integrand will bring down
extra angular factors when we apply the methods of Section
\ref{angular}. This is one case where we cannot simply rescale $v$
to obtain the correct result when an angular dependence is
introduced into the potential. With this in mind, we have
\begin{eqnarray}
\int d^{3}r\,r^{p}\left\{1 - \exp\left[-4\Lambda
V^{2}\left(\mbf{r}\right)\right]\right\} &=& 4\pi\int
dr\,r^{p+2}\left\{1 - \exp\left[-4\Lambda
\frac{\left(\mu\mu^{\prime}\right)^{2}}{r^{6}}\right]\right\}
\nonumber \\ &=& \frac{4\pi}{3}
\left[4\Lambda\left(\mu\mu^{\prime}\right)^{2}\right]^{\frac{p+3}{6}}
\int dz\,z^{p/3}\left[1 -
\exp\left(-\frac{1}{z^{2}}\right)\right],
\end{eqnarray}
where $z^{2} = r^{6}/4\Lambda\left(\mu\mu^{\prime}\right)^{2}$.
The remaining integral over $z$ can be found in tables or done by
\emph{Mathematica\/}. The result is
\begin{equation}
\int d^{3}r\,r^{p}\left\{1 - \exp\left[-4\Lambda
V^{2}\left(\mbf{r}\right)\right]\right\} =
-\frac{1}{2}\G{-\frac{p+3}{6}}\frac{4\pi}{3}
\left[4\Lambda\left(\mu\mu^{\prime}\right)^{2}\right]^{\frac{p+3}{6}},
\end{equation}
for $p < 3$. This should suffice, since we will be primarily
interested in $p=1$ and $p=2$.

Thus we have (for $p < 3$)
\begin{eqnarray}
\tilde{\mc{M}}_{p}\left(\Delta,\alpha\right) &=&
\frac{1}{2\alpha}\int_{0}^{\infty }d\Lambda\,\exp\left[ -\Lambda
\left(\alpha^{2}+\Delta^{2}\right) - 2v\sqrt{\Lambda}\right]
\times \nonumber \\ && \frac{\partial}{\partial\Lambda} \left\{
-\frac{1}{2}\G{-\frac{p+3}{6}}\frac{4\pi}{3}\frac{N}{\Omega}
\left[4\Lambda\left(\mu\mu^{\prime}\right)^{2}\right]^{\frac{p+3}{6}}
\right\} \nonumber \\ &=& -\frac{1}{2\alpha}
\left[4\left(\mu\mu^{\prime}\right)^{2}\right]^{\frac{p+3}{6}}
\frac{p+3}{12}\G{-\frac{p+3}{6}} \times \nonumber \\ &&
\frac{4\pi}{3}\frac{N}{\Omega} \int_{0}^{\infty
}d\Lambda\,\Lambda^{\frac{p-3}{6}}\exp\left[ -\Lambda \left(
\alpha ^{2}+\Delta ^{2}\right) - 2v\sqrt{\Lambda}\right] \nonumber
\\ &=& \frac{1}{\alpha}
2^{p/3}\left(\mu\mu^{\prime}\right)^{p/3}
\frac{1}{2}\G{\frac{3-p}{6}} \frac{v}{\sqrt{\pi}} \times \nonumber
\\ && \int_{0}^{\infty }d\Lambda\,\Lambda^{\frac{p-3}{6}}\exp\left[
-\Lambda \left( \alpha ^{2}+\Delta ^{2}\right) -
2v\sqrt{\Lambda}\right],
\end{eqnarray}
where in the last step we have used the fact that $\G{n+1} =
n\G{n}$ and Eq.~(\ref{v}). For $p > -3$, we have from
\emph{Mathematica\/} that
\begin{eqnarray}
\int_{0}^{\infty}
d\Lambda\,\Lambda^{\frac{p-3}{6}}\exp\left[-\Lambda A^{2} -
2v\sqrt{\Lambda}\right] &=&
\left(\frac{1}{A}\right)^{\frac{p+3}{3}}
\G{\frac{p+3}{6}}\,{}_{1}F_{1}\left(\frac{p+3}{6}; \frac{1}{2};
\frac{v^{2}}{A^{2}} \right) - \nonumber \\ &&
\frac{2v}{A^{\frac{p+6}{3}}}
\G{\frac{p+6}{6}}\,{}_{1}F_{1}\left(\frac{p+6}{6}; \frac{3}{2};
\frac{v^{2}}{A^{2}} \right),
\end{eqnarray}
where as in Chapter~\ref{sparse_no_u}, $A^{2} = \alpha^{2} +
\Delta^{2}$. It follows that for $\left| p\right| < 3$, we have
\begin{eqnarray}
\tilde{\mc{M}}_{p}\left( \Delta,\alpha \right) &=&
\frac{1}{\alpha}
2^{\frac{p-3}{3}}\left(\mu\mu^{\prime}\right)^{p/3}
\G{\frac{3-p}{6}} \frac{v}{\sqrt{\pi}} \times \nonumber
\\ && \left[ \left(\frac{1}{\alpha^{2} +
\Delta^{2}}\right)^{\frac{p+3}{6}}
\G{\frac{p+3}{6}}\,{}_{1}F_{1}\left(\frac{p+3}{6}; \frac{1}{2};
\frac{v^{2}}{\alpha^{2} + \Delta^{2}} \right) - \right. \nonumber
\\ && \left. \frac{2v}{\left(\alpha^{2} +
\Delta^{2}\right)^{\frac{p+6}{6}}}
\G{\frac{p+6}{6}}\,{}_{1}F_{1}\left(\frac{p+6}{6}; \frac{3}{2};
\frac{v^{2}}{\alpha^{2} + \Delta^{2}} \right)\right].
\end{eqnarray}

\subsection{Inverting the Laplace transform}
\label{invertLTmoments}

We know from Eq.~(\ref{hypergeodef}) that
\begin{equation}
{}_{1}F_{1}\left(a,b,z\right) = \frac{\G{b}}{\G{a}}
\sum_{j=0}^{\infty} \frac{\G{j+a}}{\G{j+b} \G{j+1}} z^{j},
\end{equation}
so
\begin{eqnarray}
\tilde{\mc{M}}_{p}\left(\Delta,\alpha\right) &=& \frac{1}{\alpha}
2^{\frac{p-3}{3}}\left(\mu\mu^{\prime}\right)^{p/3}
\G{\frac{3-p}{6}} \frac{v}{\sqrt{\pi}} \times \nonumber \\ &&
\left[ \left(\frac{1}{\alpha^{2} +
\Delta^{2}}\right)^{\frac{p+3}{6}} \G{\frac{p+3}{6}}
\frac{\G{\frac{1}{2}}}{\G{\frac{p+3}{6}}} \sum_{j=0}^{\infty}
\frac{\G{j+\frac{p+3}{6}}}{\G{j+\frac{1}{2}} \G{j+1}}
\left(\frac{v^{2}}{\alpha^{2} + \Delta^{2}}\right)^{j} - \right.
\nonumber
\\ && \left. \frac{2v}{\left(\alpha^{2} +
\Delta^{2}\right)^{\frac{p+6}{6}}} \G{\frac{p+6}{6}}
\frac{\G{\frac{3}{2}}}{\G{\frac{p+6}{6}}} \sum_{j=0}^{\infty}
\frac{\G{j+\frac{p+6}{6}}}{\G{j+\frac{3}{2}} \G{j+1}}
\left(\frac{v^{2}}{\alpha^{2} + \Delta^{2}}\right)^{j}\right]
\nonumber \\ &=& \frac{1}{\alpha}
2^{\frac{p-3}{3}}\left(\mu\mu^{\prime}\right)^{p/3}
\G{\frac{3-p}{6}} \frac{v}{\sqrt{\pi}} \times \nonumber \\ &&
\left[ \sqrt{\pi} \sum_{j=0}^{\infty}
\frac{\G{j+\frac{p+3}{6}}}{\G{j+\frac{1}{2}} \G{j+1}}
\frac{v^{2j}}{\left(\alpha^{2} + \Delta^{2}\right)^{j +
\frac{p+3}{6}}} - \right. \nonumber
\\ && \left. v\sqrt{\pi} \sum_{j=0}^{\infty}
\frac{\G{j+\frac{p+6}{6}}}{\G{j+\frac{3}{2}} \G{j+1}}
\frac{v^{2j}}{\left(\alpha^{2} + \Delta^{2}\right)^{j +
\frac{p+6}{6}}}\right] \nonumber \\ &=& \frac{1}{\alpha}
2^{\frac{p-3}{3}}\left(\mu\mu^{\prime}\right)^{p/3}
\G{\frac{3-p}{6}} v \left[ \sum_{j=0}^{\infty}
\frac{\G{j+\frac{p+3}{6}}}{\G{j+\frac{1}{2}} \G{j+1}}
\frac{v^{2j}}{\left(\alpha^{2} + \Delta^{2}\right)^{j +
\frac{p+3}{6}}} - \right. \nonumber
\\ && \left. \sum_{j=0}^{\infty}
\frac{\G{j+\frac{p+6}{6}}}{\G{j+\frac{3}{2}} \G{j+1}}
\frac{v^{2j+1}}{\left(\alpha^{2} + \Delta^{2}\right)^{j +
\frac{p+6}{6}}}\right],
\end{eqnarray}
where we have used the facts that $\G{1/2} = \sqrt{\pi}$ and
$\G{3/2} = \sqrt{\pi}/2$.

Now we note that
\begin{equation}
\sum_{j=0}^{\infty} \frac{\G{j+\frac{p+3}{6}}}{\G{j+\frac{1}{2}}
\G{j+1}} \frac{v^{2j}}{\left(\alpha^{2} + \Delta^{2}\right)^{j +
\frac{p+3}{6}}} - \sum_{j=0}^{\infty}
\frac{\G{j+\frac{p+6}{6}}}{\G{j+\frac{3}{2}}\G{j+1}}
\frac{v^{2j+1}}{\left(\alpha^{2} + \Delta^{2}\right)^{j +
\frac{p+6}{6}}} \label{diffofsums}
\end{equation}
can be rewritten as
\begin{eqnarray}
&&\sum_{j=0}^{\infty} \frac{\G{\frac{2j}{2} +
\frac{p+3}{6}}}{\G{\frac{2j}{2} + \frac{1}{2}}\G{\frac{2j}{2} +
1}} \left(-\frac{v}{\sqrt{\alpha^{2} + \Delta^{2}}}\right)^{2j}
\left(\frac{1}{\alpha^{2} + \Delta^{2}}\right)^{\frac{p+3}{6}} +
\nonumber \\ &&\sum_{j=0}^{\infty} \frac{\G{\frac{2j+1}{2} +
\frac{p+3}{6}}}{\G{\frac{2j+1}{2} + 1}\G{\frac{2j+1}{2} +
\frac{1}{2}}} \left(-\frac{v}{\sqrt{\alpha^{2} +
\Delta^{2}}}\right)^{2j+1} \left(\frac{1}{\alpha^{2} +
\Delta^{2}}\right)^{\frac{p+3}{6}},
\end{eqnarray}
and so Eq.~(\ref{diffofsums}) is equivalent to the single sum
\begin{equation}
\left(\frac{1}{\alpha^{2} +
\Delta^{2}}\right)^{\frac{p+3}{6}}\sum_{j=0}^{\infty}
\frac{\G{\frac{j}{2} + \frac{p+3}{6}}}{\G{\frac{j}{2} +
\frac{1}{2}}\G{\frac{j}{2} + 1}} \left(-\frac{v}{\sqrt{\alpha^{2}
+ \Delta^{2}}}\right)^{j}.
\end{equation}
Therefore we have
\begin{eqnarray}
\tilde{\mc{M}}_{p}\left(\Delta,\alpha\right) &=& \frac{1}{\alpha}
2^{\frac{p-3}{3}}\left(\mu\mu^{\prime}\right)^{p/3}
\G{\frac{3-p}{6}} v \times \nonumber \\ &&
\left(\frac{1}{\alpha^{2} +
\Delta^{2}}\right)^{\frac{p+3}{6}}\sum_{j=0}^{\infty}
\frac{\G{\frac{j}{2} + \frac{p+3}{6}}}{\G{\frac{j}{2} +
\frac{1}{2}}\G{\frac{j}{2} + 1}} \left(-\frac{v}{\sqrt{\alpha^{2}
+ \Delta^{2}}}\right)^{j}.
\end{eqnarray}

Using Eqs.~(\ref{before}) and (\ref{after}), we see that the
inverse Laplace transform of
\begin{equation}
\frac{1}{\alpha}\left(\frac{1}{\alpha^{2} +
\Delta^{2}}\right)^{\nu}
\end{equation}
is
\begin{equation}
\frac{t^{2\nu}}{\G{2\nu + 1}}\,
{}_{1}F_{2}\left(\nu;\nu+\frac{1}{2};\nu +
1;-\frac{1}{4}\Delta^{2}t^{2}\right).
\end{equation}
Therefore we have that
\begin{eqnarray}
\mc{M}_{p}\left(\Delta,t\right) &=&
2^{\frac{p-3}{3}}\left(\mu\mu^{\prime}\right)^{p/3}
\G{\frac{3-p}{6}} v \times \nonumber \\ && \sum_{j=0}^{\infty}
\left(-v\right)^{j} \frac{\G{\frac{j}{2} +
\frac{p+3}{6}}}{\G{\frac{j}{2} + \frac{1}{2}}\G{\frac{j}{2} + 1}
\G{j + \frac{p+3}{3} + 1}} t^{j+\frac{p+3}{3}}\times \nonumber
\\ && {}_{1}F_{2}\left(\frac{j}{2}+\frac{p+3}{6};
\frac{j}{2}+\frac{p+6}{6};\frac{j}{2}+\frac{p+9}{6};
-\frac{1}{4}\Delta^{2}t^{2}\right) \nonumber \\ &=&
2^{\frac{p-3}{3}} \G{\frac{3-p}{6}}
\left(\mu\mu^{\prime}\right)^{p/3}vt^{\frac{p+3}{3}} \times
\nonumber \\ && \sum_{j=0}^{\infty} \frac{\G{\frac{j}{2} +
\frac{p+3}{6}}}{\G{\frac{j}{2} + \frac{1}{2}}\G{\frac{j}{2} + 1}
\G{j + \frac{p+3}{3} + 1}} \left(-vt\right)^{j}\times \nonumber
\\ && {}_{1}F_{2}\left(\frac{j}{2}+\frac{p+3}{6};
\frac{j}{2}+\frac{p+6}{6},\frac{j}{2}+\frac{p+9}{6};
-\frac{1}{4}\Delta^{2}t^{2}\right).
\end{eqnarray}

\subsection{A more convenient form for $\mc{M}_{p}\left(\Delta,t\right)$}
\label{moreconvenient}

As we mentioned in Section~\ref{sparse_no_u_inverting}, it is
sometimes useful to have a result as a series in powers of the
detuning squared. Now we exploit the method used there to obtain
such a form for $\mc{M}_{p}\left(\Delta,t\right)$. For
convenience, we define
\begin{eqnarray}
\mc{B} &=& \sum_{j=0}^{\infty} \frac{\G{\frac{j+1}{2} +
\frac{p}{6}}}{\G{\frac{j+1}{2}}\G{\frac{j+2}{2}} \G{j + 2 +
\frac{p}{3}}} \left(-vt\right)^{j}\times \nonumber
\\ && {}_{1}F_{2}\left(\frac{j+1}{2}+\frac{p}{6};
\frac{j+2}{2}+\frac{p}{6},\frac{j+3}{2}+\frac{p}{6};
-\frac{1}{4}\Delta^{2}t^{2}\right),
\end{eqnarray}
so that
\begin{equation}
\mc{M}_{p}\left( \Delta,t \right) = 2^{\frac{p-3}{3}}
\G{\frac{3-p}{6}}
\left(\mu\mu^{\prime}\right)^{p/3}vt^{\frac{p+3}{3}} \mc{B}
\label{MpwithB}
\end{equation}
We now recall from Eq.~(\ref{hypergeodef}) that
\begin{equation}
{}_{1}F_{2}\left(a;b,c;z\right) = \frac{\G{b}\G{c}}{\G{a}}
\sum_{k=0}^{\infty} \frac{\G{k+a}}{\G{k+b} \G{k+c}\G{k+1}} z^{k},
\end{equation}
and hence
\begin{eqnarray}
\mc{B} &=& \sum_{j=0}^{\infty} \frac{\G{\frac{j+2}{2}+\frac{p}{6}}
\G{\frac{j+3}{2}+\frac{p}{6}}}{\G{\frac{j+1}{2}} \G{\frac{j+2}{2}}
\G{j + 2 + \frac{p}{3}}} \left(-vt\right)^{j} \times \nonumber
\\ && \sum_{k=0}^{\infty}
\frac{\G{k+\frac{j+1}{2}+\frac{p}{6}}}{\G{k+\frac{j+2}{2}+\frac{p}{6}}
\G{k+\frac{j+3}{2}+\frac{p}{6}}\G{k+1}}
\left(-\frac{1}{4}\Delta^{2}t^{2}\right)^{k}.
\end{eqnarray}
Recalling Eq.~(\ref{gamma2z}), we see that
\begin{equation}
\G{j + 2 + \frac{p}{3}} = \frac{1}{2\sqrt{\pi}}
2^{j+2+\frac{p}{3}} \G{\frac{j+2}{2} +
\frac{p}{6}}\G{\frac{j+3}{2} + \frac{p}{6}},
\end{equation}
and so
\begin{eqnarray}
\mc{B} &=&
\frac{\sqrt{\pi}}{2^{\frac{p+3}{3}}}\sum_{k=0}^{\infty}\sum_{j=0}^{\infty}
\frac{1}{\G{\frac{j+1}{2}}\G{\frac{j+2}{2}}}
\left(-\frac{vt}{2}\right)^{j} \times \nonumber
\\ &&
\frac{\G{k+\frac{j+1}{2}+\frac{p}{6}}}{\G{k+\frac{j+2}{2}+\frac{p}{6}}
\G{k+\frac{j+3}{2}+\frac{p}{6}}\G{k+1}}
\left(-\frac{1}{4}\Delta^{2}t^{2}\right)^{k} \nonumber \\ &=&
\frac{\sqrt{\pi}}{2^{\frac{p+3}{3}}}\sum_{k=0}^{\infty}
\left(\frac{-\Delta^{2}t^{2}}{4}\right)^{k} \frac{1}{\G{k+1}}
\times \nonumber
\\ && \sum_{j=0}^{\infty}
\frac{\G{\frac{j+1}{2}+k+\frac{p}{6}}}{\G{\frac{j+1}{2}}
\G{\frac{j+2}{2}} \G{\frac{j+2}{2}+k+\frac{p}{6}}
\G{\frac{j+3}{2}+k+\frac{p}{6}}} \left(-\frac{vt}{2}\right)^{j}.
\end{eqnarray}

Now we define
\begin{equation}
\mc{B}_{k}^{\mrm{odd}} = \sum_{j=0,j\,\,\mrm{odd}}^{\infty}
\frac{\G{\frac{j+1}{2}+k+\frac{p}{6}}}{\G{\frac{j+1}{2}}
\G{\frac{j+2}{2}} \G{\frac{j+2}{2}+k+\frac{p}{6}}
\G{\frac{j+3}{2}+k+\frac{p}{6}}} \left(-\frac{vt}{2}\right)^{j}
\end{equation}
and similarly,
\begin{equation}
\mc{B}_{k}^{\mrm{even}} = \sum_{j=0,j\,\,\mrm{even}}^{\infty}
\frac{\G{\frac{j+1}{2}+k+\frac{p}{6}}}{\G{\frac{j+1}{2}}
\G{\frac{j+2}{2}} \G{\frac{j+2}{2}+k+\frac{p}{6}}
\G{\frac{j+3}{2}+k+\frac{p}{6}}} \left(-\frac{vt}{2}\right)^{j},
\end{equation}
so that
\begin{equation}
\mc{B} = \frac{\sqrt{\pi}}{2^{\frac{p+3}{3}}}\sum_{k=0}^{\infty}
\left(-\frac{\Delta^{2}t^{2}}{4}\right)^{k} \frac{1}{\G{k+1}}
\left(\mc{B}_{k}^{\mrm{odd}} + \mc{B}_{k}^{\mrm{even}} \right).
\label{BitoBk}
\end{equation}
We then have (again using Eq.~(\ref{hypergeodef}))
\begin{eqnarray}
\mc{B}_{k}^{\mrm{odd}} &=& -\frac{vt}{2} \sum_{j=0}^{\infty}
\frac{\G{j+k+\frac{p}{6}+1}}{\G{j+1} \G{j+\frac{3}{2}}
\G{j+k+\frac{p}{6}+\frac{3}{2}} \G{j+k+\frac{p}{6}+2}}
\left(\frac{v^{2}t^{2}}{4}\right)^{j} \nonumber \\ &=&
-\frac{vt}{2} \frac{\G{k+\frac{p}{6}+1}}{\G{\frac{3}{2}}
\G{k+\frac{p}{6}+\frac{3}{2}} \G{k+\frac{p}{6}+2}} \times
\nonumber
\\ && {}_{1}F_{3}\left(k+\frac{p}{6}+1;
\frac{3}{2},k+\frac{p}{6}+\frac{3}{2},k+\frac{p}{6}+2;
\frac{v^{2}t^{2}}{4}\right) \nonumber
\\ &=& -\frac{vt}{\sqrt{\pi}}
\frac{1}{\left(k+\frac{p}{6}+1\right)\G{k+\frac{p}{6}+\frac{3}{2}}}
\times \nonumber \\ && {}_{1}F_{3}\left(k+\frac{p}{6}+1;
\frac{3}{2},k+\frac{p}{6}+\frac{3}{2},k+\frac{p}{6}+2;
\frac{v^{2}t^{2}}{4}\right).
\end{eqnarray}
Performing the analogous calculation for
$\mc{B}_{k}^{\mrm{even}}$, we see that
\begin{eqnarray}
\mc{B}_{k}^{\mrm{even}} &=& \sum_{j=0}^{\infty}
\frac{\G{j+k+\frac{p}{6}+\frac{1}{2}}}{\G{j+\frac{1}{2}} \G{j+1}
\G{j+k+\frac{p}{6}+1} \G{j+k+\frac{p}{6}+\frac{3}{2}}}
\left(\frac{v^{2}t^{2}}{4}\right)^{j} \nonumber \\ &=&
\frac{\G{k+\frac{p}{6}+\frac{1}{2}}}{\G{\frac{1}{2}}
\G{k+\frac{p}{6}+1} \G{k+\frac{p}{6}+\frac{3}{2}}} \times
\nonumber
\\ && {}_{1}F_{3}\left(k+\frac{p}{6}+\frac{1}{2};
\frac{1}{2},k+\frac{p}{6}+1,k+\frac{p}{6}+\frac{3}{2};
\frac{v^{2}t^{2}}{4}\right) \nonumber \\ &=& \frac{1}{\sqrt{\pi}}
\frac{1}{\left(k+\frac{p}{6}+\frac{1}{2}\right)\G{k+\frac{p}{6}+1}}
\times \nonumber \\ && {}_{1}F_{3}\left(k+\frac{p}{6}+\frac{1}{2};
\frac{1}{2},k+\frac{p}{6}+1,k+\frac{p}{6}+\frac{3}{2};
\frac{v^{2}t^{2}}{4}\right).
\end{eqnarray}

Using Eq.~(\ref{BitoBk}), we arrive at
\begin{eqnarray}
\mc{B} &=& \frac{1}{2^{\frac{p+3}{3}}}\sum_{k=0}^{\infty}
\left(-\frac{\Delta^{2}t^{2}}{4}\right)^{k} \frac{1}{\G{k+1}}
\left[ \frac{1}{\left(k+\frac{p}{6}+\frac{1}{2}\right)
\G{k+\frac{p}{6}+1}} \times \right. \nonumber
\\ && \left. {}_{1}F_{3}\left(k+\frac{p}{6}+\frac{1}{2};
\frac{1}{2},k+\frac{p}{6}+1,k+\frac{p}{6}+\frac{3}{2};
\frac{v^{2}t^{2}}{4}\right) - \right. \nonumber \\ && \left.
\frac{vt}{\left(k+\frac{p}{6}+1\right)
\G{k+\frac{p}{6}+\frac{3}{2}}} \times \right. \nonumber \\ &&
\left. {}_{1}F_{3}\left(k+\frac{p}{6}+1;
\frac{3}{2},k+\frac{p}{6}+\frac{3}{2},k+\frac{p}{6}+2;
\frac{v^{2}t^{2}}{4}\right) \right].
\end{eqnarray}

Recalling Eq.~(\ref{MpwithB}), we obtain finally
\begin{eqnarray}
\mc{M}_{p}\left(\Delta,t\right) &=& 2^{\frac{p-3}{3}}
\G{\frac{3-p}{6}}
\left(\mu\mu^{\prime}\right)^{p/3}vt^{\frac{p+3}{3}} \times
\nonumber \\ && \frac{1}{2^{\frac{p+3}{3}}}\sum_{k=0}^{\infty}
\left(-\frac{\Delta^{2}t^{2}}{4}\right)^{k} \frac{1}{\G{k+1}}
\left[ \frac{1}{\left(k+\frac{p}{6}+\frac{1}{2}\right)
\G{k+\frac{p}{6}+1}} \times \right. \nonumber
\\ && \left. {}_{1}F_{3}\left(k+\frac{p}{6}+\frac{1}{2};
\frac{1}{2},k+\frac{p}{6}+1,k+\frac{p}{6}+\frac{3}{2};
\frac{v^{2}t^{2}}{4}\right) - \right. \nonumber \\ && \left.
\frac{vt}{\left(k+\frac{p}{6}+1\right)
\G{k+\frac{p}{6}+\frac{3}{2}}} \times \right. \nonumber \\ &&
\left. {}_{1}F_{3}\left(k+\frac{p}{6}+1;
\frac{3}{2},k+\frac{p}{6}+\frac{3}{2},k+\frac{p}{6}+2;
\frac{v^{2}t^{2}}{4}\right) \right] \nonumber \\ &=& \frac{1}{4}
\G{\frac{3-p}{6}} \left(\frac{\mu\mu^{\prime}}{v}\right)^{p/3}
\left(vt\right)^{\frac{p+3}{3}} \times \nonumber
\\ && \sum_{k=0}^{\infty}
\left(-\frac{\Delta^{2}t^{2}}{4}\right)^{k} \frac{1}{\G{k+1}}
\left[ \frac{1}{\left(k+\frac{p}{6}+\frac{1}{2}\right)
\G{k+\frac{p}{6}+1}} \times \right. \nonumber
\\ && \left. {}_{1}F_{3}\left(k+\frac{p}{6}+\frac{1}{2};
\frac{1}{2},k+\frac{p}{6}+1,k+\frac{p}{6}+\frac{3}{2};
\frac{v^{2}t^{2}}{4}\right) - \right. \nonumber \\ && \left.
\frac{vt}{\left(k+\frac{p}{6}+1\right)
\G{k+\frac{p}{6}+\frac{3}{2}}} \times \right. \nonumber \\ &&
\left. {}_{1}F_{3}\left(k+\frac{p}{6}+1;
\frac{3}{2},k+\frac{p}{6}+\frac{3}{2},k+\frac{p}{6}+2;
\frac{v^{2}t^{2}}{4}\right) \right]. \label{M_p(t)}
\end{eqnarray}
It is important to recall that this result is valid for
$\left|p\right| < 3$.

The result of Eq.~(\ref{M_p(t)}) shows that the $p$th moment
$\mc{M}_{p}$ is equal to the quantity
$\left(\mu\mu^{\prime}/v\right)^{p/3}$ multiplied by a
dimensionless function. Since $\mu\mu^{\prime}/v$ is proportional
to $\Omega/N$, as can be seen from Eq.~(\ref{v}), it follows that
$\mc{M}_{p}$ has the dimensions of length to the $p$th power as it
should.

To extract a localization length $l_{p}$ from the moment
$\mc{M}_{p}$, we must first normalize the distribution $\mc{S}$ of
Eq.~(\ref{calSdeltat}). The localization length corresponding to
the $p$th moment is thus given by
\begin{equation}
l_{p}\left(\Delta,t\right) = \left[
\frac{\mc{M}_{p}\left(\Delta,t\right)}{\mc{M}_{0}\left(\Delta,t\right)}
\right]^{1/p}. \label{loclength}
\end{equation}
For each value of $p$, this expression gives a somewhat different
measurement of localization. The quantity $l_{p}$ is proportional
to $\left(\mu\mu^{\prime}/v\right)^{1/3}$, which according to
Eq.~(\ref{v}) is equal to $\left(3/4\pi^{3/2}\right)^{1/3}
\left(\Omega/N\right)^{1/3}$, or
$0.5126\,\left(\Omega/N\right)^{1/3}$. Therefore we see that
$l_{p}$ has units of length for all values of $p$, as one would
expect.

\subsection{A check on the result for $\mc{M}_{p}\left(\Delta,t\right)$}
\label{checkint}

We see from Eq.~(\ref{M_0result}) that
$\mc{M}_{0}\left(\Delta,t\right) = \left\langle
S\left(\Delta,t\right)\right\rangle$, and this provides one way to
check Eq.~(\ref{M_p(t)}). From Eq.~(\ref{M_p(t)}) we have
\begin{eqnarray}
\mc{M}_{0}\left(\Delta,t\right) &=& \frac{1}{4} \G{\frac{1}{2}} vt
\sum_{k=0}^{\infty} \left(-\frac{\Delta^{2}t^{2}}{4}\right)^{k}
\frac{1}{\G{k+1}}\times \nonumber \\ &&\left[
\frac{1}{\left(k+\frac{1}{2}\right) \G{k+1}}
{}_{1}F_{3}\left(k+\frac{1}{2}; \frac{1}{2},k+1,k+\frac{3}{2};
\frac{v^{2}t^{2}}{4}\right) - \right. \nonumber \\ && \left.
\frac{vt}{\left(k+1\right) \G{k+\frac{3}{2}}}
{}_{1}F_{3}\left(k+1; \frac{3}{2},k+\frac{3}{2},k+2;
\frac{v^{2}t^{2}}{4}\right) \right] \nonumber \\ &=&
\frac{\sqrt{\pi}}{4}vt \sum_{k=0}^{\infty}
\left(-\frac{\Delta^{2}t^{2}}{4}\right)^{k} \times \nonumber \\
&&\left\{ \frac{1}{\left(k+\frac{1}{2}\right)
\left[\G{k+1}\right]^{2}} {}_{1}F_{3}\left(k+\frac{1}{2};
\frac{1}{2},k+1,k+\frac{3}{2}; \frac{v^{2}t^{2}}{4}\right) -
\right. \nonumber \\ && \left. \frac{vt}{\G{k+\frac{3}{2}}
\G{k+2}} {}_{1}F_{3}\left(k+1; \frac{3}{2},k+\frac{3}{2},k+2;
\frac{v^{2}t^{2}}{4}\right) \right\},
\end{eqnarray}
which agrees perfectly with Eq.~(\ref{deltaexpansion}).

\subsection{Plots}
\label{localization_plots}

\begin{figure}
\resizebox{\textwidth}{!}{\includegraphics[0in,0in][8in,10in]{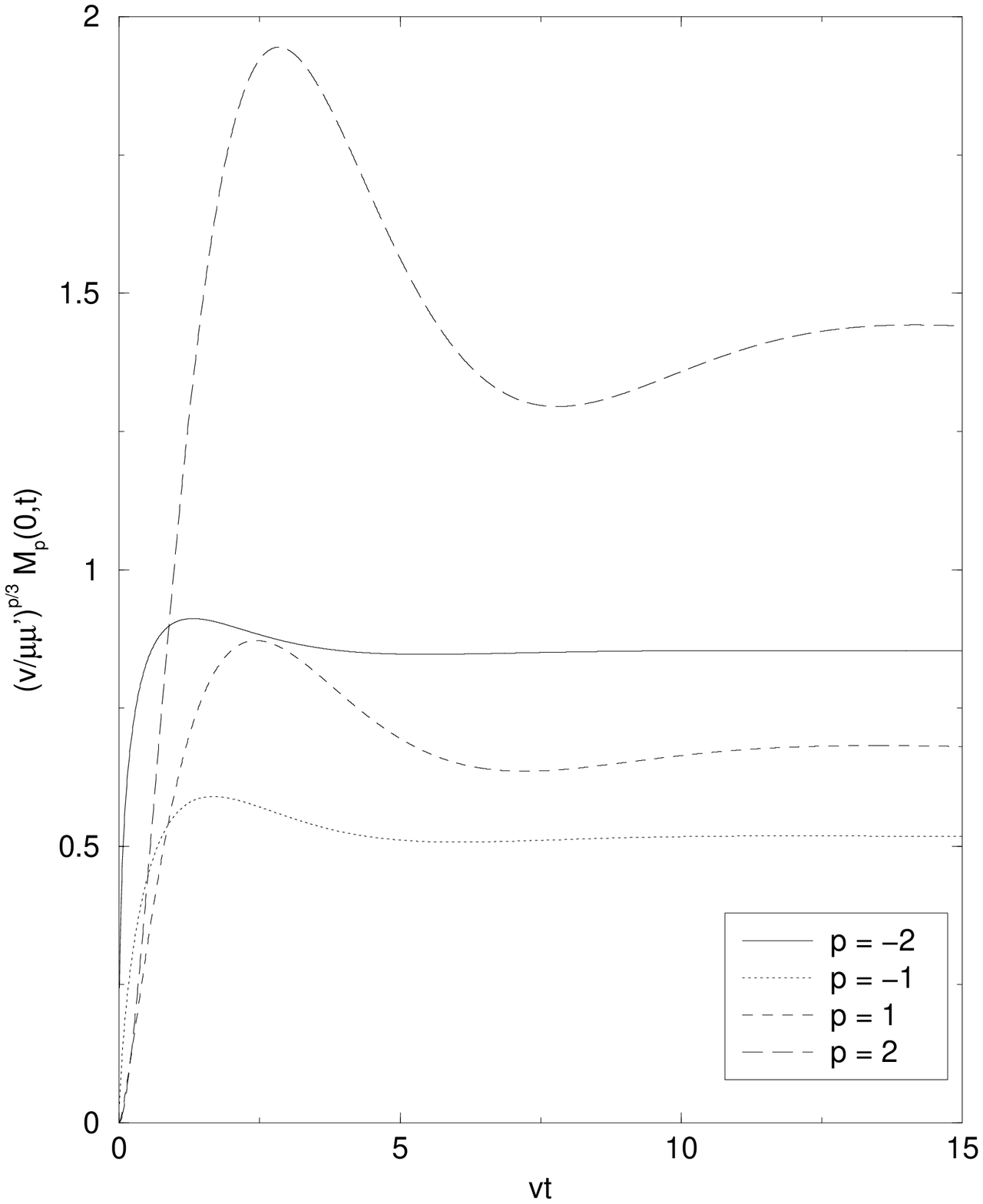}}
\caption[The moments $\mc{M}_{p}\left(0,t\right)$ for $p = -2$,
$-1$, $1$, and $2$]{Plots of the moments
$\mc{M}_{p}\left(0,t\right)$ for $p$ equal to $-2$, $-1$, $1$, and
$2$.} \label{sparse_no_u_m_delta=0v}
\end{figure}

\begin{figure}
\resizebox{\textwidth}{!}{\includegraphics[0in,0in][8in,10in]{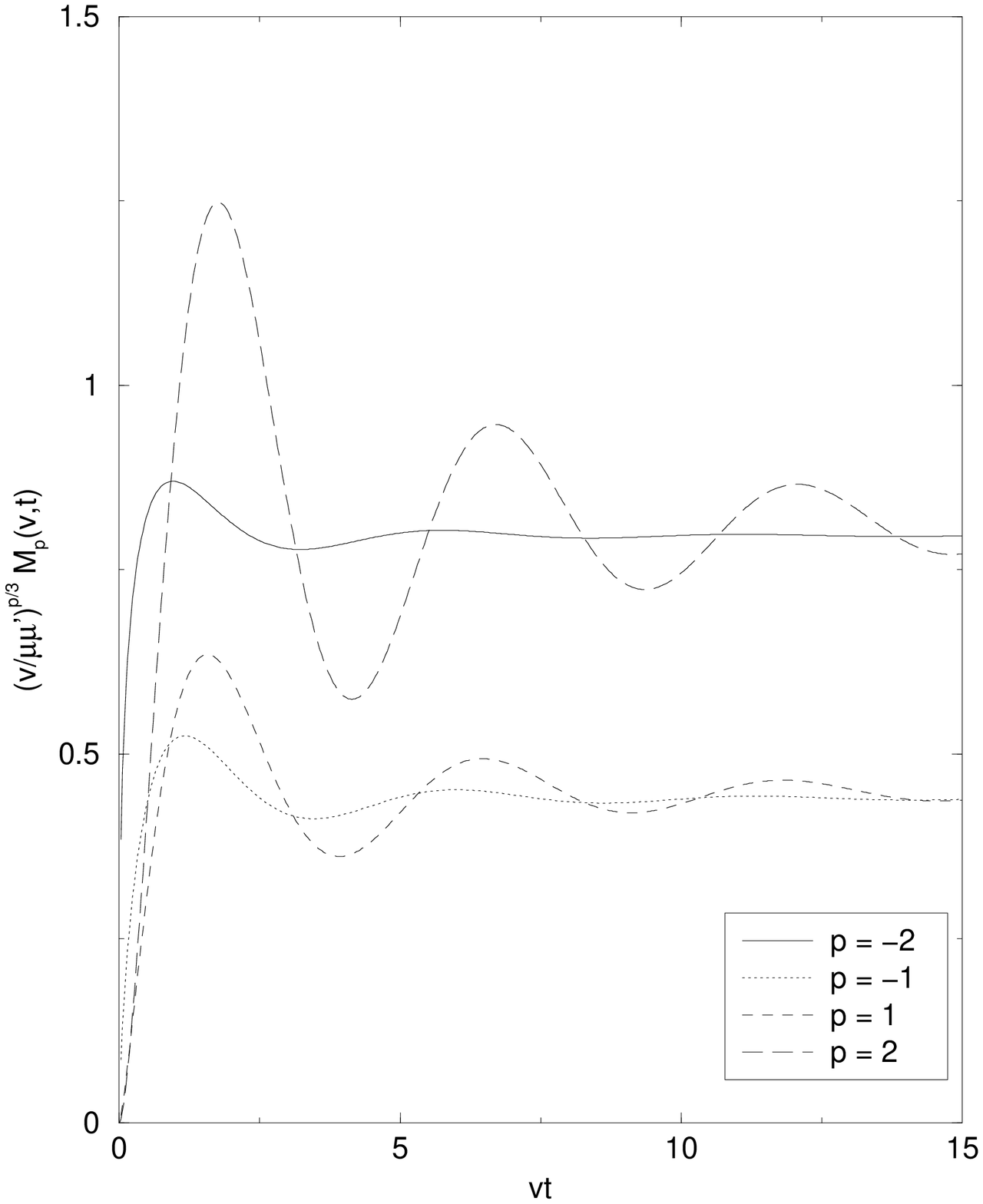}}
\caption[The moments $\mc{M}_{p}\left(v,t\right)$ for $p = -2$,
$-1$, $1$, and $2$]{Plots of the moments
$\mc{M}_{p}\left(v,t\right)$ for $p$ equal to $-2$, $-1$, $1$, and
$2$.} \label{sparse_no_u_m_delta=v}
\end{figure}

\begin{figure}
\resizebox{\textwidth}{!}{\includegraphics[0in,0in][8in,10in]{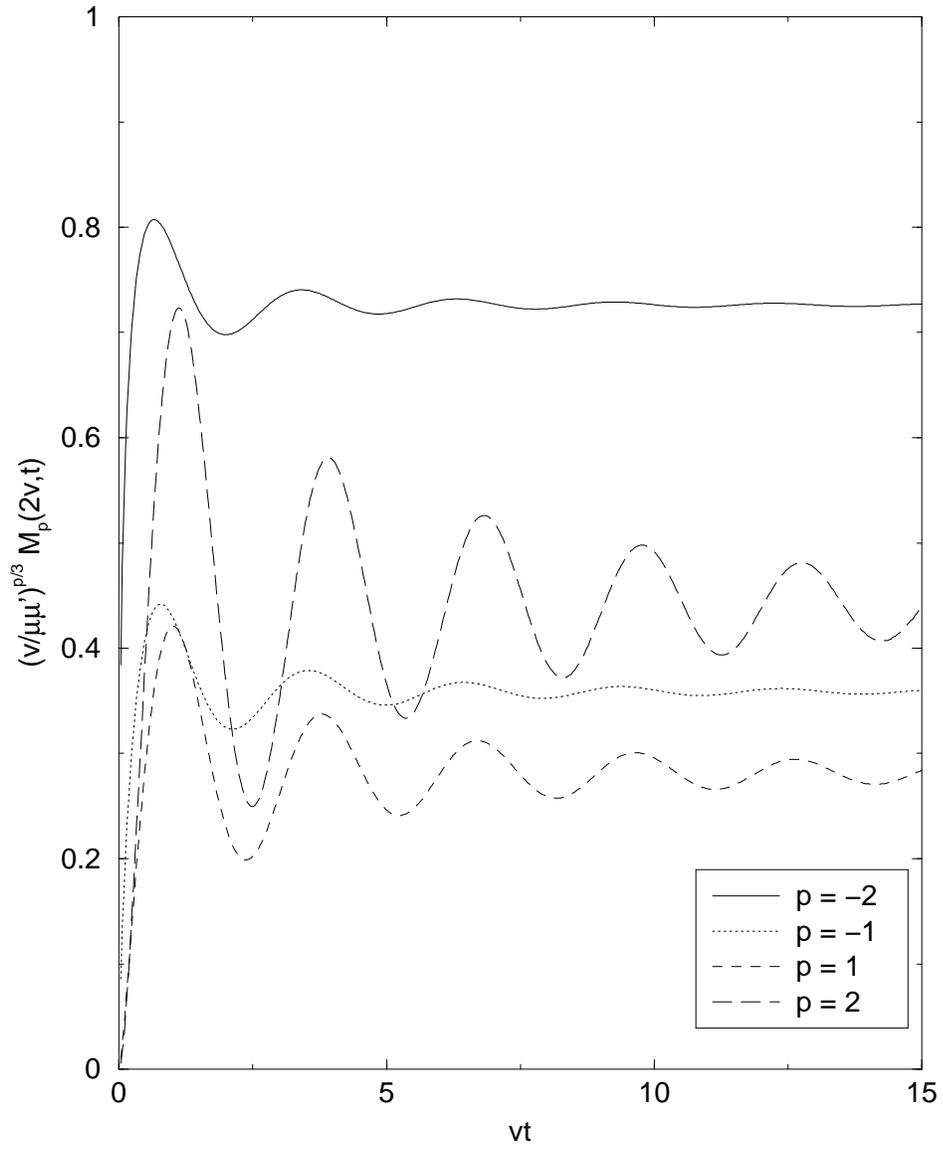}}
\caption[The moments $\mc{M}_{p}\left(2v,t\right)$ for $p = -2$,
$-1$, $1$, and $2$]{Plots of the moments
$\mc{M}_{p}\left(2v,t\right)$ for $p$ equal to $-2$, $-1$, $1$,
and $2$.} \label{sparse_no_u_m_delta=2v}
\end{figure}

Figs.~\ref{sparse_no_u_m_delta=0v} -- \ref{sparse_no_u_m_delta=2v}
give plots of the dimensionless quantities
\begin{equation}
\left(\frac{v}{\mu\mu^{\prime}}\right)^{p/3}
\mc{M}_{p}\left(\Delta,t\right),
\end{equation}
as a function of $vt$, for $p = -1$, $-1$, $1$, and $2$ and for
$\Delta = 0$, $v$, and $2v$, respectively.

\begin{figure}
\resizebox{\textwidth}{!}{\includegraphics[0in,0in][8in,10in]{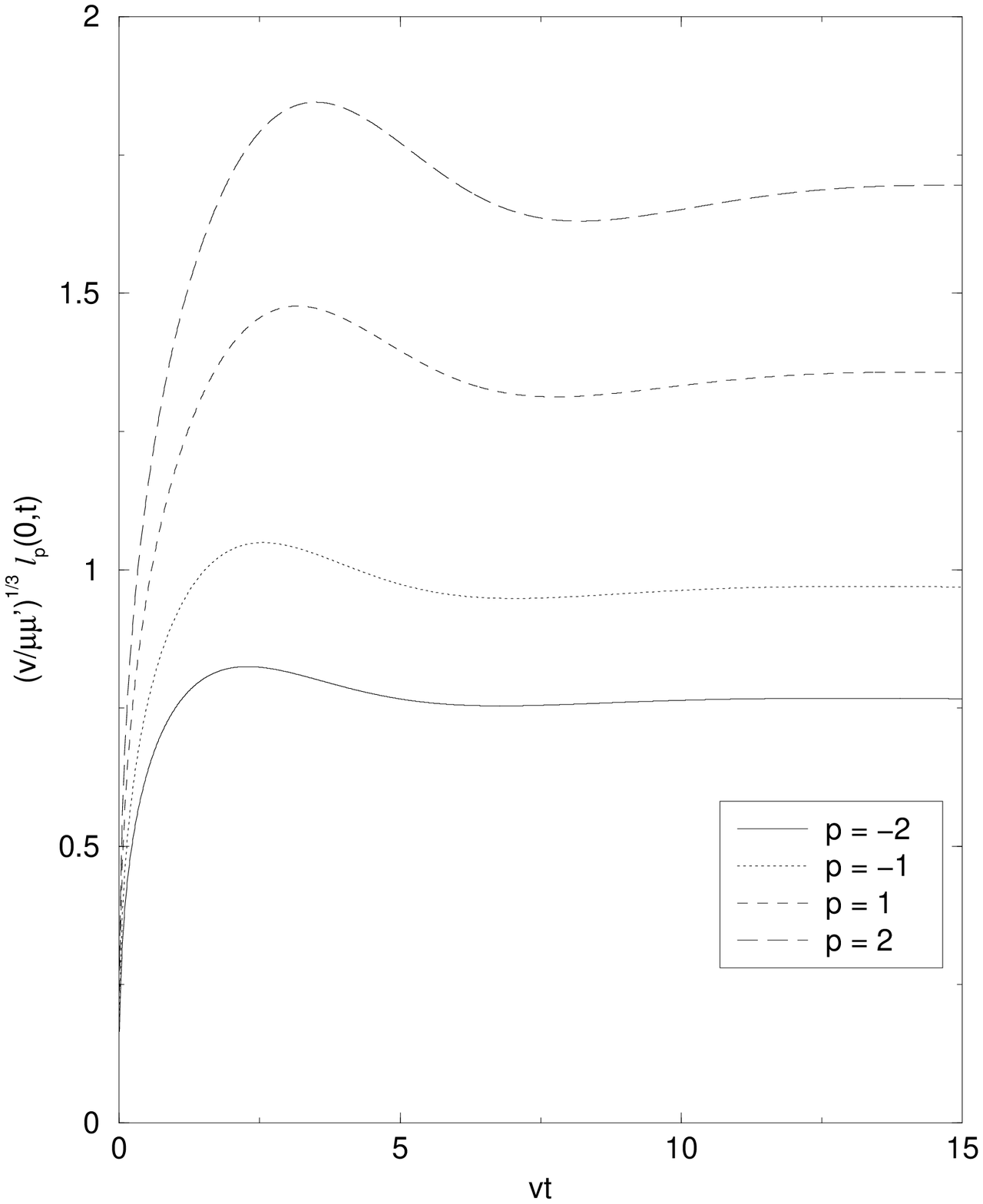}}
\caption[The localization lengths $l_{p}\left(0,t\right)$ for $p =
-2$, $-1$, $1$, and $2$]{Plots of the localization lengths
$l_{p}\left(0,t\right)$ for $p$ equal to $-2$, $-1$, $1$, and
$2$.} \label{sparse_no_u_norm_m_d0v}
\end{figure}

\begin{figure}
\resizebox{\textwidth}{!}{\includegraphics[0in,0in][8in,10in]{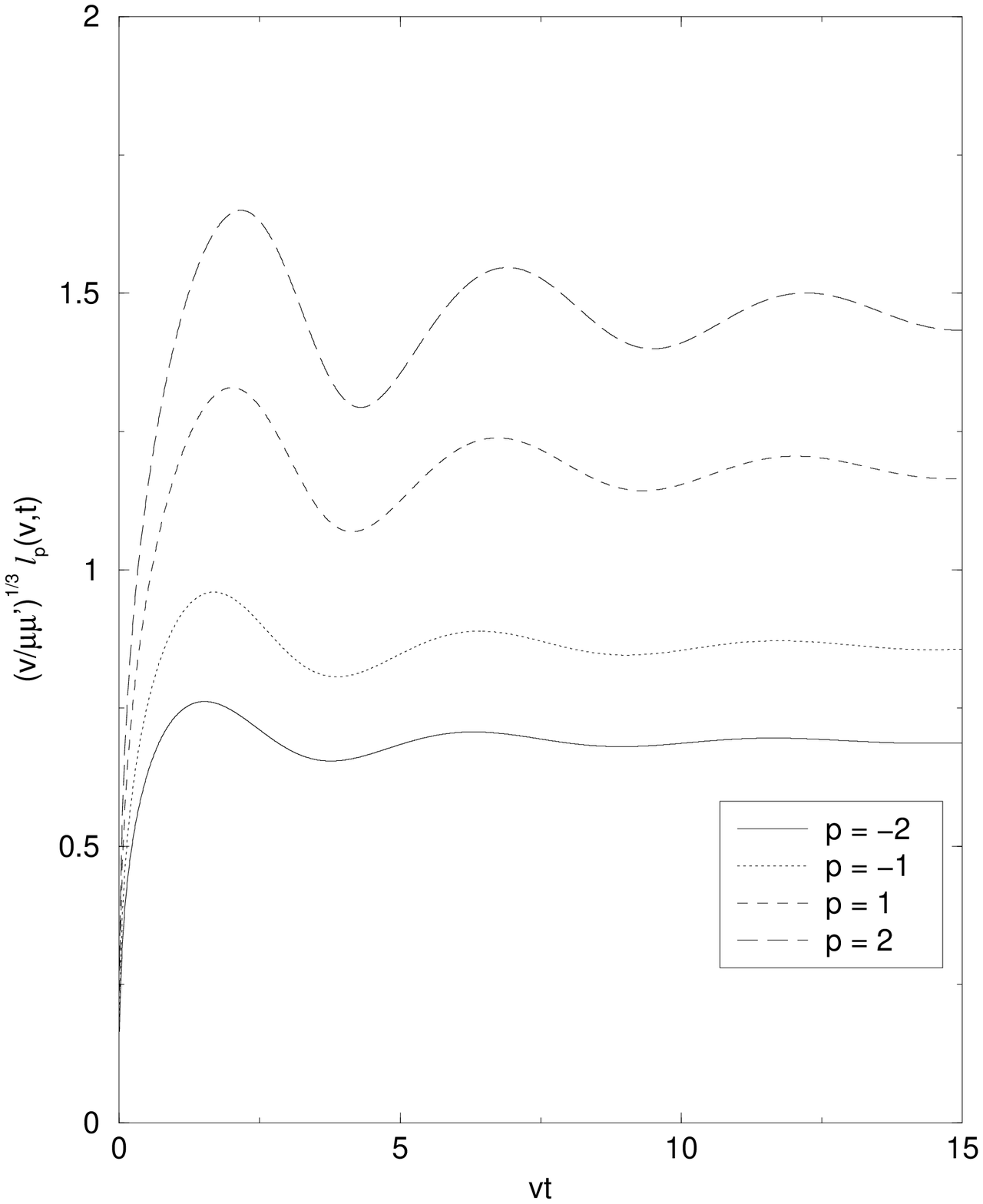}}
\caption[The localization lengths $l_{p}\left(v,t\right)$ for $p =
-2$, $-1$, $1$, and $2$]{Plots of the localization lengths
$l_{p}\left(v,t\right)$ for $p$ equal to $-2$, $-1$, $1$, and
$2$.} \label{sparse_no_u_norm_m_dv}
\end{figure}

\begin{figure}
\resizebox{\textwidth}{!}{\includegraphics[0in,0in][8in,10in]{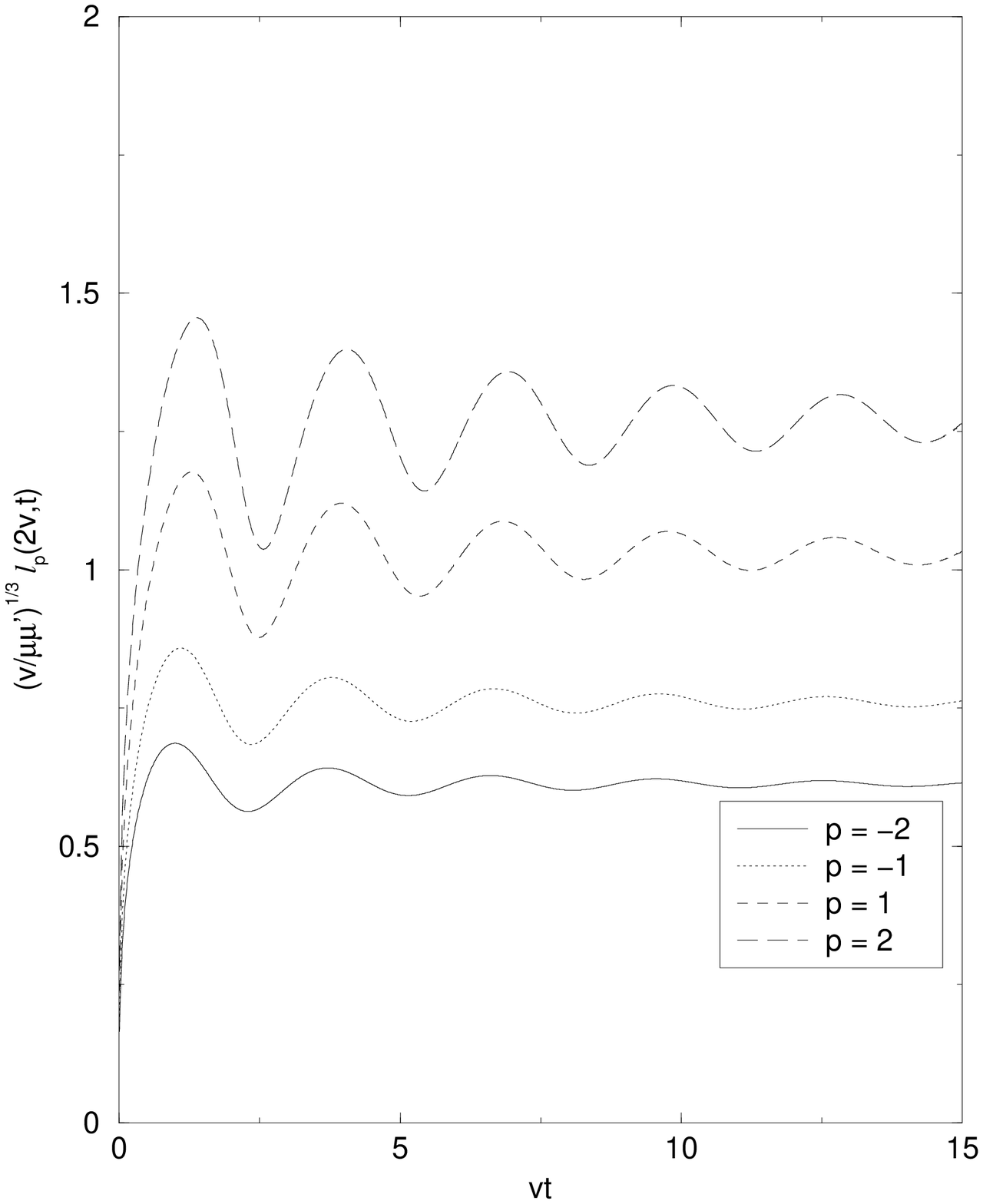}}
\caption[The localization lengths $l_{p}\left(2v,t\right)$ for $p
= -2$, $-1$, $1$, and $2$]{Plots of the localization lengths
$l_{p}\left(2v,t\right)$ for $p$ equal to $-2$, $-1$, $1$, and
$2$.} \label{sparse_no_u_norm_m_d2v}
\end{figure}

In Figs.~\ref{sparse_no_u_norm_m_d0v} --
\ref{sparse_no_u_norm_m_d2v} we see plots of the dimensionless
quantities
\begin{equation}
\left(\frac{v}{\mu\mu^{\prime}}\right)^{1/3}
l_{p}\left(\Delta,t\right),
\end{equation}
as a function of $vt$, for $\Delta = 0$, $v$, and $2v$,
respectively. We see that all the plots initially rise rapidly
like $\left(vt\right)^{1/3}$, and then oscillate around a value of
order one that depends rather weakly on the detuning $\Delta$.

\section{Summary}
\label{localization_conclusions}

We study localization in this system because we need to have some
idea of the relative importance of close pairs of atoms, as well
as some idea of the effective range of the $1/r^{3}$ interaction.
Naively, it seems that the close pairs are dominant and that the
range of the $1/r^{3}$ potential is infinite because $1/r^{3}$
diverges at $r = 0$ and yet the integral $\int d^{3}r/r^{3}$ over
all space diverges at both limits.

\chapter{Sparse $ss^{\prime}\rightarrow pp^{\prime}$ Processes
with $U$ Interaction} \markright{Chapter \arabic{chapter}: Sparse
$ss^{\prime}\rightarrow pp^{\prime}$ Processes with $U$
Interaction}
\label{sparse_with_u}

\section{Introduction}

In this section we complicate the simple model of
Chapter~\ref{sparse_no_u} by introducing the $sp\rightarrow ps$
process through the addition of an interaction potential $U$. The
$sp\rightarrow ps$ process can be viewed as an exciton propagating
through the medium \cite{Bassani1975a,Dexter1965a,Knox1963a}, as
was discussed in Ref.~\cite{Frasier2000a}.

In Section~\ref{cayleysection} we discuss the Cayley tree
approximation, starting with its justification in
Section~\ref{cayleyjustification}. In Sections~\ref{cayley_u_only}
-- \ref{cayleyUVdelta} we solve for the Laplace transform of the
probability amplitude in the case where $\Delta$ and $V$ are both
zero, the case where $V$ is nonzero but $\Delta$ is still zero,
and the case where both $\Delta$ and $V$ are nonzero,
respectively.

We consider a Hubbard band approximation in
Section~\ref{hubbardsection}. In Section~\ref{fhubbardsection} we
again solve for the Laplace transform of the probability amplitude
in the case where $\Delta$ and $V$ are both zero and in
Section~\ref{f0hubbard} we do the same for the case where $V$ is
nonzero but $\Delta$ is still zero. However, we note in this case
that the latter solution is not consistent. In the case of the
Hubbard approximation, we are also able to compute the square of
the probability amplitude in the $V = \Delta = 0$ case, as is done
in Section~\ref{hubbard_a2}.

In Section~\ref{cpasection} we consider a coherent potential
approximation (CPA). In this approximation we are able to compute
the probability amplitude in all cases. We will see in
Section~\ref{sparse_with_u_plots} that the CPA results agree best
with the results of the numerical simulations discussed in
Chapter~\ref{simulations}.

In Section~\ref{sparse_with_u_comp} we compare the results of
these three approximations. We see that they are in some ways
rather similar, but that the differences are sufficiently large
that it is easy to tell them apart, particularly when the
$sp\rightarrow ps$ process is weak. This means that one
approximation will fit the exact results better than the others,
and in the case we are studying it is the CPA. Thus, in
Section~\ref{sparse_with_u_plots} we present plots of the CPA
results in comparison with the results of the numerical
simulations discussed in Chapter~\ref{simulations}.

Finally, in Section~\ref{sparse_with_u_conclusions} we summarize
these results and conclude.

\section{The Cayley tree approximation}
\label{cayleysection}

\subsection{Justification of the Cayley tree approximation}
\label{cayleyjustification}

We start with the equations
\setcounter{tlet}{1}
\renewcommand{\theequation}{\arabic{chapter}.\arabic{equation}\alph{tlet}}
\begin{eqnarray}
 i\dot{a}_{0} &=& \Delta a_{0}+\sum_{k=1}^{N}V_{k}c_{k}, \label{azerodotU} \\
\stepcounter{tlet}\addtocounter{equation}{-1}
 i\dot{c}_{k} &=& V_{k}a_{0} + \sum_{l\neq k}U_{kl}c_{l}. \label{ckdotU}
\end{eqnarray}
\renewcommand{\theequation}{\arabic{chapter}.\arabic{equation}}
These are the same as Eqs.~(\ref{a0dot}) and (\ref{ckdot}), except
that we now allow for the $sp\rightarrow ps$ process by including
the $\sum_{l} U_{kl}c_{l}$ term in the equation for $c_{k}$. For
the purposes of compressing indices on sums and products, it is
convenient to define $U_{kl}$ to be zero when $k=l$. With this
definition, the previous equation becomes \setcounter{tlet}{1}
\renewcommand{\theequation}{\arabic{chapter}.\arabic{equation}\alph{tlet}}
\begin{eqnarray}
 i\dot{a}_{0} &=& \Delta a_{0}+\sum_{k=1}^{N}V_{k}c_{k}, \label{a0dotcayley} \\
\stepcounter{tlet}\addtocounter{equation}{-1}
 i\dot{c}_{k} &=& V_{k}a_{0} + \sum_{l=1}^{N}U_{kl}c_{l}. \label{ckdotcayley}
\end{eqnarray}
\renewcommand{\theequation}{\arabic{chapter}.\arabic{equation}}
Unfortunately the potential $U$ makes the set of equations
insoluble as they stand, so we cannot obtain an exact solution as
we did in Chapter~\ref{sparse_no_u}. However, the equations can be
solved numerically, and one can also make quite a lot of progress
by making approximations and proceeding analytically.

In developing approximations, it is useful to cast the problem in
the standard Green's function language \cite{Economou1983a}, as
discussed in Section~3 of Ref.~\cite{Celli1999a}. The Green's
function of Eq.~(\ref{ckdotcayley}) satisfies the equation
\begin{equation}
\omega G_{ln} = \delta_{ln} + \sum_{m=1}^{N}U_{lm}G_{mn},
\label{gfln}
\end{equation}
and then
\begin{equation}
a_{0}\left(\Delta,\omega\right) = \frac{i}{\omega - \Delta -
\sum_{lm}V_{l}G_{lm}V_{m}}. \label{gfv}
\end{equation}
Actually, $-ia_{0}$ is the $00$ element of the Green's function
for the entire set of $N+1$ equations.

We can derive a convenient form for the on-site Green's function
$G_{nn}$. From Eq.~(\ref{gfln}) we have, for $l = n$,
\begin{equation}
\omega G_{nn} = 1 + \sum_{m} U_{nm}G_{mn}, \label{Gnn}
\end{equation}
and if we write separately the term containing $G_{nn}$, we obtain
\begin{equation}
\omega G_{ln} = U_{ln}G_{nn} + \sum_{m\neq n} U_{lm}G_{mn},
\label{Gln}
\end{equation}
for $l\neq n$. We note here that Eqs.~(\ref{Gnn}) and (\ref{Gln})
are the analogs of Eqs.~(\ref{a0dotcayley}) and
(\ref{ckdotcayley}), respectively.

Now we imagine removing the $n$th atom. The new Green's function
for the remaining $N - 1$ atoms, which we will call $G_{lp,\left[
n\right]}$, obeys the equation
\begin{equation}
\omega G_{lp,\left[ n\right]} = \delta_{lp} + \sum_{m\neq
n}U_{lm}G_{mp,\left[n\right]}, \label{precayley1}
\end{equation}
where $l$ and $p$ are never equal to $n$. If we multiply both
sides of this equation by $U_{pn}G_{nn}$ and sum over $p$, we
obtain
\begin{equation}
\omega\sum_{p} G_{lp,\left[ n\right]} U_{pn}G_{nn} = U_{ln}G_{nn}
+ \sum_{m\neq n}U_{lm}\sum_{p} G_{mp,\left[n\right]} U_{pn}G_{nn}.
\label{precayley2}
\end{equation}
Comparing this equation with Eq.~(\ref{Gln}), we see that for
$m\neq n$
\begin{equation}
G_{mn} = \sum_{p} G_{mp,\left[ n\right]} U_{pn} G_{nn}.
\end{equation}
If we insert this result into Eq.~(\ref{Gnn}), then we have
finally that
\begin{equation}
\omega G_{nn} = 1 + \sum_{m} U_{nm} \sum_{p} G_{mp,\left[
n\right]} U_{pn} G_{nn},
\end{equation}
and hence
\begin{equation}
G_{nn} = \frac{1}{\omega - \sum_{mp} U_{nm} G_{mp,\left[n\right]}
U_{pn}}. \label{gfu}
\end{equation}
It is important to keep in mind that these Green's function
equations are exact.

The Cayley approximation
\cite{Abou-Chacra1973a,Logan1987a,Logan1985a,Logan1987b} keeps
only the diagonal $\left(l = m\right)$ terms in the sum over $l$
and $m$ in Eqs.~(\ref{gfv}) and (\ref{gfu}). It further assumes
that each $G_{ll,\left[n \right]}$ is independently distributed,
and that there are sufficiently many sites that $G_{ll,\left[ n
\right]}$ and $G_{ll}$ have the same distribution.

We choose to set $\omega = i\alpha$ and work in Laplace space
instead of Fourier space because this makes many of the integrals
that follow more obviously convergent. Up to a point, the same
analysis can be carried out with the Fourier analogs, but Laplace
transforms are needed to solve the Cayley equations exactly in
Section \ref{cayley_u_only}. We also work with the inverses of
these Green's functions, setting $a_{0}\left(i\alpha\right) =
1/f_{0}\left(\alpha\right)$ and $G_{kk}\left(i\alpha\right) =
i/f_{k}\left(\alpha\right)$, so in the Cayley approximation we
have from Eqs.~(\ref{precayley1}) and (\ref{precayley2}),
respectively \setcounter{tlet}{1}
\renewcommand{\theequation}{\arabic{chapter}.\arabic{equation}\alph{tlet}}
\begin{eqnarray}
f_{0} &=& \alpha + i\Delta +
\sum_{k=1}^{N}\frac{V_{0k}^{2}}{f_{k}}, \label{fzero} \\
\stepcounter{tlet}\addtocounter{equation}{-1}
f_{k} &=& \alpha +
\sum_{l=1}^{N}\frac{U_{kl}^{2}}{f_{l}}. \label{fk}
\end{eqnarray}
\renewcommand{\theequation}{\arabic{chapter}.\arabic{equation}}

\subsection{Cayley tree approximation with the $U$ process only}
\label{cayley_u_only}

First we consider just the $U$ process and solve Eq.~(\ref{fk}),
which is to say we look at the $sp\rightarrow ps$ process by
itself. Note that $f_{k}$ is always real and positive. The
distribution of $f_{k}$ is given by
\begin{equation}
P\left(f_{k}\right) = \int\left\langle \delta\left(
f_{k}-\alpha-\sum_{l=1}^{N}\frac{U_{kl}^{2}}{f_{l}}\right)\right\rangle
\prod_{m\neq k,m=1}^{N}P\left(f_{m}\right)\,df_{m},
\end{equation}
where for a uniform gas of volume $\Omega$,
\begin{equation}
\left\langle X\right\rangle = \frac{1}{\Omega^{N}}\int
\prod_{j=1}^{N}d^{3}r_{j}\,X.
\end{equation}
Writing the delta function as a Fourier transform, we see that
\begin{eqnarray}
P\left(f_{k}\right) &=&
\int_{-\infty}^{\infty}\frac{dq}{2\pi}\,\int \prod_{m\neq
k,m=1}^{N}P\left(f_{m}\right)df_{m}\, e^{iq\left( f_{k}-\alpha
\right)} \left\langle \exp\left(-iq\sum_{l=1}^{N}
\frac{U_{kl}^{2}}{f_{l}}\right)\right\rangle \nonumber \\ &=&
\int_{-\infty}^{\infty}\frac{dq}{2\pi}\,e^{iq\left( f_{k}-\alpha
\right)}\int \prod_{m\neq k,m=1}^{N}P\left(f_{m}\right)df_{m}
\times \nonumber \\ && \frac{1}{\Omega^{N-1}}\int \prod_{l\neq
k,l=1}^{N}d^{3}\,r_{l}
\exp\left(-iq\frac{U_{kl}^{2}}{f_{l}}\right).
\end{eqnarray}
Using the fact that $\left(\prod_{m}P_{m}\right)
\left(\prod_{l}E_{l}\right) = \prod_{m}P_{m}E_{m}$, we can write
\begin{eqnarray}
P\left(f_{k}\right) &=&
\int_{-\infty}^{\infty}\frac{dq}{2\pi}\,e^{iq\left( f_{k}-\alpha
\right)}\prod_{m\neq k,m=1}^{N}\int P\left(f_{m}\right)df_{m}\,
\frac{1}{\Omega}\int d^{3}\,r_{m}
\exp\left(-iq\frac{U_{km}^{2}}{f_{m}}\right) \nonumber \\ &=&
\int_{-\infty}^{\infty}\frac{dq}{2\pi}\,e^{iq\left( f_{k}-\alpha
\right)} \left[\int_{0}^{\infty} df_{1}\,P\left(f_{1}\right)
\frac{1}{\Omega}\int d^{3}r_{1}\,
\exp\left(-iq\frac{U_{k1}^{2}}{f_{1}}\right)\right]^{N-1}
\nonumber
\\ &=& \int_{-\infty}^{\infty}\frac{dq}{2\pi}\,e^{iq\left(
f_{k}-\alpha \right)} \times \nonumber \\ &&
\left(1-\int_{0}^{\infty} df\,P\left(f\right) \frac{1}{\Omega}\int
d^{3}r\,
\left\{1-\exp\left[-iq\frac{U^{2}\left(r\right)}{f}\right]
\right\}\right)^{N-1}, \label{Pincomplete}
\end{eqnarray}
where in the last step we have simply changed variables from
$\mbf{r}_{1}$ to $\mbf{r} = \mbf{r}_{1} - \mbf{r}_{k}$. Using the
averaging method detailed in Section~\ref{averaging}, we see
that\footnote{Of course, we must use the more elaborate averaging
method of Section~\ref{angular} if the potential $U$ is more
complicated.}
\begin{equation}
P\left(f\right) =
\int_{-\infty}^{\infty}\frac{dq}{2\pi}\,e^{iq\left( f-\alpha
\right)} \exp\left[-u\sqrt{iq}Q\left(\alpha\right)\right],
\end{equation}
where
\begin{equation}
Q\left(\alpha\right) = \int_{0}^{\infty} \frac{df}{\sqrt{f}}\,
P\left(f\right),
\end{equation}
and
\begin{equation}
u = \frac{4\pi^{3/2}}{3}\mu^{2}\frac{N}{\Omega}, \label{u}
\end{equation}
if $U$ is of the simple $\mu^{2}/r^{3}$ form with no angular
dependence, and is the equivalent of $v$ in Eq.~(\ref{fancyv})
otherwise. For $A$ and $B$ real, we have that
\begin{equation}
\int_{-\infty}^{\infty}\frac{dq}{2\pi}\, e^{iqA-\sqrt{iq}B} =
\frac{1}{2\sqrt{\pi}}\frac{B}{A^{3/2}}\exp\left(-\frac{B^{2}}{4A}\right)
\theta\left(A\right),
\end{equation}
where this integral is evaluated by the same procedure used to
evaluate the integral of Eq.~(\ref{crazyPintegral}).\footnote{In
case the reader is curious, the theta function appears here but is
omitted in Eq.~(\ref{Pv}) because $\mc{V}^{2}$ is an intrinsically
positive quantity.} With this result, we have
\begin{equation}
P\left(f\right) =
\frac{1}{2\sqrt{\pi}}\frac{uQ\left(\alpha\right)}{\left( f-\alpha
\right)^{3/2}}
\exp\left[-\frac{u^{2}Q^{2}\left(\alpha\right)}{4\left( f-\alpha
\right)}\right] \theta\left(f-\alpha\right). \label{Pfcayleynov}
\end{equation}
and so
\begin{eqnarray}
Q\left(\alpha\right) &=& \int_{\alpha}^{\infty}
\frac{df}{\sqrt{f}}\,
\frac{1}{2\sqrt{\pi}}\frac{uQ\left(\alpha\right)}{\left(f-\alpha\right)^{3/2}}
\exp\left[-\frac{u^{2}Q^{2}\left(\alpha\right)}{4\left(f-\alpha\right)}\right]
 \nonumber \\ &=&
\frac{1}{\sqrt{\alpha}}\exp\left[\frac{u^{2}Q^{2}\left
(\alpha\right)}{4\alpha} \right]
\erfc\left[\frac{uQ\left(\alpha\right)}{2\sqrt{\alpha}}\right].
\label{Qeqn}
\end{eqnarray}
We also have
\begin{eqnarray}
\left\langle\frac{1}{f}\right\rangle &=&
\int_{0}^{\infty}\frac{df}{f}\, P\left(f\right) \nonumber \\ &=&
\frac{1}{\alpha} -
\frac{\sqrt{\pi}}{2}\frac{uQ\left(\alpha\right)}{\alpha^{3/2}}
\exp\left[\frac{u^{2}Q^{2}\left(\alpha\right)}{4\alpha}\right]
\erfc\left[\frac{uQ\left(\alpha\right)}{2\sqrt{\alpha}}\right],
\end{eqnarray}
or
\begin{equation}
\left\langle\frac{1}{f}\right\rangle = \frac{1}{\alpha}\left[
1-\frac{\sqrt{\pi}}{2}uQ^{2}\left(\alpha\right) \right].
\label{fkinv}
\end{equation}
Thus we can solve numerically the transcendental relation for
$Q\left(\alpha\right)$ given in Eq.~(\ref{Qeqn}) and plug it into
Eq.~(\ref{fkinv}) to find $\left\langle 1/f\right\rangle$. It is
important to note that Eq.~(\ref{Qeqn}) defines $Q$ as an analytic
function of $\alpha$ in the half-plane where $\mrm{Re}\,\alpha
> 0$. Thus $\left\langle 1/f\right\rangle$ can be analytically
continued, in the variable $\omega=i\alpha$, from the positive
imaginary axis to the entire upper half-plane.

\subsection{Cayley tree approximation with $U$ and $V$
processes in the absence of $\Delta$}
\label{cayley_u_and_v}

Now we allow the $V$ process to be present, but we still ignore
the detuning, $\Delta$. We start with the equations
\setcounter{tlet}{1}
\renewcommand{\theequation}{\arabic{chapter}.\arabic{equation}\alph{tlet}}
\begin{eqnarray}
f_{0} &=& \alpha + \sum_{k=1}^{N}\frac{V_{0k}^{2}}{f_{k}}, \\
\stepcounter{tlet}\addtocounter{equation}{-1} f_{k} &=& \alpha +
\sum_{l=1,l\neq k}^{N}\frac{U_{kl}^{2}}{f_{l}}.
\end{eqnarray}
\renewcommand{\theequation}{\arabic{chapter}.\arabic{equation}}
Repeating the analysis we did before, we have
\begin{equation}
P_{0}\left(f_{0}\right) = \int\left\langle\delta\left(f_{0} -
\alpha - \sum_{k=1}^{N}\frac{V_{0k}^{2}}{f_{k}}
\right)\right\rangle\prod_{l=1}^{N}P\left(f_{l}\right)df_{l},
\end{equation}
which leads to the form
\begin{equation}
P_{0}\left(f_{0}\right) = \int_{-\infty}^{\infty}\frac{dq}{2\pi}\,
e^{iq\left(f_{0}-\alpha\right)} \left\{\int_{0}^{\infty}df\,
P\left(f\right)\frac{1}{\Omega}\int
d^{3}r\exp\left[-iq\frac{V^{2}\left(r\right)}{f}\right]
\right\}^{N}.
\end{equation}
At this point it is clear from the work we did to obtain
Eq.~(\ref{Pfcayleynov}) from (\ref{Pincomplete}) that this
equation leads to
\begin{equation}
P_{0}\left(f_{0}\right) = \frac{1}{2\sqrt{\pi}}
\frac{vQ\left(\alpha\right)}{\left(f_{0}-\alpha\right)^{3/2}}
\exp\left[
-\frac{v^{2}Q^{2}\left(\alpha\right)}{4\left(f_{0}-\alpha\right)}
\right]\theta\left(f_{0}-\alpha\right),
\end{equation}
where $Q\left(\alpha\right)$ is still given by Eq.~(\ref{Qeqn}),
and depends on $u$. Therefore we obtain the result
\begin{eqnarray}
\left\langle\frac{1}{f_{0}}\right\rangle &=&
\int_{0}^{\infty}\frac{df_{0}}{f_{0}}\,P_{0}\left(f_{0}\right)
\nonumber \\ &=& \frac{1}{\alpha}-\frac{\sqrt{\pi}}{2}
\frac{vQ\left(\alpha\right)}{\alpha^{3/2}}
\exp\left[\frac{v^{2}Q^{2}\left(\alpha\right)}{4\alpha}\right]
\erfc\left[ \frac{vQ\left(\alpha\right)}{2\sqrt{\alpha}} \right],
\label{fzerowithu}
\end{eqnarray}
which again can be analytically continued.

\subsection{Cayley tree approximation with $U$ and $V$
processes in the presence of $\Delta$}
\label{cayleyUVdelta}

If we further consider the effect of a detuning, $\Delta$, then we
have the full set of Eqs.~(\ref{fzero}) and (\ref{fk}). The
equation for $f_{k}$ is unchanged from the previous case, but
because the equation for $f_{0}$ is now complex we have to
consider a distribution in the two variables $f_{0}^{R} =
\mrm{Re}\,f_{0}$ and $f_{0}^{I} = \mrm{Im}\,f_{0}$. We have
\begin{equation}
P_{0}\left(f_{0}^{R},f_{0}^{I}\right) =
\int\left\langle\delta\left(f_{0}^{R} - \alpha -
\sum_{k=1}^{N}\frac{V_{0k}^{2}}{f_{k}}
\right)\delta\left(f_{0}^{I} - \Delta
\right)\right\rangle\prod_{l=1}^{N}P\left(f_{l}\right)df_{l}.
\end{equation}
Clearly the delta function involving $f_{0}^{I}$ does not
participate in the averaging or in the integrations over $f_{l}$.
Therefore we can treat the part involving $f_{0}^{R}$ just as we
did before, and the $f_{0}^{I}$ part will just be carried along.
Hence we find
\begin{equation}
P_{0}\left(f_{0}^{R},f_{0}^{I}\right) = \frac{1}{2\sqrt{\pi}}
\frac{vQ\left(\alpha\right)}{\left(f_{0}^{R}-\alpha\right)^{3/2}}
\exp\left[
-\frac{v^{2}Q^{2}\left(\alpha\right)}{4\left(f_{0}^{R}-\alpha\right)}
\right]\theta\left(f_{0}^{R}-\alpha\right)\delta\left(f_{0}^{I} -
\Delta\right). \label{P0f0}
\end{equation}
It follows, then, that
\begin{eqnarray}
\left\langle\frac{1}{f_{0}}\right\rangle &=&
\int_{0}^{\infty}df_{0}^{R}\int_{-\infty}^{\infty}df_{0}^{I}\,
\frac{1}{f_{0}^{R} + i f_{0}^{I}}
P_{0}\left(f_{0}^{R},f_{0}^{I}\right) \nonumber \\ &=&
\frac{1}{\alpha + i\Delta}-\frac{\sqrt{\pi}}{2}
\frac{vQ\left(\alpha\right)}{\left(\alpha + i\Delta\right)^{3/2}}
\exp\left[ \frac{v^{2}Q^{2}\left(\alpha\right)}{4\left(\alpha +
i\Delta\right)}\right] \erfc\left[
\frac{vQ\left(\alpha\right)}{2\sqrt{\alpha + i\Delta}} \right].
\label{1/f0cayley}
\end{eqnarray}

\section{The Hubbard approximation}
\label{hubbardsection}

It is also possible to make approximations other than the one we
have chosen. For example, instead of working with Eq.~(\ref{fk}),
one can instead choose to work with
\begin{equation}
f_{k} = \alpha + \sum_{l=1}^{N}\frac{U_{kl}^{2}}{f_{k}},
\label{fhubbard}
\end{equation}
which gives explicitly
\begin{equation}
f_{k} = \frac{1}{2}\left(\alpha +
\sqrt{\alpha^{2}+4\sum_{l}U_{kl}^{2}}\right). \label{hubbardband}
\end{equation}
This approximation corresponds to a Hubbard band.

\subsection{Computation of $\left\langle 1/f\right\rangle$}
\label{fhubbardsection}

It is possible to average Eq.~(\ref{hubbardband}), although it
requires a little more manipulation than the averaging methods we
have applied previously because of the presence of the square
root. We first note that
\begin{eqnarray}
\frac{1}{f_{k}} &=& \frac{2}{\alpha +
\sqrt{\alpha^{2}+4\mc{U}^{2}}} \nonumber \\ &=&
\frac{1}{2\mc{U}^{2}}\left( \sqrt{\alpha^{2} + 4\mc{U}^{2}} -
\alpha\right), \label{fkunavg}
\end{eqnarray}
where
\begin{equation}
\mc{U}^{2} = \sum_{l}U_{kl}^{2}.
\end{equation}
Interestingly enough, the Laplace transform in Eq.~(\ref{fkunavg})
has a known inverse. We see from Eq.~(2.1.35) of
Ref.~\cite{Roberts1966a} that the inverse Laplace transform of
Eq.~(\ref{fkunavg}) is
\begin{equation}
\frac{1}{\mc{U}t}J_{1}\left( 2\mc{U}t\right).
\label{inversetransform}
\end{equation}
We will use Eq.~(\ref{inversetransform}) in
Section~\ref{hubbard_a2} to compute $\left\langle\left|
a_{0}\left(\Delta = 0,t\right) \right|^{2}\right\rangle$ in the
case where only the $U$ process is present.

We now note that
\begin{equation}
\sqrt{x^{2} + y} - x =
\frac{y}{2}\int_{0}^{1}\frac{d\lambda}{\sqrt{\alpha^{2} + \lambda
y}}, \label{lambdaintegral}
\end{equation}
so we can rewrite Eq.~(\ref{fkunavg}) as
\begin{equation}
\frac{1}{f_{k}} = \int_{0}^{1}\frac{d\lambda}{\sqrt{\alpha^{2} +
4\lambda\mc{U}^{2}}}.
\end{equation}
We introduce yet another integral by observing that
\begin{equation}
\frac{1}{\sqrt{\alpha^{2} + C\lambda\mc{U}^{2}}} =
\frac{2}{\sqrt{\pi}}\int_{0}^{\infty}d\beta\,
\exp\left[-\beta^{2}\left(\alpha^{2} +
C\lambda\mc{U}^{2}\right)\right], \label{beta2integral}
\end{equation}
if $\mrm{Re}\,\left(C\lambda\right) > 0$, so
\begin{equation}
\frac{1}{f_{k}} = \frac{2}{\sqrt{\pi}}\int_{0}^{\infty}d\beta\,
\int_{0}^{1}d\lambda\, \exp\left[-\beta^{2}\left(\alpha^{2} +
4\lambda\mc{U}^{2}\right)\right].
\end{equation}
From our previous experience with averaging over $\mc{U}$, we know
that
\begin{equation}
\left\langle\frac{1}{f}\right\rangle =
\frac{2}{\sqrt{\pi}}\int_{0}^{\infty}d\beta\,
\int_{0}^{1}d\lambda\, \exp\left(-\beta^{2}\alpha^{2} +
2u\beta\sqrt{\lambda}\right).
\end{equation}
Here we write $\left\langle 1/f\right\rangle$ in place of
$\left\langle 1/f_{k}\right\rangle$ because the averages of all
the $f_{k}$ are equal. The integral over $\beta$ is familiar to
us, since it is the type of integral that shows up repeatedly in
this averaging method. We have
\begin{eqnarray}
\left\langle\frac{1}{f}\right\rangle &=& \frac{2}{\sqrt{\pi}}
\int_{0}^{1}d\lambda\,
\frac{\sqrt{\pi}}{2\alpha}\exp\left(\frac{u^{2}\lambda}{\alpha^{2}}\right)
\erfc\left(\frac{u\sqrt{\lambda}}{\alpha}\right) \nonumber \\ &=&
\frac{1}{\alpha} \int_{0}^{1}d\lambda\,
\exp\left(\frac{u^{2}\lambda}{\alpha^{2}}\right)
\erfc\left(\frac{u\sqrt{\lambda}}{\alpha}\right).
\end{eqnarray}
The remaining integral over $\lambda$ can be done by parts, with
the result that
\begin{equation}
\left\langle\frac{1}{f}\right\rangle = \frac{2}{u\sqrt{\pi}} +
\frac{\alpha}{u^{2}}\left[\exp\left(\frac{u^{2}}{\alpha^{2}}\right)
\erfc\left(\frac{u}{\alpha}\right)-1\right]. \label{hubbard1/f}
\end{equation}

We can also compute $\left\langle 1/f\right\rangle$ by first
computing the distribution $P_{\mrm{H}}\left(f_{k}\right)$. It
will turn out that knowing $P_{\mrm{H}}\left(f_{k}\right)$ will
aid greatly in the computation of $1/f_{0}$ for this
approximation, as we will see in Section~\ref{f0hubbard}. Because
in this case $f_{k}$ is explicitly given in terms of $\mc{U}^{2}$
through Eq.~(\ref{hubbardband}), we have
\begin{equation}
P_{\mrm{H}}\left(f_{k}\right) = \left\langle \delta\left(f_{k} -
\frac{\alpha}{2} - \frac{1}{2}\sqrt{\alpha^{2} +
4\mc{U}^{2}}\right) \right\rangle.
\end{equation}
This average is easy to compute using the analog of Eq.~(\ref{Pv})
for the potential $U$. We have
\begin{eqnarray}
P_{\mrm{H}}\left(f_{k}\right) &=& \int_{0}^{\infty}
d\left(\mc{U}^{2}\right)\, \delta\left(f_{k} - \frac{\alpha}{2} -
\frac{1}{2}\sqrt{\alpha^{2} + 4\mc{U}^{2}}\right)
\frac{1}{2\sqrt{\pi}} \frac{u}{\mc{U}^{3}}
\exp\left(-\frac{u^{2}}{4\mc{U}^{2}}\right) \nonumber \\ &=&
\frac{1}{2\sqrt{\pi}}
\frac{2f-\alpha}{\left[f\left(f-\alpha\right)\right]^{3/2}}
\exp\left[-\frac{u^{2}}{4f\left(f-\alpha\right)}\right]
\theta\left(f-\alpha\right), \label{Pfkhubbard}
\end{eqnarray}
where we have used the fact that
\begin{equation}
\delta\left(f_{k} - \frac{\alpha}{2} - \frac{1}{2}\sqrt{\alpha^{2}
+ 4\mc{U}^{2}}\right) = \left(2f-\alpha\right)
\delta\left[\mc{U}^{2}-f\left(f-\alpha\right)\right].
\end{equation}
The theta function is present in Eq.~(\ref{Pfkhubbard}) because
$\mc{U}^{2}$ and $f$ are positive definite quantities, and hence
$\delta\left[\mc{U}^{2}-f\left(f-\alpha\right)\right]$ cannot be
nonzero unless $f-\alpha > 0$.

Using the distribution of Eq.~(\ref{Pfkhubbard}) we can compute
\begin{eqnarray}
\left\langle \frac{1}{f}\right\rangle &=& \int_{0}^{\infty}
\frac{df}{f}\,P_{\mrm{H}}\left(f\right) \nonumber \\ &=&
\int_{0}^{\infty} \frac{df}{f}\,\frac{1}{2\sqrt{\pi}}
\frac{2f-\alpha}{\left[f\left(f-\alpha\right)\right]^{3/2}}
\exp\left[-\frac{u^{2}}{4f\left(f-\alpha\right)}\right]
\theta\left(f-\alpha\right) \nonumber \\ &=& \frac{1}{2\sqrt{\pi}}
\int_{\alpha}^{\infty} \frac{df}{f}\,
\frac{2f-\alpha}{\left[f\left(f-\alpha\right)\right]^{3/2}}
\exp\left[-\frac{u^{2}}{4f\left(f-\alpha\right)}\right].
\end{eqnarray}
This integral is very difficult to compute analytically in its
present form, even by a computer package such as
\emph{Mathematica}. Therefore we change variables to $\chi =
f\left(f-\alpha\right)$. Then $d\chi = \left(2f -
\alpha\right)df$, and because $f$ must be positive we have $f =
\alpha + \sqrt{\alpha^{2} + 4\chi}$. It follows that
\begin{equation}
\left\langle \frac{1}{f}\right\rangle = \frac{1}{2\sqrt{\pi}}
\int_{0}^{\infty} \frac{d\chi}{\chi^{3/2}}\,\frac{1}{\alpha +
\sqrt{\alpha^{2} + 4\chi}} \exp\left(-\frac{u^{2}}{4\chi}\right).
\label{chitransf}
\end{equation}
This integral can now be done by \emph{Mathematica\/}, and it
gives exactly the result of Eq.~(\ref{hubbard1/f}).

\subsection{Computation of $\left\langle 1/f_{0}\right\rangle$}
\label{f0hubbard}

Going back to Eq.~(\ref{fzero}), we see that
\begin{equation}
\frac{1}{f_{0}} = \left(\alpha + i\Delta + \sum_{k}
\frac{V_{k}^{2}}{f_{k}}\right)^{-1}.
\end{equation}
Since this is the same equation we considered for the Cayley
approximation in Section~\ref{cayleyUVdelta} we see that the
result for $P_{0,\mrm{H}}\left(f_{0}^{R},f_{0}^{I}\right)$ is very
similar to Eq.~(\ref{P0f0}). Specifically, we have
\begin{equation}
P_{0,\mrm{H}}\left(f_{0}^{R},f_{0}^{I}\right) =
\frac{1}{2\sqrt{\pi}}
\frac{vQ_{\mrm{H}}\left(\alpha\right)}{\left(f_{0}^{R}-\alpha\right)^{3/2}}
\exp\left[ -\frac{v^{2}
Q_{\mrm{H}}^{2}\left(\alpha\right)}{4\left(f_{0}^{R}-\alpha\right)}
\right]\theta\left(f_{0}^{R}-\alpha\right)\delta\left(f_{0}^{I} -
\Delta\right),
\end{equation}
where
\begin{equation}
Q_{\mrm{H}}\left(\alpha\right) = \int_{0}^{\infty}
\frac{df}{\sqrt{f}}\,P_{\mrm{H}}\left(f\right).
\label{Qeqnhubbard}
\end{equation}
It then follows from Eq.~(\ref{1/f0cayley}) that
\begin{equation}
\left\langle\frac{1}{f_{0}}\right\rangle = \frac{1}{\alpha +
i\Delta}-\frac{\sqrt{\pi}}{2}
\frac{vQ_{\mrm{H}}\left(\alpha\right)}{\left(\alpha +
i\Delta\right)^{3/2}} \exp\left[
\frac{v^{2}Q_{\mrm{H}}^{2}\left(\alpha\right)}{4\left(\alpha +
i\Delta\right)}\right] \erfc\left[
\frac{vQ_{\mrm{H}}\left(\alpha\right)}{2\sqrt{\alpha + i\Delta}}
\right]. \label{1/f0hubbard}
\end{equation}

One might think that because $P_{\mrm{H}}\left(f\right)$ is known
explicitly in this case, instead of involving transcendental
quantities as Eq.~(\ref{Pfcayleynov}) does, it should be possible
to compute the function $Q_{\mrm{H}}\left(\alpha\right)$
explicitly as well. Unfortunately the integral of
Eq.~(\ref{Qeqnhubbard}) cannot be done analytically. If we make
the same change of variable that we did to obtain
Eq.~(\ref{chitransf}), however, then Eq.~(\ref{Qeqnhubbard})
becomes
\begin{equation}
Q_{\mrm{H}}\left(\alpha\right) = \frac{1}{2\sqrt{\pi}}
\int_{0}^{\infty} \frac{d\chi}{\chi^{3/2}}\,\frac{1}{\sqrt{\alpha
+ \sqrt{\alpha^{2} + 4\chi}}}
\exp\left(-\frac{u^{2}}{4\chi}\right),
\end{equation}
and this integral can be done numerically for a given value of
$\alpha$.

Thus $Q_{\mrm{H}}\left(\alpha\right)$ can be computed numerically,
and the result substituted into Eq.~(\ref{1/f0hubbard}) to obtain
$\left\langle 1/f_{0} \right\rangle$.

Although we have shown how to compute $\left\langle 1/f_{0}
\right\rangle$ for the Hubbard model using Eqs.~(\ref{fzero}) and
(\ref{hubbardband}), we note that these equations are not
consistent. Consistency demands that $\left\langle
1/f_{0}\right\rangle$ reduces to $\left\langle 1/f\right\rangle$
when $u = v$ and $\Delta = 0$, and this is clearly not the case
with Eqs.~(\ref{1/f0hubbard}) and (\ref{fhubbard}). In fact, one
can see this inconsistency immediately from Eqs.~(\ref{fzero}) and
(\ref{hubbardband}). If we set $\mc{U}^{2} = \mc{V}^{2}$ and
$\Delta = 0$ in these equations, then we have \setcounter{tlet}{1}
\renewcommand{\theequation}{\arabic{chapter}.\arabic{equation}\alph{tlet}}
\begin{eqnarray}
f_{0} &=& \alpha + \sum_{k=1}^{N}\frac{V_{0k}^{2}}{f_{k}}, \\
\stepcounter{tlet}\addtocounter{equation}{-1} f_{k} &=& \alpha +
\frac{\mc{V}^{2}}{f_{k}}.
\end{eqnarray}
\renewcommand{\theequation}{\arabic{chapter}.\arabic{equation}}
We see that even in this limit, $f_{0}$ and $f_{k}$ are still
treated differently in this approximation. Therefore the Hubbard
approximation is inconsistent, and to be safe one should disregard
the result for $\left\langle 1/f_{0}\right\rangle$ and only
consider the result for $\left\langle 1/f\right\rangle$ of
Section~\ref{fhubbardsection} as valid.

\subsection{Computation of $\left\langle\left|\, a_{0}\left( \Delta,t\right)
\right|^{2}\right\rangle$ with the $U$ process only at resonance}
\label{hubbard_a2}

We saw in Eq.~(\ref{inversetransform}) that at resonance
\begin{equation}
a_{0}\left(0,t\right) = \frac{1}{\mc{U}t}
J_{1}\left(2\mc{U}t\right),
\end{equation}
and so
\begin{equation}
\left| a_{0}\left(0,t\right)\right|^{2} =
\frac{1}{\mc{U}^{2}t^{2}} J_{1}^{2}\left(2\mc{U}t\right).
\end{equation}
According to Eq.~(10.1.1) of Ref~\cite{Roberts1966a}, this
function has the inverse Laplace transform
\begin{eqnarray}
\frac{1}{\alpha} - \tilde{S}\left(0,\alpha\right) &=&
\frac{1}{\pi\mc{U}}\int_{0}^{\pi}d\theta\,
\left(1+\cos\theta\right) \left[
\sqrt{\frac{\alpha^{2}}{4\mc{U}^{2}} + 2 - 2\cos\theta} -
\frac{\alpha}{2\mc{U}}\right] \nonumber \\ &=&
\frac{1}{2\pi\mc{U}^{2}}\int_{0}^{\pi}d\theta\,
\left(1+\cos\theta\right) \left[ \sqrt{\alpha^{2} +
8\mc{U}^{2}\left(1 - \cos\theta\right)} - \alpha\right].
\end{eqnarray}
Now we again apply Eq.~(\ref{lambdaintegral}), with the result
that
\begin{eqnarray}
\tilde{S}\left(0,\alpha\right) &=& \frac{1}{\alpha} -
\frac{1}{2\pi\mc{U}^{2}}\int_{0}^{\pi}d\theta\,
\left(1+\cos\theta\right) 4\left(1-\cos\theta\right)\mc{U}^{2}
\int_{0}^{1}\frac{d\lambda}{\sqrt{\alpha^{2} +
8\lambda\mc{U}^{2}\left(1-\cos\theta\right)}} \nonumber \\ &=&
\frac{1}{\alpha} - \frac{2}{\pi}\int_{0}^{\pi}d\theta\,
\sin^{2}\theta \int_{0}^{1}\frac{d\lambda}{\sqrt{\alpha^{2} +
8\lambda\mc{U}^{2}\left(1-\cos\theta\right)}}.
\end{eqnarray}
At this point, we apply Eq.~(\ref{beta2integral}) to find that
\begin{equation}
\tilde{S}\left(0,\alpha\right) = \frac{1}{\alpha} -
\frac{4}{\pi^{3/2}}\int_{0}^{\pi}d\theta\, \sin^{2}\theta
\int_{0}^{1}d\lambda \int_{0}^{\infty}d\beta\,
e^{-\beta\left[\alpha^{2} + 8\lambda\mc{U}^{2}
\left(1-\cos\theta\right)\right]}.
\end{equation}
This expression can now be averaged. We find that
\begin{equation}
\left\langle\tilde{S}\left(0,\alpha\right)\right\rangle =
\frac{1}{\alpha} - \frac{4}{\pi^{3/2}}\int_{0}^{\pi}d\theta\,
\sin^{2}\theta \int_{0}^{1}d\lambda\, \int_{0}^{\infty}d\beta\,
e^{-\beta\alpha^{2} - 2\sqrt{2}u\sqrt{\beta\lambda
\left(1-\cos\theta\right)}}.
\end{equation}
The integration over $\beta$ can be performed at this point, with
the result that
\begin{eqnarray}
\left\langle\tilde{S}\left(0,\alpha\right)\right\rangle &=&
\frac{1}{\alpha} - \frac{4}{\pi^{3/2}}\int_{0}^{\pi}d\theta\,
\sin^{2}\theta \int_{0}^{1}d\lambda\,
\frac{\sqrt{\pi}}{2}\frac{1}{\alpha} \times \nonumber \\ &&
\exp\left[\frac{2\lambda
u^{2}}{\alpha^{2}}\left(1-\cos\theta\right)\right]
\erfc\left[\frac{u}{\alpha}\sqrt{2\lambda
\left(1-\cos\theta\right)}\right] \nonumber \\ &=&
\frac{1}{\alpha} -
\frac{2}{\pi}\frac{1}{\alpha}\int_{0}^{\pi}d\theta\,
\sin^{2}\theta \int_{0}^{1}d\lambda\, \exp\left[\frac{2\lambda
u^{2}}{\alpha^{2}}\left(1-\cos\theta\right)\right] \times
\nonumber \\ && \erfc\left[\frac{u}{\alpha}\sqrt{2\lambda
\left(1-\cos\theta\right)}\right].
\end{eqnarray}
The $\lambda$ integral can be done by parts. The result is
\begin{eqnarray}
\left\langle\tilde{S}\left(0,\alpha\right)\right\rangle &=&
\frac{1}{\alpha} -
\frac{2}{\pi}\frac{1}{\alpha}\int_{0}^{\pi}d\theta\,
\sin^{2}\theta \left\{
\frac{\alpha^{2}}{2u^{2}\left(1-\cos\theta\right)} \exp\left[
\frac{2u^{2}}{\alpha^{2}}\left( 1-\cos\theta\right)\right] \right.
\times \nonumber \\ && \left. \erfc\left[
\frac{u}{\alpha}\sqrt{2\left(1-\cos\theta\right)}\right] -
\frac{\alpha^{2}}{2u^{2}\left(1-\cos\theta\right)} +
\sqrt{\frac{2}{\pi}}\frac{\alpha}{u\sqrt{1-\cos\theta}} \right\}
\nonumber \\ &=& \frac{1}{\alpha} -
\frac{1}{\pi}\frac{\alpha}{u^{2}}\int_{0}^{\pi}d\theta\,
\frac{\sin^{2}\theta}{1-\cos\theta} \exp\left[
\frac{2u^{2}}{\alpha^{2}}\left( 1-\cos\theta\right)\right]
\erfc\left[
\frac{u}{\alpha}\sqrt{2\left(1-\cos\theta\right)}\right] +
\nonumber \\ &&
\frac{1}{\pi}\frac{\alpha}{u^{2}}\int_{0}^{\pi}d\theta\,
\left(1+\cos\theta\right) -
\left(\frac{2}{\pi}\right)^{3/2}\frac{1}{u}\int_{0}^{\pi}d\theta\,
\frac{\sin^{2}\theta}{\sqrt{1-\cos\theta}}.
\end{eqnarray}
The last two terms are standard integrals, and we have that
\begin{eqnarray}
\left\langle\tilde{S}\left(0,\alpha\right)\right\rangle &=&
\frac{1}{\alpha} -
\frac{1}{\pi}\frac{\alpha}{u^{2}}\int_{0}^{\pi}d\theta\,
\left(1+\cos\theta\right) \exp\left[
\frac{2u^{2}}{\alpha^{2}}\left( 1-\cos\theta\right)\right]
\erfc\left[
\frac{u}{\alpha}\sqrt{2\left(1-\cos\theta\right)}\right] +
\nonumber
\\ && \frac{\alpha}{u^{2}} - \frac{16}{3\pi^{3/2}}\frac{1}{u}.
\label{a0thetaintegral}
\end{eqnarray}
Now we consider the remaining integral
\begin{equation}
\Lambda = \frac{1}{\pi}\frac{\alpha}{u^{2}}\int_{0}^{\pi}d\theta\,
\left(1+\cos\theta\right) \exp\left[
\frac{2u^{2}}{\alpha^{2}}\left( 1-\cos\theta\right)\right]
\erfc\left[
\frac{u}{\alpha}\sqrt{2\left(1-\cos\theta\right)}\right].
\end{equation}
First we define $\phi = \theta/2$, so that $d\theta = 2\, d\phi$.
Thus $1+\cos\theta = 2\cos^{2}\phi$ and similarly $1-\cos\theta =
2\sin^{2}\phi$. In addition, since $0\leq\theta\leq\pi$, it
follows that $0\leq\phi\leq\pi/2$, and hence both $\cos\phi$ and
$\sin\phi$ are positive quantities. With this substitution we have
\begin{equation}
\Lambda = \frac{4}{\pi}\frac{\alpha}{u^{2}}\int_{0}^{\pi/2}d\phi\,
\cos^{2}\phi \exp\left(
\frac{4u^{2}}{\alpha^{2}}\sin^{2}\phi\right) \erfc\left(
\frac{2u}{\alpha}\sin\phi\right).
\end{equation}
Using Eq.~(\ref{expzerfcz}), we see that
\begin{eqnarray}
\Lambda &=&
\frac{4}{\pi}\frac{\alpha}{u^{2}}\int_{0}^{\pi/2}d\phi\,
\cos^{2}\phi \sum_{n=0}^{\infty}
\frac{1}{\G{\frac{n+2}{2}}}\left(-\frac{2u}{\alpha}\sin\phi\right)^{n}
\nonumber \\ &=&
\frac{4}{\pi}\frac{\alpha}{u^{2}}\sum_{n=0}^{\infty}
\frac{1}{\G{\frac{n+2}{2}}}\left(-\frac{2u}{\alpha}\right)^{n}
\int_{0}^{\pi/2}d\phi\, \cos^{2}\phi \sin^{n}\phi.
\end{eqnarray}
The integral over $\phi$ is a standard one, and we find that
\begin{eqnarray}
\Lambda &=&
\frac{1}{\sqrt{\pi}}\frac{\alpha}{u^{2}}\sum_{n=0}^{\infty}
\frac{\G{\frac{n+1}{2}}}{\G{\frac{n+2}{2}}\G{\frac{n+4}{2}}}
\left(-\frac{2u}{\alpha}\right)^{n} \nonumber \\ &=&
\frac{\alpha}{u^{2}} -\frac{16}{3\pi^{3/2}}\frac{1}{u} +
\frac{1}{\alpha} +
\frac{1}{\sqrt{\pi}}\frac{\alpha}{u^{2}}\sum_{n=3}^{\infty}
\frac{\G{\frac{n+1}{2}}}{\G{\frac{n+2}{2}}\G{\frac{n+4}{2}}}
\left(-\frac{2u}{\alpha}\right)^{n} \nonumber \\ &=&
\frac{\alpha}{u^{2}} -\frac{16}{3\pi^{3/2}}\frac{1}{u} +
\frac{1}{\alpha} -
\frac{8}{\sqrt{\pi}}\frac{u}{\alpha^{2}}\sum_{n=0}^{\infty}
\frac{\G{\frac{n+4}{2}}}{\G{\frac{n+5}{2}}\G{\frac{n+7}{2}}}
\left(-\frac{2u}{\alpha}\right)^{n}.
\end{eqnarray}
Substituting this back into Eq.~(\ref{a0thetaintegral}), we see
that
\begin{equation}
\left\langle\tilde{S}\left(0,\alpha\right)\right\rangle =
\frac{8}{\sqrt{\pi}}\frac{u}{\alpha^{2}}\sum_{n=0}^{\infty}
\frac{\G{\frac{n+4}{2}}}{\G{\frac{n+5}{2}}\G{\frac{n+7}{2}}}
\left(-\frac{2u}{\alpha}\right)^{n}.
\end{equation}

At this point we can invert this Laplace transform term by term,
using the fact that the inverse Laplace transform of
$\alpha^{-\nu}$ is $t^{\nu-1}/\G{\nu}$. We finally arrive at
\begin{equation}
\left\langle S\left(0,t\right)\right\rangle =
\frac{8}{\sqrt{\pi}}ut\sum_{n=0}^{\infty}
\frac{\G{\frac{n+4}{2}}}{\G{\frac{n+5}{2}}\G{\frac{n+7}{2}}\G{n+2}}
\left(-2ut\right)^{n}.
\end{equation}
The even terms in this sum reduce to
\begin{eqnarray}
S_{\mrm{H},\,\mrm{even}} &=& \frac{8}{\sqrt{\pi}}ut
\sum_{n=0}^{\infty}
\frac{\G{\frac{2n+4}{2}}}{\G{\frac{2n+5}{2}}\G{\frac{2n+7}{2}}\G{2n+2}}
\left(-2ut\right)^{2n} \nonumber \\ &=& 4ut \sum_{n=0}^{\infty}
\frac{\G{n+2}}{\G{n+\frac{5}{2}}\G{n+\frac{7}{2}}\G{n+1}\G{n+\frac{3}{2}}}
\left(4u^{2}t^{2}\right)^{n} \nonumber \\ &=&
\frac{256}{45\pi^{3/2}} ut\,
{}_{1}F_{3}\left(2;\frac{3}{2},\frac{5}{2},\frac{7}{2};u^{2}t^{2}\right),
\end{eqnarray}
where in the third step we have used Eq.~(\ref{gamma2z}) and in
the final step we have used Eq.~(\ref{hypergeodef}). Similarly,
the odd terms in this sum reduce to
\begin{eqnarray}
S_{\mrm{H},\,\mrm{odd}} &=& \frac{8}{\sqrt{\pi}}ut
\sum_{n=0}^{\infty}
\frac{\G{\frac{2n+5}{2}}}{\G{\frac{2n+6}{2}}\G{\frac{2n+8}{2}}\G{2n+3}}
\left(-2ut\right)^{2n+1} \nonumber \\ &=& -4u^{2}t^{2}
\sum_{n=0}^{\infty}
\frac{\G{n+\frac{5}{2}}}{\G{n+3}\G{n+4}\G{n+\frac{3}{2}}\G{n+2}}
\left(u^{2}t^{2}\right)^{n} \nonumber \\ &=& -4
\sum_{n=1}^{\infty}
\frac{\G{n+\frac{3}{2}}}{\G{n+2}\G{n+3}\G{n+\frac{1}{2}}\G{n+1}}
\left(u^{2}t^{2}\right)^{n} \nonumber \\ &=& 1 -
{}_{1}F_{3}\left(\frac{3}{2};\frac{1}{2},2,3;u^{2}t^{2} \right) ,
\end{eqnarray}
and so we have finally that
\begin{equation}
\left\langle S\left(0,t\right)\right\rangle = 1 +
\frac{256}{45\pi^{3/2}} ut\,
{}_{1}F_{3}\left(2;\frac{3}{2},\frac{5}{2},\frac{7}{2};u^{2}t^{2}\right)
- {}_{1}F_{3}\left(\frac{3}{2};\frac{1}{2},2,3;u^{2}t^{2} \right).
\label{hubbardsignalresult}
\end{equation}

\begin{figure}
\resizebox{\textwidth}{!}{\includegraphics[0in,0in][8in,10in]{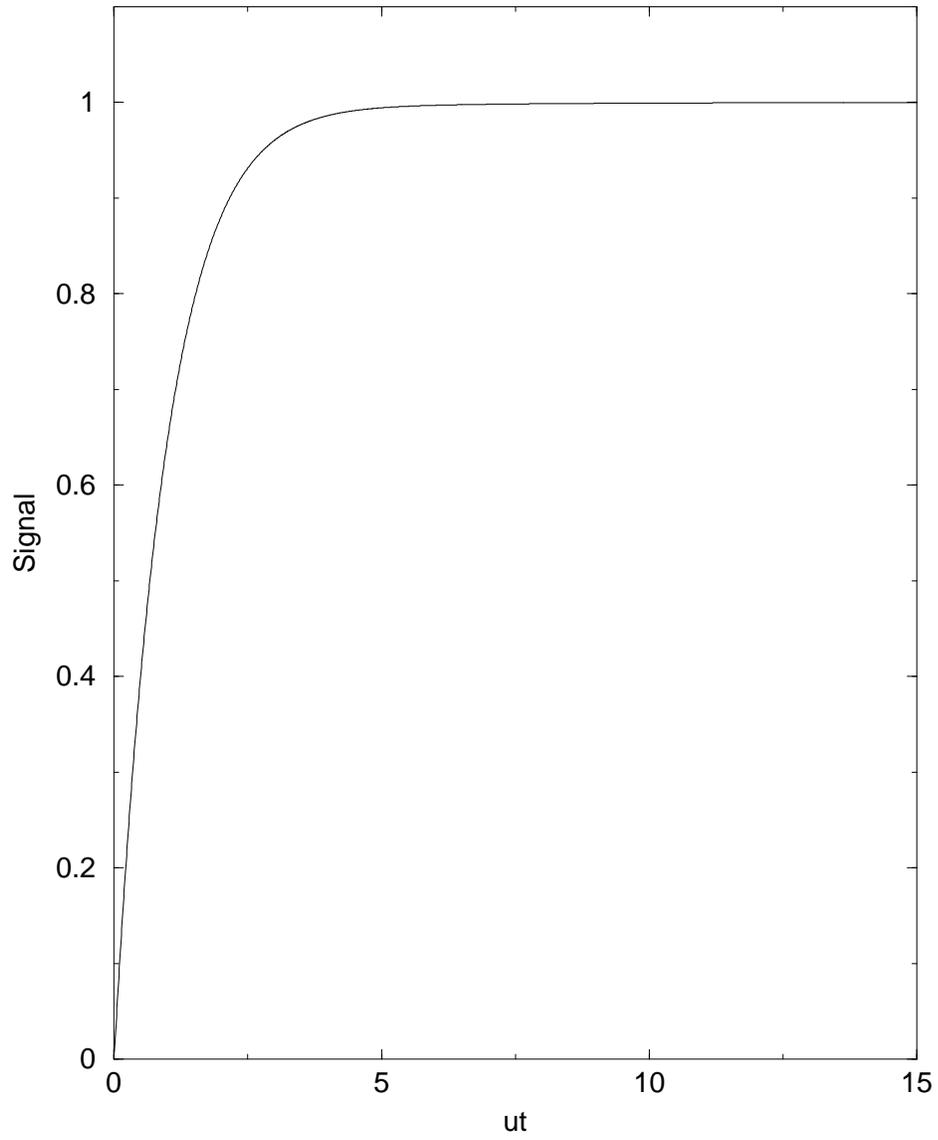}}
\caption[Averaged signal as a function of time for the Hubbard
approximation]{A plot of the averaged signal as a function of time
for $u=v$ and $\Delta = 0$ in the Hubbard approximation.}
\label{sparse_with_u_hubbard_sig}
\end{figure}

Figure~\ref{sparse_with_u_hubbard_sig} shows a plot of this result
as a function of time. Looking back to
Fig.~\ref{introduction_expsig}, we see that this signal saturates
too quickly to provide a good fit to the experimental data. In
fact, looking ahead to Fig.~\ref{simulation_u=v_signal}, we see
that it reaches saturation too quickly to favorably compare with
the numerical simulation results for $u = 0$ at resonance.

\section{The coherent potential approximation}
\label{cpasection}

Instead of making the Hubbard substitution of
Eq.~(\ref{fhubbard}), we can replace $1/f_{l}$ by its average in
Eq.~(\ref{fk}), obtaining
\begin{equation}
f_{k} = \alpha + \sum_{l=1}^{N}U_{kl}^{2}
\left\langle\frac{1}{f_{l}}\right\rangle. \label{fcpa}
\end{equation}
This corresponds to a coherent potential approximation (CPA).

\subsection{Computation of $\left\langle 1/f\right\rangle$}
\label{fcpasection}

We have from Eq.~(\ref{fcpa}) that
\begin{equation}
\frac{1}{f_{k}} = \frac{1}{\alpha + \sum_{l=1}^{N}U_{kl}^{2}
\left\langle\frac{1}{f}\right\rangle}.
\end{equation}
We again note that $\left\langle 1/f_{k}\right\rangle$ is
independent of $k$, so we drop the subscript and we see that
\begin{equation}
\left\langle \frac{1}{f}\right\rangle = \left\langle
\frac{1}{\alpha + \mc{U}^{2}\left\langle
\frac{1}{f}\right\rangle}\right\rangle. \label{startwithme}
\end{equation}
The averaging procedure of Section~\ref{averaging} is by now
second nature, and we see that
\begin{eqnarray}
\left\langle \frac{1}{f}\right\rangle &=& \left\langle
\int_{0}^{\infty}d\beta\, \exp\left[ -\beta \left(\alpha +
\mc{U}^{2}\left\langle \frac{1}{f}\right\rangle\right) \right]
\right\rangle \nonumber \\ &=& \int_{0}^{\infty}d\beta\,
\exp\left( -\beta\alpha - u\sqrt{\beta\left\langle
\frac{1}{f}\right\rangle}\right) \nonumber \\ &=& \frac{1}{\alpha}
- \frac{\sqrt{\pi}}{2}\frac{u}{\alpha^{3/2}} \sqrt{\left\langle
\frac{1}{f}\right\rangle}\, \exp\left(\frac{u^{2}}{4\alpha}
\left\langle \frac{1}{f}\right\rangle\right) \erfc\left(
\frac{u}{2\sqrt{\alpha}}\sqrt{\left\langle
\frac{1}{f}\right\rangle}\right). \label{endwithme}
\end{eqnarray}
This is an implicit equation for $\left\langle 1/f\right\rangle$,
which can be rearranged to give
\begin{equation}
\alpha\left\langle\frac{1}{f}\right\rangle +
\frac{\sqrt{\pi}}{2}\frac{u}{\sqrt{\alpha}}
\sqrt{\left\langle\frac{1}{f}\right\rangle}\,
\exp\left(\frac{u^{2}}{4\alpha}
\left\langle\frac{1}{f}\right\rangle\right)
\erfc\left(\frac{u}{2\sqrt{\alpha}}
\sqrt{\left\langle\frac{1}{f}\right\rangle}\right) = 1.
\label{1/fCPA}
\end{equation}
Eq.~(\ref{1/fCPA}) can then be solved numerically for
$\left\langle 1/f\right\rangle$.

\subsection{Computation of $\left\langle 1/f_{0}\right\rangle$}
\label{f0cpasection}

Just as we saw when considering the Hubbard approximation, the
modification of Eq.~(\ref{fk}) to Eq.~(\ref{fcpa}) makes the
latter equation inconsistent with Eq.~(\ref{fzero}). It is
possible to use Eq.~(\ref{1/fCPA}) coupled with Eq.~(\ref{fzero})
to obtain an expression for $\left\langle 1/f_{0}\right\rangle$,
but it does not reduce to Eq.~(\ref{1/fCPA}) when we take the
limit $\Delta = 0$ and $u = v$. In this case, however, we can
modify Eq.~(\ref{fzero}) to
\begin{equation}
f_{0} = \alpha + i\Delta + \mc{V}^{2} \left\langle
\frac{1}{f}\right\rangle, \label{fzeroCPA}
\end{equation}
so that it and Eq.~(\ref{fcpa}) are again consistent.

Proceeding as we did in going from Eq.~(\ref{startwithme}) to
\ref{endwithme}, we have
\begin{eqnarray}
\left\langle\frac{1}{f_{0}}\right\rangle &=& \left\langle
\frac{1}{\alpha + i\Delta + \mc{V}^{2} \left\langle
\frac{1}{f}\right\rangle} \right\rangle \nonumber \\ &=&
\left\langle \int_{0}^{\infty} d\beta\, \exp\left[-\beta
\left(\alpha + i\Delta + \mc{V}^{2} \left\langle
\frac{1}{f}\right\rangle\right)\right]\right\rangle \nonumber \\
&=& \int_{0}^{\infty}d\beta\, \exp\left[-\beta \left(\alpha +
i\Delta\right) - v\sqrt{\beta\left\langle
\frac{1}{f}\right\rangle}\right],
\end{eqnarray}
and hence
\begin{eqnarray}
\left\langle\frac{1}{f_{0}}\right\rangle &=& \frac{1}{\alpha +
i\Delta} - \frac{\sqrt{\pi}}{2} \frac{v}{\left(\alpha +
i\Delta\right)^{3/2}}\sqrt{\left\langle
\frac{1}{f}\right\rangle}\,\times \nonumber \\ &&
\exp\left[\frac{v^{2}}{4\left(\alpha + i\Delta
\right)}\left\langle \frac{1}{f}\right\rangle\right]
\erfc\left(\frac{v}{2\sqrt{\alpha + i\Delta}}\sqrt{\left\langle
\frac{1}{f}\right\rangle}\right). \label{finalCPAresult}
\end{eqnarray}

\section{Comparison of the Cayley, Hubbard, and CPA approximations}
\label{sparse_with_u_comp}

We have examined these approximations and found that the results
that agree best with the simulations are those obtained using the
CPA approximation. In fact, as we will see in
Section~\ref{sparse_with_u_plots}, the CPA results agree
strikingly well with the simulations. This is somewhat surprising,
since in the Cayley equations (\ref{fzero}) and (\ref{fk}) the
on-site Green's functions for the $U$ problem are allowed to vary
from site to site, whereas in Eqs.~(\ref{fhubbard}) and
(\ref{fcpa}) they are not. However, the Cayley tree approximation
is exact in a one-dimensional system with nearest neighbor
interactions and becomes increasingly inaccurate as the
dimensionality of the problem and the range of the interaction
increase. The CPA, on the other hand, is a mean field
approximation, and it becomes accurate for higher dimensional
problems and long-range interactions.

\begin{figure}
\resizebox{\textwidth}{!}{\includegraphics[0in,0in][8in,10in]{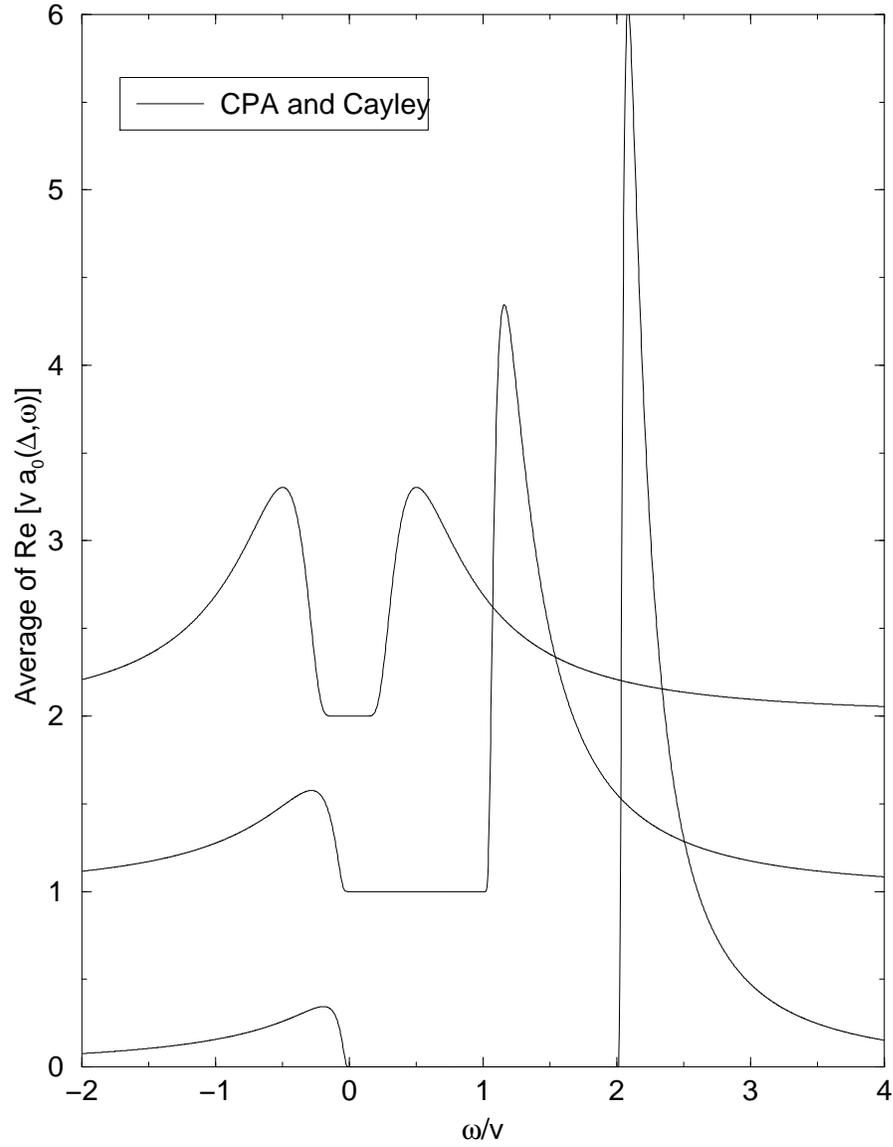}}
\caption[Real part of $\left\langle
a_{0}\left(\Delta,\omega\right)\right\rangle$ for $u = 0$]{Plots
of $\mrm{Re}\,\left\langle
\mrm{Re}\,a_{0}\left(\Delta,\omega\right)\right\rangle$ for $u =
0$ and, from left to right, $\Delta = 0$, $v$, and $2v$. The
$\Delta = v$ curve has been shifted vertically upward by $1$, and
the $\Delta = 2v$ curve by $2$. The results are exact in this
case.} \label{sparse_with_u_comp_u=0v}
\end{figure}

We see in Fig.~\ref{sparse_with_u_comp_u=0v} plots of
$\mrm{Re}\,\left\langle
a_{0}\left(\Delta,\omega\right)\right\rangle$ for $u = 0$ and
several values of $\Delta$. The Cayley and CPA approximations
coincide in this case. The exact result for $u = 0$ can actually
be obtained analytically by the methods for computing
$\left\langle\tilde{S}\left(\Delta,\alpha\right)\right\rangle$
used in Ref.~\cite{Frasier1999a} and described in detail in
Chapter~\ref{sparse_no_u}. We reproduce here the result of
Eq.~(\ref{a0exact}), which is
\begin{equation}
\left\langle \tilde{a}_{0}\left(\Delta,\alpha\right)\right\rangle
= \frac{1}{\alpha+i\Delta} -
\frac{\sqrt{\pi}}{2}\frac{v}{\left(\alpha+i\Delta\right)^{3/2}
\sqrt{\alpha}}
\exp\left[\frac{v^{2}}{4\alpha\left(\alpha+i\Delta\right)}\right]
{\mathrm{erfc}}\left[\frac{v}{2\sqrt{\alpha\left(\alpha +
i\Delta\right)}}\right]. \label{a0alpha}
\end{equation}
As one would expect, this agrees with Eq.~(\ref{1/f0cayley}) for
$Q\left(\alpha\right) = 1/\sqrt{\alpha}$ and with
Eq.~(\ref{finalCPAresult}) for $\left\langle 1/f\right\rangle =
1/\alpha$.

We note the ``gap'' that is present in each of the curves, and we
also note that the width of the gap increases with $\Delta$. This
effect can be seen analytically from Eq.~(\ref{a0alpha}). We start
by converting from Laplace to Fourier space by making the
substitution $\alpha = \omega/i$, so that we have
\begin{equation}
\left\langle a_{0}\left(\Delta,\omega\right)\right\rangle =
\frac{i}{\omega-\Delta} +
\frac{\sqrt{\pi}}{2}\frac{v}{\left(\omega-\Delta\right)^{3/2}
\sqrt{\omega}}
\exp\left[-\frac{v^{2}}{4\omega\left(\omega-\Delta\right)}\right]
\erfc\left[i\frac{v}{2\sqrt{\omega\left(\omega -
\Delta\right)}}\right].
\end{equation}
Using Eq.~(\ref{expzerfcz}), we see that
\begin{equation}
\mrm{Re}\,\left\langle
a_{0}\left(\Delta,\omega\right)\right\rangle = \mrm{Re}\,\left\{
\frac{\sqrt{\pi}}{2}\frac{v}{\left(\omega-\Delta\right)^{3/2}
\sqrt{\omega}}
\exp\left[-\frac{v^{2}}{4\omega\left(\omega-\Delta\right)}\right]\right\},
\end{equation}
which reduces to
\begin{equation}
\mrm{Re}\,\left\langle
a_{0}\left(\Delta,\omega\right)\right\rangle =
\frac{\sqrt{\pi}}{2}\frac{v}{\left(\omega-\Delta\right)^{3/2}
\sqrt{\omega}}
\exp\left[-\frac{v^{2}}{4\omega\left(\omega-\Delta\right)}\right],
\label{rea0omegafull}
\end{equation}
if $\omega > \Delta$, and zero otherwise. This explains the gaps
seen in the $\Delta = v$ and $\Delta = 2v$ curves. For $\Delta =
0$, Eq.~(\ref{rea0omegafull}) becomes
\begin{equation}
\mrm{Re}\,\left\langle
a_{0}\left(\Delta,\omega\right)\right\rangle =
\frac{\sqrt{\pi}}{2}\frac{v}{\omega^{2}}
\exp\left(-\frac{v^{2}}{4\omega^{2}}\right),
\end{equation}
which produces the quasigap around $\omega = 0$ seen in the
resonance curve.

\begin{figure}
\resizebox{\textwidth}{!}{\includegraphics[0in,0in][8in,10in]{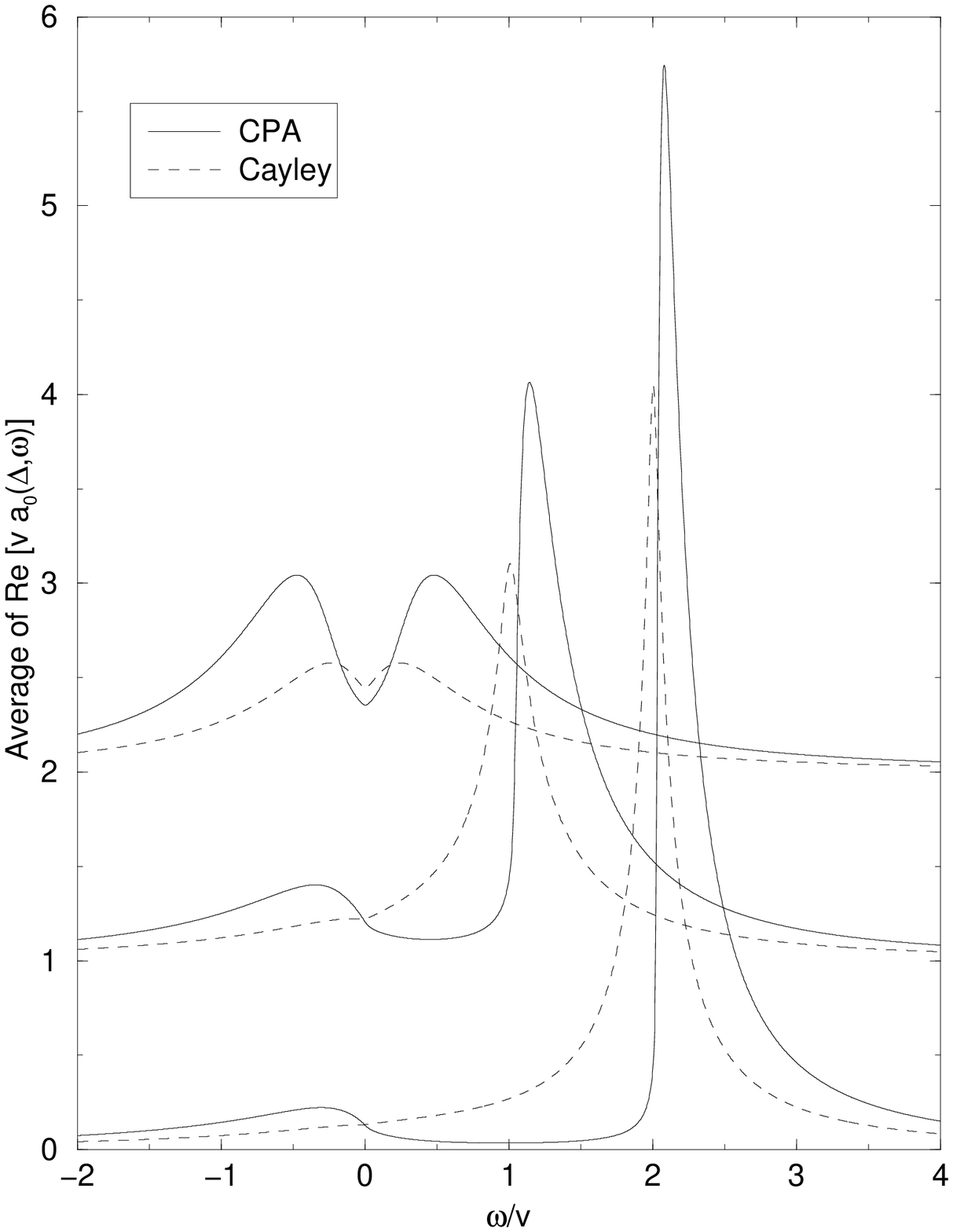}}
\caption[Real part of $\left\langle
a_{0}\left(\Delta,\omega\right)\right\rangle$ for $u =
0.25v$]{Plots of $\mrm{Re}\,\left\langle
a_{0}\left(\Delta,\omega\right)\right\rangle$ for $u = 0.25v$ and,
from left to right, $\Delta = 0$, $v$, and $2v$. The $\Delta = v$
curve has been shifted vertically upward by $1$, and the $\Delta =
2v$ curve by $2$.} \label{sparse_with_u_comp_u=0.25v}
\end{figure}

In Fig.~\ref{sparse_with_u_comp_u=0.25v}, which presents plots of
$\mrm{Re}\,\left\langle a_{0}\left(\Delta,t\right)\right\rangle$
for $u = 0.25v$ and several values of $\Delta$, we begin to see
differences between the CPA and Cayley approximations. We see that
the CPA results retain the gap structure of the $u = 0$ case
somewhat, while the only vestiges of the gap remaining in the
Cayley results are the dimple in the $\Delta = 0$ curve at
$\omega/v = 0$ and a barely perceptible bend in the $\Delta = v$
curve at $\omega/v = 0$. The off-resonant CPA results are also
noticeably asymmetric around $\omega = \Delta$.

\begin{figure}
\resizebox{\textwidth}{!}{\includegraphics[0in,0in][8in,10in]{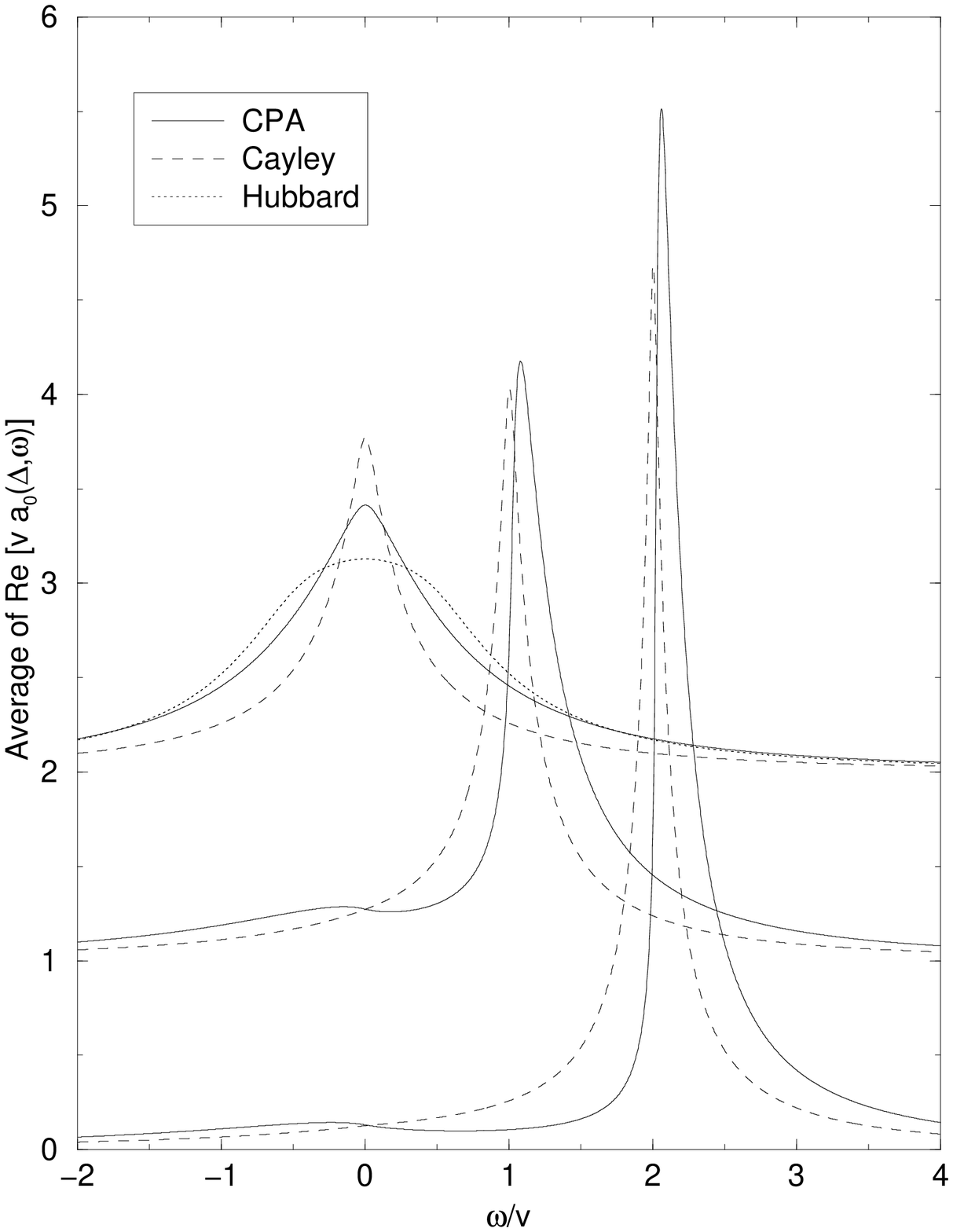}}
\caption[Real part of $\left\langle
a_{0}\left(\Delta,\omega\right)\right\rangle$ for $u = v$]{Plots
of $\mrm{Re}\,\left\langle
a_{0}\left(\Delta,\omega\right)\right\rangle$ for $u = v$ and,
from left to right, $\Delta = 0$, $v$, and $2v$. The $\Delta = v$
curve has been shifted vertically upward by $1$, and the $\Delta =
2v$ curve by $2$.} \label{sparse_with_u_comp_u=v}
\end{figure}

In Fig.~\ref{sparse_with_u_comp_u=v} we see plots for $u = v$, and
many of the trends of Fig.~\ref{sparse_with_u_comp_u=0.25v}
continue. The off-resonant CPA results continue to retain some
memory of the gap and are still asymmetric around $\omega =
\Delta$. The gap structure is no longer evident in the
on-resonance curve. For $\Delta = 0$ the result for the Hubbard
approximation is also shown.

\begin{figure}
\resizebox{\textwidth}{!}{\includegraphics[0in,0in][8in,10in]{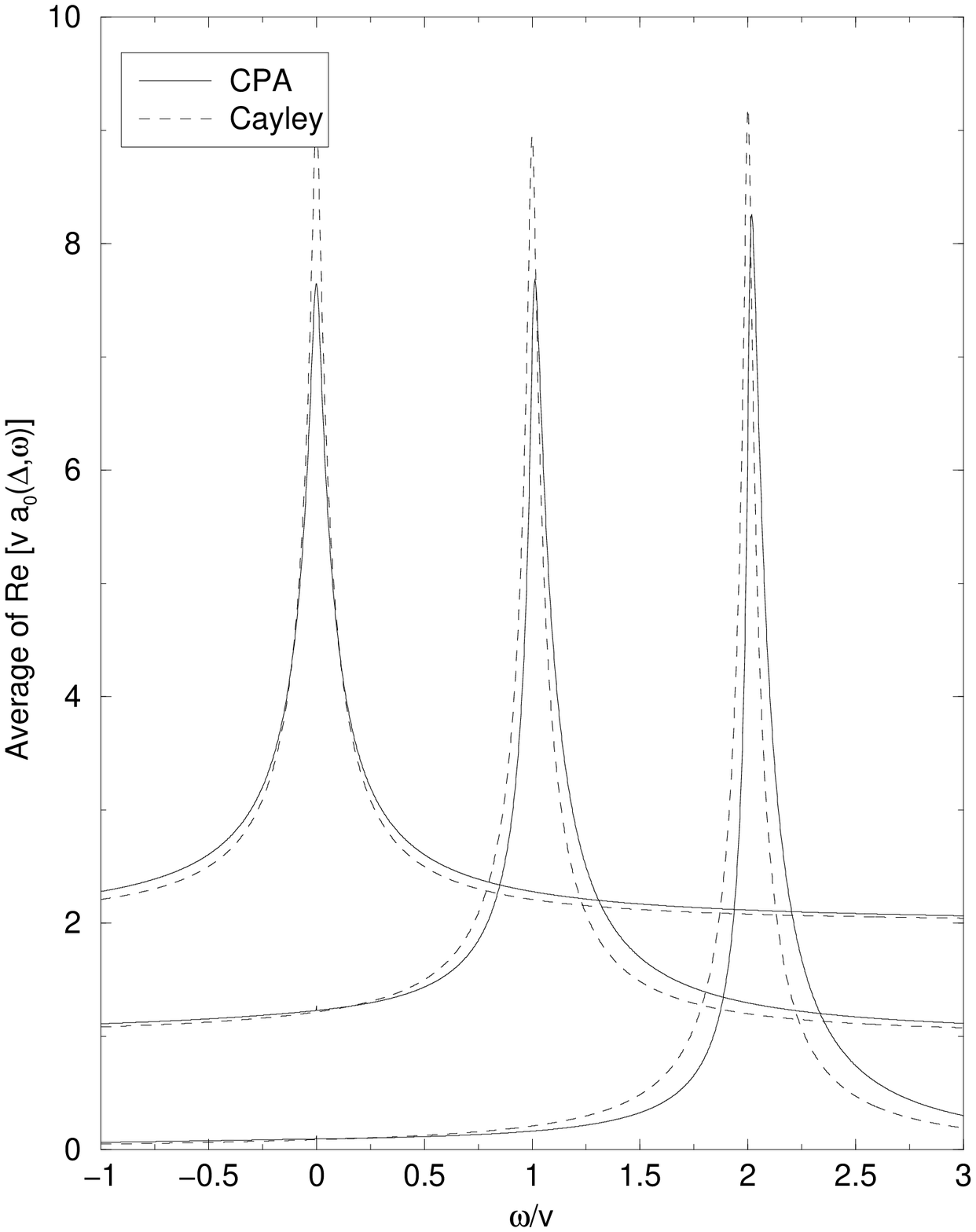}}
\caption[Real part of $\left\langle
a_{0}\left(\Delta,\omega\right)\right\rangle$ for $u = 4v$]{Plots
of $\mrm{Re}\,\left\langle
a_{0}\left(\Delta,\omega\right)\right\rangle$ for $u = 4v$ and,
from left to right, $\Delta = 0$, $v$, and $2v$. The $\Delta = v$
curve has been shifted vertically upward by $1$, and the $\Delta =
2v$ curve by $2$.} \label{sparse_with_u_comp_u=4v}
\end{figure}

We see that the CPA and Cayley results look the most similar in
Fig.~\ref{sparse_with_u_comp_u=4v}, where $u = 4v$. However, upon
careful examination one notes that the CPA results are still
slightly asymmetric about $\omega = \Delta$. The peaks are also
slightly wider for the CPA results.

\section{Plots}
\label{sparse_with_u_plots}

\subsection{The averaged amplitude $\left\langle a_{0} \right\rangle$
as a function of time}

\begin{figure}
\resizebox{0.5\textwidth}{!}{\includegraphics[0in,0in][8in,10in]{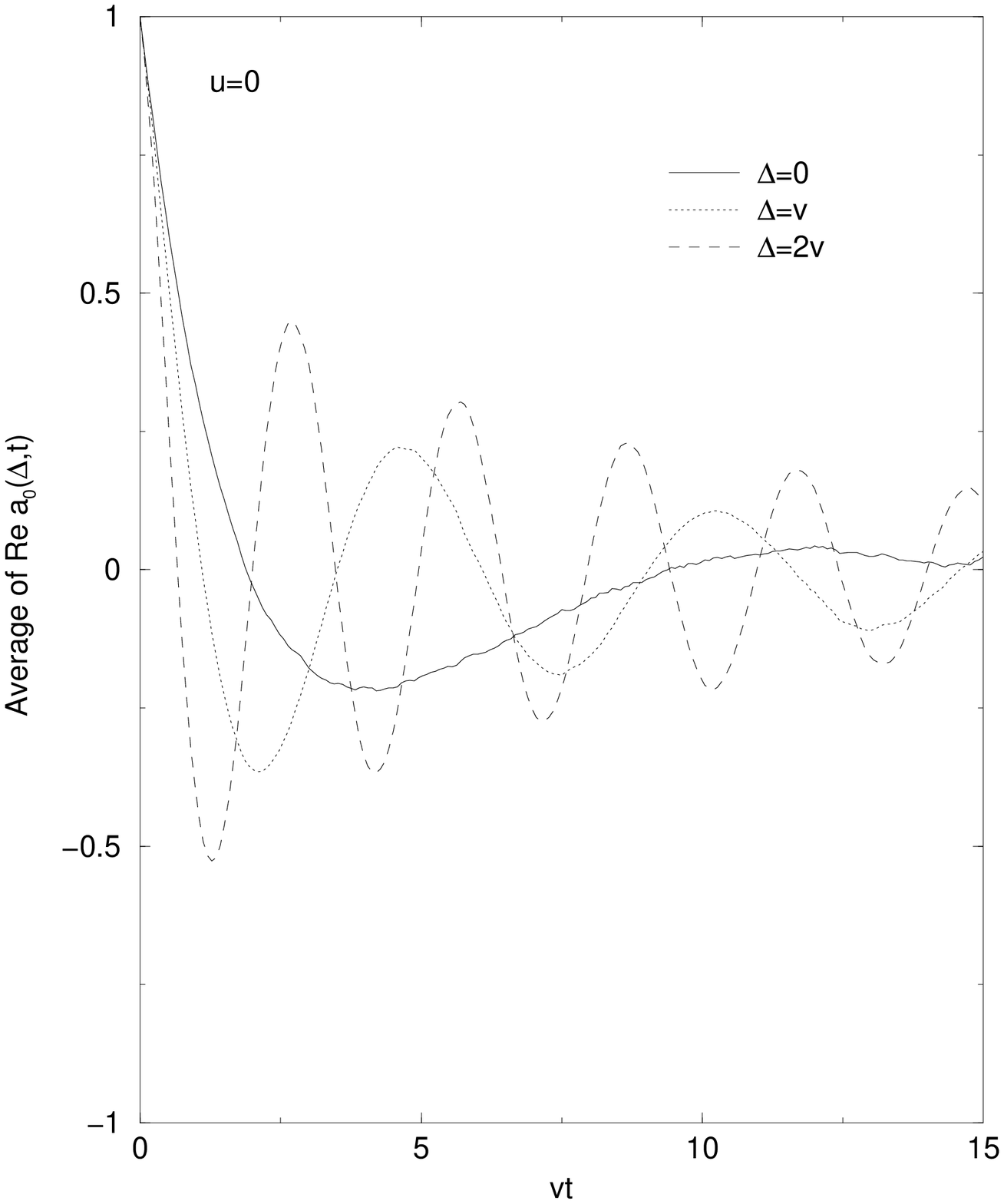}}
\resizebox{0.5\textwidth}{!}{\includegraphics[0in,0in][8in,10in]{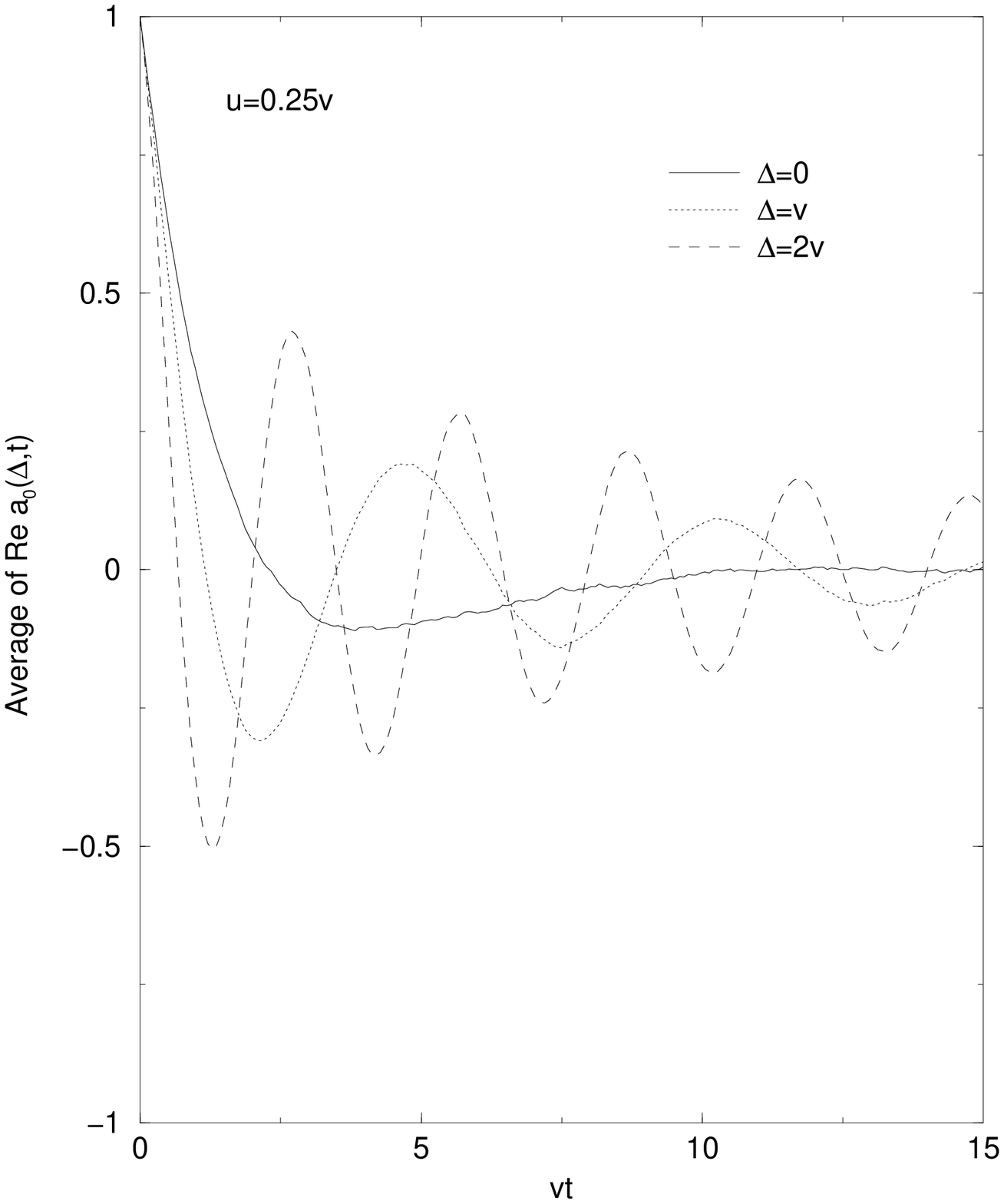}}
\\
\resizebox{0.5\textwidth}{!}{\includegraphics[0in,0in][8in,10in]{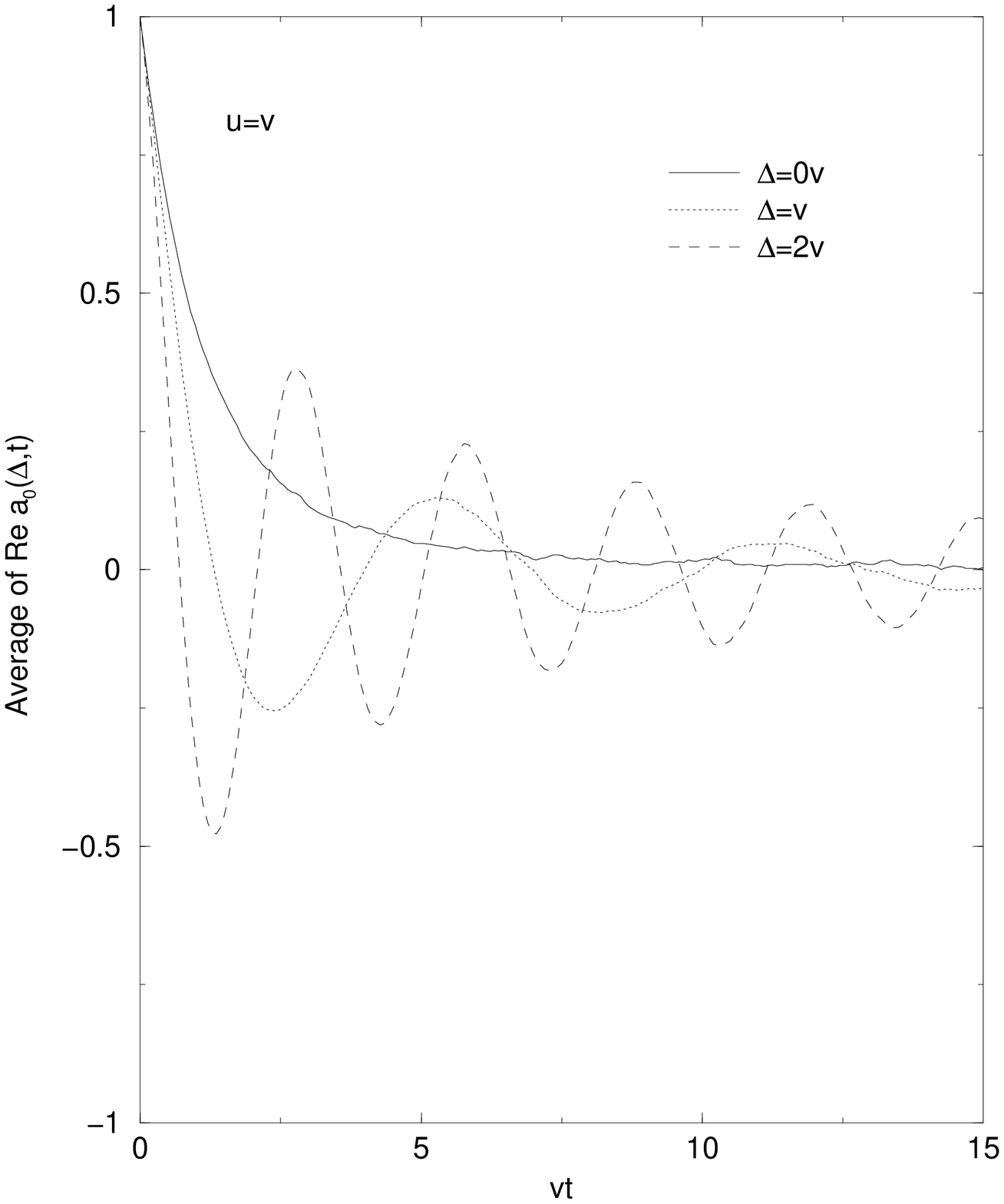}}
\resizebox{0.5\textwidth}{!}{\includegraphics[0in,0in][8in,10in]{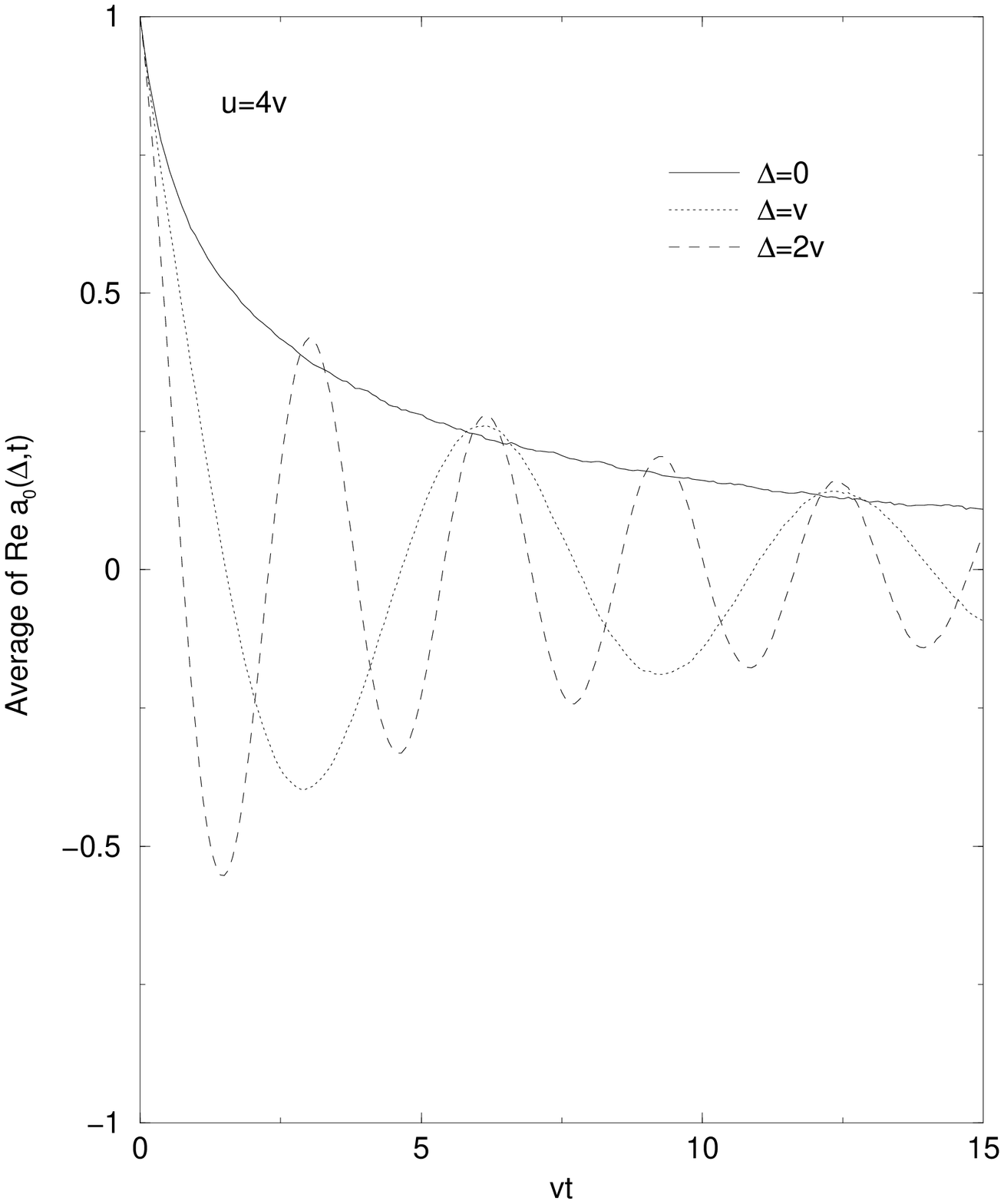}}
\caption[Real part of $\left\langle a_{0}\left(\Delta,t\right)
\right\rangle$]{Plots of the real part of $\left\langle
a_{0}\left(\Delta,t\right) \right\rangle$ for $u=0$, $u=0.25v$,
$u=v$, and $u=4v$.} \label{sparse_with_u_reala0}
\end{figure}

\begin{figure}
\resizebox{0.5\textwidth}{!}{\includegraphics[0in,0in][8in,10in]{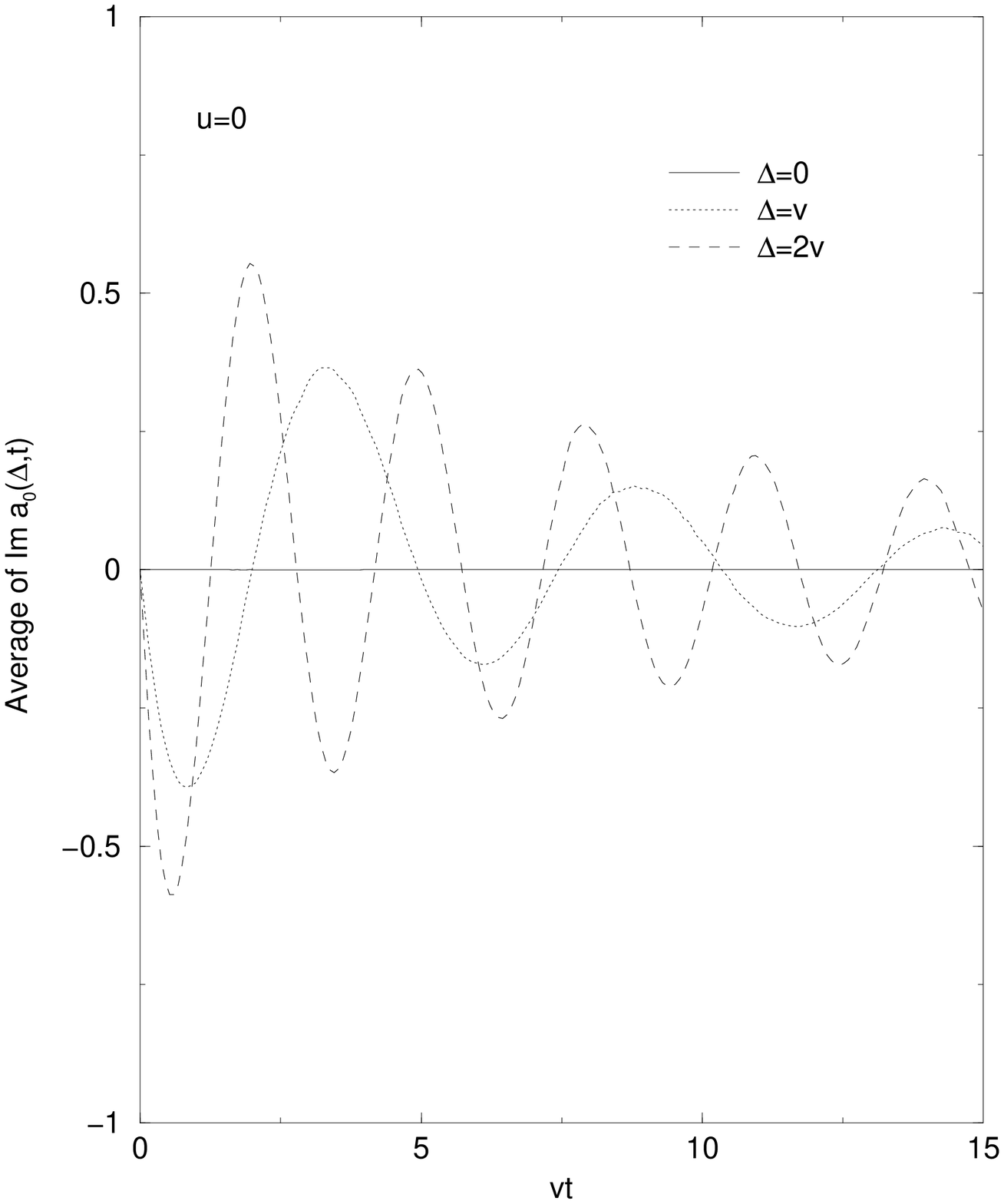}}
\resizebox{0.5\textwidth}{!}{\includegraphics[0in,0in][8in,10in]{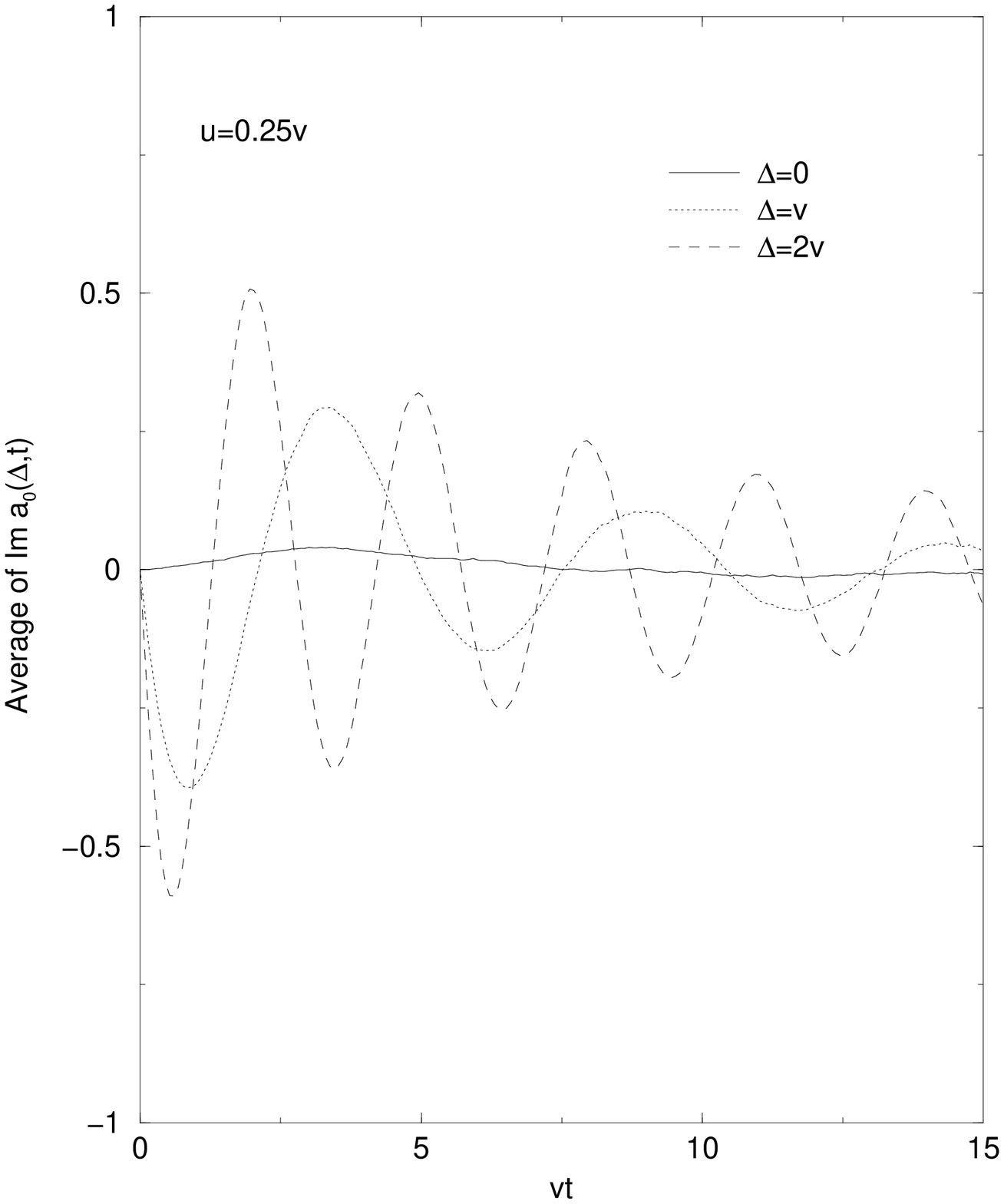}}
\\
\resizebox{0.5\textwidth}{!}{\includegraphics[0in,0in][8in,10in]{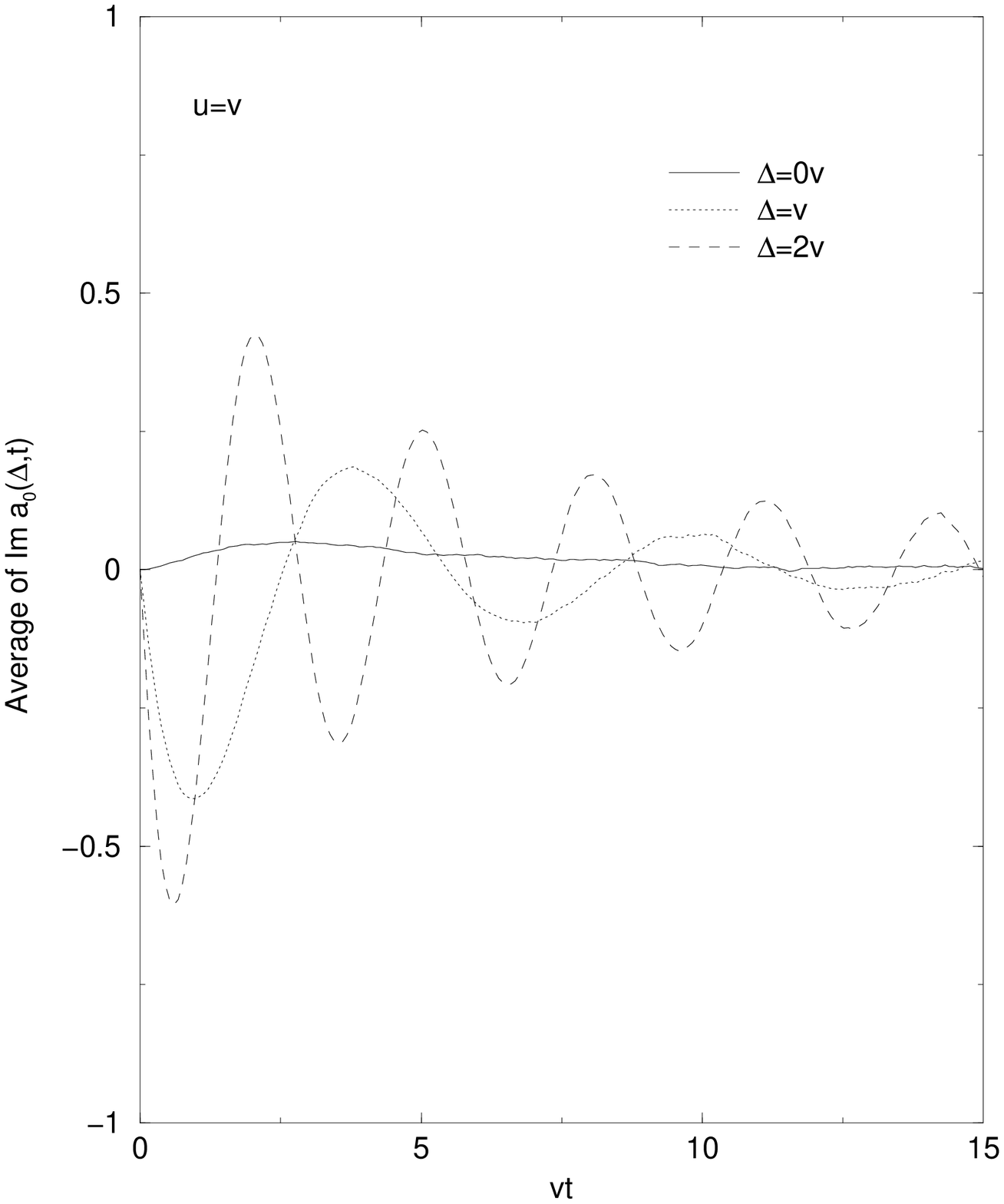}}
\resizebox{0.5\textwidth}{!}{\includegraphics[0in,0in][8in,10in]{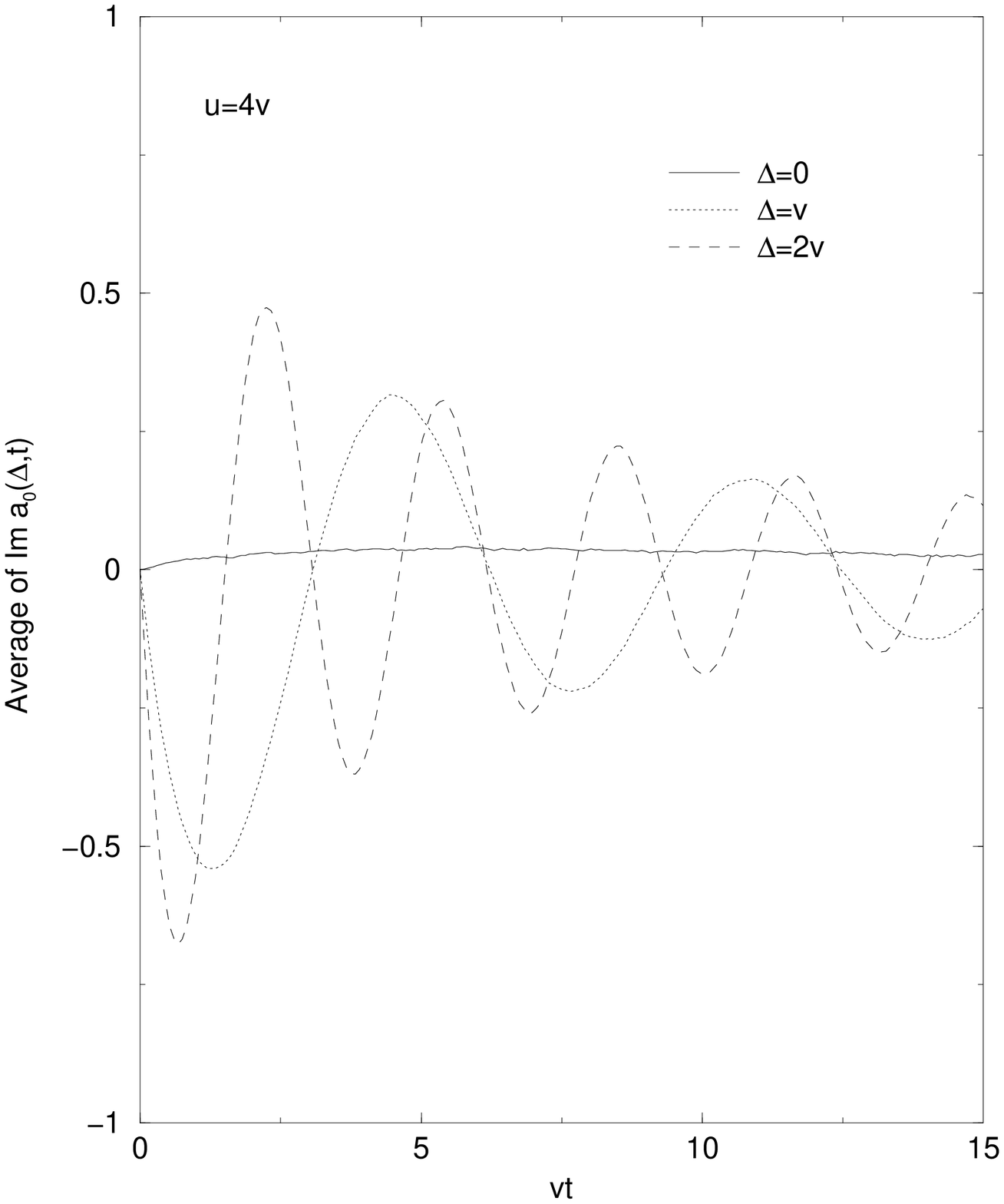}}
\caption[Imaginary part of $\left\langle
a_{0}\left(\Delta,t\right) \right\rangle$]{Plots of the imaginary
part of $\left\langle a_{0}\left(\Delta,t\right) \right\rangle$
for $u=0$, $u=0.25v$, $u=v$, and $u=4v$.}
\label{sparse_with_u_imaga0}
\end{figure}

Figs.~\ref{sparse_with_u_reala0} and \ref{sparse_with_u_imaga0}
display the numerical simulations of ${\mathrm{Re}}\left\langle
a_{0}\left(\Delta,t\right)\right\rangle$ and
${\mathrm{Im}}\left\langle
a_{0}\left(\Delta,t\right)\right\rangle$, respectively, for
$u/v=0,0.25,1$, and $4$. We recall that $\left\langle
a_{0}\left(\Delta,t\right)\right\rangle $ is the probability
amplitude that the $ss^{\prime}\rightarrow pp^{\prime}$ has not
occurred during the time $t$. The striking feature of these graphs
is that, off-resonance, oscillations with an approximate period
$\Delta$ persist for a long time, not only when the exciton does
not propagate $\left( u = 0 \right)$, but even for $u = 4v$.

The graphs for $u = 0$ can be obtained analytically by the methods
for computing $\left\langle S\left(\Delta,t\right)\right\rangle$
used in Ref.~\cite{Frasier1999a} and described in detail in
Chapter~\ref{sparse_no_u}. In fact, we have obtained $\left\langle
a_{0}\left(\Delta,t\right)\right\rangle$ by expanding
Eq.~(\ref{a0alpha}) (or, equivalently, Eq.~(\ref{a0exact})) in
powers of $1/\alpha$ and inverting the Laplace transform term by
term. As in Chapter~\ref{sparse_no_u}, various formulas involving
special functions can also be obtained. The exact analytical
results in this case (not shown here) agree with the numerical
simulations about as well as in
Fig.~\ref{simulation_exact_signal}. The initial dependence on time
is given by
\begin{equation}
\left\langle a_{0}\left(\Delta,t\right)\right\rangle \approx 1 -
\frac{\sqrt{\pi }}{2}vt - i\Delta t  \label{a0lin}
\end{equation}
Looking at the graphs, it seems that the initial decrease of
$\mrm{Re}\,\left\langle a_{0}\left(\Delta,t\right)\right\rangle$
depends on $u$, but this perception only indicates that the range
of validity of the linear approximation Eq.~(\ref{a0lin}) is
small. On the other hand, the graphs of $1 - \left\langle \left|
a_{0}\right|^{2}\right\rangle$ as a function of time in
Figs.~\ref{simulation_exact_signal}--\ref{simulation_u=4v_signal}
look linear and independent of $u$ over a wide range of $vt$. Note
also that $\left| \left\langle
a_{0}\left(\Delta,t\right)\right\rangle \right|^{2} \approx 1 -
\sqrt{\pi}vt$ decreases initially twice as fast as $\left\langle
\left| a_{0}\left(\Delta,t\right)\right|^{2}\right\rangle \approx
1 - \left(\sqrt{\pi}/2\right)vt$. Of course, it must be true that
the mod square of the average is smaller than or equal to the
average of the square, but the above result shows that $\left|
\left\langle a_{0}\left(\Delta,t\right)\right\rangle \right|^{2}$
is not even a fair approximation to $\left\langle \left|
a_{0}\left(\Delta,t\right)\right|^{2}\right\rangle$.

\subsection{The averaged amplitude $\left\langle a_{0} \right\rangle$
as a function of frequency}
\label{a0freq}

\begin{figure}
\resizebox{\textwidth}{!}{\includegraphics[0in,0in][8in,10in]{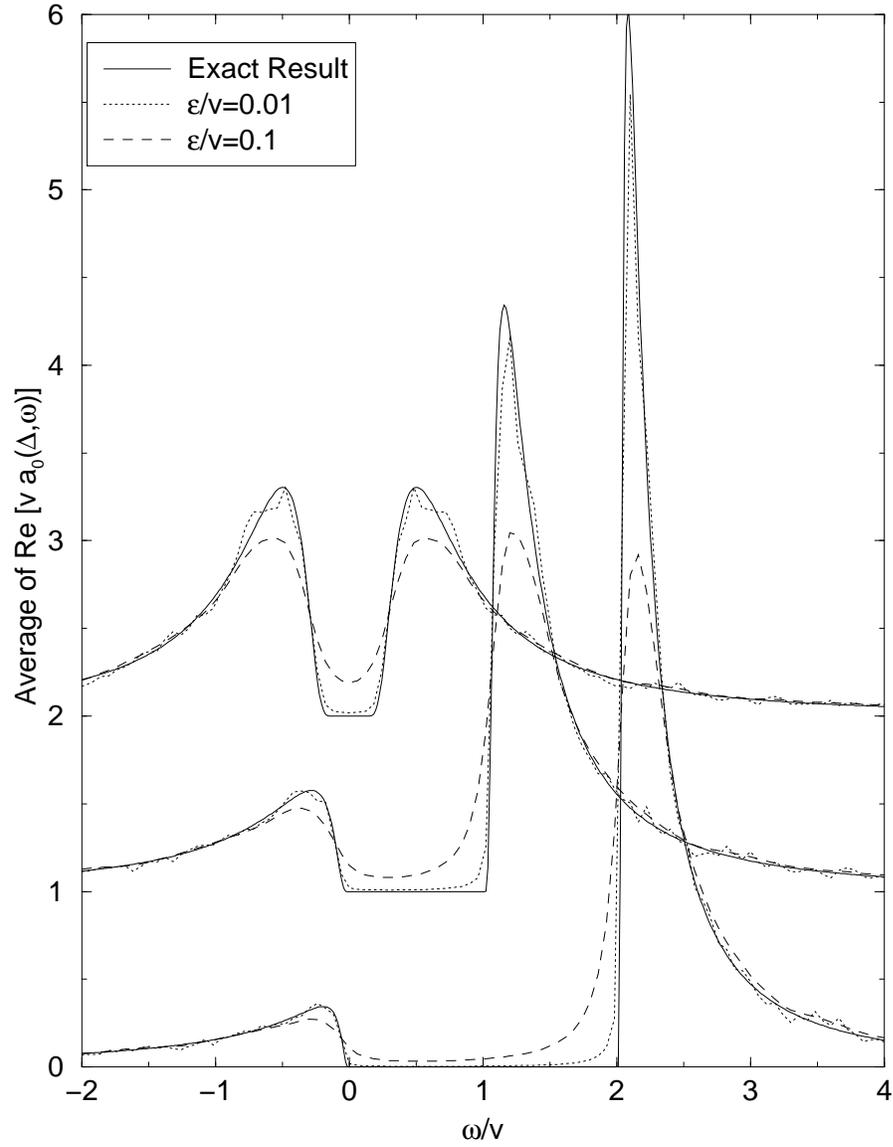}}
\caption[Real part of $\left\langle
a_{0}\left(\Delta,\omega\right)\right\rangle$ for $u=0$ ---
comparison with simulation]{A plot of the real part of
$\left\langle a_{0}\left(\Delta,\omega\right)\right\rangle$ for
$u=0$ and, from left to right, $\Delta = 0$, $v$, and $2v$. To aid
the eye in separating the curves, the $\Delta = 0$ curve has been
vertically offset by $2$ and the $\Delta = v$ curve by $1$.}
\label{sparse_with_u_a0w_u=0v}
\end{figure}

\begin{figure}
\resizebox{\textwidth}{!}{\includegraphics[0in,0in][8in,10in]{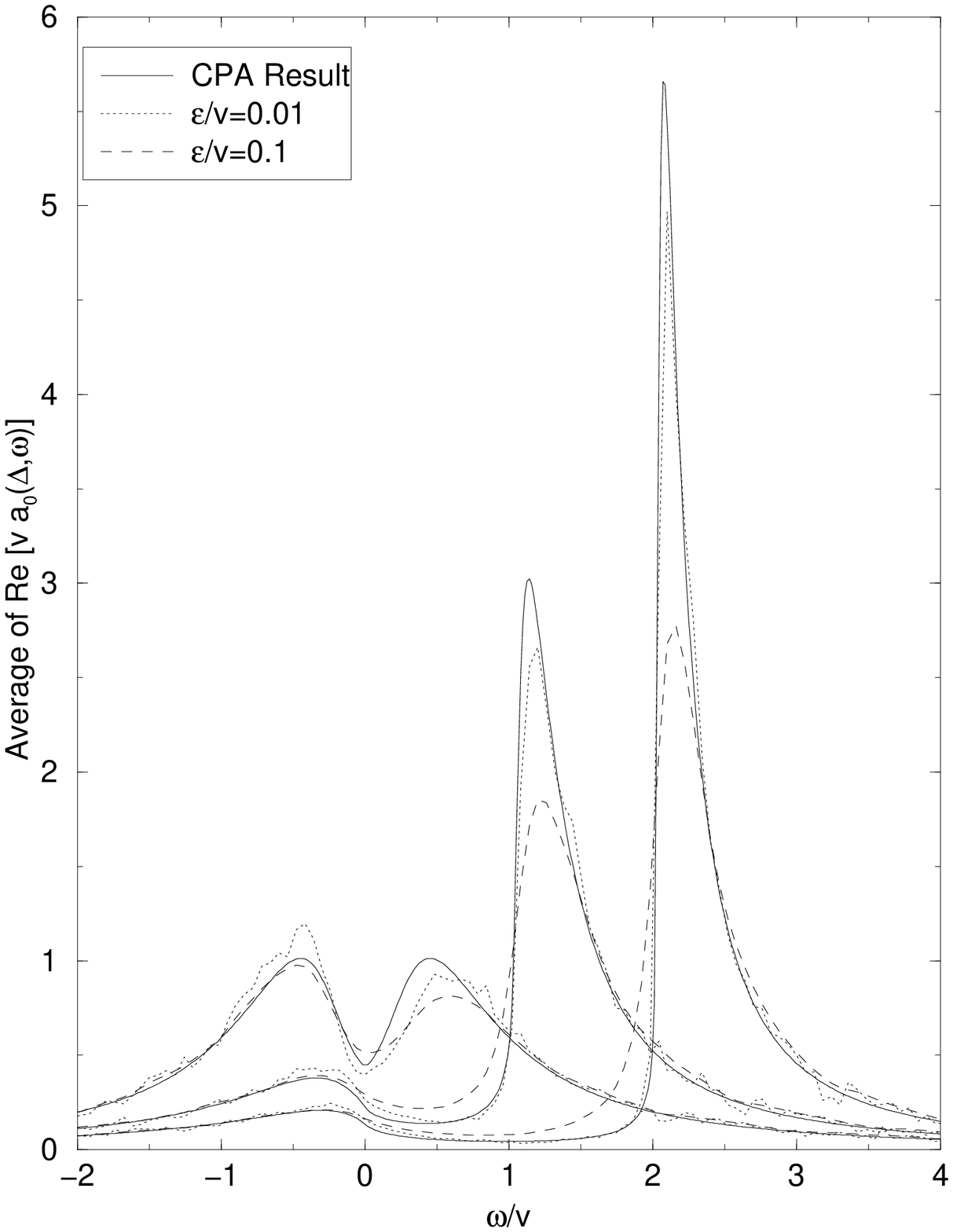}}
\caption[Real part of $\left\langle
a_{0}\left(\Delta,\omega\right)\right\rangle$ for $u=0.25v$ ---
comparison with simulation]{A plot of the real part of
$\left\langle a_{0}\left(\Delta,\omega\right)\right\rangle$ for
$u=0.25v$ and, from left to right, $\Delta = 0$, $v$, and $2v$.}
\label{sparse_with_u_a0w_u=0.25v}
\end{figure}

\begin{figure}
\resizebox{\textwidth}{!}{\includegraphics[0in,0in][8in,10in]{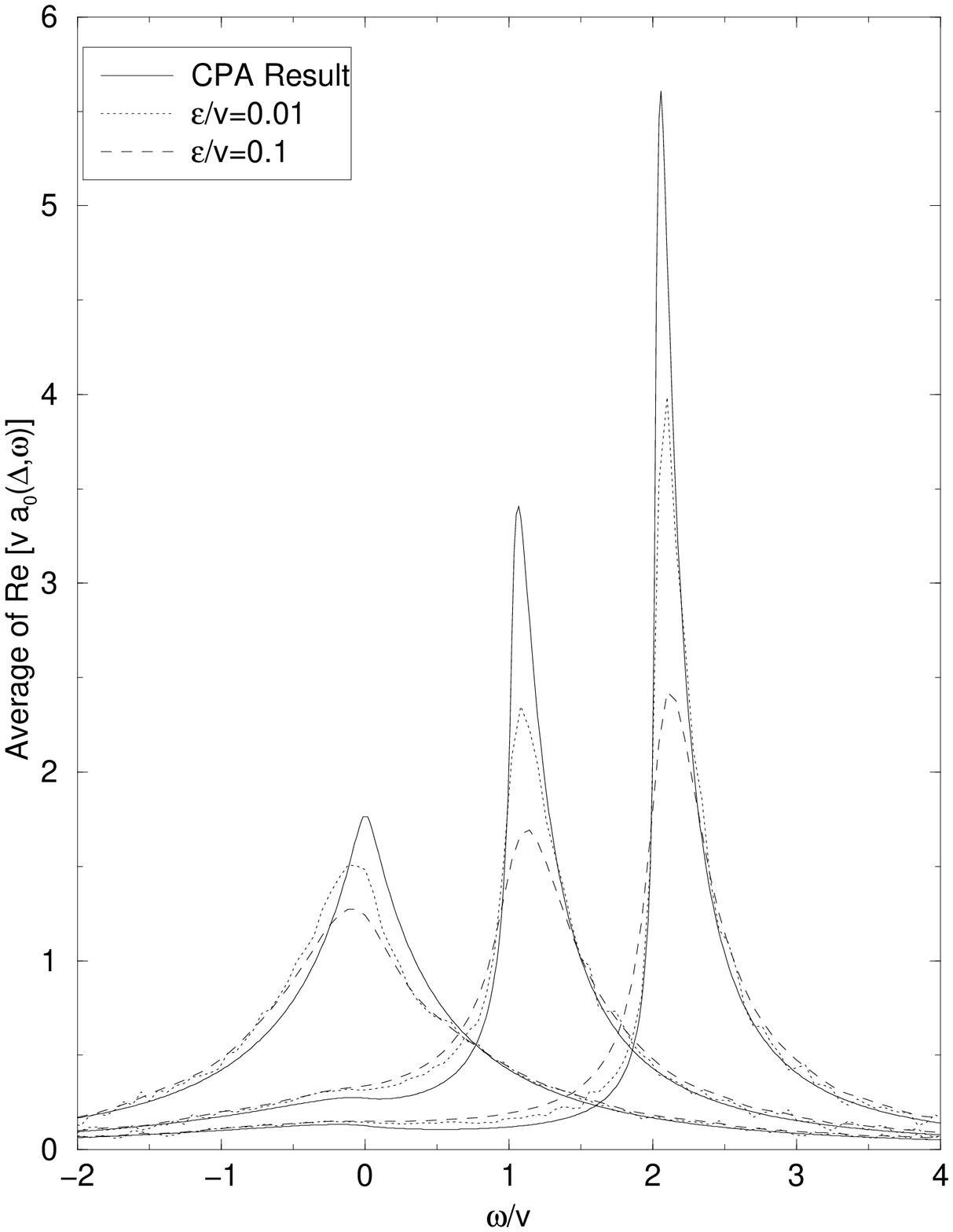}}
\caption[Real part of $\left\langle
a_{0}\left(\Delta,\omega\right)\right\rangle$ for $u=v$ ---
comparison with simulation]{A plot of the real part of
$\left\langle a_{0}\left(\Delta,\omega\right)\right\rangle$ for
$u=v$ and, from left to right, $\Delta = 0$, $v$, and $2v$.}
\label{sparse_with_u_a0w_u=v}
\end{figure}

\begin{figure}
\resizebox{\textwidth}{!}{\includegraphics[0in,0in][8in,10in]{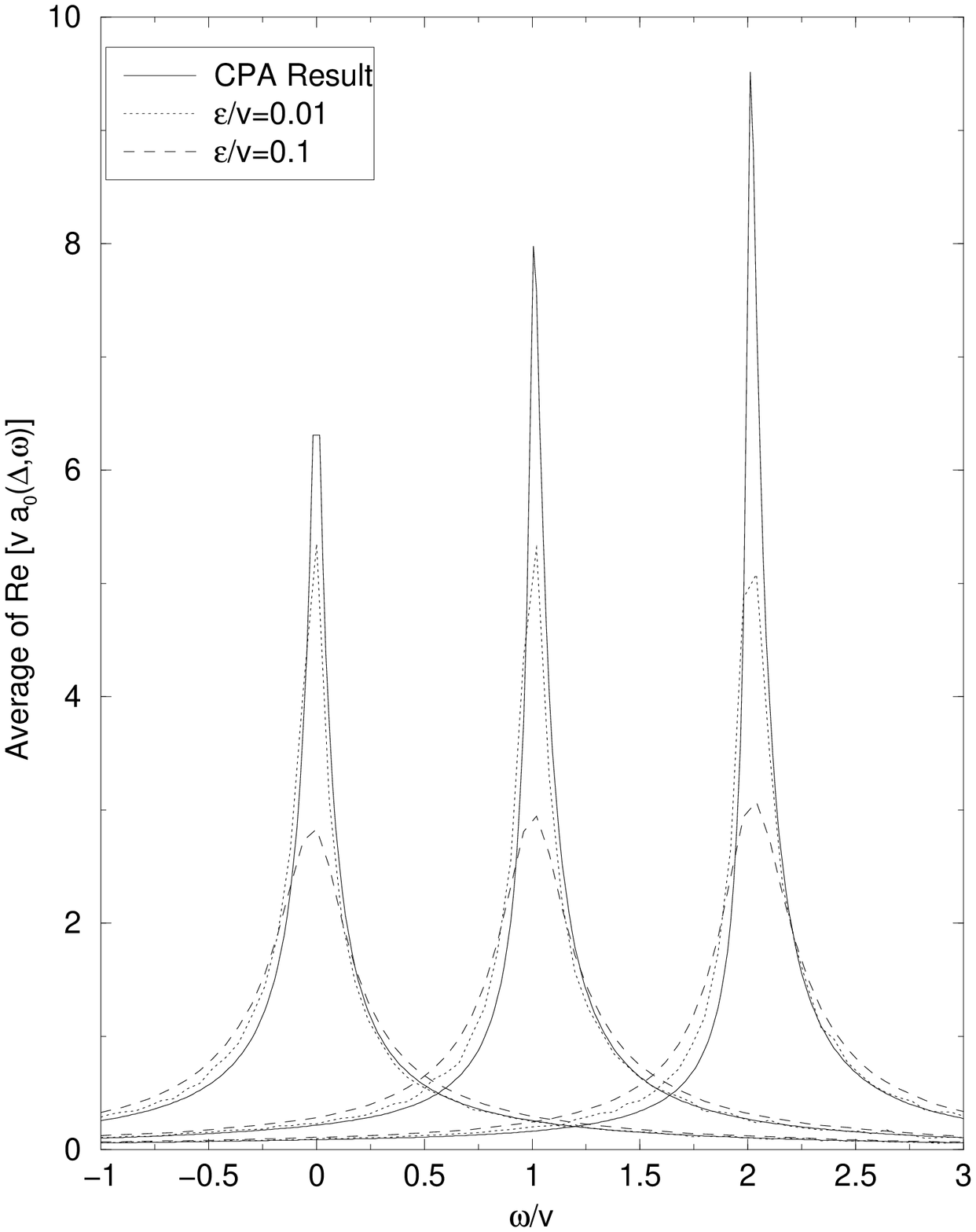}}
\caption[Real part of $\left\langle
a_{0}\left(\Delta,\omega\right)\right\rangle$ for $u=4v$ ---
comparison with simulation]{A plot of the real part of
$\left\langle a_{0}\left(\Delta,\omega\right)\right\rangle$ for
$u=4v$ and, from left to right, $\Delta = 0$, $v$, and $2v$.}
\label{sparse_with_u_a0w_u=4v}
\end{figure}

Figs.~\ref{sparse_with_u_a0w_u=0v}--\ref{sparse_with_u_a0w_u=4v}
present graphs of ${\mathrm{Re}}\left\langle
a_{0}\left(\Delta,\omega \right)\right\rangle $ that correspond to
the graphs of $\left\langle
a_{0}\left(\Delta,t\right)\right\rangle $ in
Figs.~\ref{sparse_with_u_reala0} and \ref{sparse_with_u_imaga0}.
For $u=v$ and $\Delta = 0$, ${\mathrm{Re}}\left\langle
a_{0}\left(\Delta,\omega\right)\right\rangle$ is the exciton
spectral density (times $\pi$); more generally, it is an
excitation spectral density for the resonant process
$ss^{\prime}\rightarrow pp^{\prime}$. Each figure gives a
comparison of the numerical simulation with the CPA approximation,
which in particular for $u=0$ reduces correctly to the real part
of the exact result of Eq.~(\ref{a0alpha}), with $\alpha =
\omega/i$. We compare spectral densities, rather than time graphs,
in part because obtaining $\left\langle
a_{0}\left(\Delta,t\right)\right\rangle$ for the CPA approximation
is laborious, and in part because the comparison of spectral
densities is instructive, as we now discuss.

In the notation of Section~\ref{descriptionofmethod}, the spectral
density is obtained numerically from Eq.~(\ref{azeroeqn}) as
\begin{equation}
\mrm{Re}\,a_{0}\left( \Delta,\omega +i\varepsilon \right) =
\mrm{Im}\sum_{m=0}^{N} \frac{\beta_{m0}\beta_{0m}}{\omega
-\lambda_{m}+i\varepsilon}
\end{equation}
in the limits $\varepsilon \rightarrow 0$ and $N\rightarrow
\infty$. In practice it is difficult to go beyond $N = 10^{3}$,
and one must choose $\varepsilon$ to be fairly large to obtain a
smooth plot for the spectral density where the eigenvalues are
sparse, even after averaging over $10^{4}$ realizations. On the
other hand, the distribution of eigenvalues for a random dipolar
gas is highly structured near $\omega = 0$, and if $\varepsilon$
is too large this structure will be missed. Rather than spending
much time to go to large $N$ and refine the numerical technique,
we set $N = 100$, plot the results for $\varepsilon/v = 0.1$ and
$0.01$, and let the eye extrapolate to the correct limit. It is
remarkable that the CPA approximation agrees best with the
smallest value of $\varepsilon/v$, except for wiggles that are
usually in regions of sparse eigenvalues. This is exactly how the
exact spectral density is expected to compare with a calculation
for finite $N$. A finite $\varepsilon$ can also be viewed as a
simulation of the decoherence introduced by atomic motions, or of
losses to other channels, such as the decay of the excited atomic
states due to blackbody radiation as described in Chapter~5 of
Ref.~\cite{Gallagher1994a}. Thus, $\varepsilon$ is almost
equivalent to the $\gamma$ of the Lorentzian broadening model of
Ref.~\cite{Frasier1999a}, the difference being that in
Ref.~\cite{Frasier1999a} $i\gamma$ is added to $\omega$ for the
Fourier transform of Eq.~(\ref{ckdot}) only, while here
$i\varepsilon$ is added to every $\omega$ that appears. As
$\varepsilon$ increases, the highly structured spectral densities
of the ideal frozen gas become increasingly similar to Lorentzian
lines. Because of unavoidable losses and incoherences, the
experimental spectral densities will resemble the results of the
simulations for some finite $\varepsilon$.

In Fig.~\ref{sparse_with_u_a0w_u=0v}, where results are presented
for $u = 0$, the spectral density is quite different from a
Lorentzian. At $\Delta = 0$ there is a quasi-gap around $\omega =
0$. As $\Delta$ increases the gap extends from $\omega = 0$ to
$\omega = \Delta$, as was described in
Section~\ref{sparse_with_u_comp}.

In Fig.~\ref{sparse_with_u_a0w_u=0.25v} are presented results for
$u = 0.25v$. We see that the sharp features of the $u=0$ spectrum
are rounded off and the gap is beginning to be filled, but is
still recognizable.

Results for $u=v$ are presented in
Fig.~\ref{sparse_with_u_a0w_u=v}. We note that the spectrum at
$\Delta = 0$ has the appearance of a single, asymmetric line. At
$\Delta = v$ and $\Delta = 2v$ a shoulder on the left of the line
is the remnant of the gap of Fig.~\ref{sparse_with_u_a0w_u=0v}.

Fig.~\ref{sparse_with_u_a0w_u=4v} contains results for $u = 4v$.
We see that there is a single line for any value of $\Delta$,
which becomes increasingly sharp as $u$ increases. The lineshape
is not Lorentzian, but even a small additional broadening will
make it nearly so, as indicated by the calculation for
$\varepsilon/v = 0.01$. This sharp line leads to the damped
sinusoidal oscillations seen in the $u=4v$ graphs of
Figs.~\ref{sparse_with_u_reala0} and \ref{sparse_with_u_imaga0}.
The general phenomenon of line narrowing can be discussed using
the simple Lorentzian model for the exciton band, in which
Eq.~(1b) of Ref.~\cite{Frasier1999a} is replaced by
\begin{equation}
i\dot{c_{k}} = V_{k}a_{0} - i\gamma c_{k}.
\end{equation}
Solving the equations in this limit, one finds a pole at
\begin{equation}
\omega = -i\frac{\gamma\sum_{k}V_{k}^{2}}{\gamma^{2} + \Delta^{2}}
+ \Delta,
\end{equation}
which gives a decreasing width as $u$ increases, since $\gamma$ is
proportional to $u$. One cannot simply replace $\sum_{k}V_{k}^{2}$
by its average, which does not even exist. Instead, one can
Fourier transform the average to find the time dependence
\begin{equation}
\exp\left(-i\Delta t - \sqrt{\gamma_{eq}t}\right),
\end{equation}
where
\begin{equation}
\gamma_{eq} = \frac{v^{2}\gamma}{\gamma^{2} + \Delta^{2}},
\end{equation}
and $v$ has been defined in Eq.~(\ref{v}). A similar argument was
applied in Ref.~\cite{Frasier1999a} to find the time dependence of
$\left\langle\left|a_{0}\right|^{2}\right\rangle$.

From the comparison with the simulations, it is apparent that a
shortcoming of the CPA approximation is that it gives a symmetric
spectral density at resonance $\left(\Delta =0\right)$, while the
correct spectral density is slightly asymmetric, except for $u=0$.
Correspondingly, $\left\langle
a_{0}\left(\Delta,t\right)\right\rangle $ is not purely real at
resonance, as can be seen in Fig.~\ref{sparse_with_u_imaga0}. For
$\Delta \neq 0$, the spectral density at $\omega < \Delta$ is
underestimated by the CPA approximation.

Clearly, the asymmetry of the spectral density at $\Delta =0$
comes from the terms with $l\neq m$ in the denominator of
Eq.~(\ref{gfv}), which are neglected in the CPA approximation. It
is surprising that these terms have such small effect on
$\left\langle a_{0}\left(\Delta,\omega\right)\right\rangle$ at
$\Delta = 0$. At finite $\Delta$, the underestimate of
${\mathrm{Re}}\left\langle
a_{0}\left(\Delta,\omega\right)\right\rangle $ for $\omega <
\Delta$ is probably due to the neglect of these same terms. We
have already shown how to include these terms in the CPA
formulation \cite{Celli1999a}, but in practice, when the simple
CPA is not adequate one can just as well use the numerical
simulations.

\section{Summary}
\label{sparse_with_u_conclusions}

In this chapter we have extended the simple model of
Chapter~\ref{sparse_no_u} by allowing the $sp\rightarrow ps$
process to take place. Although this complication no longer allows
a direct analytical solution of the problem, we consider three
approximations, each of which allows us to make some progress. We
compare these approximations with the results of numerical
simulations, which are carried out for a finite number of atoms by
the methods described later in Chapter~\ref{simulations} of this
thesis.

The results of the Cayley tree approximation are not found to give
the best results for the problem we are considering here, but they
are still extremely worthwhile results. They extend the Cayley
tree results of Ref.~\cite{Abou-Chacra1973a}, which are used
extensively, for example, in the study of dipolar
liquids~\cite{Chandler1982a,Hall1985a,Hoye1981a,Hoye1982a,Logan1987a,
Logan1987b,Thompson1982a}.

The results of the Hubbard approximation are somewhat
disappointing, since in this case we can obtain an explicit
expression for the exciton spectrum $\left( 1/f_{l} \right)$ but
we are not able to obtain a consistent result when the effects of
the $ss^{\prime}\rightarrow pp^{\prime}$ process are included.
However, in this approximation we can obtain the average of
$\left| a_{0} \right|^{2}$ as a function of time when the $V$
process is absent, and perhaps in the future this will lead to a
more general technique that will allow this quantity to be
computed in the other approximations as well.

The CPA approximation is found to agree best with the results of
the numerical simulations, and we are able to carry our
computations in this approximation just as far as in the Cayley
tree approximation. It seems at first strange that the CPA
approximation gives better results than the Cayley, since in the
Cayley equations (\ref{fzero}) and (\ref{fk}) the on-site Green's
functions for the $U$ problem are allowed to vary from site to
site, whereas in Eqs.~(\ref{fhubbard}) and (\ref{fcpa}) they are
not. However, the Cayley tree approximation is exact in a
one-dimensional system with nearest neighbor interactions and
becomes increasingly inaccurate as the dimensionality of the
problem and the range of the interaction increase. The CPA, on the
other hand, is a mean field approximation, and it becomes accurate
for higher dimensional problems and long-range interactions. It
may also happen that, when the off-diagonal terms of
Eqs.~(\ref{gfv}) and (\ref{gfu}) are included, the Cayley is found
to be the better approximation.

\chapter{Analogy With Spin Glasses}
\markright{Chapter \arabic{chapter}: An Analogy With Spin Glasses}
\label{cesiumappendix}

\section{Introduction}

In this chapter, we examine a correspondence between the atomic
system we are seeking to model and spin glasses. We start in
Section~\ref{Rbspinglass} by considering the rubidium experiments
of Anderson \emph{et al.\/}~\cite{Anderson1996a,Anderson1998a} and
Lowell \emph{et al.\/}~\cite{Lowell1998a}. In
Section~\ref{Rbspinglass1} we derive in a rather straightforward
manner a Hamiltonian that describes these experiments as a system
of two interpenetrating spin glasses. We also determine the
corresponding equations of motion. In Section~\ref{reduction} we
illustrate how the results of Section~\ref{Rbspinglass1} reduce to
those of Chapter~\ref{sparse_with_u} in the appropriate limit. We
derive a new, and seemingly more cumbersome Hamiltonian in
Section~\ref{Rbspinglass2} that also describes the rubidium
experiments and is entirely equivalent to the Hamiltonian of
Section~\ref{Rbspinglass1}. The method used to generate the new
Hamiltonian is more generally applicable, however, as we
demonstrate in Section~\ref{Csspinglass} by using the new method
to construct a spin Hamiltonian that describes the cesium
experiments of Mourachko \emph{et
al.\/}~\cite{Mourachko1998a,Mourachko1999a}. Finally, we conclude
in Section~\ref{spinglasssummary}.

\section{The rubidium system as a spin glass}
\label{Rbspinglass}

\subsection{Spin glass Hamiltonian and equations of motion}
\label{Rbspinglass1}

We consider $N$ atoms at positions $\mbf{r}_{k}$ which are
initially in state $s$, and $N^{\prime}$ atoms at positions
$\mbf{r}_{k^{\prime}}$ which are initially in state $s^{\prime}$.
We represent the $s$ and $p$ states at $\mbf{r}_{k}$ with the down
and up states of an effective spin $\mbf{\sigma}_{k}$,
respectively, and similarly represent the $s^{\prime}$ and
$p^{\prime}$ states at $\mbf{r}_{k^{\prime}}$ with a spin
$\mbf{\sigma}_{k^{\prime}}$. The Hamiltonian that describes this
system with the $ss^{\prime}\rightarrow pp^{\prime}$ and
$sp\rightarrow ps$ processes mediated by dipole-dipole interaction
potentials $V$ and $U$, respectively, is then
\begin{eqnarray}
H &=& \sum_{k=1}^{N}\left[ \varepsilon_{s} + \left(
\varepsilon_{p} - \varepsilon_{s}\right)
\sigma_{k}^{+}\sigma_{k}^{-}\right] +
\sum_{k^{\prime}=1}^{N^{\prime}} \left[ \varepsilon_{s}^{\prime} +
\left( \varepsilon_{p}^{\prime} - \varepsilon_{s}^{\prime} \right)
\sigma_{k^{\prime}}^{+}\sigma_{k^{\prime}}^{-} \right] + \nonumber
\\ && \sum_{k,k^{\prime}} V_{kk^{\prime}}\left[ \sigma_{k}^{+}
\sigma_{k^{\prime}}^{+} +
\sigma_{k}^{-}\sigma_{k^{\prime}}^{-}\right] + \sum_{k,l\neq k}
U_{kl}\sigma_{k}^{+}\sigma_{l}^{-},  \label{H}
\end{eqnarray}
where
\begin{equation}
V_{kk^{\prime}} = \frac{\mu\mu^{\prime}}{\left| \mbf{r}_{k} -
\mbf{r}_{k^{\prime}}\right|^{3}}, \label{yetanotherV}
\end{equation}
and
\begin{equation}
U_{kl} = \frac{\mu^{2}}{\left| \mbf{r}_{k} -
\mbf{r}_{l}\right|^{3}}, \label{yetanotherU}
\end{equation}
if $l\neq k$ and $U_{kl} = 0$ otherwise.\footnote{Of course, the
$1/r^{3}$ potentials of Eqs.~(\ref{yetanotherV}) and
(\ref{yetanotherU}) are just the simplest examples of
dipole-dipole interaction potentials. We are free to choose more
complicated potentials if we so desire.} This is precisely the
Hamiltonian for two interpenetrating spin glasses that interact
via dipole-dipole potentials.

For the experiments under consideration $\mu \approx
4\mu^{\prime}$, and the $sp\rightarrow ps$ coupling $\mu^{2}$,
which leads to ``spin diffusion,'' is larger than the
$ss^{\prime}\rightarrow pp^{\prime}$ coupling $\mu\mu^{\prime}$
that is responsible for the resonant energy transfer.

We can write down basic equations of motion such as
\begin{eqnarray}
i\frac{d}{dt}\sigma _{k^{\prime}}^{+} &=&
\Delta\sigma_{k^{\prime}}^{+} + \sigma_{k^{\prime}}^{z}
\sum_{k=1}^{N} V_{k^{\prime}k}\sigma_{k}^{-}, \label{s'+} \\
i\frac{d}{dt}\sigma_{k}^{+} &=& \sigma_{k}^{z}\sum_{k^{\prime}}
V_{kk^{\prime}}\sigma_{k^{\prime}}^{-} + \sigma_{k}^{z}\sum_{l\neq
k} U_{kl}\sigma_{l}^{+}, \label{s+}
\end{eqnarray}
and
\begin{equation}
i\frac{d}{dt}\sigma_{k^{\prime}}^{z} = 2\sum_{k=1}^{N}
V_{k^{\prime}k} \left(\sigma_{k}^{+}\sigma _{k^{\prime}}^{+} -
\sigma_{k}^{-}\sigma_{k^{\prime}}^{-}\right), \label{sz}
\end{equation}
where $\Delta $ is the detuning from resonance.\footnote{To obtain
these expressions we take the equations of motion that come about
naturally from the Hamiltonian in Eq.~(\ref{H}) and introduce a
phase $\exp\left[ -i\left(\epsilon_{p} - \epsilon_{s}\right)
t\right]$ into the definitions of $\sigma_{k}^{+}$ and
$\sigma_{k^{\prime}}^{+}$.} From these we can obtain
\begin{eqnarray}
\frac{1}{2}\frac{d^{2}}{dt^{2}}\sigma_{k^{\prime}}^{z} &=&
-\sum_{kl^{\prime}} V_{k^{\prime}k}V_{kl^{\prime}} \sigma_{k}^{z}
\left( \sigma_{k^{\prime}}^{-}\sigma_{l^{\prime}}^{+} +
\sigma_{k^{\prime}}^{+}\sigma_{l^{\prime}}^{-}\right) -\sum_{kl}
V_{k^{\prime}k}V_{lk^{\prime}}\sigma_{k^{\prime}}^{z} \left(
\sigma_{k}^{-}\sigma_{l}^{+} + \sigma_{k}^{+}\sigma_{l}^{-}\right)
+  \nonumber \\  && -\sum_{k,l\neq k}
V_{k^{\prime}k}U_{kl}\sigma_{k}^{z}
\left(\sigma_{k^{\prime}}^{+}\sigma_{l}^{+} +
\sigma_{k^{\prime}}^{-}\sigma_{l}^{-}\right) - \Delta \sum_{k}
V_{k^{\prime}k}\left( \sigma_{k^{\prime}}^{+} \sigma_{k}^{+} +
\sigma_{k^{\prime}}^{-}\sigma_{k}^{-}\right). \label{s2'}
\end{eqnarray}

These differential equations are rather difficult to work with,
but we have already discussed one possible approximation scheme in
Ref.~\cite{Frasier1999a}.

\subsection{Reduction to the sparse limit}
\label{reduction}

We now show explicitly how the general spin-variable formalism of
Section~\ref{Rbspinglass1} relates to the coefficients $a_{0}$ and
$c_{k}$ of Eqs.~(\ref{azerodotU}) and (\ref{ckdotU}) when there is
only a single atom that can be in the $s^{\prime}$, $p^{\prime}$
pair of states (i.e., when only a single spin variable of the type
$\sigma_{k^{\prime}}$ is present.) The key to the correspondence
is that in this case the initial state $\left| i\right\rangle$,
which consists of all spins down, evolves to
\begin{equation}
\left| t\right\rangle = \left( a_{0}\left(\Delta,t\right) +
\sum_{q}c_{q}\left(\Delta,t\right)\sigma_{k^{\prime}}^{+}\sigma_{q}^{+}\right)
\left| i\right\rangle,
\end{equation}
at time $t$, where the spin variables now denote time-independent
Schr\"{o}dinger operators. With the understanding that
$\mbf{r}_{k^{\prime}} = 0$, we have:
\begin{equation}
\left\langle i \left| \sigma_{k^{\prime}}^{z}\left(t\right)\right|
i \right\rangle = \left\langle t\left|
\sigma_{k^{\prime}}^{z}\right| t\right\rangle = 1 - 2\left|
a_{0}\left(\Delta,t\right)\right| ^{2},
\end{equation}
\begin{equation}
\left\langle t\left| \sigma_{k}^{z} \right| t \right\rangle =
2\left| c_{k}\left(\Delta,t\right)\right|^{2} - 1,
\end{equation}
and
\begin{equation}
\left\langle t \left| \sigma_{k^{\prime}}^{+}\sigma_{k}^{+}\right|
t \right\rangle =
c_{k}^{\dagger}\left(\Delta,t\right)a_{0}\left(\Delta,t\right).
\end{equation}

It is then clear that the expectation value of Eq.~(\ref{sz}) is
equivalent to
\begin{equation}
i\frac{d}{dt}\left| a_{0}\left(\Delta,t\right)\right|^{2} =
\sum_{k} V_{k}\left[
a_{0}^{\dagger}\left(\Delta,t\right)c_{k}\left(\Delta,t\right) -
c_{k}^{\dagger}\left(\Delta,t\right)a_{0}\left(\Delta,t\right)
\right],
\end{equation}
where $V_{k}$ is short for $V_{k^{\prime}k}$. This equation is
itself a consequence of Eq.~(\ref{azerodotU}). Similarly, the
equations for $\left\langle t \left| \sigma_{k}^{z} \right| t
\right\rangle$ and $\left\langle t \left|
\sigma_{k^{\prime}}^{+}\sigma_{k}^{+}\right| t \right\rangle$ are
equivalent to the equations for $\left|
c_{k}\left(\Delta,t\right)\right|^{2}$ and
$c_{k}\left(\Delta,t\right)^{\dagger}a_{0}\left(\Delta,t\right)$
that follow from Eqs.~(\ref{azerodotU}) and (\ref{ckdotU}).

\subsection{Another spin glass Hamiltonian for the rubidium system}
\label{Rbspinglass2}

In the rubidium experiments of Anderson \emph{et
al.\/}~\cite{Anderson1996a,Anderson1998a} and Lowell \emph{et
al.\/}~\cite{Lowell1998a} there are the four states $s$, $p$,
$s^{\prime}$, and $p^{\prime}$. Therefore we can define a column
vector of length four for each site, so that if an atom at site
$\Lambda$ is in the $s$, $p$, $s^{\prime}$, or $p^{\prime}$ state,
then the column vector corresponding to that site is $\left(
0,0,0,1\right)^{T}$, $\left( 0,0,1,0\right)^{T}$, $\left(
0,1,0,0\right)^{T}$, or $\left( 1,0,0,0\right)^{T}$, respectively.
The Hamiltonian for this system can then be written
\begin{eqnarray}
H_{\mrm{Rb}} &=& \sum_{\Lambda} \epsilon_{\Lambda} +
\sum_{\Lambda_{1},\Lambda_{2}} V_{\Lambda_{1}\Lambda_{2}} \left[
\left(\sigma^{+}_{sp}\right)_{\Lambda_{1}}
\left(\sigma^{+}_{s^{\prime}p^{\prime}}\right)_{\Lambda_{2}} +
\mrm{h.c.} \right] + \nonumber \\ &&
\sum_{\Lambda_{1},\Lambda_{2}} U_{\Lambda_{1}\Lambda_{2}} \left[
\left(\sigma^{+}_{sp}\right)_{\Lambda_{1}}
\left(\sigma^{-}_{sp}\right)_{\Lambda_{2}} + \mrm{h.c.}\right],
\label{RbHamiltonian}
\end{eqnarray}
where $\epsilon$, $\sigma^{\pm}_{sp}$, and
$\sigma^{\pm}_{s^{\prime}p^{\prime}}$ are the $4\times 4$ matrices
\begin{equation}
\epsilon = \left(
\begin{array}{cccc}
\epsilon_{p^{\prime}} & 0 & 0 & 0 \\ 0 & \epsilon_{s^{\prime}} & 0
& 0 \\ 0 & 0 & \epsilon_{p} & 0 \\ 0 & 0 & 0 & \epsilon_{s} \\
\end{array}
\right),
\end{equation}
\begin{equation}
\sigma^{\pm}_{sp} = \left(
\begin{array}{cc}
\begin{array}{cc} 0 & 0 \\ 0 & 0 \end{array} &
\begin{array}{cc} 0 & 0 \\ 0 & 0 \end{array} \\
\begin{array}{cc} 0 & 0 \\ 0 & 0 \end{array} &
\sigma^{\pm}
\end{array}
\right),
\end{equation}
and
\begin{equation}
\sigma^{\pm}_{s^{\prime}p^{\prime}} = \left(
\begin{array}{cc}
\sigma^{\pm} &
\begin{array}{cc} 0 & 0 \\ 0 & 0 \end{array} \\
\begin{array}{cc} 0 & 0 \\ 0 & 0 \end{array} &
\begin{array}{cc} 0 & 0 \\ 0 & 0 \end{array}
\end{array}
\right),
\end{equation}
and the subscripts on these matrices indicate on which site they
are to be applied. The quantities $V_{\Lambda_{1}\Lambda_{2}}$ and
$U_{\Lambda_{1}\Lambda_{2}}$ are defined to be zero if
$\Lambda_{1} = \Lambda_{2}$, and are the interaction potentials
\begin{equation}
V_{\Lambda_{1}\Lambda_{2}} = \frac{\mu\mu^{\prime}}{\left|
\mbf{r}_{\Lambda_{1}} - \mbf{r}_{\Lambda_{2}}\right|^{3}},
\label{appV}
\end{equation}
and
\begin{equation}
U_{\Lambda_{1}\Lambda_{2}} = \frac{\mu^{2}}{\left|
\mbf{r}_{\Lambda_{1}} - \mbf{r}_{\Lambda_{2}}\right|^{3}},
\label{appU}
\end{equation}
otherwise. In the interest of completeness, we note that we could
also include the $s^{\prime}p^{\prime}\rightarrow
p^{\prime}s^{\prime}$ interaction by adding a term
\begin{equation}
\sum_{\Lambda_{1},\Lambda_{2}} W_{\Lambda_{1}\Lambda_{2}} \left[
\left(\sigma^{+}_{s^{\prime}p^{\prime}}\right)_{\Lambda_{1}}
\left(\sigma^{-}_{s^{\prime}p^{\prime}}\right)_{\Lambda_{2}} +
\mrm{h.c.}\right]
\end{equation}
to the Hamiltonian of Eq.~(\ref{RbHamiltonian}).

The Hamiltonian of Eq.~(\ref{RbHamiltonian}) is entirely
equivalent to the Hamiltonian of Eq.~(\ref{H}). The fact that
Eq.~(\ref{H}) contains one set of spins corresponding to the
primed sites and another set of spins corresponding to the
unprimed sites is seen to arise from the fact that
Eq.~(\ref{RbHamiltonian}) can be viewed as containing two distinct
types of operators. One type operates only on the $s^{\prime}$ and
$p^{\prime}$ states and hence only on primed sites, while the
other acts only on the $s$ and $p$ states and hence only on the
unprimed sites. Eq.~(\ref{RbHamiltonian}) appears more complicated
than Eq.~(\ref{H}), but it actually embodies the general technique
for constructing spin Hamiltonians for these systems, as we will
see in Section~\ref{Csspinglass}.

\section{Spin glass model for the cesium system}
\label{Csspinglass}

In the cesium experiments of Mourachko \emph{et
al.\/}~\cite{Mourachko1998a,Mourachko1999a}, there are three
states that contribute to the experimental results. These are the
$23s$, $23p$, and the $24s$ states, which we refer to as the $s$,
$p$, and $s^{\prime}$ states, respectively. The processes
$pp\rightarrow ss^{\prime}$, $ps\rightarrow sp$, and
$ps^{\prime}\rightarrow s^{\prime}p$ can occur. The $pp\rightarrow
ss^{\prime}$ process can be tuned in and out of resonance via the
Stark shift by application of an electric field, while the
$ps\rightarrow sp$ and $ps^{\prime}\rightarrow s^{\prime}p$
processes are always resonant. In the case of the cesium
experiments, the $ps\rightarrow sp$ and $ps^{\prime}\rightarrow
s^{\prime}p$ processes are of comparable strength and both must be
considered.

Just as we did in Section~\ref{Rbspinglass}, we define a column
vector of length three for each site. If an atom at site $\Lambda$
is in the $s$, $p$, or $s^{\prime}$ state, then the column vector
corresponding to that site is $\left(0,0,1\right)^{T}$,
$\left(0,1,0\right)^{T}$, or $\left(1,0,0\right)^{T}$,
respectively. The Hamiltonian for this system is then
\begin{eqnarray}
H_{\mrm{Cs}} &=& \sum_{\Lambda} \epsilon_{\Lambda} + \nonumber \\
&& \sum_{\Lambda_{1},\Lambda_{2}} V_{\Lambda_{1}\Lambda_{2}}
\left[ \left(\sigma^{+}_{ps^{\prime}}\right)_{\Lambda_{1}}
\left(\sigma^{-}_{ps}\right)_{\Lambda_{2}} + \mrm{h.c.}\right] +
\sum_{\Lambda_{1},\Lambda_{2}} U_{\Lambda_{1}\Lambda_{2}} \left[
\left(\sigma^{+}_{ps^{\prime}}\right)_{\Lambda_{1}}
\left(\sigma^{-}_{ps^{\prime}}\right)_{\Lambda_{2}} +
\mrm{h.c.}\right] + \nonumber
\\ && \sum_{\Lambda_{1},\Lambda_{2}}
W_{\Lambda_{1}\Lambda_{2}} \left[
\left(\sigma^{-}_{ps}\right)_{\Lambda_{1}}
\left(\sigma^{+}_{ps}\right)_{\Lambda_{2}} + \mrm{h.c.}\right],
\label{Hcs}
\end{eqnarray}
where $\epsilon$, $\sigma^{\pm}_{ps^{\prime}}$, and
$\sigma^{\pm}_{ps^{\prime}}$ are the $3\times 3$ matrices
\begin{equation}
\epsilon = \left(
\begin{array}{ccc}
\epsilon_{s^{\prime}} & 0 & 0 \\ 0 & \epsilon_{p} & 0 \\ 0 & 0 &
\epsilon_{s}
\end{array}
\right),
\end{equation}
\begin{equation}
\sigma^{\pm}_{ps^{\prime}} = \left(
\begin{array}{cc}
\sigma^{\pm} & \begin{array}{c} 0 \\ 0 \end{array} \\
\begin{array}{cc} 0 & 0 \end{array} & 0
\end{array}
\right),
\end{equation}
and
\begin{equation}
\sigma^{\pm}_{ps} = \left(
\begin{array}{cc}
0 & \begin{array}{cc} 0 & 0 \end{array} \\
\begin{array}{c} 0 \\ 0 \end{array} & \sigma^{\pm}
\end{array}
\right),
\end{equation}
and the interaction potentials $V_{\Lambda_{1}\Lambda_{2}}$,
$U_{\Lambda_{1}\Lambda_{2}}$, and $W_{\Lambda_{1}\Lambda_{2}}$ are
zero if $\Lambda_{1} = \Lambda_{2}$ and otherwise are the
potentials corresponding to the $pp\rightarrow ss^{\prime}$,
$ps^{\prime}\rightarrow s^{\prime}p$, and $ps\rightarrow sp$
processes, respectively.

The equations of motion can be determined in this case as well,
simply by computing commutators.

\section{Summary}
\label{spinglasssummary}

We have seen in this chapter how the frozen dipolar gases studied
by Anderson \emph{et al.\/}~\cite{Anderson1996a,Anderson1998a} and
Lowell \emph{et al.\/}~\cite{Lowell1998a}, as well as Mourachko
\emph{et al.\/}~\cite{Mourachko1998a,Mourachko1999a}, are
completely equivalent to systems of interpenetrating spin glasses.
Casting the problem in this form has the advantage that it becomes
possible to at least write down equations that go beyond the
sparse model of Chapters~\ref{sparse_no_u} and
\ref{sparse_with_u}.

\chapter{The Effect of Spin}
\markright{Chapter \arabic{chapter}: The Effect of Spin}
\label{spin}

\section{Introduction}

Throughout most of Chapter~\ref{sparse_no_u}, we used an
interaction potential that varied as $1/r^{3}$, with no angular
dependence. In Section~\ref{angular} we discussed how to deal with
interaction potentials that have a more complicated dependence on
$r$, depend on the angular variables, or both. In this chapter we
determine exactly what the spin-dependent interatomic potentials
are for the $ss^{\prime}\rightarrow pp^{\prime}$ and
$sp\rightarrow ps$ processes in the experiment we are
modeling~\cite{Anderson1996a,Anderson1998a,Lowell1998a}. The
interaction potential depends directly on the $m_{j}$ values for
the states involved, where $m_{j}$ is the projection of the total
angular momentum in the direction of the applied electric field.
For states with $l = 0$, $m_{j}$ is just the projection of the
outer electron's spin. For states with $l = 1$, the spin-orbit
coupling separates the states with $j = 3/2$ from the states with
$j = 1/2$. Further, we assume that for $j = 3/2$, the applied
electric field splits the states with $\left|m_{j}\right| = 3/2$
from those with $\left|m_{j}\right| = 1/2$. As is illustrated in
Fig.~\ref{introduction_states}, this is the case for the $34p$
states of Rb under the conditions of the experiments in which we
are interested. Then, for each of the states we consider,
$\left|m_{j}\right|$ is fixed and $m_{j}$ can only take on the
values $\pm\left|m_{j}\right|$, so that it is effectively
equivalent to a spin of $1/2$. When we refer to spin effects in
this chapter we mean, more precisely, the effects due to $m_{j}$
which are derived from the spin-orbit coupling.

We start in Section~\ref{matrixelements} by examining the matrix
elements of the dipole-dipole interaction in more detail, and
taking into account the effect of the various atomic spin states.
The techniques we use for doing this are inspired by
Ref.~\cite{Movre1977a}, which we found to be extremely helpful in
that it pointed us in the right direction when we began our study
of spin effects in the frozen system. In
Section~\ref{dipoleinteraction} we recall the form of the
dipole-dipole interaction potential and in Section~\ref{dme} we
begin the calculation of the dipolar matrix elements with the
effects of the atomic states taken into consideration. This
computation requires a change of basis of the atomic spin states,
as explained in Section~\ref{cob}. In Section~\ref{me} the
laborious procedure necessary to compute the individual matrix
elements is explained, and the results are presented in tabular
form.

Next, in Section~\ref{determiningthepotential} we attempt to
determine the effective interaction potential for the resonant and
nonresonant processes. We start in Section~\ref{statesandprobamps}
by determining the states and probability amplitudes of the system
when the complications due to the presence of atomic spin are
taken into account. We proceed in Section~\ref{equationsofmotion}
to discuss the corresponding equations of motion. We find, as
expected, that in general the quantity $m_{j}$ is not conserved.
Therefore the number of interacting states for a system of $N + 1$
atoms is $2^{N+1}$, whereas it is only $N + 1$ when the effects of
spin are neglected. However, when states of $\left|m_{j}\right| =
3/2$ are involved it is still possible to obtain exact formulas as
we did in Chapter~\ref{sparse_no_u}, provided that the quantity
$V$ is replaced by an effective interaction. We refer to this case
as the separable case, for reasons that will become apparent in
Section~\ref{separability}. Section~\ref{diracmatrices} concerns
the properties of Dirac matrices, which are instrumental in our
discussion of separability and its applicability to averaging in
Section~\ref{separability}. In Section~\ref{averagingwithspin} we
determine the effective interaction potential in the separable
case. Finally, we summarize and conclude in
Section~\ref{spin_conclusions}.

\section{Matrix elements for the $ss^{\prime}\rightarrow pp^{\prime}$ and
$sp\rightarrow ps$ processes}
\label{matrixelements}

\subsection{The dipole-dipole interaction}
\label{dipoleinteraction}

We consider two atoms, the first with its nucleus located at
$\mbf{x}_{\mrm{nuc}}$ and outer electron located at
$\mbf{x}_{\mrm{elec}}$, and the second with nucleus located at
$\mbf{x}_{\mrm{nuc}}^{\prime}$ and outer electron located at
$\mbf{x}_{\mrm{elec}}^{\prime}$. Next we define
\begin{equation}
\mbf{R} = \mbf{x}_{\mrm{nuc}} - \mbf{x}_{\mrm{nuc}}^{\prime},
\end{equation}
\begin{equation}
\mbf{r} = \mbf{x}_{\mrm{elec}} - \mbf{x}_{\mrm{nuc}},
\end{equation}
and
\begin{equation}
\mbf{r^{\prime}} = \mbf{x}_{\mrm{elec}}^{\prime} -
\mbf{x}_{\mrm{nuc}}^{\prime}.
\end{equation}
The potential for the electrostatic interaction between the two
atoms is then, in atomic units,
\begin{equation}
U\left(\mbf{r},\mbf{r}^{\prime},\mbf{R}\right) = \frac{1}{R} +
\frac{1}{\left| \mbf{R}+\mbf{r^{\prime}-\mbf{r}}\right|} -
\frac{1}{\left| \mbf{R}+\mbf{r^{\prime}}\right|} - \frac{1}{\left|
\mbf{R}-\mbf{r}\right|}. \label{UVdWfull}
\end{equation}
The first two terms on the right hand side represent the
interaction between the two nuclei and the interaction between the
two electrons, respectively. The last two terms represent the
interactions between a nucleus and the electron that belongs to
the other atom.

As is shown on page~283 of Ref.~\cite{Schwabl1995a} and in more
detail on page~260 of Ref.~\cite{Schiff1960a}, expanding
Eq.~(\ref{UVdWfull}) and keeping terms only to leading order in
$r/R$ and $r^{\prime}/R$ gives
\begin{equation}
U\left(\mbf{r},\mbf{r}^{\prime},\mbf{R}\right) =
\frac{\mbf{r}\cdot\mbf{r}^{\prime} -
3\left(\hat{\mbf{R}}\cdot\mbf{r}\right)
\left(\hat{\mbf{R}}\cdot\mbf{r}^{\prime}\right)}{R^{3}}.
\label{UVdW}
\end{equation}
This is the well known result for the dipole-dipole potential
between two atoms.

It is important to note that, although Eq.~(\ref{UVdW}) is the
same potential used in the derivation of the van der Waals
interaction, the effects we study are transient effects that take
place after the atoms have been allowed to interact only a very
short time. Therefore they are quite different from the
steady-state van der Waals effect.

\subsection{Dipolar matrix elements}
\label{dme}

We will later be averaging over the positions of the atoms, so we
only want to determine the angular dependence of the matrix
elements.  Thus the $1/R^{3}$ factor can be ignored until
averaging, and we need only consider matrix elements of the type
\begin{equation}
\left\langle p\left| \left\langle p^{\prime}\left|
\,\mbf{r}\cdot\mbf{r}^{\prime} -
3\left(\mbf{r}\cdot\hat{\mbf{R}}\right)
\left(\mbf{r}^{\prime}\cdot\hat{\mbf{R}}\right)
\right|s^{\prime}\right\rangle \right| s\right\rangle,
\label{ss'pp'me}
\end{equation}
or
\begin{equation}
\left\langle p\left| \left\langle s\left|
\,\mbf{r}\cdot\mbf{r}^{\prime} -
3\left(\mbf{r}\cdot\hat{\mbf{R}}\right)
\left(\mbf{r}^{\prime}\cdot\hat{\mbf{R}}\right)
\right|p\right\rangle \right| s\right\rangle. \label{sppsme}
\end{equation}
Here we note that when we write the states $\left|a\right\rangle
\left|b\right\rangle$ and $\left\langle c\right| \left\langle
d\right|$ we really mean $\left|a\right\rangle \otimes
\left|b\right\rangle$ and $\left\langle c\right| \otimes
\left\langle d\right|$, respectively. In Section~\ref{me}, where
we deal with products of a great many states and the tensor
products are more urgently called for, we will write the tensor
product symbols out explicitly. For now, however, since we are
only dealing with products of two states at a time, we omit them
to save space and make the notation a little simpler.

We now recall that, in the experiments we are modeling, $s$,
$s^{\prime}$, $p$, and $p^{\prime}$ refer to the $25s$, $33s$,
$24p_{1/2}$, and $34p_{3/2}$ states, respectively. We also recall
that there are really two distinct resonances in the
$ss^{\prime}\rightarrow pp^{\prime}$ interaction, corresponding to
whether $\left| m_{j}\right| = 1/2$ or $\left| m_{j}\right| = 3/2$
for the $p^{\prime}$ atom. The complication for the interaction
potential enters through the $p$ and $p^{\prime}$ states. Clearly
the matrix element in Eq.~(\ref{sppsme}) will be different for the
case where the $p$ state on the left has $m_{j} = 1/2$ and the one
on the right has $m_{j} = -1/2$ than for the case where both $p$
states have $m_{j} = 1/2$, because the $p$ state with $m_{j} =
1/2$ will involve a different mixture of the $p_{x}$, $p_{y}$, and
$p_{z}$ states than the state where $m_{j} = -1/2$. The matrix
element in Eq.~(\ref{ss'pp'me}) will similarly depend on the
$m_{j}$ values of the $p$ and $p^{\prime}$ states, although there
is some simplification because we are able to consider the
$p^{\prime}$ states with $\left| m_{j}\right| = 1/2$ and those
with $\left| m_{j}\right| = 3/2$ separately.

For simplicity, we define
\begin{equation}
\rho = \mbf{r}\cdot\mbf{r}^{\prime} -
3\left(\mbf{r}\cdot\hat{\mbf{R}}\right)
\left(\mbf{r}^{\prime}\cdot\hat{\mbf{R}}\right),
\end{equation}
and we note that
\begin{equation}
\rho = xx^{\prime} + yy^{\prime} + zz^{\prime} -
3\left(x\hat{R}_{x} + y\hat{R}_{y} + z\hat{R}_{z} \right)
\left(x^{\prime}\hat{R}_{x} + y^{\prime}\hat{R}_{y} +
z^{\prime}\hat{R}_{z} \right),
\end{equation}
where $\hat{R}_{x} = \sin\theta\cos\phi$, $\hat{R}_{y} =
\sin\theta\sin\phi$, and $\hat{R}_{z} = \cos\theta$.

Consider for a moment the matrix element
\begin{equation}
\left\langle p_{a} \left| \left\langle p_{b}^{\prime} \left|
xx^{\prime} \right| s^{\prime} \right\rangle \right| s
\right\rangle = \left\langle p_{a} \left| x \right| s
\right\rangle \left\langle p_{b}^{\prime} \left| x^{\prime}
\right| s^{\prime} \right\rangle,
\end{equation}
where $a$ and $b$ can be $x$, $y$, or $z$. We can apply the usual
parity argument to determine that $\left\langle p_{a} \left| x
\right| s \right\rangle$ is nonzero only if $a = x$, and in that
case is equal to the dipole moment $\mu$. Similarly, $\left\langle
p_{b}^{\prime} \left| x^{\prime} \right| s^{\prime} \right\rangle$
is nonzero only if $b = x$, and in that case is equal to the
dipole moment $\mu^{\prime}$. Thus we have
\begin{eqnarray}
\left\langle p_{a} \left| \left\langle p_{b}^{\prime} \left|
xx^{\prime} \right| s^{\prime} \right\rangle \right| s
\right\rangle &=& \left\langle p_{a} \left| x \right| s
\right\rangle \left\langle p_{b}^{\prime} \left| x^{\prime}
\right| s^{\prime} \right\rangle \nonumber \\ &=& \mu\mu^{\prime}
\delta_{ax}\delta_{bx}.
\end{eqnarray}
Analogously,
\begin{eqnarray}
\left\langle p_{b} \left| \left\langle s \left| xx^{\prime}
\right| p_{a} \right\rangle \right| s \right\rangle &=&
\left\langle p_{b} \left| x \right| s \right\rangle \left\langle s
\left| x^{\prime} \right| p_{a} \right\rangle \nonumber \\ &=&
\mu^{2}\delta_{ax}\delta_{bx}.
\end{eqnarray}

In fact, the matrix elements we consider are a little more
complicated than this. The atoms have a total spin $\mbf{J} =
\mbf{L} + \mbf{S}$. The potential $\rho$ changes $\mbf{L}$, but
does not at all affect $\mbf{S}$ and therefore cannot change $s$
or $m_{s}$. Therefore the matrix elements we really consider are
of the form
\begin{eqnarray}
\left\langle p_{a},\alpha\left| \left\langle p^{\prime}_{b},\beta
\left| xx^{\prime} \right| s^{\prime},\gamma\right\rangle \right|
s,\epsilon \right\rangle &=& \left\langle p_{a},\alpha \left| x
\right| s,\epsilon \right\rangle \left\langle p_{b}^{\prime},\beta
\left| x^{\prime} \right| s^{\prime},\gamma \right\rangle
\nonumber
\\ &=& \mu\mu^{\prime} \delta_{ax}\delta_{bx}
\delta_{\alpha\epsilon}\delta_{\beta\gamma}, \label{ss'xx'pp'}
\end{eqnarray}
where, for example, $\left| s,\alpha\right\rangle$ is short for
$\left| s,m_{s}=\alpha \right\rangle$. Analogously,
\begin{eqnarray}
\left\langle p_{a},\alpha\left| \left\langle s,\beta \left|
xx^{\prime} \right| p_{b},\gamma \right\rangle\right| s,\epsilon
\right\rangle &=& \left\langle p_{a},\alpha \left| x \right|
s,\epsilon \right\rangle \left\langle s,\beta \left| x^{\prime}
\right| p_{b},\gamma \right\rangle \nonumber \\ &=&
\mu^{2}\delta_{ax}\delta_{bx}
\delta_{\alpha\epsilon}\delta_{\beta\gamma}. \label{spxx'ps}
\end{eqnarray}

Using Eqs.~(\ref{ss'xx'pp'}) and (\ref{spxx'ps}), it is trivial to
see that
\begin{equation}
\left\langle p_{a},\alpha\left| \left\langle p_{b}^{\prime},\beta
\left| \rho \right| s^{\prime},\gamma\right\rangle \right|
s,\epsilon \right\rangle = \mu\mu^{\prime} \left(\delta_{ab} -
3\hat{R}_{a}\hat{R}_{b}\right)
\delta_{\alpha\epsilon}\delta_{\beta\gamma}, \label{ss'me}
\end{equation}
and
\begin{equation}
\left\langle p_{a},\alpha\left| \left\langle s,\beta \left| \rho
\right| p_{b},\gamma \right\rangle\right| s,\epsilon \right\rangle
= \mu^{2}\left(\delta_{ab} - 3\hat{R}_{a}\hat{R}_{b}\right)
\delta_{\alpha\epsilon}\delta_{\beta\gamma}. \label{spme}
\end{equation}

\subsection{A change of basis}
\label{cob}

We can easily write out the states $s$, $s^{\prime}$, $p$, and
$p^{\prime}$ in terms of states of the form $\left|
j,l,m_{j}\right\rangle$. For example, the state $s$ denotes the
pair of states $\left| j=1/2,l=0,m_{j}=1/2\right\rangle$ and
$\left| j=1/2,l=0,m_{j}=-1/2\right\rangle$, while the state
$s^{\prime}$ denotes the pair of states $\left|
j^{\prime}=1/2,l^{\prime}=0,m_{j}=1/2\right\rangle$ and $\left|
j^{\prime}=1/2,l^{\prime}=0,m_{j}=-1/2\right\rangle$. Similarly,
$p$ denotes the pair of states $\left|
j=1/2,l=1,m_{j}=1/2\right\rangle$ and $\left|
j=1/2,l=1,m_{j}=-1/2\right\rangle$. Recalling that the
$p^{\prime}$ states with $\left| m_{j}\right| = 3/2$ are treated
separately from those with $\left| m_{j}\right| = 1/2$, we see
that $p^{\prime}$ denotes either the pair of states $\left|
j^{\prime}=3/2,l^{\prime}=1,m_{j}=3/2\right\rangle$ and $\left|
j^{\prime}=3/2,l^{\prime}=1,m_{j}=-3/2\right\rangle$, or the pair
of states $\left|
j^{\prime}=3/2,l^{\prime}=1,m_{j}=1/2\right\rangle$ and $\left|
j^{\prime}=3/2,l^{\prime}=1,m_{j}=-1/2\right\rangle$. In order to
use Eqs.~(\ref{ss'me}) and (\ref{spme}), we have to write the $p$
and $p^{\prime}$ states as linear combinations of states of the
form $\left| p_{a},m_{s}\right\rangle$. This is easily done by
consulting a table of Clebsch-Gordan coefficients, as can be found
on pages~76--77 of Ref.~\cite{Condon1964a}. Using such a table, we
determine the decompositions in Table~\ref{cgtable}.

\begin{table}
\begin{center}
\begin{tabular}{||c|c||}
\hline \emph{State} & \emph{Decomposition} \\ \hline\hline

$\left| j=\frac{3}{2},m_{j} = \frac{3}{2}\right\rangle$ &
$-\frac{1}{\sqrt{2}}\left(\left|p_{x},\frac{1}{2}\right\rangle + i
\left|p_{y},\frac{1}{2}\right\rangle\right)$ \\

$\left |j=\frac{3}{2},m_{j} = -\frac{3}{2}\right\rangle$ &
$\frac{1}{\sqrt{2}}\left(\left|p_{x},-\frac{1}{2}\right\rangle - i
\left|p_{y},-\frac{1}{2}\right\rangle\right)$ \\ \hline

$\left| j=\frac{3}{2},m_{j} = \frac{1}{2}\right\rangle$ &
$\sqrt{\frac{2}{3}}\left| p_{z},\frac{1}{2}\right\rangle -
\frac{1}{\sqrt{6}}\left(\left|p_{x},-\frac{1}{2}\right\rangle + i
\left|p_{y},-\frac{1}{2}\right\rangle\right)$ \\

$\left| j=\frac{3}{2},m_{j} = -\frac{1}{2}\right\rangle$ &
$\sqrt{\frac{2}{3}}\left| p_{z},-\frac{1}{2}\right\rangle +
\frac{1}{\sqrt{6}}\left(\left|p_{x},\frac{1}{2}\right\rangle - i
\left|p_{y},\frac{1}{2}\right\rangle\right)$ \\ \hline

$\left| j=\frac{1}{2},m_{j} = \frac{1}{2}\right\rangle$ &
$-\frac{1}{\sqrt{3}}\left| p_{z},\frac{1}{2}\right\rangle -
\frac{1}{\sqrt{3}}\left(\left|p_{x},-\frac{1}{2}\right\rangle + i
\left|p_{y},-\frac{1}{2}\right\rangle\right)$ \\

$\left| j=\frac{1}{2},m_{j} = -\frac{1}{2}\right\rangle$ &
$\frac{1}{\sqrt{3}}\left| p_{z},-\frac{1}{2}\right\rangle -
\frac{1}{\sqrt{3}}\left(\left|p_{x},\frac{1}{2}\right\rangle - i
\left|p_{y},\frac{1}{2}\right\rangle\right)$ \\ \hline

\end{tabular}
\caption[Decomposition of $p$ and $p^{\prime}$ states]{The
decomposition of all $p$ and $p^{\prime}$ states into linear
combinations of states of the form $\left|
p_{a},m_{s}\right\rangle$.} \label{cgtable}
\end{center}
\end{table}

\subsection{The $ss^{\prime}\rightarrow
pp^{\prime}$ and $sp\rightarrow ps$ matrix elements}
\label{me}

With the decomposition of Table~\ref{cgtable} and the rules of
Eqs.~(\ref{ss'me}) and (\ref{spme}) in hand, it is now possible to
compute the matrix elements we need. I will first show how to
compute one matrix element, as an example, then present all the
matrix elements in table form.

Consider the matrix element
\begin{equation}
\left\langle p,m_{j}=-\frac{1}{2}\right|\left\langle
p^{\prime},m_{j}=\frac{3}{2}\right| \frac{\rho}{\mu\mu^{\prime}}
\left| s^{\prime},m_{j}=\frac{1}{2}\right\rangle \left|
s,m_{j}=-\frac{1}{2}\right\rangle.
\end{equation}
We have from Table~\ref{cgtable} that
\begin{equation}
\left| p^{\prime},m_{j}=\frac{3}{2}\right\rangle =
-\frac{1}{\sqrt{2}}\left(\left|p^{\prime}_{x},\frac{1}{2}\right\rangle
+ i \left|p^{\prime}_{y},\frac{1}{2}\right\rangle\right),
\end{equation}
and
\begin{equation}
\left| p,m_{j}=-\frac{1}{2}\right\rangle =
\frac{1}{\sqrt{3}}\left| p_{z},-\frac{1}{2}\right\rangle -
\frac{1}{\sqrt{3}}\left(\left|p_{x},\frac{1}{2}\right\rangle - i
\left|p_{y},\frac{1}{2}\right\rangle\right).
\end{equation}
Dropping the occurrences of ``$m_{j}=$'' that appear to save
space, we see that
\begin{eqnarray}
\left\langle p,-\frac{1}{2}\right|\left\langle
p^{\prime},\frac{3}{2}\right| &=& \left[ \frac{1}{\sqrt{3}}\left|
p_{z},-\frac{1}{2}\right\rangle -
\frac{1}{\sqrt{3}}\left(\left|p_{x},\frac{1}{2}\right\rangle - i
\left|p_{y},\frac{1}{2}\right\rangle\right)\right]^{\dagger}
\times \nonumber \\ && \left[
-\frac{1}{\sqrt{2}}\left(\left|p^{\prime}_{x},\frac{1}{2}\right\rangle
+ i
\left|p^{\prime}_{y},\frac{1}{2}\right\rangle\right)\right]^{\dagger}
\nonumber
\\ &=& \left[
\frac{1}{\sqrt{3}}\left\langle p_{z},-\frac{1}{2}\right| -
\frac{1}{\sqrt{3}}\left(\left\langle p_{x},\frac{1}{2}\right| + i
\left\langle p_{y},\frac{1}{2}\right| \right)\right] \times
\nonumber
\\ && \left[
-\frac{1}{\sqrt{2}}\left(\left\langle
p^{\prime}_{x},\frac{1}{2}\right| - i \left\langle
p^{\prime}_{y},\frac{1}{2}\right|\right)\right] \nonumber \\ &=&
\frac{1}{\sqrt{6}}\left\{ -\left\langle
p_{z},-\frac{1}{2}\right|\left\langle
p^{\prime}_{x},\frac{1}{2}\right| + \left\langle
p_{x},\frac{1}{2}\right|\left\langle
p^{\prime}_{x},\frac{1}{2}\right| + i\left\langle
p_{y},\frac{1}{2}\right|\left\langle
p^{\prime}_{x},\frac{1}{2}\right| + \right. \nonumber \\ && \left.
i\left\langle p_{z},-\frac{1}{2}\right|\left\langle
p^{\prime}_{y},\frac{1}{2}\right| - i\left\langle
p_{x},\frac{1}{2}\right|\left\langle
p^{\prime}_{y},\frac{1}{2}\right| + \left\langle
p_{y},\frac{1}{2}\right|\left\langle
p^{\prime}_{y},\frac{1}{2}\right| \,\right\}. \label{longpp'}
\end{eqnarray}
Now we simply apply Eq.~(\ref{ss'me}) to each matrix element
corresponding to a term in the sum of Eq.~(\ref{longpp'}).  For
example, the first term in the sum corresponds to the matrix
element
\begin{equation}
-\frac{1}{\sqrt{6}}\left\langle
p_{z},-\frac{1}{2}\right|\left\langle
p^{\prime}_{x},\frac{1}{2}\right| \frac{\rho}{\mu\mu^{\prime}}
\left| s^{\prime},\frac{1}{2}\right\rangle\left|
s,-\frac{1}{2}\right\rangle,
\end{equation}
which according to Eq.~(\ref{ss'me}) is equal to
\begin{equation}
-\frac{1}{\sqrt{6}}\left(-3\hat{R}_{x}\hat{R}_{z}\right) =
\sqrt{\frac{3}{2}}\sin\theta\cos\theta\cos\phi.
\end{equation}
Computing the other terms similarly and summing them, we determine
one of the matrix elements for the $ss^{\prime}\rightarrow
pp^{\prime}$ process.

The process is straightforward, albeit laborious, and one obtains
the matrix elements presented in the Tables~\ref{ss'pp'3/2table}
-- \ref{sppstable}.

\begin{table}
\begin{center}
\begin{tabular}{||c||c|c|c|c||}
\hline

$\rho/\left(\mu\mu^{\prime}\right)$ & $\left|
s^{\prime},\frac{1}{2}\right\rangle\left|
s,\frac{1}{2}\right\rangle$ & $\left|
s^{\prime},\frac{1}{2}\right\rangle\left|
s,-\frac{1}{2}\right\rangle$ & $\left|
s^{\prime},-\frac{1}{2}\right\rangle\left|
s,\frac{1}{2}\right\rangle$ & $\left|
s^{\prime},-\frac{1}{2}\right\rangle\left|
s,-\frac{1}{2}\right\rangle$
\\ \hline\hline

$\left\langle p,\frac{1}{2}\right|\left\langle
p^{\prime},\frac{3}{2}\right|$ &
$-\sqrt{\frac{3}{2}}\cos\theta\sin\theta e^{i\phi}$ &
$-\sqrt{\frac{3}{2}}\sin^{2}\theta e^{2i\phi}$ & $0$ & $0$
\\ \hline

$\left\langle p,-\frac{1}{2}\right|\left\langle
p^{\prime},\frac{3}{2}\right|$ &
$-\frac{1}{\sqrt{6}}\left(1-3\cos^{2}\theta\right)$ &
$\sqrt{\frac{3}{2}}\cos\theta\sin\theta e^{i\phi}$ & $0$ & $0$
\\ \hline

$\left\langle p,\frac{1}{2}\right|\left\langle
p^{\prime},-\frac{3}{2}\right|$ & $0$ & $0$ &
$\sqrt{\frac{3}{2}}\cos\theta\sin\theta e^{-i\phi}$ &
$\frac{1}{\sqrt{6}}\left(1-3\cos^{2}\theta\right)$
\\ \hline

$\left\langle p,-\frac{1}{2}\right|\left\langle
p^{\prime},-\frac{3}{2}\right|$ & $0$ & $0$ &
$\sqrt{\frac{3}{2}}\sin^{2}\theta e^{-2i\phi}$ &
$-\sqrt{\frac{3}{2}}\cos\theta\sin\theta e^{-i\phi}$
\\ \hline
\end{tabular}
\caption[Matrix elements for the $ss^{\prime}\rightarrow
pp^{\prime}$ process ($\left| m_{j}\right| = 3/2$)]{The matrix
elements $\left\langle p,m_{j}\left|\left\langle
p^{\prime},m_{j}\left| \rho/\left(\mu\mu^{\prime}\right)\right|
s^{\prime},m_{j}\right\rangle\right| s,m_{j}\right\rangle$ for the
$ss^{\prime}\rightarrow pp^{\prime}$ process when $\left|
m_{j}\right| = 3/2$ for the $p^{\prime}$ atom.}
\label{ss'pp'3/2table}
\end{center}
\end{table}

\begin{table}
\begin{center}
\begin{tabular}{||c||c|c|c|c||}
\hline

$\rho/\left(\mu\mu^{\prime}\right)$ & $\left|
s^{\prime},\frac{1}{2}\right\rangle\left|
s,\frac{1}{2}\right\rangle$ & $\left|
s^{\prime},\frac{1}{2}\right\rangle\left|
s,-\frac{1}{2}\right\rangle$ & $\left|
s^{\prime},-\frac{1}{2}\right\rangle\left|
s,\frac{1}{2}\right\rangle$ & $\left|
s^{\prime},-\frac{1}{2}\right\rangle\left|
s,-\frac{1}{2}\right\rangle$
\\ \hline\hline

$\left\langle p,\frac{1}{2}\right|\left\langle
p^{\prime},\frac{1}{2}\right|$ &
$-\frac{\sqrt{2}}{3}\left(1-3\cos^{2}\theta\right)$ &
$\sqrt{2}\cos\theta\sin\theta e^{i\phi}$ &
$-\frac{1}{\sqrt{2}}\cos\theta\sin\theta e^{i\phi}$ &
$-\frac{1}{\sqrt{2}}\sin^{2}\theta e^{2i\phi}$
\\ \hline

$\left\langle p,-\frac{1}{2}\right|\left\langle
p^{\prime},\frac{1}{2}\right|$ & $\sqrt{2}\cos\theta\sin\theta
e^{-i\phi}$ & $\frac{\sqrt{2}}{3}\left(1-3\cos^{2}\theta\right)$ &
$-\frac{1}{3\sqrt{2}}\left(1-3\cos^{2}\theta\right)$ &
$\frac{1}{\sqrt{2}}\cos\theta\sin\theta e^{i\phi}$
\\ \hline

$\left\langle p,\frac{1}{2}\right|\left\langle
p^{\prime},-\frac{1}{2}\right|$ &
$\frac{1}{\sqrt{2}}\cos\theta\sin\theta e^{-i\phi}$ &
$\frac{1}{3\sqrt{2}}\left(1-3\cos^{2}\theta\right)$ &
$-\frac{\sqrt{2}}{3}\left(1-3\cos^{2}\theta\right)$ &
$\sqrt{2}\cos\theta\sin\theta e^{i\phi}$
\\ \hline

$\left\langle p,-\frac{1}{2}\right|\left\langle
p^{\prime},-\frac{1}{2}\right|$ &
$\frac{1}{\sqrt{2}}\sin^{2}\theta e^{-2i\phi}$ &
$-\frac{1}{\sqrt{2}}\cos\theta\sin\theta e^{-i\phi}$ &
$\sqrt{2}\cos\theta\sin\theta e^{-i\phi}$ &
$\frac{\sqrt{2}}{3}\left(1-3\cos^{2}\theta\right)$
\\ \hline
\end{tabular}
\caption[Matrix elements for the $ss^{\prime}\rightarrow
pp^{\prime}$ process ($\left| m_{j}\right| = 1/2$)]{The matrix
elements $\left\langle p,m_{j}\left|\left\langle
p^{\prime},m_{j}\left| \rho/\left(\mu\mu^{\prime}\right) \right|
s^{\prime},m_{j}\right\rangle\right| s,m_{j}\right\rangle$ for the
$ss^{\prime}\rightarrow pp^{\prime}$ process when $\left|
m_{j}\right| = 1/2$ for the $p^{\prime}$ atom.}
\label{ss'pp'1/2table}
\end{center}
\end{table}

\begin{table}
\begin{center}
\begin{tabular}{||c||c|c|c|c||}
\hline

$\rho/\mu^{2}$ & $\left|p,\frac{1}{2}\right\rangle\left|
s,\frac{1}{2}\right\rangle$ &
$\left|p,\frac{1}{2}\right\rangle\left|
s,-\frac{1}{2}\right\rangle$ &
$\left|p,-\frac{1}{2}\right\rangle\left|
s,\frac{1}{2}\right\rangle$ &
$\left|p,-\frac{1}{2}\right\rangle\left|
s,-\frac{1}{2}\right\rangle$
\\ \hline\hline

$\left\langle p,\frac{1}{2}\right|\left\langle
s,\frac{1}{2}\right|$ &
$\frac{1}{3}\left(1-3\cos^{2}\theta\right)$ &
$-\cos\theta\sin\theta e^{-i\phi}$ & $-\cos\theta\sin\theta
e^{-i\phi}$ & $-\sin^{2}\theta e^{-2i\phi}$
\\ \hline

$\left\langle p,\frac{1}{2}\right|\left\langle
s,-\frac{1}{2}\right|$ & $-\cos\theta\sin\theta e^{i\phi}$ &
$-\frac{1}{3}\left(1-3\cos^{2}\theta\right)$ &
$-\frac{1}{3}\left(1-3\cos^{2}\theta\right)$ &
$\cos\theta\sin\theta e^{-i\phi}$
\\ \hline

$\left\langle p,-\frac{1}{2}\right|\left\langle
s,\frac{1}{2}\right|$ & $-\cos\theta\sin\theta e^{i\phi}$ &
$-\frac{1}{3}\left(1-3\cos^{2}\theta\right)$ &
$-\frac{1}{3}\left(1-3\cos^{2}\theta\right)$ &
$\cos\theta\sin\theta e^{-i\phi}$
\\ \hline

$\left\langle p,-\frac{1}{2}\right|\left\langle
s,-\frac{1}{2}\right|$ & $-\sin^{2}\theta e^{2i\phi}$ &
$\cos\theta\sin\theta e^{i\phi}$ & $\cos\theta\sin\theta
e^{i\phi}$ & $\frac{1}{3}\left(1-3\cos^{2}\theta\right)$
\\ \hline
\end{tabular}
\caption[Matrix elements for the $sp\rightarrow ps$ process]{The
matrix elements $\left\langle p,m_{j}\left|\left\langle
s,m_{j}\left| \rho/\left(\mu^{2}\right)\right|
p,m_{j}\right\rangle\right| s,m_{j}\right\rangle$ for the
$sp\rightarrow ps$ process.} \label{sppstable}
\end{center}
\end{table}

\section{Determining the potential}
\label{determiningthepotential}

\subsection{States and probability amplitudes with spin included}
\label{statesandprobamps}

In the initial state, the system is labeled by an index $\left|
s^{\prime},m_{j}\right\rangle$ representing the state of the
$s^{\prime}$ atom and a set of $N$ indices $\left\{\left|
s,m_{j}\right\rangle\right\}$ representing the states of the $N$
$s$ atoms. In the absence of an applied magnetic field, the
initial values of these indices are random. The state
corresponding to the configuration where the primed atom is in the
$s^{\prime}$ state with $m_{j}=\alpha_{0}$ and the unprimed atoms
are all in the $s$ state with $\left\{m_{j}\right\} =
\left\{\alpha_{0};\alpha_{1},\ldots ,\alpha_{N}\right\}$ is
\begin{equation}
\left| s^{\prime},\alpha_{0}\right\rangle \otimes \left|
s,\alpha_{1}\right\rangle\otimes\cdots\otimes\left|
s,\alpha_{N}\right\rangle. \label{astate}
\end{equation}
Of course, because the $s$ and $s^{\prime}$ states have total spin
$1/2$, all $\alpha_{k}$ are required to have the value $1/2$ or
$-1/2$. The probability amplitude corresponding to this state is
\begin{equation}
a_{\left\{\alpha_{0};\alpha_{1},\ldots,\alpha_{N}\right\}}\left(t\right)
= \left\{ \left\langle s,\alpha_{N}\right| \otimes \cdots \otimes
\left\langle s,\alpha_{1}\right| \otimes \left\langle
s^{\prime},\alpha_{0}\right| \right\} \left|
\psi\left(t\right)\right\rangle, \label{awithspin}
\end{equation}
where $\left| \psi\left(t\right) \right\rangle$ is the state of
the system at time $t$.

Because the system is allowed to make $ss^{\prime}\rightarrow
pp^{\prime}$ transitions, we must also consider states in which
the primed atom is in the state $p^{\prime}$ and a single unprimed
atom is in the state $p$. This state is labeled by the indices
$\left| p^{\prime},m_{j}\right\rangle$ and $\left|
p,m_{j}\right\rangle$, together with the indices $\left\{
\left|s,m_{j} \right\rangle\right\}$ corresponding to the $s$
atoms not involved in the  $ss^{\prime}\rightarrow pp^{\prime}$
transition. The state corresponding to the configuration where the
primed atom is in the $p^{\prime}$ state with $m_{j}=\beta_{0}$
and all the unprimed atoms except for the $k$th are in the $s$
state while the $k$th unprimed atom is in the $p$ state with
$m_{j}=\beta_{k}$ is
\begin{equation}
\left| p^{\prime},\beta_{0}\right\rangle \otimes \left|
s,\beta_{1}\right\rangle \otimes \cdots \otimes \left|
s,\beta_{k-1}\right\rangle \otimes \left| p,\beta_{k}\right\rangle
\otimes \left| s,\beta_{k+1}\right\rangle \otimes \cdots \otimes
\left| s,\beta_{N}\right\rangle. \label{ckstate}
\end{equation}
The quantities $\beta_{1},\ldots
,\beta_{k-1},\beta_{k+1},\ldots,\beta_{N}$ are restricted to the
values $\pm 1/2$ because the $s$ state has total spin $1/2$.
Because the $p$ state also has total spin $1/2$, $\beta_{k}$ is
similarly restricted to the values $\pm 1/2$. The $p^{\prime}$
state has total spin $3/2$, and the quantity $\beta_{0}$ is
restricted to the values $\pm 3/2$ or $\pm 1/2$, depending on
whether we are considering the $\left| m_{j}\right| = 3/2$ or
$\left| m_{j}\right| = 1/2$ resonance, respectively. Similarly to
Eq.~(\ref{awithspin}), the probability amplitude corresponding to
the state in Eq.~(\ref{ckstate}) is
\begin{equation}
c_{k;\left\{\beta_{0};\beta_{1},\ldots,
\beta_{N}\right\}}\left(t\right) = \left\{ \left\langle
s,\beta_{N}\right| \otimes \cdots \otimes \left\langle
s,\beta_{k+1}\right| \otimes \left\langle p,\beta_{k}\right|
\otimes \left\langle s,\beta_{k-1}\right| \otimes \cdots \otimes
\left\langle p^{\prime},\beta_{0}\right| \right\} \left|
\psi\left(t\right)\right\rangle, \label{ckwithspin}
\end{equation}
where $\left| \psi\left(t\right) \right\rangle$ is again the state
of the system at time $t$.

\subsection{Equations of motion}
\label{equationsofmotion}

In the interest of brevity, we denote the subscript lists
\begin{equation}
\left\{ \alpha_{0};\alpha_{1},\ldots,\alpha_{N}\right\},
\end{equation}
and
\begin{equation}
\left\{ \gamma_{0};\beta_{1},\ldots,
\beta_{k-1},\gamma_{k},\beta_{k+1},\ldots,\beta_{N} \right\}
\end{equation}
by $A$ and $B_{k}\left(\gamma_{0},\gamma_{k}\right)$,
respectively. We label the $0$th and $k$th components of
$B_{k}\left(\gamma_{0},\gamma_{k}\right)$ with a different symbol
than the rest because we will sum over
$B_{k}\left(\gamma_{0},\gamma_{k}\right)$, and it will turn out
that the $B_{k}\left(\gamma_{0},\gamma_{k}\right)$ term in the sum
corresponds to the $s^{\prime}$ atom at the origin and the $s$
atom at position $\mbf{r}_{k}$ making the $ss^{\prime}\rightarrow
pp^{\prime}$ transition. Thus the use of a different symbol serves
as a reminder of which atoms are involved in the transition.

The equations of motion for the $a$ and $c$ coefficients of
Eqs.~(\ref{awithspin}) and Eqs.~(\ref{ckwithspin}) in Fourier
space are then
\setcounter{tlet}{1}
\renewcommand{\theequation}{\arabic{chapter}.\arabic{equation}\alph{tlet}}
\begin{eqnarray}
\left(\omega - \Delta\right) a_{A} &=&
i\prod_{m=0}^{N}\delta_{\alpha_{m},\alpha_{m}^{\left(i\right)}} +
\sum_{k=1}^{N}\sum_{B_{k}\left(\gamma_{0},\gamma_{k}\right)} V_{k}
M_{A,B_{k}\left(\gamma_{0},\gamma_{k}\right)}^{\left(k\right)}
c_{k;B_{k}\left(\gamma_{0},\gamma_{k}\right)}, \label{a_A}
\\ \stepcounter{tlet}\addtocounter{equation}{-1} \omega
c_{k;B_{k}\left(\gamma_{0},\gamma_{k}\right)} &=&
\sum_{A^{\prime}} V_{k}
M_{B_{k}\left(\gamma_{0},\gamma_{k}\right),A^{\prime}}^{\left(k\right)}
a_{A^{\prime}}, \label{c_B}
\end{eqnarray}
\renewcommand{\theequation}{\arabic{chapter}.\arabic{equation}}
where $\alpha_{m}^{\left(i\right)}$ is the initial $m_{j}$ value
for the $m$th atom and
\begin{equation}
V_{k} = \frac{\mu\mu^{\prime}}{r_{k}^{3}}
\end{equation}
is just the radial part of the interaction potential. Here we have
defined $A^{\prime} = \left\{\alpha_{0}^{\prime};
\alpha_{1}^{\prime},\ldots,\alpha_{N}^{\prime}\right\}$ to be
another set of indices subject to the same constraints as $A$. The
matrix
$M_{B_{k}\left(\gamma_{0},\gamma_{k}\right),A}^{\left(k\right)}$
is a $2^{N+1}\times 2^{N+1}$ matrix that is constructed from the
$4\times 4$ matrix formed by Table~\ref{ss'pp'3/2table} or
\ref{ss'pp'1/2table}, depending on whether we are considering the
case where the $p^{\prime}$ atom has $\left| m_{j}\right| = 3/2$
or $\left| m_{j}\right| = 1/2$. Explicitly, if the set of
\begin{equation}
m_{\left\{\beta,\beta^{\prime}\right\},\left\{\alpha,\alpha^{\prime}\right\}}
\left(\theta_{k},\phi_{k}\right)
 = \left\langle p,\beta\left|\left\langle
p^{\prime},\beta^{\prime}\left|
\frac{\rho}{\mu\mu^{\prime}}\right|
s^{\prime},\alpha^{\prime}\right\rangle\right|
s,\alpha\right\rangle \label{mdefn}
\end{equation}
are the elements belonging to the $4\times 4$ matrix that
corresponds to the value of $\left| m_{j}\right|$ that we want for
the $p^{\prime}$ state, then
\begin{equation}
M_{B_{k}\left(\gamma_{0},\gamma_{k}\right),A}^{\left(k\right)}
=
m_{\left\{\gamma_{0},\gamma_{k}\right\},\left\{\alpha_{0},\alpha_{k}\right\}}
\left(\theta_{k},\phi_{k}\right) \prod_{l=1,l\neq k}^{N}
\delta_{\alpha_{l}\beta_{l}}. \label{Mdefn}
\end{equation}
We see immediately from Eqs.~(\ref{Mdefn}) and (\ref{mdefn}) that
\begin{equation}
M_{A,B_{k}\left(\gamma_{0},\gamma_{k}\right)}^{\left(k\right)} =
\left[
M_{B_{k}\left(\gamma_{0},\gamma_{k}\right),A}^{\left(k\right)}
\right]^{\ast}. \label{switchstar}
\end{equation}

Multiplying both sides of Eq.~(\ref{c_B}) by $V_{k} M_{A,B_{k}
\left(\gamma_{0},\gamma_{k}\right)}^{\left(k\right)}$ and summing
over $k$ and $B_{k}\left(\gamma_{0},\gamma_{k}\right)$, we see
that
\begin{equation}
\omega
\sum_{k=1}^{N}\sum_{B_{k}\left(\gamma_{0},\gamma_{k}\right)} V_{k}
M_{A,B_{k}\left(\gamma_{0},\gamma_{k}\right)}^{\left(k\right)}
c_{k;B_{k}\left(\gamma_{0},\gamma_{k}\right)} =
\sum_{k=1}^{N}\sum_{A^{\prime}} V_{k}^{2}
P_{A,A^{\prime}}^{\left(k\right)} a_{A^{\prime}},
\label{thisresult}
\end{equation}
where
\begin{equation}
P_{A,A^{\prime}}^{\left(k\right)} =
\sum_{B_{k}\left(\gamma_{0},\gamma_{k}\right)}
M_{A,B_{k}\left(\gamma_{0},\gamma_{k}\right)}^{\left(k\right)}
M_{B_{k}\left(\gamma_{0},\gamma_{k}\right),A^{\prime}}^{\left(k\right)}.
\label{Pdefnlong}
\end{equation}
Using Eq.~(\ref{switchstar}) we see that, in terms of matrices,
Eq.~(\ref{Pdefnlong}) is equivalent to
\begin{equation}
P^{\left(k\right)} = \left[ M^{\left(k\right)} \right]^{\dagger}
M^{\left(k\right)}.
\end{equation}
Plugging the result of Eq.~(\ref{thisresult}) into
Eq.~(\ref{a_A}), we see that
\begin{equation}
\omega\left(\omega - \Delta\right) a_{A} =
i\omega\prod_{m=0}^{N}\delta_{\alpha_{m},\alpha_{m}^{\left(i\right)}}
+ \sum_{k=1}^{N}\sum_{A^{\prime}} V_{k}^{2}
P_{A,A^{\prime}}^{\left(k\right)} a_{A^{\prime}}. \label{aeqn1}
\end{equation}
For $N+1$ atoms, this represents $2^{N+1}$ equations.

The quantity $a_{A}$ actually depends on the two sets of indices
$A$ and $A^{\left(i\right)} = \left\{
\alpha_{0}^{\left(i\right)};\alpha_{1}^{\left(i\right)},\ldots
,\alpha_{N}^{\left(i\right)}\right\}$, as is clear from the first
term on the right hand side of Eq.~(\ref{aeqn1}). Therefore we can
consider $a$ to be a matrix, and we can write
\begin{equation}
\omega\left(\omega - \Delta\right) a = i\omega + \sum_{k=1}^{N}
V_{k}^{2} P^{\left(k\right)} a.
\end{equation}
At this point we solve this equation formally and switch to
Laplace space by making the substitution $\omega = i\alpha$. We
find
\begin{eqnarray}
a &=& \frac{\alpha}{\alpha\left(\alpha + i\Delta\right) +
\sum_{k=1}^{N} V_{k}^{2} P^{\left(k\right)}} \nonumber
\\ &=& \alpha\int_{0}^{\infty}d\beta\,
\exp\left[-\beta\alpha\left(\alpha + i\Delta\right)\right]
\exp\left[ -\beta \sum_{k=1}^{N} V_{k}^{2}
P^{\left(k\right)}\right]. \label{prearta}
\end{eqnarray}

Because Eq.~(\ref{c_B}) is coupled to Eq.~(\ref{a_A}),
$c_{k,B_{k}\left(\gamma_{0},\gamma_{k}\right)}$ also depends on
the set of indices $A^{\left(i\right)}$. Thus $c_{k}$ can also be
considered a matrix, and we can rewrite Eqs.~(\ref{a_A}) and
(\ref{c_B}) in the matrix form
\setcounter{tlet}{1}
\renewcommand{\theequation}{\arabic{chapter}.\arabic{equation}\alph{tlet}}
\begin{eqnarray}
\left(\omega - \Delta\right) a &=& i + \sum_{k=1}^{N} V_{k}
\left[M^{\left(k\right)}\right]^{\dagger} c_{k}, \label{amatrix}
\\ \stepcounter{tlet}\addtocounter{equation}{-1} \omega
c_{k} &=& V_{k} M^{\left(k\right)} a. \label{cmatrix}
\end{eqnarray}
\renewcommand{\theequation}{\arabic{chapter}.\arabic{equation}}

We note that these equations correspond to Eqs.~(\ref{a0dot}) and
(\ref{ckdot}), and so we hope to obtain an expression for
$\tilde{S}\left(\Delta,\alpha\right)$ analogous to
Eq.~(\ref{Stild}). We have to exercise a little caution, however,
since Eqs.~(\ref{a0dot}) and (\ref{ckdot}) are equations for
scalar quantities while Eqs.~(\ref{amatrix}) and (\ref{cmatrix})
are matrix equations. First consider Eq.~(\ref{prearta}) in the
rational form we had before we introduced the integral over
$\beta$. Because the matrix $\sum_{k=1}^{N} V_{k}^{2}
P^{\left(k\right)}$ is Hermitian we can diagonalize it. If we let
$\left| \lambda \right\rangle$ be the eigenstate of
$\sum_{k=1}^{N} V_{k}^{2} P^{\left(k\right)}$ with eigenvalue
$\lambda$, then
\begin{equation}
\left\langle \lambda \left| a \right| \lambda \right\rangle =
\frac{\alpha}{\alpha\left(\alpha + i\Delta\right) + \lambda^{2}}.
\end{equation}
The quantity $\left\langle \lambda \left| a \right| \lambda
\right\rangle$ is just a scalar, so we can certainly invert this
Laplace transform, take the mod square, and Laplace transform
again just as we did in Section~\ref{basicequations}. In this way
we find that
\begin{equation}
\left\langle\lambda\left|\tilde{S}\left(\Delta,\alpha\right)
\right|\lambda\right\rangle = \frac{1}{2\alpha} \left[1 -
\frac{\alpha^{2} + \Delta^{2}}{\alpha^{2} + \Delta^{2} +
4\lambda^{2}}\right].
\end{equation}
We could do this for any of the eigenvectors
$\left|\lambda\right\rangle$, and so it follows that
\begin{eqnarray}
\tilde{S}\left(\Delta,\alpha\right) &=& \frac{1}{2\alpha} \left[1
- \frac{\alpha^{2} + \Delta^{2}}{\alpha^{2} + \Delta^{2} +
4\sum_{k=1}^{N} V_{k}^{2}P^{\left(k\right)}}\right] \nonumber \\
&=& \frac{1}{2\alpha} - \frac{\alpha^{2} + \Delta^{2}}{2\alpha}
\int_{0}^{\infty} d\beta\,\exp\left[-\beta\left(\alpha^{2} +
\Delta^{2}\right)\right] \exp\left[-4\beta\sum_{k=1}^{N} V_{k}^{2}
P^{\left(k\right)}\right], \label{preaadagger}
\end{eqnarray}
which looks very much like Eq.~(\ref{Stild}).

To obtain results that can be compared with the experiments, we
also want to average over initial spin configurations and sum over
final spin configurations. Thus we want to compute the quantity
\begin{eqnarray}
\frac{1}{2^{N+1}}\Tr\,\tilde{S}\left(\Delta,\alpha\right) &=&
\frac{1}{2\alpha} - \frac{\alpha^{2} + \Delta^{2}}{2\alpha}
\int_{0}^{\infty} d\beta\,\exp\left[-\beta\left(\alpha^{2} +
\Delta^{2}\right)\right] \times \nonumber \\ &&
\frac{1}{2^{N+1}}\Tr\,\exp\left[-4\beta\sum_{k=1}^{N} V_{k}^{2}
P^{\left(k\right)}\right]. \label{aadagger}
\end{eqnarray}

\subsection{Dirac matrices}
\label{diracmatrices}

Before we go any further with Eq.~(\ref{aadagger}), we first
examine some properties of $4\times 4$ matrices. It will turn out
that this will aid us greatly in carrying out the average of
Eq.~(\ref{aadagger}).

We know from pages~211--213 of Ref.~\cite{Arfken1985a} that any
$4\times 4$ matrix can be written as a unique linear combination
of the sixteen Dirac matrices. These matrices are of the form
$\sigma_{\alpha}\otimes\sigma_{\beta}$, where $\alpha$ and $\beta$
range from zero to three. Here we define $\sigma_{\alpha}$ to be
the $2\times 2$ unit matrix if $\alpha = 0$ and the usual Pauli
matrix otherwise. Thus, if we have any $4\times 4$ matrix $Q$ then
we can write
\begin{equation}
Q = \sum_{\alpha=0}^{3}\sum_{\beta=0}^{3}
q_{\alpha\beta}\,\sigma_{\alpha}\otimes\sigma_{\beta},
\label{sumeqn}
\end{equation}
for some coefficients $q_{\alpha\beta}$. These coefficients are in
fact easy to determine, since
\begin{equation}
q_{\alpha\beta} = \frac{1}{4}
\Tr\,\left[Q\left(\sigma_{\alpha}\otimes\sigma_{\beta}\right)\right].
\label{traceeqn}
\end{equation}
Eq.~(\ref{traceeqn}) follows easily from the well-known properties
of the Pauli matrices and the facts that
\begin{equation}
\left(A\otimes B\right)\left(C\otimes D\right) =
\left(AC\right)\otimes\left(BD\right),
\label{tensorproductproduct}
\end{equation}
and
\begin{equation}
\Tr\,\left(A\otimes B\right) =
\left(\Tr\,A\right)\left(\Tr\,B\right).
\end{equation}

Let $M_{ss^{\prime}\rightarrow pp^{\prime},\frac{3}{2}}$,
$M_{ss^{\prime}\rightarrow pp^{\prime},\frac{1}{2}}$, and
$M_{sp\rightarrow ps}$ be the $4\times 4$ matrices determined by
Tables~\ref{ss'pp'3/2table} -- \ref{sppstable}. Then define
\begin{equation}
P_{ss^{\prime}\rightarrow pp^{\prime},\frac{3}{2}} =
M_{ss^{\prime}\rightarrow pp^{\prime},\frac{3}{2}}^{\dagger}
M_{ss^{\prime}\rightarrow pp^{\prime},\frac{3}{2}},
\end{equation}
and similarly for $P_{ss^{\prime}\rightarrow
pp^{\prime},\frac{1}{2}}$ and $P_{sp\rightarrow ps}$. We will see
in Section~\ref{separability} that these three matrices are very
important when it comes to averaging Eq.~(\ref{aadagger}).
Therefore we present the three $P$ matrices in
Tables~\ref{Pss'pp'3/2}--\ref{Pspps}.

\begin{table}
\begin{center}
\begin{tabular}{||c||c|c|c|c||}
\hline

$P_{ss^{\prime}\rightarrow pp^{\prime},3/2}$ & $\left|
s^{\prime},\frac{1}{2}\right\rangle\left|
s,\frac{1}{2}\right\rangle$ & $\left|
s^{\prime},\frac{1}{2}\right\rangle\left|
s,-\frac{1}{2}\right\rangle$ & $\left|
s^{\prime},-\frac{1}{2}\right\rangle\left|
s,\frac{1}{2}\right\rangle$ & $\left|
s^{\prime},-\frac{1}{2}\right\rangle\left|
s,-\frac{1}{2}\right\rangle$
\\ \hline\hline

$\left\langle s,\frac{1}{2}\right|\left\langle
s^{\prime},\frac{1}{2}\right|$ & $\frac{3}{2}\sin^{2}\theta$ &
$-\frac{1}{2}\sin\left(2\theta\right)e^{-i\phi}$ & $0$ & $0$
\\ \hline

$\left\langle s,-\frac{1}{2}\right|\left\langle
s^{\prime},\frac{1}{2}\right|$ &
$-\frac{1}{2}\sin\left(2\theta\right)e^{i\phi}$ &
$\frac{1}{12}\left[ 5+3\cos\left(2\theta\right) \right]$ & $0$ &
$0$
\\ \hline

$\left\langle s,\frac{1}{2}\right|\left\langle
s^{\prime},-\frac{1}{2}\right|$ & $0$ & $0$ & $\frac{1}{12}\left[
5+3\cos\left(2\theta\right) \right]$ &
$\frac{1}{2}\sin\left(2\theta\right) e^{-i\phi}$
\\ \hline

$\left\langle s,-\frac{1}{2}\right|\left\langle
s^{\prime},-\frac{1}{2}\right|$ & $0$ & $0$ &
$\frac{1}{2}\sin\left(2\theta\right) e^{i\phi}$ &
$\frac{3}{2}\sin^{2}\theta$
\\ \hline
\end{tabular}
\caption[The $P$ matrix for the $ss^{\prime}\rightarrow
pp^{\prime}$ process $\left( \left| m_{j}\right|=3/2 \right)$]{The
$P$ matrix for the $ss^{\prime}\rightarrow pp^{\prime}$ process
when $\left| m_{j}\right|=3/2$.} \label{Pss'pp'3/2}
\end{center}
\end{table}

\begin{table}
\begin{center}
\begin{tabular}{||c||c|c|c|c||}
\hline

$P_{ss^{\prime}\rightarrow pp^{\prime},1/2}$ & $\left|
s^{\prime},\frac{1}{2}\right\rangle\left|
s,\frac{1}{2}\right\rangle$ & $\left|
s^{\prime},\frac{1}{2}\right\rangle\left|
s,-\frac{1}{2}\right\rangle$ & $\left|
s^{\prime},-\frac{1}{2}\right\rangle\left|
s,\frac{1}{2}\right\rangle$ & $\left|
s^{\prime},-\frac{1}{2}\right\rangle\left|
s,-\frac{1}{2}\right\rangle$
\\ \hline\hline

$\left\langle s,\frac{1}{2}\right|\left\langle
s^{\prime},\frac{1}{2}\right|$ & $\frac{1}{36}\left[
29+3\cos\left(2\theta\right) \right]$ &
$-\frac{1}{6}\sin\left(2\theta\right) e^{-i\phi}$ &
$-\frac{1}{3}\sin\left(2\theta\right) e^{-i\phi}$ &
$-\frac{2}{3}\sin^{2}\theta e^{-2i\phi}$
\\ \hline

$\left\langle s,-\frac{1}{2}\right|\left\langle
s^{\prime},\frac{1}{2}\right|$ &
$-\frac{1}{6}\sin\left(2\theta\right) e^{i\phi}$ &
$\frac{5}{36}\left[ 5+3\cos\left(2\theta\right) \right]$ &
$\frac{1}{9}\left[ 5+3\cos\left(2\theta\right) \right]$ &
$\frac{1}{3}\sin\left(2\theta\right) e^{-i\phi}$
\\ \hline

$\left\langle s,\frac{1}{2}\right|\left\langle
s^{\prime},-\frac{1}{2}\right|$ &
$-\frac{1}{3}\sin\left(2\theta\right) e^{i\phi}$ &
$\frac{1}{9}\left[ 5+3\cos\left(2\theta\right) \right]$ &
$\frac{5}{36}\left[ 5+3\cos\left(2\theta\right) \right]$ &
$\frac{1}{6}\sin\left(2\theta\right) e^{-i\phi}$
\\ \hline

$\left\langle s,-\frac{1}{2}\right|\left\langle
s^{\prime},-\frac{1}{2}\right|$ & $-\frac{2}{3}\sin^{2}\theta
e^{2i\phi}$ & $\frac{1}{3}\sin\left(2\theta\right) e^{i\phi}$ &
$\frac{1}{6}\sin\left(2\theta\right) e^{i\phi}$ &
$\frac{1}{36}\left[ 29+3\cos\left(2\theta\right) \right]$
\\ \hline
\end{tabular}
\caption[The $P$ matrix for the $ss^{\prime}\rightarrow
pp^{\prime}$ process $\left( \left| m_{j}\right|=1/2 \right)$]{The
$P$ matrix for the $ss^{\prime}\rightarrow pp^{\prime}$ process
when $\left| m_{j}\right|=1/2$.} \label{Pss'pp'1/2}
\end{center}
\end{table}

\begin{table}
\begin{center}
\begin{tabular}{||c||c|c|c|c||}
\hline

$P_{sp\rightarrow ps}$ & $\left|p,\frac{1}{2}\right\rangle\left|
s,\frac{1}{2}\right\rangle$ &
$\left|p,\frac{1}{2}\right\rangle\left|
s,-\frac{1}{2}\right\rangle$ &
$\left|p,-\frac{1}{2}\right\rangle\left|
s,\frac{1}{2}\right\rangle$ &
$\left|p,-\frac{1}{2}\right\rangle\left|
s,-\frac{1}{2}\right\rangle$
\\ \hline\hline

$\left\langle s,\frac{1}{2}\right|\left\langle
p,\frac{1}{2}\right|$ & $\frac{1}{9}\left[
7-3\cos\left(2\theta\right) \right]$ &
$-\frac{1}{3}\sin\left(2\theta\right)e^{-i\phi}$ &
$-\frac{1}{3}\sin\left(2\theta\right)e^{-i\phi}$ &
$-\frac{2}{3}\sin^{2}\theta e^{-2i\phi}$
\\ \hline

$\left\langle s,-\frac{1}{2}\right|\left\langle
p,\frac{1}{2}\right|$ &
$-\frac{1}{3}\sin\left(2\theta\right)e^{i\phi}$ &
$\frac{1}{9}\left[ 5+3\cos\left(2\theta\right) \right]$ &
$\frac{1}{9}\left[ 5+3\cos\left(2\theta\right) \right]$ &
$\frac{1}{3}\sin\left(2\theta\right) e^{-i\phi}$
\\ \hline

$\left\langle s,\frac{1}{2}\right|\left\langle
p,-\frac{1}{2}\right|$ &
$-\frac{1}{3}\sin\left(2\theta\right)e^{i\phi}$ &
$\frac{1}{9}\left[ 5+3\cos\left(2\theta\right) \right]$ &
$\frac{1}{9}\left[ 5+3\cos\left(2\theta\right) \right]$ &
$\frac{1}{3}\sin\left(2\theta\right) e^{-i\phi}$
\\ \hline

$\left\langle s,-\frac{1}{2}\right|\left\langle
p,-\frac{1}{2}\right|$ & $-\frac{2}{3}\sin^{2}\theta e^{2i\phi}$ &
$\frac{1}{3}\sin\left(2\theta\right) e^{i\phi}$ &
$\frac{1}{3}\sin\left(2\theta\right) e^{i\phi}$ &
$\frac{1}{9}\left[ 7-3\cos\left(2\theta\right) \right]$
\\ \hline
\end{tabular}
\caption[The $P$ matrix for the $sp\rightarrow ps$ process]{The
$P$ matrix for the $sp\rightarrow ps$ process.} \label{Pspps}
\end{center}
\end{table}

We will also see in Section~\ref{separability} that in order to
obtain the average of Eq.~(\ref{aadagger}) we will need the
decompositions of the $P$ matrices into linear combinations of the
Dirac matrices. We have seen that we can use Eqs.~(\ref{sumeqn})
and (\ref{traceeqn}) to determine these linear combinations, and
in this way we obtain
Tables~\ref{ss'pp'3/2tracetable}--\ref{sppstracetable}.

\begin{table}
\begin{center}
\begin{tabular}{||c|c||}
\hline

\emph{Dirac Matrix} & \emph{Projection}
\\ \hline\hline

$\sigma_{0}\otimes\sigma_{0}$ & $\frac{1}{12}\left[7 -
3\cos\left(2\theta\right)\right]$
\\ \hline

$\sigma_{0}\otimes\sigma_{1}$ & $0$
\\ \hline

$\sigma_{0}\otimes\sigma_{2}$ & $0$
\\ \hline

$\sigma_{0}\otimes\sigma_{3}$ & $0$
\\ \hline

$\sigma_{1}\otimes\sigma_{0}$ & $0$
\\ \hline

$\sigma_{1}\otimes\sigma_{1}$ & $0$
\\ \hline

$\sigma_{1}\otimes\sigma_{2}$ & $0$
\\ \hline

$\sigma_{1}\otimes\sigma_{3}$ & $0$
\\ \hline

$\sigma_{2}\otimes\sigma_{0}$ & $0$
\\ \hline

$\sigma_{2}\otimes\sigma_{1}$ & $0$
\\ \hline

$\sigma_{2}\otimes\sigma_{2}$ & $0$
\\ \hline

$\sigma_{2}\otimes\sigma_{3}$ & $0$
\\ \hline

$\sigma_{3}\otimes\sigma_{0}$ & $0$
\\ \hline

$\sigma_{3}\otimes\sigma_{1}$ &
$-\frac{1}{2}\sin\left(2\theta\right)\cos\phi$
\\ \hline

$\sigma_{3}\otimes\sigma_{2}$ &
$-\frac{1}{2}\sin\left(2\theta\right)\sin\phi$
\\ \hline

$\sigma_{3}\otimes\sigma_{3}$ & $\frac{1}{6}\left[1 -
3\cos\left(2\theta\right)\right]$
\\ \hline
\end{tabular}
\caption[Projections of $P$ onto the Dirac matrices for the
$ss^{\prime}\rightarrow pp^{\prime}$ process
$\left(\left|m_{j}\right| = 3/2\right)$]{Projections of the
$4\times 4$ matrix $P$ onto the Dirac matrices for the
$ss^{\prime}\rightarrow pp^{\prime}$ process when $\left|
m_{j}\right| = 3/2$ for the $p^{\prime}$ atom.}
\label{ss'pp'3/2tracetable}
\end{center}
\end{table}

\begin{table}
\begin{center}
\begin{tabular}{||c|c||}
\hline

\emph{Dirac Matrix} & \emph{Projection}
\\ \hline\hline

$\sigma_{0}\otimes\sigma_{0}$ & $\frac{1}{4}\left[3 +
\cos\left(2\theta\right)\right]$
\\ \hline

$\sigma_{0}\otimes\sigma_{1}$ & $0$
\\ \hline

$\sigma_{0}\otimes\sigma_{2}$ & $0$
\\ \hline

$\sigma_{0}\otimes\sigma_{3}$ & $0$
\\ \hline

$\sigma_{1}\otimes\sigma_{0}$ & $0$
\\ \hline

$\sigma_{1}\otimes\sigma_{1}$ & $\frac{1}{18}\left[5 +
3\cos\left(2\theta\right) -
6\sin^{2}\theta\cos\left(2\phi\right)\right]$
\\ \hline

$\sigma_{1}\otimes\sigma_{2}$ &
$-\frac{1}{3}\sin^{2}\theta\sin\left(2\phi\right)$
\\ \hline

$\sigma_{1}\otimes\sigma_{3}$ &
$-\frac{1}{3}\sin\left(2\theta\right)\cos\phi$
\\ \hline

$\sigma_{2}\otimes\sigma_{0}$ & $0$
\\ \hline

$\sigma_{2}\otimes\sigma_{1}$ &
$-\frac{1}{3}\sin^{2}\theta\sin\left(2\phi\right)$
\\ \hline

$\sigma_{2}\otimes\sigma_{2}$ & $\frac{1}{18}\left[5 +
3\cos\left(2\theta\right) +
6\sin^{2}\theta\cos\left(2\phi\right)\right]$
\\ \hline

$\sigma_{2}\otimes\sigma_{3}$ &
$-\frac{1}{3}\sin\left(2\theta\right)\sin\phi$
\\ \hline

$\sigma_{3}\otimes\sigma_{0}$ & $0$
\\ \hline

$\sigma_{3}\otimes\sigma_{1}$ &
$-\frac{1}{6}\sin\left(2\theta\right)\cos\phi$
\\ \hline

$\sigma_{3}\otimes\sigma_{2}$ &
$-\frac{1}{6}\sin\left(2\theta\right)\sin\phi$
\\ \hline

$\sigma_{3}\otimes\sigma_{3}$ & $\frac{1}{18}\left[1 -
3\cos\left(2\theta\right)\right]$
\\ \hline
\end{tabular}
\caption[Projections of $P$ onto the Dirac matrices for the
$ss^{\prime}\rightarrow pp^{\prime}$ process
$\left(\left|m_{j}\right| = 1/2\right)$]{Projections of the
$4\times 4$ matrix $P$ onto the Dirac matrices for the
$ss^{\prime}\rightarrow pp^{\prime}$ process when $\left|
m_{j}\right| = 1/2$ for the $p^{\prime}$ atom.}
\label{ss'pp'1/2tracetable}
\end{center}
\end{table}

\begin{table}
\begin{center}
\begin{tabular}{||c|c||}
\hline

\emph{Dirac Matrix} & \emph{Projection}
\\ \hline\hline

$\sigma_{0}\otimes\sigma_{0}$ & $\frac{2}{3}$
\\ \hline

$\sigma_{0}\otimes\sigma_{1}$ & $0$
\\ \hline

$\sigma_{0}\otimes\sigma_{2}$ & $0$
\\ \hline

$\sigma_{0}\otimes\sigma_{3}$ & $0$
\\ \hline

$\sigma_{1}\otimes\sigma_{0}$ & $0$
\\ \hline

$\sigma_{1}\otimes\sigma_{1}$ & $\frac{1}{18}\left[5 +
3\cos\left(2\theta\right) -
6\cos\left(2\phi\right)\sin^{2}\theta\right]$
\\ \hline

$\sigma_{1}\otimes\sigma_{2}$ &
$-\frac{1}{3}\sin^{2}\theta\sin\left( 2\phi\right)$
\\ \hline

$\sigma_{1}\otimes\sigma_{3}$ &
$-\frac{1}{3}\sin\left(2\theta\right)\cos\phi$
\\ \hline

$\sigma_{2}\otimes\sigma_{0}$ & $0$
\\ \hline

$\sigma_{2}\otimes\sigma_{1}$ &
$-\frac{1}{3}\sin^{2}\theta\sin\left( 2\phi\right)$
\\ \hline

$\sigma_{2}\otimes\sigma_{2}$ & $\frac{1}{18}\left[5 +
3\cos\left(2\theta\right) +
6\cos\left(2\phi\right)\sin^{2}\theta\right]$
\\ \hline

$\sigma_{2}\otimes\sigma_{3}$ &
$-\frac{1}{3}\sin\left(2\theta\right)\sin\phi$
\\ \hline

$\sigma_{3}\otimes\sigma_{0}$ & $0$
\\ \hline

$\sigma_{3}\otimes\sigma_{1}$ &
$-\frac{1}{3}\sin\left(2\theta\right)\cos\phi$
\\ \hline

$\sigma_{3}\otimes\sigma_{2}$ &
$-\frac{1}{3}\sin\left(2\theta\right)\sin\phi$
\\ \hline

$\sigma_{3}\otimes\sigma_{3}$ &
$\frac{1}{9}\left[1-3\cos\left(2\theta\right)\right]$
\\ \hline
\end{tabular}
\caption[Projections of $P$ onto the Dirac matrices for the
$sp\rightarrow ps$ process]{Projections of the $4\times 4$ matrix
$P$ onto the Dirac matrices for the $sp\rightarrow ps$ process.}
\label{sppstracetable}
\end{center}
\end{table}

\subsection{Separability}
\label{separability}

At this point we would like to write
\begin{equation}
\exp\left[ -\beta\sum_{k}V_{k}^{2} P^{\left(k\right)}\right] =
\prod_{k} \exp\left[-\beta V_{k}^{2} P^{\left(k\right)}\right],
\label{wishfulthinking}
\end{equation}
so that we can perform our usual averaging procedure as described
in Section~\ref{averaging}. Unfortunately,
Eq.~(\ref{wishfulthinking}) does not hold unless the commutator of
$P^{\left(k\right)}$ with $P^{\left(l\right)}$ is a $c$-number. We
can use Tables~\ref{ss'pp'3/2tracetable}--\ref{sppstracetable} to
check if this is the case for our three potentials.

Because it is the simplest, we first consider the potential for
the $ss^{\prime}\rightarrow pp^{\prime}$ process when $\left|
m_{j}\right| = 3/2$. Using Table~\ref{ss'pp'3/2tracetable} we see
that
\begin{eqnarray}
P^{\left(k\right)} &=&
p_{00}^{\left(k\right)}\left\{\sigma_{0}^{\left(0\right)}\otimes
\sigma_{0}^{\left(k\right)}\right\} +
p_{31}^{\left(k\right)}\left\{\sigma_{3}^{\left(0\right)}\otimes
\sigma_{1}^{\left(k\right)}\right\} + \nonumber \\ &&
p_{32}^{\left(k\right)}\left\{\sigma_{3}^{\left(0\right)}\otimes
\sigma_{2}^{\left(k\right)}\right\} +
p_{33}^{\left(k\right)}\left\{\sigma_{3}^{\left(0\right)}\otimes
\sigma_{3}^{\left(k\right)}\right\},
\end{eqnarray}
for some coefficients $p_{00}^{\left(k\right)}$,
$p_{31}^{\left(k\right)}$, $p_{32}^{\left(k\right)}$, and
$p_{33}^{\left(k\right)}$ that depend on $\theta_{k}$ and
$\phi_{k}$. We also define
\begin{equation}
\left\{\sigma_{\alpha}^{\left(0\right)}\otimes
\sigma_{\beta}^{\left(k\right)}\right\} =
\sigma_{\alpha}^{\left(0\right)}\otimes
\sigma_{0}^{\left(1\right)}\otimes\cdots\otimes
\sigma_{0}^{\left(k-1\right)}\otimes
\sigma_{\beta}^{\left(k\right)}\otimes
\sigma_{0}^{\left(k+1\right)}\otimes\cdots\otimes
\sigma_{0}^{\left(N\right)}.
\end{equation}
Similarly, we have
\begin{eqnarray}
P^{\left(l\right)} &=&
p_{00}^{\left(l\right)}\left\{\sigma_{0}^{\left(0\right)}\otimes
\sigma_{0}^{\left(l\right)}\right\} +
p_{31}^{\left(l\right)}\left\{\sigma_{3}^{\left(0\right)}\otimes
\sigma_{1}^{\left(l\right)}\right\} + \nonumber \\ &&
p_{32}^{\left(l\right)}\left\{\sigma_{3}^{\left(0\right)}\otimes
\sigma_{2}^{\left(l\right)}\right\} +
p_{33}^{\left(l\right)}\left\{\sigma_{3}^{\left(0\right)}\otimes
\sigma_{3}^{\left(l\right)}\right\}. \label{Pss'pp'3/2expanded}
\end{eqnarray}

Because $\sigma_{0}$ is just the $2\times 2$ identity matrix, it
follows that $\left\{ \sigma_{0}^{\left(0\right)}\otimes
\sigma_{0}^{\left(k\right)} \right\}$ is just the $2^{N+1}\times
2^{N+1}$ identity matrix. Thus $\left\{
\sigma_{0}^{\left(0\right)}\otimes \sigma_{0}^{\left(k\right)}
\right\}$ will commute with anything. Now consider the product
\begin{equation}
\Pi = \left\{\sigma_{3}^{\left(0\right)}\otimes
\sigma_{\alpha}^{\left(k\right)}\right\}
\left\{\sigma_{3}^{\left(0\right)}\otimes
\sigma_{\beta}^{\left(l\right)}\right\}. \label{Piprod}
\end{equation}
Although the result does not depend on it, we assume for
convenience that $k \lneq l$. We also expand each of the terms in
Eq.~(\ref{Piprod}) and apply Eq.~(\ref{tensorproductproduct}). We
find that
\begin{eqnarray}
\Pi &=& \left[\,\sigma_{3}^{\left(0\right)}\otimes
\sigma_{0}^{\left(1\right)}\otimes\cdots\otimes
\sigma_{0}^{\left(k-1\right)}\otimes
\sigma_{\alpha}^{\left(k\right)}\otimes
\sigma_{0}^{\left(k+1\right)}\otimes\cdots\otimes
\sigma_{0}^{\left(N\right)}\right] \times \nonumber \\ &&
\left[\,\sigma_{3}^{\left(0\right)}\otimes
\sigma_{0}^{\left(1\right)}\otimes\cdots\otimes
\sigma_{0}^{\left(l-1\right)}\otimes
\sigma_{\beta}^{\left(l\right)}\otimes
\sigma_{0}^{\left(l+1\right)}\otimes\cdots\otimes
\sigma_{0}^{\left(N\right)}\right] \nonumber \\ &=&
\left(\sigma_{3}^{\left(0\right)}
\sigma_{3}^{\left(0\right)}\right)\otimes
\sigma_{0}^{\left(1\right)}\otimes\cdots\otimes
\sigma_{0}^{\left(k-1\right)}\otimes
\sigma_{\alpha}^{\left(k\right)}\otimes
\sigma_{0}^{\left(k+1\right)}\otimes\cdots\otimes \nonumber \\ &&
\sigma_{0}^{\left(l-1\right)}\otimes
\sigma_{\beta}^{\left(l\right)}\otimes
\sigma_{0}^{\left(l+1\right)}\otimes\cdots\otimes
\sigma_{0}^{\left(N\right)},
\end{eqnarray}
since $\sigma_{0}$ is just the identity matrix. Note that if we
had considered instead the product
\begin{equation}
\left\{\sigma_{3}^{\left(0\right)}\otimes
\sigma_{\beta}^{\left(l\right)}\right\}
\left\{\sigma_{3}^{\left(0\right)}\otimes
\sigma_{\alpha}^{\left(k\right)}\right\},
\end{equation}
then we would have obtained the same result. Therefore each term
in $P^{\left(k\right)}$ commutes with each of the terms in
$P^{\left(l\right)}$. It follows that $P^{\left(k\right)}$
commutes with $P^{\left(l\right)}$, and hence we can perform the
decomposition of Eq.~(\ref{wishfulthinking}) in this case.

Now consider the potential for the $ss^{\prime}\rightarrow
pp^{\prime}$ process when $\left| m_{j}\right| = 1/2$. Using
Table~\ref{ss'pp'1/2tracetable} we see that
\begin{eqnarray}
P^{\left(k\right)} &=& p_{00}^{\left(k\right)}
\left\{\sigma_{0}^{\left(0\right)}\otimes\sigma_{0}^{\left(k\right)}\right\}
+ p_{11}^{\left(k\right)}
\left\{\sigma_{1}^{\left(0\right)}\otimes\sigma_{1}^{\left(k\right)}\right\}
+ p_{12}^{\left(k\right)}
\left\{\sigma_{1}^{\left(0\right)}\otimes\sigma_{2}^{\left(k\right)}\right\}
+ p_{13}^{\left(k\right)}
\left\{\sigma_{1}^{\left(0\right)}\otimes\sigma_{3}^{\left(k\right)}\right\}
+ \nonumber \\ && p_{21}^{\left(k\right)}
\left\{\sigma_{2}^{\left(0\right)}\otimes\sigma_{1}^{\left(k\right)}\right\}
+ p_{22}^{\left(k\right)}
\left\{\sigma_{2}^{\left(0\right)}\otimes\sigma_{2}^{\left(k\right)}\right\}
+ p_{23}^{\left(k\right)}
\left\{\sigma_{2}^{\left(0\right)}\otimes\sigma_{3}^{\left(k\right)}\right\}
+ p_{31}^{\left(k\right)}
\left\{\sigma_{3}^{\left(0\right)}\otimes\sigma_{1}^{\left(k\right)}\right\}
+ \nonumber \\ && p_{32}^{\left(k\right)}
\left\{\sigma_{3}^{\left(0\right)}\otimes\sigma_{2}^{\left(k\right)}\right\}
+ p_{33}^{\left(k\right)}
\left\{\sigma_{3}^{\left(0\right)}\otimes\sigma_{3}^{\left(k\right)}\right\},
\end{eqnarray}
and similarly for $P^{\left(l\right)}$. Examining the product
\begin{equation}
\Pi =
\left\{\sigma_{1}^{\left(0\right)}\otimes\sigma_{3}^{\left(k\right)}\right\}
\left\{\sigma_{2}^{\left(0\right)}\otimes\sigma_{1}^{\left(l\right)}\right\},
\end{equation}
we see that
\begin{eqnarray}
\Pi &=& \left(\sigma_{1}^{\left(0\right)}
\sigma_{2}^{\left(0\right)}\right)\otimes
\sigma_{0}^{\left(1\right)}\otimes\cdots\otimes
\sigma_{0}^{\left(k-1\right)}\otimes
\sigma_{3}^{\left(k\right)}\otimes
\sigma_{0}^{\left(k+1\right)}\otimes\cdots\otimes \nonumber \\ &&
\sigma_{0}^{\left(l-1\right)}\otimes
\sigma_{1}^{\left(l\right)}\otimes
\sigma_{0}^{\left(l+1\right)}\otimes\cdots\otimes
\sigma_{0}^{\left(N\right)},
\end{eqnarray}
while if we had multiplied the factors in the opposite order we
would have obtained
\begin{eqnarray}
&& \left(\sigma_{2}^{\left(0\right)}
\sigma_{1}^{\left(0\right)}\right)\otimes
\sigma_{0}^{\left(1\right)}\otimes\cdots\otimes
\sigma_{0}^{\left(k-1\right)}\otimes
\sigma_{3}^{\left(k\right)}\otimes
\sigma_{0}^{\left(k+1\right)}\otimes\cdots\otimes \nonumber \\ &&
\sigma_{0}^{\left(l-1\right)}\otimes
\sigma_{1}^{\left(l\right)}\otimes
\sigma_{0}^{\left(l+1\right)}\otimes\cdots\otimes
\sigma_{0}^{\left(N\right)}.
\end{eqnarray}
Since $\sigma_{1}^{\left(0\right)}\sigma_{2}^{\left(0\right)} =
i\sigma_{3}^{\left(0\right)}$, while
$\sigma_{2}^{\left(0\right)}\sigma_{1}^{\left(0\right)} =
-i\sigma_{3}^{\left(0\right)}$, we see that the commutator of
$\left\{\sigma_{1}^{\left(0\right)}\otimes\sigma_{3}^{\left(k\right)}\right\}$
and
$\left\{\sigma_{2}^{\left(0\right)}\otimes\sigma_{1}^{\left(l\right)}\right\}$
is not a $c$-number. Because the set of $\left\{
p_{\alpha\beta}^{\left(k\right)} \right\}$ is not the same as the
set of $\left\{ p_{\alpha\beta}^{\left(l\right)} \right\}$, it
follows that the commutator of $P^{\left(k\right)}$ and
$P^{\left(l\right)}$ cannot be a $c$-number. Hence we cannot
perform the separation of Eq.~(\ref{wishfulthinking}) in this
case.

Considering the potential for the $sp\rightarrow ps$ process and
consulting Table~\ref{ss'pp'1/2tracetable}, we see that the
coefficients of $\sigma_{1}\otimes\sigma_{3}$ and
$\sigma_{2}\otimes\sigma_{1}$ are both nonzero. Therefore the
argument of the previous paragraph will hold here too, and again
we cannot perform the separation of Eq.~(\ref{wishfulthinking}).

\subsection{Averaging the $ss^{\prime}\rightarrow pp^{\prime}$
process when $\left|m_{j}\right| = 3/2$ for the $p^{\prime}$ atom}
\label{averagingwithspin}

In this case we can perform the separation of
Eq.~(\ref{wishfulthinking}), and so we see from
Eq.~(\ref{aadagger}) that the averaging problem reduces the
computation of the average of
\begin{equation}
\frac{1}{2^{N+1}}\Tr\,\prod_{k=1}^{N} \exp\left[ -\beta V_{k}^{2}
P^{\left(k\right)}\right]. \label{separateproduct}
\end{equation}
Using Table~\ref{ss'pp'3/2tracetable}, we can write
\begin{eqnarray}
\exp\left[ -\beta V_{k}^{2} P^{\left(k\right)}\right] &=&
\exp\left[ -\beta v_{00}^{\left(k\right)} \left\{
\sigma_{0}^{\left(0\right)}\otimes\sigma_{0}^{\left(k\right)}\right\}
-\beta \sum_{b=1}^{3} v_{3b}^{\left(k\right)} \left\{
\sigma_{3}^{\left(0\right)}\otimes\sigma_{b}^{\left(k\right)}\right\}
\right] \nonumber \\ &=& \exp\left[ -\beta v_{00}^{\left(k\right)}
\right] \exp\left[ -\beta \sum_{b=1}^{3} v_{3b}^{\left(k\right)}
\left\{
\sigma_{3}^{\left(0\right)}\otimes\sigma_{b}^{\left(k\right)}\right\}
\right], \label{expintermediate}
\end{eqnarray}
since $\left\{
\sigma_{0}^{\left(0\right)}\otimes\sigma_{0}^{\left(k\right)}\right\}$
is just the $2^{N+1}\times 2^{N+1}$ identity matrix, which
certainly commutes with $\left\{
\sigma_{3}^{\left(0\right)}\otimes\sigma_{b}^{\left(k\right)}\right\}$.
Here $v_{\alpha\beta}^{\left(k\right)}$ depends on $\mbf{r}_{k}$,
and specifically
\begin{equation}
v_{\alpha\beta}^{\left(k\right)} =
\frac{\mu\mu^{\prime}}{r_{k}^{3}}
p_{\alpha\beta}^{\left(k\right)},
\end{equation}
where $p_{\alpha\beta}^{\left(k\right)}$ is the same quantity as
in Eq.~(\ref{Pss'pp'3/2expanded}).

If we define
\begin{equation}
\mbf{v}^{\left(k\right)} = \sum_{b=1}^{3} v_{3b}^{\left(k\right)}
\, \hat{\mbf{e}}_{b}
\end{equation}
and
\begin{equation}
\mbf{\sigma}^{\left(k\right)} = \sum_{b=1}^{3}
\sigma_{3b}^{\left(k\right)} \, \hat{\mbf{e}}_{b},
\end{equation}
then we can rewrite Eq.~(\ref{expintermediate}) as
\begin{equation}
\exp\left[ -\beta V_{k}^{2} P^{\left(k\right)}\right] = \exp\left[
-\beta v_{00}^{\left(k\right)}\right] \exp\left( \left\{
\sigma_{3}^{\left(0\right)} \otimes \left[-\beta
\mbf{v}^{\left(k\right)} \cdot
\mbf{\sigma}^{\left(k\right)}\right]\right\} \right).
\end{equation}
Now we note that
\begin{eqnarray}
\exp\left( \sigma_{3} \otimes A\right) &=& \sum_{j=0}^{\infty}
\frac{1}{j!} \left( \sigma_{3} \otimes A \right)^{j} \nonumber \\
&=& \sum_{j=0}^{\infty} \frac{1}{j!} \left( \sigma_{3} \right)^{j}
\otimes \left(A\right)^{j} \nonumber \\ &=& \sum_{k=0}^{\infty}
\frac{\sigma_{0} \otimes \left(A\right)^{2k}}{\left(2k\right)!} +
\sum_{l=0}^{\infty} \frac{\sigma_{3} \otimes
\left(A\right)^{2l+1}}{\left(2l+1\right)!} \nonumber \\ &=&
\sigma_{0} \otimes \cosh A + \sigma_{3} \otimes \sinh A,
\label{exp3A}
\end{eqnarray}
so that
\begin{equation}
\exp\left( \left\{ \sigma_{3}^{\left(0\right)} \otimes
\left[-\beta \mbf{v}^{\left(k\right)} \cdot
\mbf{\sigma}^{\left(k\right)}\right]\right\} \right) =
\sigma_{0}^{\left(0\right)} \otimes \cosh\left[
-\beta\,\Sigma^{\left(k\right)}\right] +
\sigma_{3}^{\left(0\right)} \otimes \sinh\left[
-\beta\,\Sigma^{\left(k\right)}\right],
\end{equation}
where to save space we define
\begin{equation}
\Sigma^{\left(k\right)} = \sigma_{0}^{\left(1\right)}
\otimes\cdots\otimes \sigma_{0}^{\left(k-1\right)} \otimes
\left[\mbf{v}^{\left(k\right)}\cdot\mbf{\sigma}^{\left(k\right)}\right]
\otimes \sigma_{0}^{\left(k+1\right)} \otimes\cdots\otimes
\sigma_{0}^{\left(N\right)}.
\end{equation}

A derivation very similar to that of Eq.~(\ref{exp3A}) shows that
\begin{equation}
\cosh\left( \sigma_{0} \otimes A\right) = \sigma_{0} \otimes \cosh
A,
\end{equation}
and
\begin{equation}
\sinh\left( \sigma_{0} \otimes A\right) = \sigma_{0} \otimes \sinh
A,
\end{equation}
and by repeated application of these last two relations we finally
arrive at
\begin{eqnarray}
\exp\left( \left\{ \sigma_{3}^{\left(0\right)} \otimes
\left[-\beta \mbf{v}^{\left(k\right)} \cdot
\mbf{\sigma}^{\left(k\right)}\right]\right\} \right) &=& \left\{
\sigma_{0}^{\left(0\right)} \otimes
\cosh\left[-\beta\mbf{v}^{\left(k\right)}\cdot
\mbf{\sigma}^{\left(k\right)}\right] \right\} - \nonumber
\\ && \left\{ \sigma_{3}^{\left(0\right)} \otimes
\sinh\left[-\beta\mbf{v}^{\left(k\right)}
\cdot\mbf{\sigma}^{\left(k\right)}\right] \right\}.
\end{eqnarray}

From page 166 of Ref.~\cite{Sakurai1994a} we have the familiar
result that
\begin{equation}
\left(\mbf{v}^{\left(k\right)}\cdot\mbf{\sigma}^{\left(k\right)}\right)^{n}
= \left\{
\begin{array}{ll}
1 & \mbox{if $n$ is even,} \\
\mbf{v}^{\left(k\right)}\cdot\mbf{\sigma}^{\left(k\right)} &
\mbox{if $n$ is odd.}
\end{array}
\right.
\end{equation}
From this it is easy to show that
\begin{equation}
\cosh\left[-\beta\mbf{v}^{\left(k\right)}\cdot
\mbf{\sigma}^{\left(k\right)}\right] = \cosh\left[ \beta
v^{\left(k\right)}\right],
\end{equation}
and
\begin{equation}
\sinh\left[-\beta\mbf{v}^{\left(k\right)}\cdot
\mbf{\sigma}^{\left(k\right)}\right] =
-\hat{\mbf{v}}^{\left(k\right)}\cdot \mbf{\sigma}^{\left(k\right)}
\sinh\left[ \beta v^{\left(k\right)}\right].
\end{equation}
Thus we see that
\begin{eqnarray}
\exp\left( \left\{ \sigma_{3}^{\left(0\right)} \otimes
\left[-\beta \mbf{v}^{\left(k\right)} \cdot
\mbf{\sigma}^{\left(k\right)}\right]\right\} \right) &=&
\cosh\left[\beta v^{\left(k\right)}\right] - \nonumber
\\ && \sinh\left[\beta
v^{\left(k\right)} \right] \left\{ \sigma_{3}^{\left(0\right)}
\otimes \hat{\mbf{v}}^{\left(k\right)}
\cdot\mbf{\sigma}^{\left(k\right)}\right\},
\end{eqnarray}
and hence
\begin{eqnarray}
\exp\left[ -\beta V_{k}^{2} P^{\left(k\right)}\right] &=&
\exp\left[ -\beta v_{00}^{\left(k\right)}\right] \cosh\left[\beta
v^{\left(k\right)}\right] - \nonumber
\\ && \exp\left[ -\beta v_{00}^{\left(k\right)}\right] \sinh\left[\beta
v^{\left(k\right)} \right] \left\{ \sigma_{3}^{\left(0\right)}
\otimes \hat{\mbf{v}}^{\left(k\right)}
\cdot\mbf{\sigma}^{\left(k\right)}\right\}. \label{fourterms}
\end{eqnarray}

We now note that because $\Tr\,\sigma_{a} = 0$ if $a$ is equal to
$1$, $2$, or $3$, it follows that when we expand the product
\begin{equation}
\left[ a_{00}^{\left(k\right)}
\left\{\sigma_{0}^{\left(0\right)}\otimes
\sigma_{0}^{\left(k\right)}\right\} + \left\{
\sigma_{3}^{\left(0\right)}\otimes
\mbf{a}^{\left(k\right)}\cdot\mbf{\sigma}^{\left(k\right)}
\right\} \right] \left[ a_{00}^{\left(l\right)}\left\{
\sigma_{0}^{\left(0\right)}\otimes\sigma_{0}^{\left(l\right)}
\right\} + \left\{ \sigma_{3}^{\left(0\right)}\otimes
\mbf{a}^{\left(k\right)}\cdot\mbf{\sigma}^{\left(l\right)}
\right\} \right]
\end{equation}
and take the trace, only the term
\begin{equation}
a_{00}^{\left(k\right)}a_{00}^{\left(l\right)}
\Tr\,\left\{\sigma_{0}^{\left(0\right)}\otimes
\sigma_{0}^{\left(k\right)}\right\} = 2^{N+1}
a_{00}^{\left(k\right)}a_{00}^{\left(l\right)}
\end{equation}
survives. This together with Eqs.~(\ref{separateproduct}) and
(\ref{fourterms}) implies that
\begin{equation}
\frac{1}{2^{N+1}}\Tr\,\prod_{k=1}^{N} \exp\left[ -\beta V_{k}^{2}
P^{\left(k\right)}\right] = \prod_{k=1}^{N} \exp\left[ -\beta
v_{00}^{\left(k\right)}\right] \cosh\left[ \beta
v^{\left(k\right)}\right].
\end{equation}

We can now change variables so that
\begin{equation}
\left\langle \prod_{k=1}^{N} \exp\left[ -\beta
v_{00}^{\left(k\right)}\right] \cosh\left[ \beta
v^{\left(k\right)}\right] \right\rangle
\end{equation}
is equal to a product of $N$ copies of
\begin{equation}
\int_{0}^{\infty} dr_{k}\,r_{k}^{2} \int_{0}^{\pi}
d\theta_{k}\,\sin\theta_{k} \int_{0}^{2\pi} d\phi_{k} \exp\left[
-\beta v_{00}^{\left(k\right)}\right] \cosh\left[ \beta
v^{\left(k\right)}\right].
\end{equation}
This integral, in turn, is equal to the sum of the integrals
\begin{equation}
\frac{1}{2\Omega}\int_{0}^{\infty} dr_{k}\,r_{k}^{2}
\int_{0}^{\pi} d\theta_{k}\,\sin\theta_{k} \int_{0}^{2\pi}
d\phi_{k} \exp\left\{ -\beta \left[v_{00}^{\left(k\right)} -
v^{\left(k\right)}\right]\right\}, \label{integral1}
\end{equation}
and
\begin{equation}
\frac{1}{2\Omega}\int_{0}^{\infty} dr_{k}\,r_{k}^{2}
\int_{0}^{\pi} d\theta_{k}\,\sin\theta_{k} \int_{0}^{2\pi}
d\phi_{k} \exp\left\{ -\beta \left[v_{00}^{\left(k\right)} +
v^{\left(k\right)}\right]\right\}. \label{integral2}
\end{equation}
Once again using the coefficients of
Table~\ref{ss'pp'3/2tracetable}, we find that
\begin{equation}
v_{00}^{\left(k\right)} = \frac{\mu\mu^{\prime}}{r_{k}^{3}}
\frac{1}{12}\left[ 7 - 3\cos\left(2\theta_{k}\right) \right],
\end{equation}
and
\begin{equation}
v^{\left(k\right)} = \frac{\mu\mu^{\prime}}{r_{k}^{3}}
\frac{1}{3\sqrt{3}}\sqrt{5 - 3\cos\left(2\theta_{k}\right)}.
\end{equation}

The integrals of Eqs.~(\ref{integral1}) and (\ref{integral2}) are
now of a familiar type, and they can be done using the methods
explained in Sections~\ref{averaging} and \ref{angular}.
Unfortunately the angular integrals have to be done numerically,
but the result is that\footnote{The astute reader will be relieved
to know that the quantities $v_{00}^{\left(k\right)} +
v^{\left(k\right)}$ and $v_{00}^{\left(k\right)} -
v^{\left(k\right)}$ are both positive for all values of
$\theta_{k}$ and $\phi_{k}$, and so there is no difficulty in
applying the methods of Sections~\ref{averaging} and \ref{angular}
here.}
\begin{equation}
\frac{1}{\Omega}\int_{0}^{\infty} dr_{k}\,r_{k}^{2} \int_{0}^{\pi}
d\theta_{k}\,\sin\theta_{k} \int_{0}^{2\pi} d\phi_{k} \exp\left[
-\beta v_{00}^{\left(k\right)}\right] \cosh\left[ \beta
v^{\left(k\right)}\right] = 1 - 5.18061
\frac{\mu\mu^{\prime}}{\Omega} \sqrt{\beta}.
\end{equation}
It follows that
\begin{equation}
\frac{1}{2^{N+1}}\Tr\,\prod_{k=1}^{N} \exp\left[ -\beta V_{k}^{2}
P^{\left(k\right)}\right] = \exp\left(
-v_{spin}\sqrt{\beta}\right),
\end{equation}
where
\begin{equation}
v_{\mrm{spin}} = 5.18061\, \mu\mu^{\prime}\frac{N}{\Omega}.
\end{equation}

Recalling Eq.~(\ref{v}), we see that when we used the simple
$\mu\mu^{\prime}/r^{3}$ interaction potential of
Chapter~\ref{sparse_no_u}, we obtained
\begin{equation}
v = \frac{4\pi^{3/2}}{3} \mu\mu^{\prime} \frac{N}{\Omega} =
7.42444\, \mu\mu^{\prime} \frac{N}{\Omega}.
\end{equation}
Hence we arrive at
\begin{equation}
v_{\mrm{spin}} = 0.697778\, v. \label{vspin}
\end{equation}

Inserting our averaged result back into Eq.~(\ref{aadagger}), we
see that
\begin{equation}
\left\langle \frac{1}{2^{N+1}}\Tr\,\tilde{S}\left(\alpha\right)
\right\rangle = \frac{1}{2\alpha} - \frac{\alpha^{2} +
\Delta^{2}}{2\alpha} \int_{0}^{\infty} d\beta\,\exp\left[
-\beta\left(\alpha^{2} + \Delta^{2}\right) -
2v_{\mrm{spin}}\sqrt{\beta} \right]. \label{aadaggeravg}
\end{equation}
Comparing with Eq.~(\ref{Stld2}), one immediately sees that
Eq.~(\ref{aadaggeravg}) gives the same results as the signal
computed in Chapter~\ref{sparse_no_u} but with the quantity
$v_{\mrm{spin}}$ appearing in place of $v$.

\section{Summary}
\label{spin_conclusions}

In this chapter we have considered the detailed effects of spin on
the system we are studying. We were only able to fully average the
$ss^{\prime}\rightarrow pp^{\prime}$ process when
$\left|m_{j}\right| = 3/2$ for the $p^{\prime}$ atom, but we can
see from Fig.~\ref{introduction_explineshape} that the case where
$\left|m_{j}\right| = 1/2$ for the $p^{\prime}$ atom is not very
different. In the case that we could treat fully, we found that
the effect of the spin was to change the value of the parameter
$v$ by a factor of $0.697778$ from the simple case where we
neglect the effects of spin and consider an angular independent
$1/r^{3}$ interaction potential.

The consideration of the effects of spin becomes extremely
important when a magnetic field is applied to the system. It was
reported by Renn \emph{et al.\/}~\cite{Renn1994a} that the
introduction of a magnetic field greatly proliferates the number
of nondegenerate states present in a room temperature Rydberg gas.
One can imagine applying a magnetic field that is parallel to the
detuning electric field and of sufficient strength to lift the
$m_{j}$ degeneracy. The form of the $ss^{\prime}\rightarrow
pp^{\prime}$ interaction would then be given by the appropriate
matrix element of Table~\ref{ss'pp'3/2table} or
\ref{ss'pp'1/2table}, and the form of the $sp\rightarrow ps$
interaction would similarly be found in Table~\ref{sppstable}.
Thus we see that this is one possible scheme for selecting the
angular form of the interaction potential.

\chapter{Numerical Simulations}
\markright{Chapter \arabic{chapter}: Numerical Simulations}
\label{simulations}

\section{Introduction}

In this chapter we discuss numerical simulations of the system we
are considering. We start by describing in detail how the
simulations are carried out, then we present the results,
including comparisons of the simulations with the experimental
data.

\section{Review of theoretical model}
\label{theoryreview}

Before discussing exactly how the simulations are carried out, we
review the basic elements of the theory presented at the beginning
of Sections~\ref{basicequations} and \ref{cayleyjustification} of
this thesis. We consider one atom at the origin, initially in the
state $s^{\prime}$, in interaction with a gas of $s$ atoms through
a resonant $ss^{\prime}\rightarrow pp^{\prime}$ dipole-dipole
interaction $V$. We also allow for interaction among the unprimed
atoms via an $sp\rightarrow ps$ process mediated by another
dipole-dipole potential $U$. This system is described by
Eqs.~(\ref{azerodotU}) and (\ref{ckdotU}), which are
\setcounter{tlet}{1}
\renewcommand{\theequation}{\arabic{chapter}.\arabic{equation}\alph{tlet}}
\begin{eqnarray}
 i~\dot{a}_{0} &=& \Delta a_{0} + \sum_{k=1}^{N}V_{k}c_{k}, \label{1a} \\
\stepcounter{tlet}\addtocounter{equation}{-1}
 i~\dot{c}_{k} &=& V_{k}a_{0} + \sum_{l=1}^{N}U_{kl}c_{l}.
 \label{1b}
\end{eqnarray}
\renewcommand{\theequation}{\arabic{chapter}.\arabic{equation}}
Here $a_{0}\left(\Delta,t\right)$ is the amplitude of the state in
which the atom at the origin is in state $s^{\prime}$ and all
other atoms are in state $s$, while $c_{k}\left(\Delta,t\right)$
is the amplitude of the state in which the atom at the origin is
in state $p^{\prime}$ and the atom at $\mbf{r}_{k}$ is in state
$p$, while all the others remain in state $s$. The quantities
$V_{k}$ and $U_{kl}$ are the interaction potentials and $\Delta =
\epsilon_{p^{\prime}} +
\epsilon_{p}-\epsilon_{s^{\prime}}-\epsilon_{s}$ is the detuning
from resonance.

The simplest possibilities for the dipole-dipole interactions $V$
and $U$ are
\begin{equation}
V_{k} = -\frac{\mu\mu^{\prime}}{r_{k}^{3}}, \label{Vsimsimple}
\end{equation}
and
\begin{equation}
U_{kl} = -\frac{\mu^{2}}{r_{kl}^{3}}. \label{Usimsimple}
\end{equation}
Here $\mu$ is the dipole matrix element connecting the $s$ and $p$
states, and $\mu^{\prime}$ is the dipole matrix element connecting
the $s^{\prime}$ and $p^{\prime}$ states.  We also define $U_{ll}
= 0$, so that the restriction $l\neq k$ is automatic in sums of
the type appearing in Eq.~(\ref{1b}). The atoms are assumed to be
sufficiently cold that during the time scale of interest they move
only a very small fraction of their separation, and therefore
$V_{k}$ and $U_{kl}$ can be taken to be independent of time.

As was stated in Section~\ref{aligneddipoles}, the induced dipoles
in the magneto-optical trap become aligned if the applied electric
field is sufficiently large that the Stark effect has passed from
the quadratic to the linear regime. This is the case considered,
for example, in Refs.~\cite{Jaksch2000a} and \cite{Santos2000a}.
For the numerical simulation results of this thesis we use the
slightly more complicated dipole-dipole interaction potential
corresponding to this case, which is
\begin{equation}
V_{k} = -\frac{2\mu\mu^{\prime}}{r_{k}^{3}}
P_{2}\left(\cos\theta_{k}\right), \label{Vsim}
\end{equation}
where $\theta_{k}$ is the angle between $\mbf{r}_{k}$ and the
applied electric field. Similarly, $U_{kl}$ is of the form
\begin{equation}
U_{kl} = -\frac{2\mu^{2}}{r_{kl}^{3}}
P_{2}\left(\cos\theta_{kl}\right), \label{Usim}
\end{equation}
for $l\neq k$, and $U_{kk}=0$. These formulas for $V$ and $U$ are
also valid, in some cases, in the quadratic Stark regime if a
magnetic field parallel to the electric field is used to align the
total angular momentum of the atoms, as we now discuss more fully.

We will show below that, using interactions such as
Eqs.~(\ref{Vsimsimple}) and (\ref{Usimsimple}), or
Eqs.~(\ref{Vsim}) and (\ref{Usim}), we are able to reduce the
computation of any quantity to the diagonalization of an
$\left(N+1\right)\times\left(N+1\right)$ matrix. On the other
hand, we saw in Chapter~\ref{spin} that in the quadratic Stark
regime the actual interaction potentials depend on the quantum
numbers $m_{j}$ of the atoms involved. A complete consideration of
the spins would require the diagonalization of a $2^{N+1}\times
2^{N+1}$ matrix in this case, which is computationally prohibitive
to say the least. The problem can be reduced again to
$\mc{O}\left(N\times N\right)$ by applying a magnetic field that
is parallel to the electric field and of sufficient strength to
lift the $m_{j}$ degeneracy. The $V$ interaction is then given by
the appropriate matrix element of Table~\ref{ss'pp'3/2table} or
\ref{ss'pp'1/2table}, and the $U$ interaction is similarly found
in Table~\ref{sppstable}. We see from these tables that for many
matrix elements the angular dependence is the same as in
Eqs.~(\ref{Vsim}) and (\ref{Usim}), which can therefore be taken
as representative. We have also shown in
Section~\ref{averagingwithspin} that, when a $p^{\prime}$ state
with $\left|m_{j}\right| = 3/2$ is involved, each $s$ atom
interacts separately with the $s^{\prime}$ atom and, as the system
evolves, the average number of $p^{\prime}$ states is the same as
for a simple spin-independent interaction, but with an effective
$V_{k}$. The resulting effective interaction energy
$v_{\mrm{spin}}$ is given by Eq.~(\ref{vspin}) to be $0.698\,v$,
where $v$ is the effective interaction energy for the interaction
potential of Eq.~(\ref{Vsimsimple}) and is defined explicitly in
Eq.~(\ref{v}). For comparison, the effective interaction energy
for the interaction potential of Eq.~(\ref{Vsim}) is given by
Eq.~(\ref{valigned}) to be $0.770\,v$. Thus we see that the exact
angular form of the interaction potential for $V$ does not greatly
affect the results. The numerical simulations reported in this
chapter show that the same is true for the angular form of $U$.

\section{Description of method}
\label{descriptionofmethod}

If we consider a column vector $C\left( \Delta,t\right) = \left(
a_{0}\left( \Delta,t\right), c_{1}\left( \Delta,t\right),
c_{2}\left( \Delta,t\right),\ldots , c_{N}\left( \Delta,t\right)
\right)^{T}$, then clearly we can construct a matrix $H$ such that
the set of $N + 1$ coupled differential equations of
Eqs.~(\ref{1a}) and (\ref{1b}) can be written in the form
$i~\dot{C} = HC$. If we let $\left\{ \lambda_{m}\right\}$ and
$\left\{ \theta_{m}\right\}$ be the eigenvalues and eigenvectors
(represented as column vectors) of the real Hermitian matrix $H$,
respectively, and if we further define $e_{n}$ to be the column
vector of length $N+1$ with the $n$th entry set equal to one and
all others set equal to zero, then we can write $e_{n} =
\sum_{m=0}^{N}\beta_{mn}\theta_{m}$ and $\theta_{m} =
\sum_{n=0}^{N}\beta_{nm}e_{n}$, where $\beta_{mn} = \theta_{m}^{T}
e_{n}$. The system starts off in the state described by $e_{0}$,
so it follows that
\begin{eqnarray}
C\left( \Delta,t\right) &=& \exp\left( -iHt\right)e_{0} \nonumber
\\ &=&
\sum_{m=0}^{N}\beta_{m0}\exp\left(-i\lambda_{m}t\right)\theta_{m}
\nonumber \\ &=& \sum_{m=0}^{N}\sum_{n=0}^{N}\beta_{m0}\beta_{nm}
\exp\left( -i\lambda_{m}t\right) e_{n}.
\end{eqnarray}
The quantity $a_{0}\left( \Delta,t\right)$ is just the $e_{0}$
component of this expression, so
\begin{equation}
a_{0}\left( \Delta,t\right) = e_{0}^{T} C\left( \Delta,t\right) =
\sum_{m=0}^{N} \beta_{m0}\beta_{0m}\exp\left(
-i\lambda_{m}t\right). \label{azeroeqn}
\end{equation}
Thus we see that, in this form, solving for $a_{0}$ at a given
time becomes just a matter of diagonalizing the matrix $H$.

To carry out the numerical simulations, we first generate a set of
random positions for the unprimed atoms. For the results presented
in this Sections~\ref{signalsimulationresults} and
\ref{widthsimulationresults}, these random positions are uniformly
distributed so that the results can be more easily compared with
the analytical results of previous chapters. However, when
proceeding numerically it is just as easy to compute random
positions for a gas blob with any given density profile. For the
experiments we are
modeling~\cite{Anderson1996a,Anderson1998a,Lowell1998a}, the
density profile is a Gaussian with an aspect ratio of
approximately $1:1:5$~\cite{Gallagher2000a}, and so this is what
we use to generate the results of Section~\ref{compexpdata}. Once
the atomic positions are generated, they are used to compute the
interaction potentials $V_{k}$ and $U_{kl}$, and these are in turn
used to compute the matrix $H$. We then diagonalize $H$ and
construct the quantity $a_{0}$ at a set of values of $t$ using
Eq.~(\ref{azeroeqn}). We do this for the same set of times for
many random distributions of the unprimed atoms and thereby obtain
an average of the quantities $a_{0}\left(\Delta,t\right)$ and
$\left| a_{0}\left(\Delta,t\right)\right|^{2}$ at a given set of
times over many configurations of the system. The signal measured
by the experiments is then proportional to the average of $1 -
\left| a_{0}\left(\Delta,t\right)\right|^{2}$. We can further
obtain plots of the linewidth as a function of time by generating
signal versus time curves for many values of the detuning,
$\Delta$, and then determining for each value of time which of
these signals is equal to one half the value of the resonance
signal. One can also directly generate Fourier transforms from
Eq.~(\ref{azeroeqn}), as we did in Section~\ref{a0freq}.

By comparing the results of the simulations to the analytical
results of Chapter~\ref{sparse_no_u} we find that an average over
$10^{4}$ realizations, each consisting of a single $s^{\prime}$
atom and $100$ $s$ atoms, is sufficient, as can be seen in
Fig.~\ref{simulation_exact_signal}. These are the parameters we
have used for all the simulation plots presented in this thesis,
although we have run the simulations for as many as $800$ $s$
atoms.

The simulations are written in \emph{Fortran 90}, and the
computation process is greatly accelerated by the use of a
\emph{Beowulf} cluster of thirteen \emph{Linux} machines. Using
this cluster, the computation can be carried out in parallel, as
we now describe. A master program is started on one of the
machines. The master then starts up a slave program on each
machine in the cluster. The master initializes the slaves by
sending them seeds for their random number generators.\footnote{It
is vital that the master control the random number generator
seeds, since each slave program must be seeded differently so that
each generates a different set of realizations for the atomic
positions.} The slaves then generate random distributions for the
atoms, compute whatever quantity is required, and send the result
back to the master. The master collects and averages the results
returned by the slaves and saves the result to a data file. The
master also takes care of any interaction with the user. The
passing of messages and data between the master and slave
processes is handled through the software program \emph{Parallel
Virtual Machine} (\emph{PVM})~\cite{Geist1994a}.

We use the (parallel) random number generation routines
\emph{gasdev} and \emph{ran1} described in Refs.~\cite{Press1992a}
and~\cite{Press1996a} to generate the random distributions of
atoms. The routine \emph{gasdev} generates Gaussian deviates with
unit standard deviation, while \emph{ran1} generates uniform
deviates between zero and one. The computation of eigenvalues and
eigenvectors for the matrix $H$ corresponding to each realization
is performed using the \emph{ssyevx} routine of the \emph{LAPACK}
package, which is described in Ref.~\cite{Anderson1999a}.

\section{The effect of many $s^{\prime}$ atoms}

\subsection{The chemical argument}
\label{chemicalargument}

Considering the $ss^{\prime}\rightarrow pp^{\prime}$ process as a
chemical reaction, we have the rate equation
\begin{equation}
\frac{dN_{p^{\prime}}}{dt} = k\left( N_{s}N_{s^{\prime}} -
N_{p}N_{p^{\prime}} \right),
\end{equation}
where $k$ is a constant with the units of inverse time, which is
equivalent to energy since we take $\hbar$ equal to unity. There
are similar rate equations for $N_{p}$, $N_{s}$, and
$N_{s^{\prime}}$, but it is sufficient to consider only one of
these because of the following constraints.

We know that $N_{p} = N_{p^{\prime}}$, since if there is a $p$ or
$p^{\prime}$ atom present it can only have been produced as a
result of a pair of atoms in the $s$ and $s^{\prime}$ states
making the $ss^{\prime}\rightarrow pp^{\prime}$ transition. We
also know that $N = N_{s} + N_{p} = N_{s} + N_{p^{\prime}}$ and
$N^{\prime} = N_{s^{\prime}} + N_{p^{\prime}}$, since primed and
unprimed atoms can change from $s$ to $p$ states and vice versa,
but can never be created or destroyed. Therefore we have
\begin{eqnarray}
\frac{dN_{p^{\prime}}}{dt} &=& k\left[ \left(N -
N_{p}\right)\left(N^{\prime} - N_{p^{\prime}}\right) -
N_{p}N_{p^{\prime}} \right] \nonumber \\ &=& k\left[ \left(N -
N_{p^{\prime}}\right)\left(N^{\prime} - N_{p^{\prime}}\right) -
N_{p^{\prime}}N_{p^{\prime}} \right] \nonumber \\ &=& k \left[
NN^{\prime} - \left(N + N^{\prime}\right)N_{p^{\prime}} \right].
\end{eqnarray}
With the initial condition $N_{p^{\prime}} = 0$ at $t=0$, this
gives
\begin{equation}
N_{p^{\prime}} = \frac{NN^{\prime}}{N + N^{\prime}} \left[ 1 -
e^{-\left(N + N^{\prime}\right) kt} \right],
\label{chemicalsolution}
\end{equation}
and at small times
\begin{equation}
N_{p^{\prime}} = NN^{\prime}kt. \label{initialbehaviorchemsoln}
\end{equation}

\subsection{Extrapolation to many $s^{\prime}$ atoms}

In our case, the reaction rate is not constant and the approach to
equilibrium is not accurately described by
Eq.~(\ref{chemicalsolution}). However, the result we have from
Chapter~\ref{sparse_no_u} in the limit $N^{\prime} \ll N$ starts
off at small times as
\begin{equation}
N_{p^{\prime}} = C \mu\mu^{\prime}\frac{N}{\Omega} N^{\prime} t,
\label{N_p'}
\end{equation}
where $C$ is a numerical constant that does not concern us here.
This expression is fine for all values of $N$ and $N^{\prime}$,
and in particular satisfies the requirement that $N_{p} =
N_{p^{\prime}}$. It is of the same form as
Eq.~(\ref{initialbehaviorchemsoln}) and shows that at small times
we have
\begin{equation}
k = C\frac{\mu\mu^{\prime}}{\Omega}. \label{ratecoeff}
\end{equation}
Furthermore, we can write
\begin{equation}
N_{p^{\prime}} = \frac{NN^{\prime}}{N + N^{\prime}} F\left(
\mu\mu^{\prime}\frac{N + N^{\prime}}{\Omega}t,\ldots \right),
\end{equation}
where the dots indicate that the function $F$ depends on other
dimensionless ratios that characterize the system but do not
involve $t$. The prefactor $NN^{\prime}/\left(N +
N^{\prime}\right)$ and the characteristic energy
$\mu\mu^{\prime}\left(N + N^{\prime}\right)/\Omega$ are suggested
by the chemical argument of Section~\ref{chemicalargument}, since
these factors are explicitly seen in Eq.~(\ref{chemicalsolution})
if we keep in mind Eq.~(\ref{ratecoeff}). The full function
$F\left( x,\ldots\right)$ is computable only in the dilute limits
$N^{\prime} \ll N$ or $N \ll N^{\prime}$, but we know in general
that it starts off linearly in $x$.

For large values of $t$ we have, again for $N^{\prime} \ll N$,
\begin{equation}
N_{p^{\prime}} = N^{\prime} f\left(
\frac{\Delta}{\mu\mu^{\prime}\frac{N}{\Omega}},
\frac{\mu}{\mu^{\prime}}\right), \label{longt}
\end{equation}
where $f\left(y,z\right)$ is some function that can be computed
numerically using the methods of
Section~\ref{descriptionofmethod}. This result cannot be correct
for all values of $N/N^{\prime}$. In fact, for $N \ll N^{\prime}$
we can compute $N_{p}$ and we find
\begin{equation}
N_{p} = N f\left(
\frac{\Delta}{\mu\mu^{\prime}\frac{N^{\prime}}{\Omega}},
\frac{\mu^{\prime}}{\mu}\right). \label{longtobverse}
\end{equation}
Since $N_{p^{\prime}}$ is equal to $N_{p}$, we see that this
(correct) result is quite different from what one would obtain
using Eq.~(\ref{longt}), which is incorrect in this limit.
Interpolating between the limits given by Eqs.~(\ref{longt}) and
(\ref{longtobverse}), we postulate that for large $t$
\begin{equation}
N_{p^{\prime}} = \frac{NN^{\prime}}{N+N^{\prime}} f\left(
\frac{\Delta}{\mu\mu^{\prime}\frac{N+N^{\prime}}{\Omega}},
\frac{\mu}{\mu^{\prime}} \right).
\end{equation}

We can join this result smoothly to the small $t$ result of
Eq.~(\ref{N_p'}) by postulating a general result of the form
\begin{equation}
N_{p^{\prime}} = \frac{NN^{\prime}}{N+N^{\prime}} F\left(
\mu\mu^{\prime}\frac{N+N^{\prime}}{\Omega}t,
\frac{\Delta}{\mu\mu^{\prime}\frac{N+N^{\prime}}{\Omega}},
\frac{N}{N+N^{\prime}} \frac{\mu}{\mu^{\prime}} +
\frac{N^{\prime}}{N+N^{\prime}} \frac{\mu^{\prime}}{\mu} \right),
\end{equation}
where the function $F\left(x,y,z\right)$ is proportional to $x$
for small $x$ and approaches $f\left(y,z\right)$ for large $x$. We
have also used the same scaling for the quantities $t$ and
$\Delta$, so that $xy = \Delta t$.

With this approach we can use the function $F\left(x,y,z\right)$,
which can be computed numerically in the dilute limit, as a
reasonable guess for the functional dependence on the suitable
dimensionless variables for any value of $N^{\prime}/N$.

The combination
\begin{equation}
\frac{N}{N+N^{\prime}}\frac{\mu}{\mu^{\prime}} +
\frac{N^{\prime}}{N+N^{\prime}}\frac{\mu^{\prime}}{\mu}
\label{strangequantity}
\end{equation}
is just an interpolation between the correct results in the limits
of small $N$ and small $N^{\prime}$. It is reasonable, however,
since the quantities
\begin{equation}
N\mu^{2}\frac{N}{\Omega},
\end{equation}
\begin{equation}
N^{\prime}\left(\mu^{\prime}\right)^{2}\frac{N^{\prime}}{\Omega},
\end{equation}
and
\begin{equation}
N^{\prime}\mu\mu^{\prime}\frac{N}{\Omega}
\end{equation}
give the effective $U$ for the $sp\rightarrow ps$ interaction for
a fraction $N^{\prime}/\left(N+N^{\prime}\right)$ of the atoms,
the effective $U^{\prime}$ for the
$s^{\prime}p^{\prime}\rightarrow p^{\prime}s^{\prime}$ interaction
for a fraction $N/\left(N+N^{\prime}\right)$ of the atoms, and the
$V$ interaction for the $ss^{\prime}\rightarrow pp^{\prime}$
process for all the atoms. The quantity of
Eq.~(\ref{strangequantity}) is obtained by taking the ratio
\begin{equation}
\frac{N\mu^{2}\frac{N}{\Omega} +
N^{\prime}\left(\mu^{\prime}\right)^{2}\frac{N^{\prime}}{\Omega}}{N^{\prime}
\mu\mu^{\prime}\frac{N}{\Omega}}.
\end{equation}

\section{Results of simulations for the signal}
\label{signalsimulationresults}

For the results of this section, we well as those of
Section~\ref{widthsimulationresults}, we use the interaction
potentials of Eqs.~(\ref{Vsim}) and (\ref{Usim}). We also use a
uniform distribution to generate the random atomic positions, so
that the results can be more easily compared with the analytical
results of previous chapters.

\begin{figure}
\resizebox{\textwidth}{!}{\includegraphics[0in,0in][8in,10in]{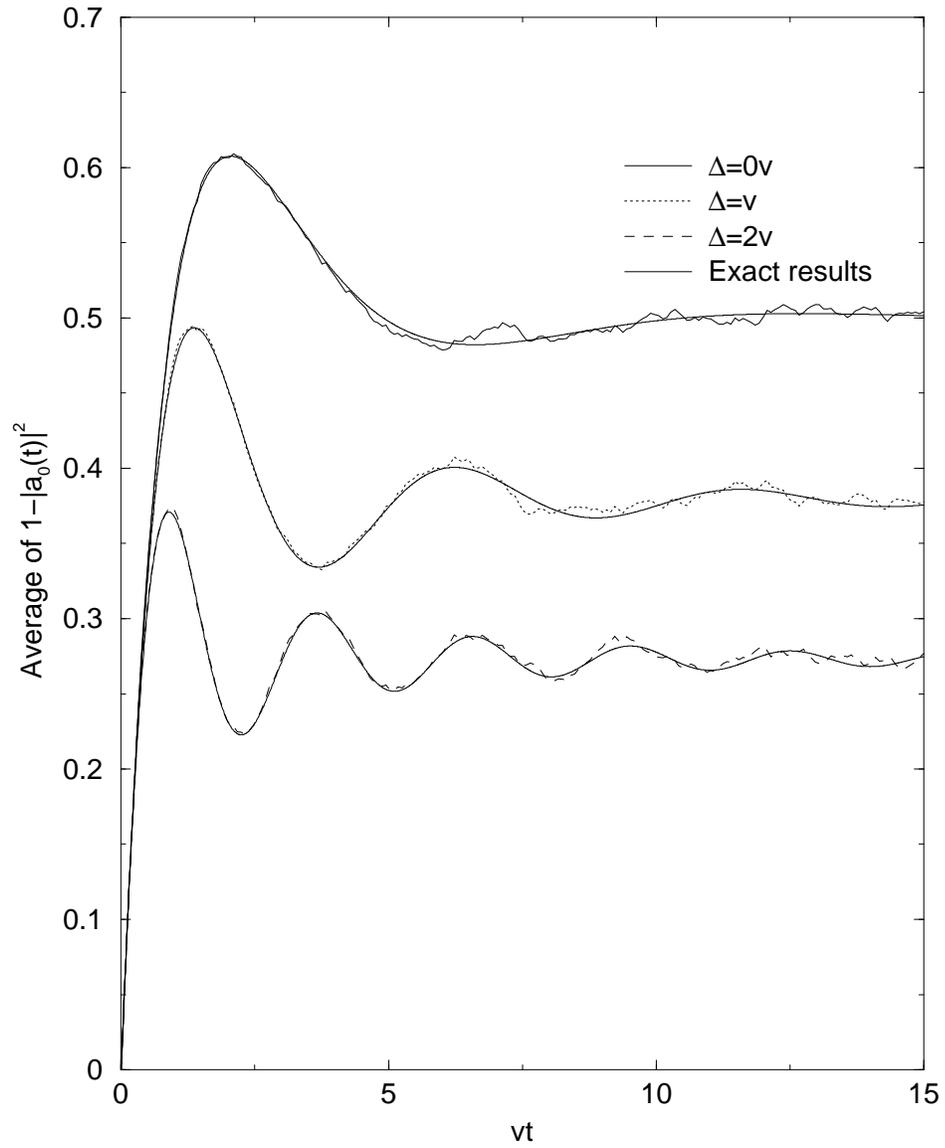}}
\caption[Comparison of simulation with exact results]{Plots of
$1-\left\langle\left|a_{0}\left(
\Delta,t\right)\right|^{2}\right\rangle$ for $u=0$ and several
values of the detuning $\Delta$. The exact results of
Chapter~\ref{sparse_no_u} are shown for comparison.}
\label{simulation_exact_signal}
\end{figure}

\begin{figure}
\resizebox{\textwidth}{!}{\includegraphics[0in,0in][8in,10in]{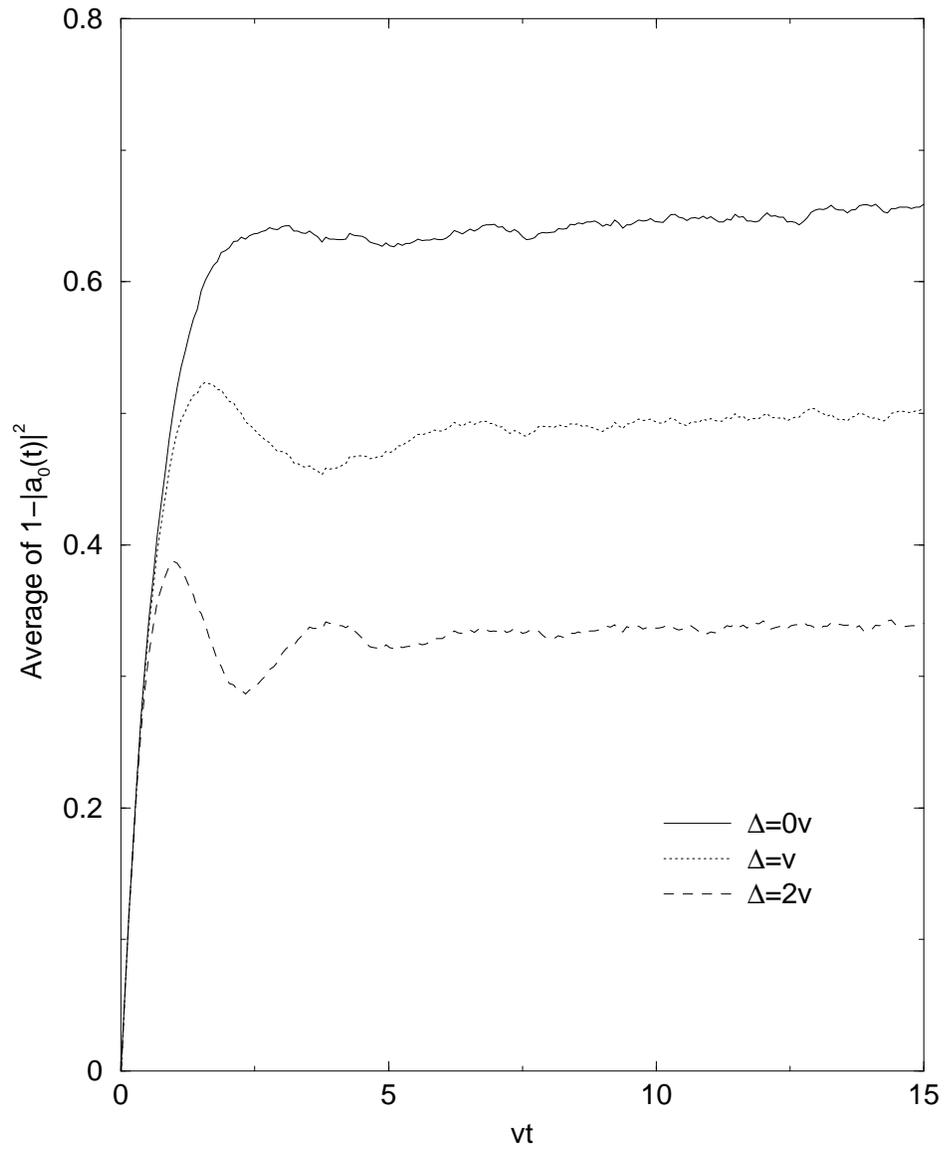}}
\caption[Averaged signal as a function of time for
$u=0.25v$]{Plots of $1-\left\langle\left|a_{0}\left(\Delta,
t\right)\right|^{2}\right\rangle$ for $u=0.25v$ and several values
of the detuning $\Delta$.} \label{simulation_u=0.25v_signal}
\end{figure}

\begin{figure}
\resizebox{\textwidth}{!}{\includegraphics[0in,0in][8in,10in]{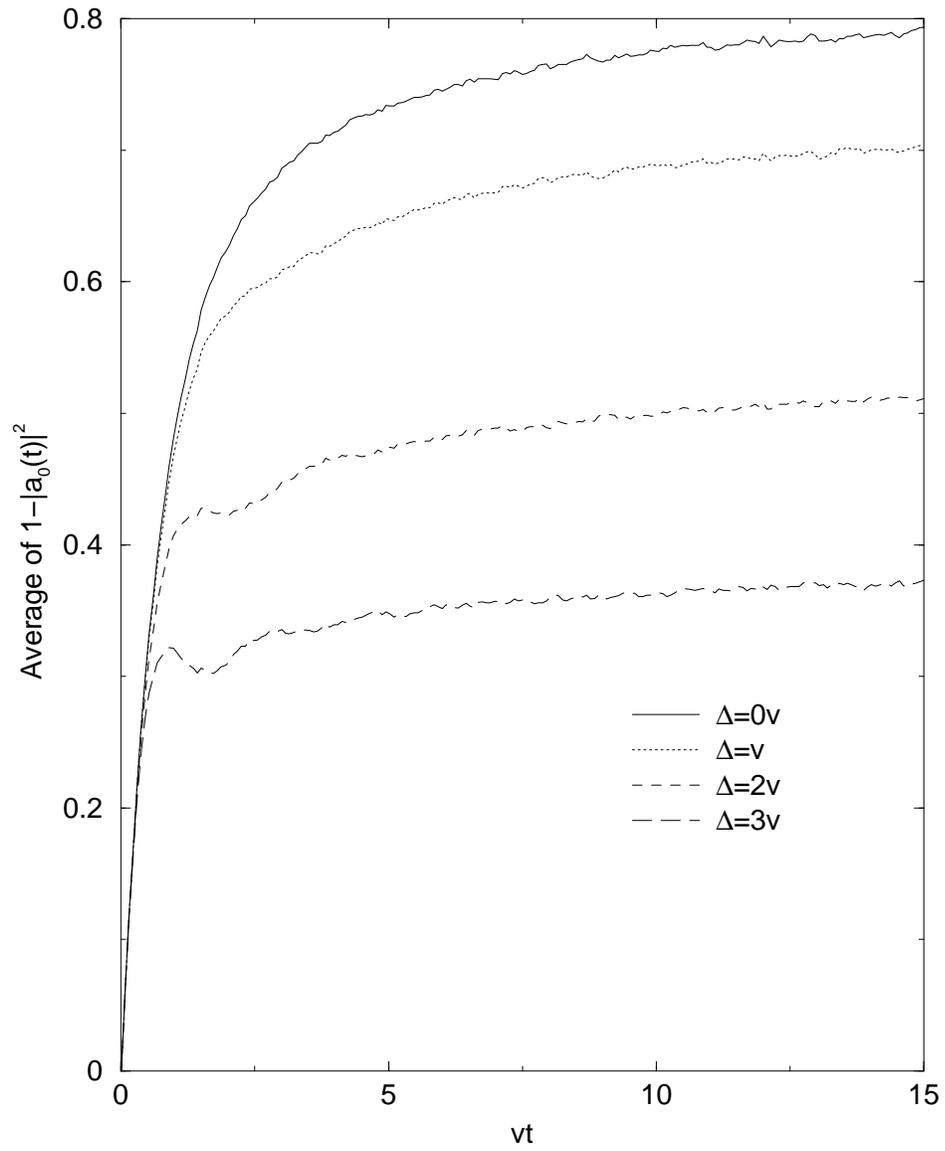}}
\caption[Averaged signal as a function of time for $u=v$]{Plots of
$1-\left\langle\left|a_{0}\left(
\Delta,t\right)\right|^{2}\right\rangle$ for $u=v$ and several
values of the detuning $\Delta$.} \label{simulation_u=v_signal}
\end{figure}

\begin{figure}
\resizebox{\textwidth}{!}{\includegraphics[0in,0in][8in,10in]{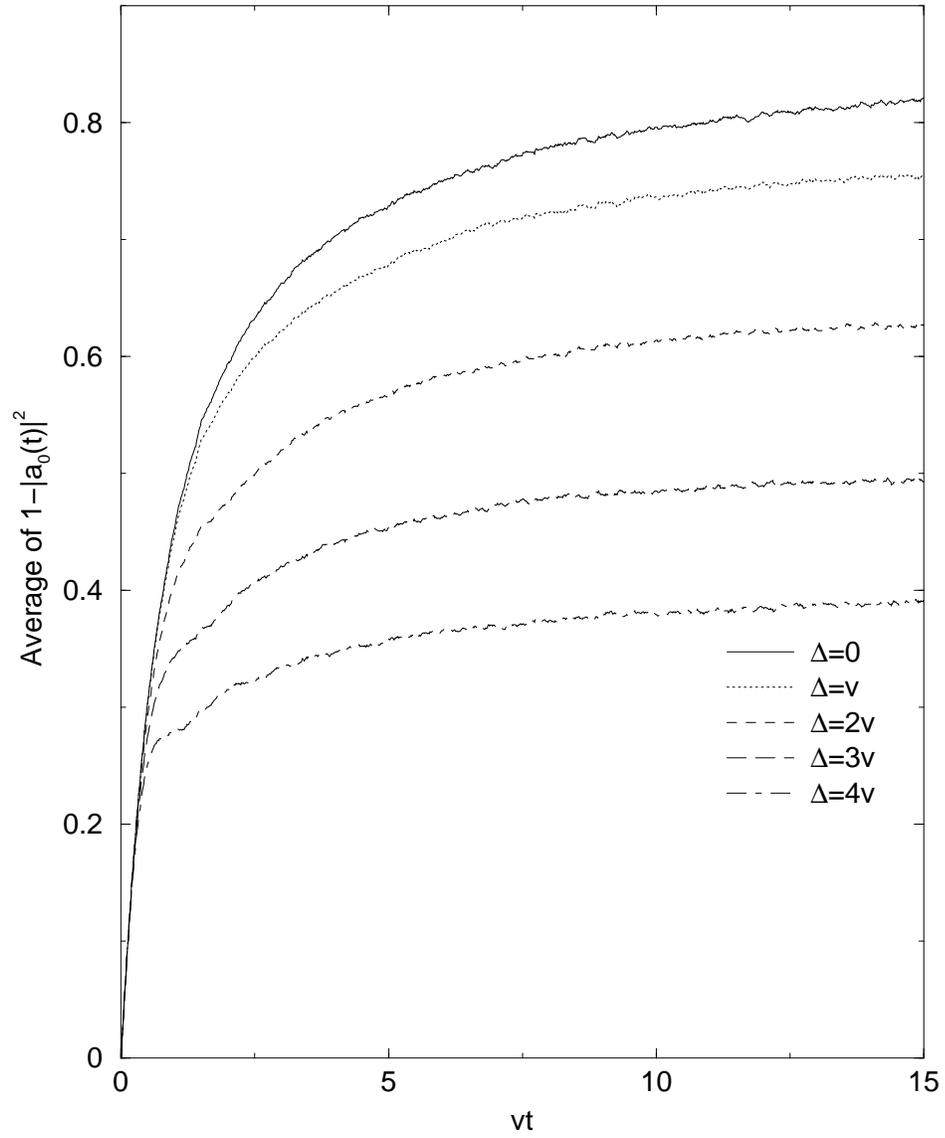}}
\caption[Averaged signal as a function of time for $u=2v$]{Plots
of $1-\left\langle\left|a_{0}\left(
\Delta,t\right)\right|^{2}\right\rangle$ for $u=2v$ and several
values of the detuning $\Delta$.} \label{simulation_u=2v_signal}
\end{figure}

\begin{figure}
\resizebox{\textwidth}{!}{\includegraphics[0in,0in][8in,10in]{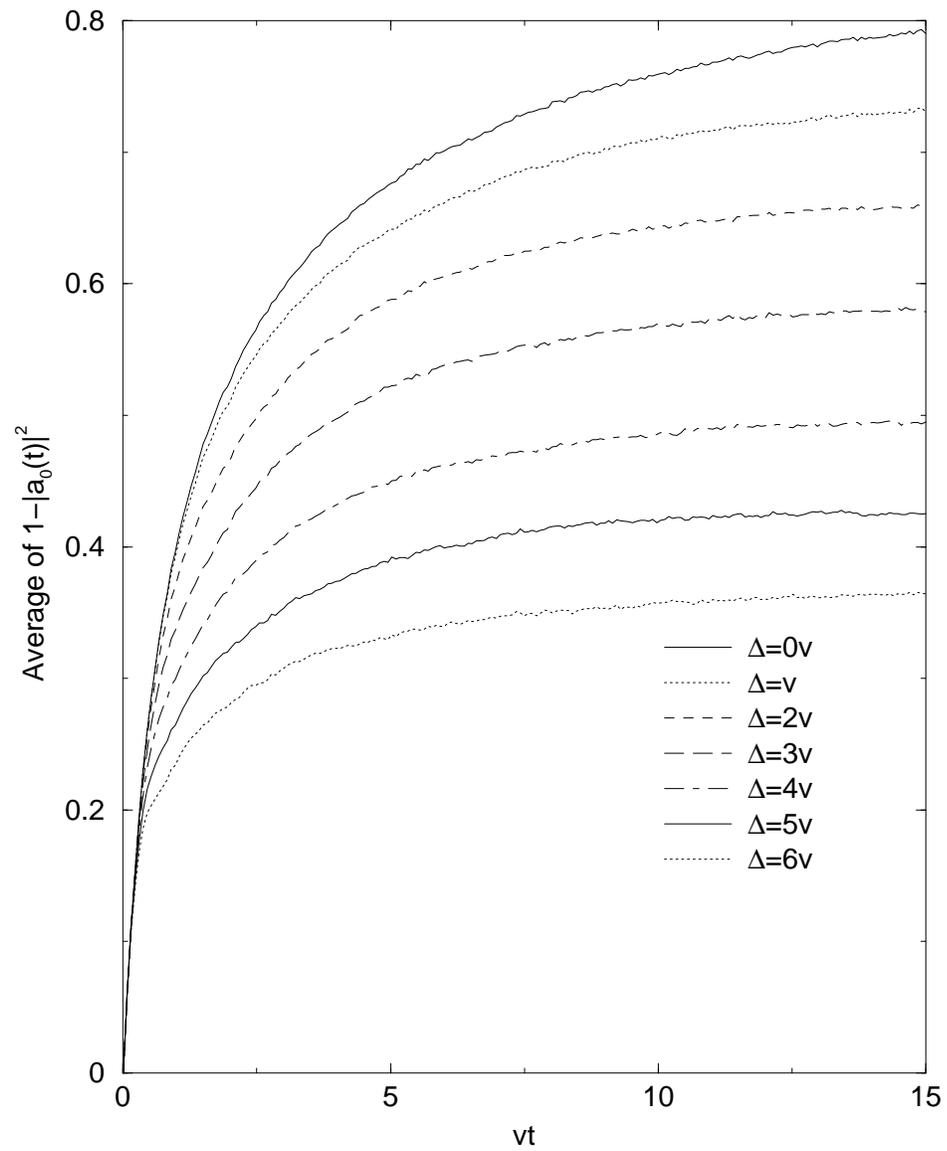}}
\caption[Averaged signal as a function of time for $u=4v$]{Plots
of $1-\left\langle\left|a_{0}\left(
\Delta,t\right)\right|^{2}\right\rangle$ for $u=4v$ and several
values of the detuning $\Delta$.} \label{simulation_u=4v_signal}
\end{figure}

Figs.~\ref{simulation_exact_signal}--\ref{simulation_u=4v_signal}
display the numerical simulations of $1-\left\langle \left|
a_{0}\right| ^{2}\right\rangle $, corresponding to the
experimental signal $N_{p^{\prime }}/N^{\prime }$, as a function
of $vt$ for $u/v=0$, $1/4$, $1$, $2$, and $4$, and enough values
of the detuning $\Delta$ to give an idea of the resonance width in
each case. The FWHM at large $t$ can be directly deduced from the
intercepts of the graphs with the right edge of the figure, as is
discussed more fully below.

In addition, Fig.~\ref{simulation_exact_signal} shows the exact
results for $u=0$ that were discussed in
Chapter~\ref{sparse_no_u}. Comparison with the numerical
simulations, which were carried out for $N = 100$, shows that the
infinite system is already well simulated, but that noticeable
wiggles remain after averaging over $10^{4}$ realizations. The
reason for this persistence of fluctuations is that most
realizations do not look at all like the average. In fact, the
variance with realization is comparable to the average. In
particular, for small $t$ the average is given by $\left(\sqrt{\pi
}/2\right)vt$ while each realization is quadratic in $t$, as
already discussed in Ref.~\cite{Frasier1999a}. It can be seen from
the graphs that the linear rise law,
$\left(\sqrt{\pi}/2\right)vt$, is independent both of $\Delta$ and
of $u$ and holds for a substantial range of $vt$.

\section{Results of simulations for the width}
\label{widthsimulationresults}

The time-dependent resonance lineshape is obtained from the
ordinates of the curves in
Figs.~\ref{simulation_exact_signal}--\ref{simulation_u=4v_signal}
at a given $vt$. In general the lineshape is not Lorentzian, but
we never find a split line with a minimum at resonance as reported
in Ref.~\cite{Mourachko1998a}. Thus it is fair to characterize the
line by its FWHM. This cannot be directly compared with an
experimental width, which is usually the FWHM of a Lorentzian
fitted to a noisy experimental line, but it is indicative of
trends. For a more accurate comparison with experiment, one should
of course convolute the calculated lineshape with the broadening
due to other effects.

\begin{figure}
\resizebox{\textwidth}{!}{\includegraphics[0in,0in][8in,10in]{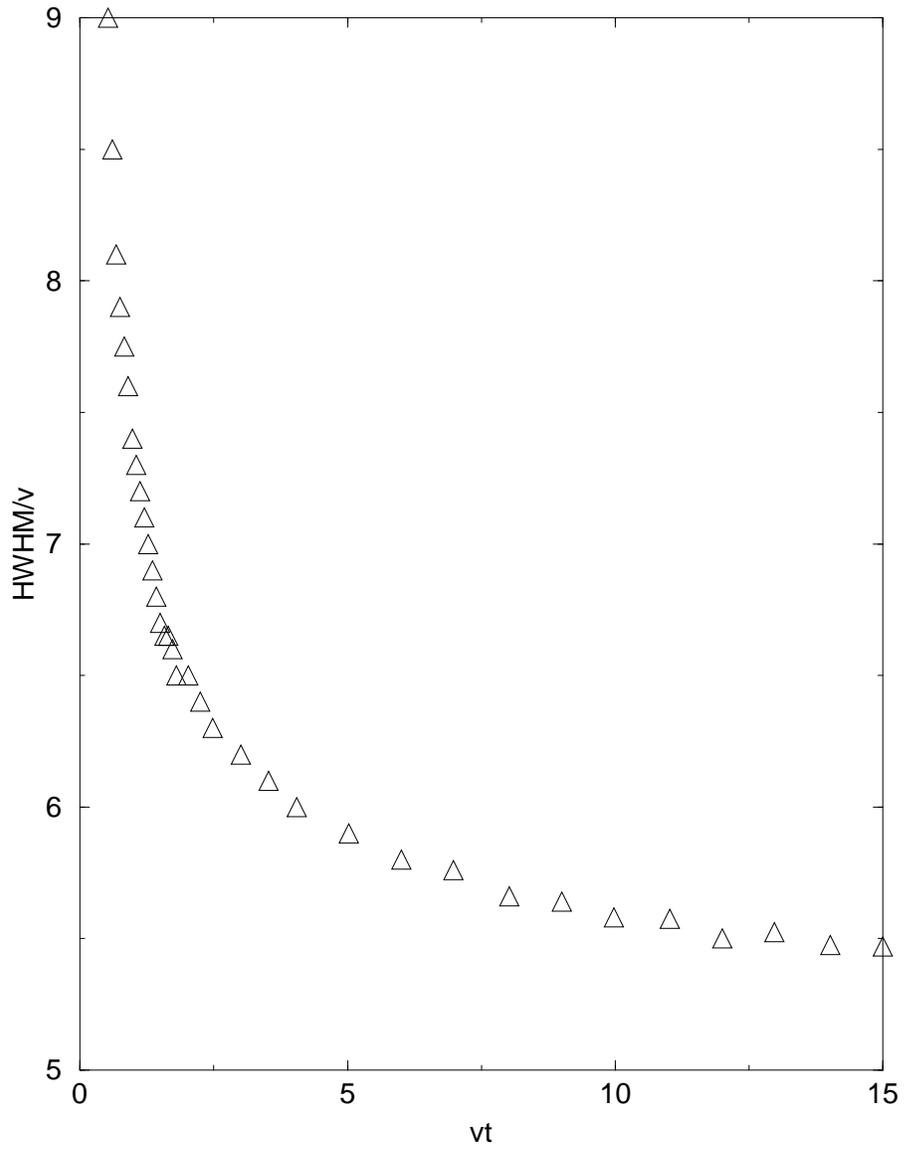}}
\caption[Width as a function of time for $u=4v$]{A plot of the
width as a function of time for $u = 4v$.}
\label{simulation_widthvstime}
\end{figure}

Fig.~\ref{simulation_widthvstime} shows one half of the FWHM, $w$,
as a function of $vt$ for $u=4v$. One sees the typical decrease of
$w$ from a universal $t^{-1}$ behavior at small $vt$ towards an
asymptotic value, which is almost reached at $vt = 15$. The
$t^{-1}$ law, which results simply from the uncertainty principle
\cite{Anderson1998a,Thomson1990a}, has already been discussed in
Ref.~\cite{Frasier1999a} where on the basis of a small $\Delta$
expansion it was estimated that $w=\sqrt{12}/t$ for small $t$.
This $t^{-1}$ behavior was observed experimentally by Renn
\emph{et al.\/}~\cite{Thomson1990a} in a different system, and was
later observed by Anderson \emph{et
al.\/}~\cite{Anderson1998a,Anderson1996a} in the Rb system.

\begin{figure}
\resizebox{\textwidth}{!}{\includegraphics[0in,0in][8in,10in]{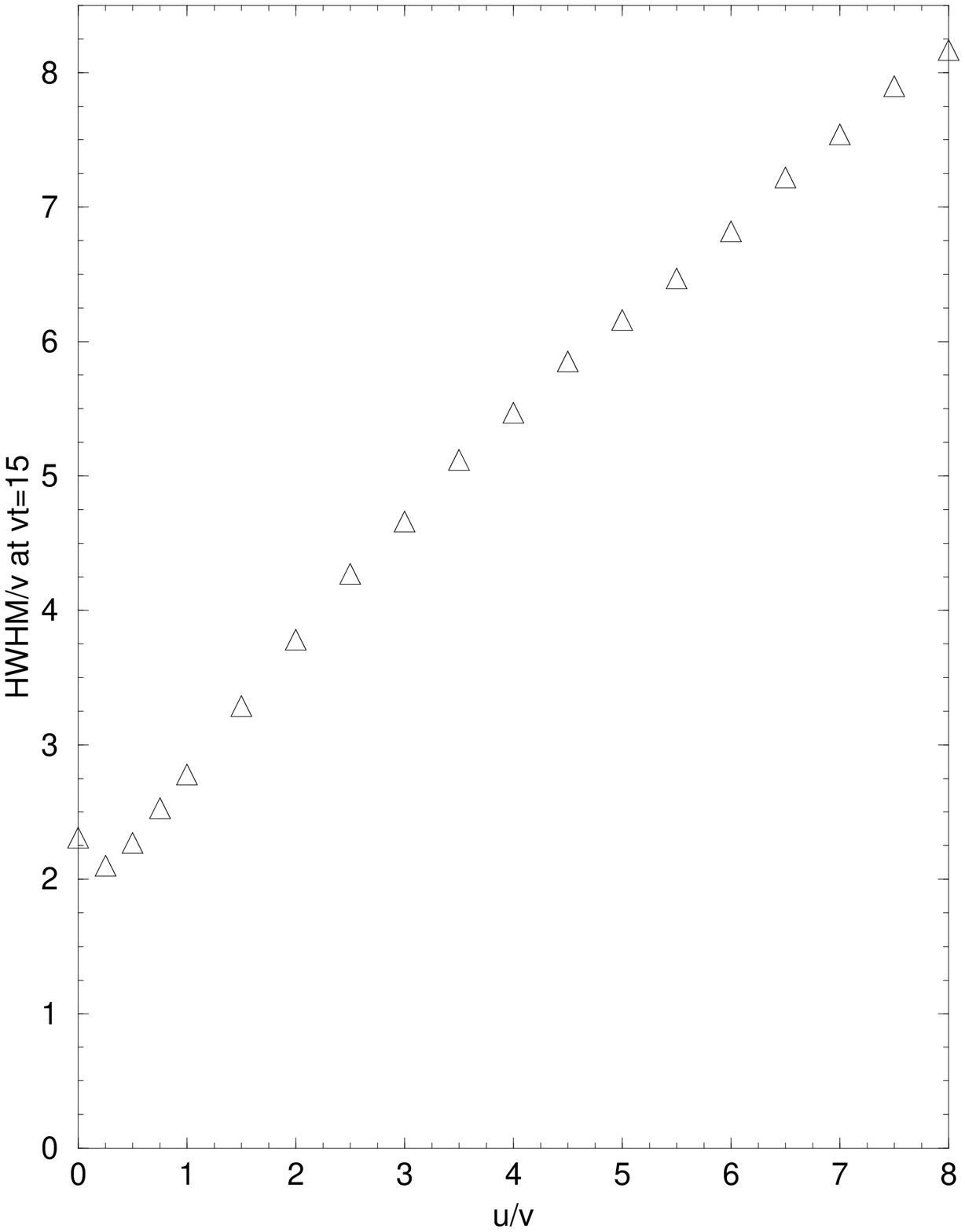}}
\caption[Width as a function of $u/v$]{A plot of the width at $vt
= 15$ as a function of the ratio $u/v$.}
\label{simulation_widthvsu}
\end{figure}

The near-asymptotic value of the FWHW at $vt=15$ is shown in
Fig.~\ref{simulation_widthvsu} as a function of $u/v$. Overall,
the graph is roughly linear except at small $u/v$, although a
careful examination shows that there are really two linear regimes
with somewhat different slopes that meet at $u/v \approx 3.5$.
There is an unexpected minimum around $u=v/4$, which is not a
numerical artifact, but can be seen directly from
Figs.~\ref{simulation_exact_signal}--\ref{simulation_u=4v_signal}
by comparing the plot for $u=v/4$ with those for $u=0$ and $u=v$.
What happens is that the resonance line always grows taller and
wider as $u$ increases, but for small $u$ its height (the signal
at $\Delta=0$) grows faster than its width.

\section{Comparison with experimental data}
\label{compexpdata}

For the results in this section, we use the interaction potentials
of Eqs.~(\ref{Vsim}) and (\ref{Usim}). Because it corresponds more
closely to the experimental situation, we also use a Gaussian
distribution with aspect ratio $1:1:5$ to generate the random
atomic positions.

\begin{figure}
\resizebox{\textwidth}{!}{\includegraphics[0in,0in][8in,10in]{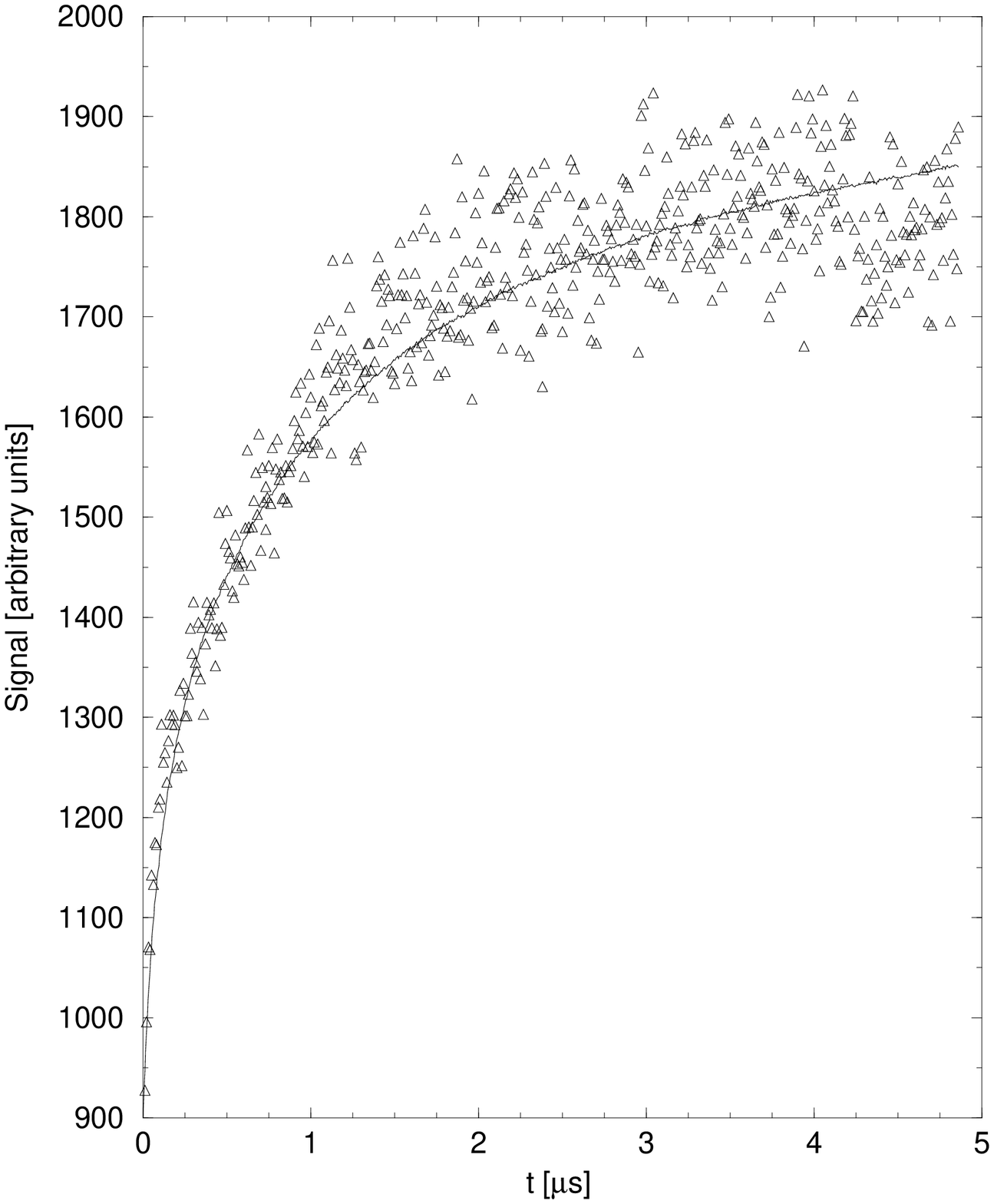}}
\caption[Experimental signal as a function of time --- larger
density]{The experimental signal as a function of time
(triangles), together with a signal curve computed using the
numerical simulations and fit to the data. The fit parameters are
$u = 12v$ and $N/\Omega = 2.25\times 10^{9}$ $\mrm{cm}^{-3}$.}
\label{simulation_expsig_tall}
\end{figure}

\begin{figure}
\resizebox{\textwidth}{!}{\includegraphics[0in,0in][8in,10in]{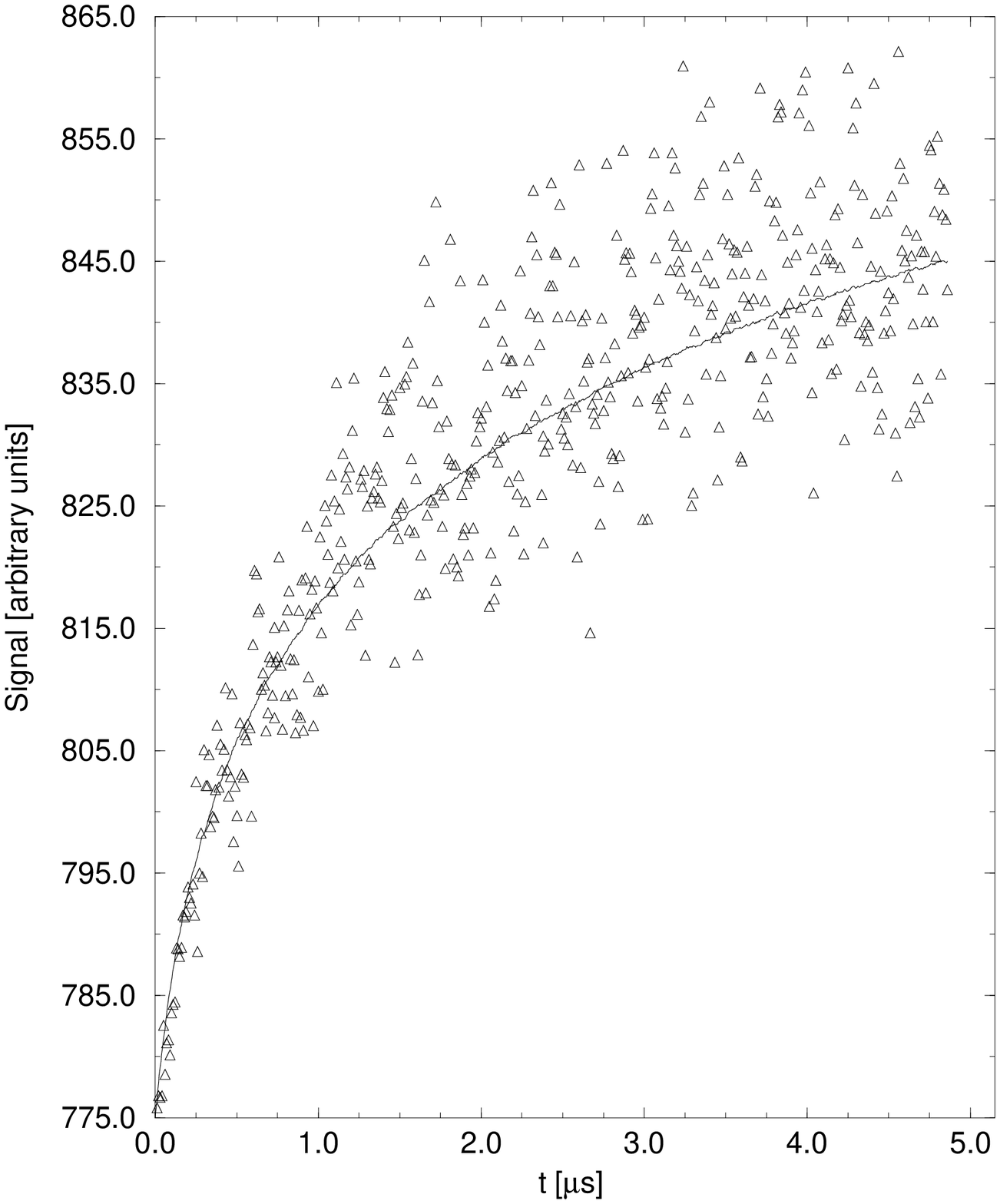}}
\caption[Experimental signal as a function of time --- smaller
density]{The experimental signal as a function of time
(triangles), together with a signal curve computed using the
numerical simulations and fit to the data. The fit parameters are
$u = 12v$ and $N/\Omega = 6.70\times 10^{8}$ $\mrm{cm}^{-3}$.}
\label{simulation_expsig_short}
\end{figure}

Figs.~\ref{simulation_expsig_tall} and
\ref{simulation_expsig_short} show the experimental signal as a
function of time, together with a signal computed using the
numerical simulations and fit to the data. The experimental data
for both graphs are the data of Fig.~3.8 on page~71 of
Ref.~\cite{Lowell1998a}, but we show here all the data points.

The fits are made by first determining the vertical coordinates
where the experimental signal starts and stops. The difference
between the two gives the vertical scale of the experimental
signal, which is necessary because as one can see the experimental
signal does not simply start at zero and rise to some value less
than one. For example, in Fig.~\ref{simulation_expsig_tall} we
assumed that the experimental signal started at $900$ and ended at
$1850$ so that the vertical scale was $950$. The initial slope is
then determined by fitting the first several points to a straight
line. For Fig.~\ref{simulation_expsig_tall} we obtained an initial
slope of $4107$ $\mu\mrm{s}^{-1}$. We then have
\begin{equation}
\frac{\sqrt{\pi}}{2}\frac{v}{\hbar} = \mbox{$3.83115$
$\mu\mrm{s}^{-1}$},
\end{equation}
which gives a value for the parameter $v$ and hence the density
$N/\Omega$. This fixes the horizontal scale, and so all that
remains is to adjust the parameter $u/v$ until we find a good fit.

\begin{figure}
\resizebox{\textwidth}{!}{\includegraphics[0in,0in][8in,10in]{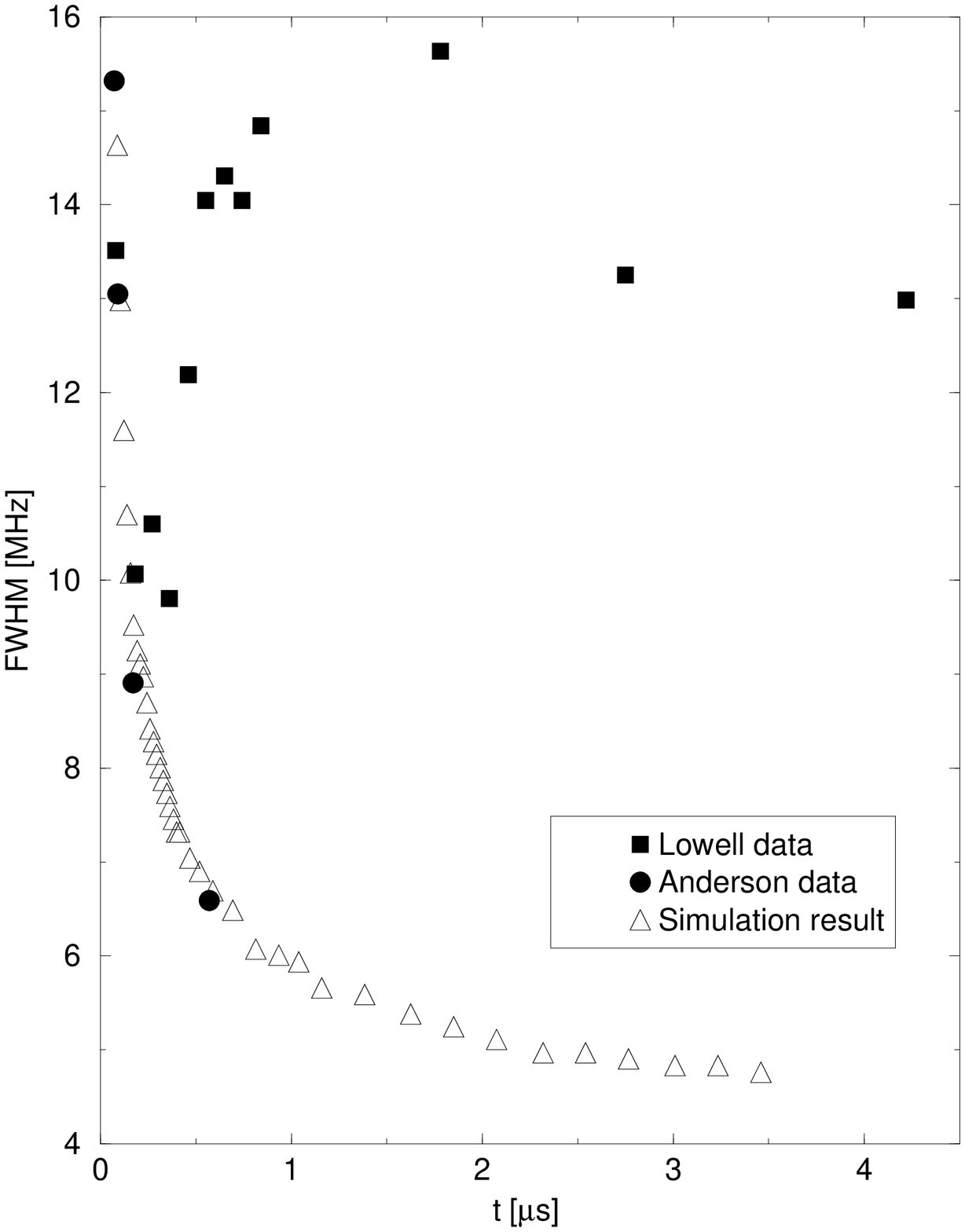}}
\caption[Experimental width as a function of time]{The
experimental width as a function of time. The circles are the data
of Anderson \emph{et al.\/}~\cite{Anderson1996a,Anderson1998a},
while the squares are the data of Lowell~\cite{Lowell1998a}. The
triangles are a width curve computed using the numerical
simulations and fit to the data. The fit parameters are $u = 4v$
and $N/\Omega = 2\times 10^{9}$ $\mrm{cm}^{-3}$.}
\label{simulation_expwidth}
\end{figure}

Fig.~\ref{simulation_expwidth} shows the experimental width as a
function of time. The data of Anderson \emph{et al.\/} are taken
from Fig.~4-18 on page~149 of Ref.~\cite{Anderson1996a}, while the
data of Lowell \emph{et al.\/} are taken from Fig.~3.10 on page~75
of Ref.~\cite{Lowell1998a}. The discrepancy between the two sets
of data has not yet been resolved by the experimentalists.
Although the result of the simulations fit Anderson's data quite
well for the parameters $u = 4v$ and $N/\Omega = 2\times 10^{9}$
$\mrm{cm}^{-3}$, it is important to note that these data are for
rather small times, and are within or very close to the region
where the universal $t^{-1}$ behavior discussed in
Section~\ref{widthsimulationresults} is present. Thus for a wide
range of values of the parameter $u$ there will exist densities
$N/\Omega$ that produce a good fit to the Anderson data.

\section{Summary}

We have seen in this chapter how to develop numerical simulations
to solve Eqs.~(\ref{1a}) and (\ref{1b}) for any quantity we want
and average the result over many realizations of the system.
Although Eqs.~(\ref{1a}) and (\ref{1b}) apply to the sparse case
where we begin initially with a single $s^{\prime}$ atom in a sea
of $s$ atoms, we have also discussed in this chapter how to extend
these results to the nonsparse case. The simulation results agree
very well with the exact analytical results of
Chapter~\ref{sparse_no_u}, as well as with the approximate
analytical results of Chapter~\ref{sparse_with_u}. The results of
the simulations can also be compared with the experimental data of
Anderson \emph{et al.}~\cite{Anderson1996a,Anderson1998a} and
Lowell \emph{et al.}~\cite{Lowell1998a}, and we find that the
agreement is again quite good.

\chapter{Final Thoughts}
\markright{Chapter \arabic{chapter}: Final Thoughts}
\label{finalthoughts}

Although we have focused chiefly on resonant dipole-dipole
interactions among frozen Rydberg atoms in this thesis, there are
many connections of this work with related experiments, as well as
more general applications of the techniques we have developed.

In this thesis we have always considered the atoms to be frozen in
place. It is true that the atoms are sufficiently cold that they
move only a small fraction of their average interatomic spacing
during the course of the experiment, but close pairs of atoms can
be separated by distances considerably smaller than this, and
hence the effects of interatomic motion can be considered.
Certainly this would be a worthwhile endeavor, since one would
ideally like to eventually have a theory that reduces to the
frozen results of this thesis at low temperatures and, as the
temperature increases, eventually makes the transition to the
high-temperature regime where the results are dominated by binary
atomic collisions. Some ideas for the inclusion of finite
temperature effects can be gleaned from the extensive literature
on dipolar
liquids~\cite{Chandler1982a,Hall1985a,Hoye1981a,Hoye1982a,Logan1987a,
Logan1987b,Thompson1982a}.

Dipole-dipole interactions among frozen Rydberg atoms may one day
be useful for quantum computing, as methods for producing
entanglement in such systems have already been studied
theoretically~\cite{Brennan2000a,Jaksch2000a,Lukin2000a} as well
as experimentally~\cite{Maitre1997a}. This method for quantum
computing has advantages over others that have been proposed.
Entanglement of photon pairs, for example, has the disadvantage
that the production of such pairs is not very efficient.

Experiments have been done recently that illustrate the creation
of ultracold plasmas using Rydberg
atoms~\cite{Killian1999a,Kulin2000a}, and even the spontaneous
evolution of Rydberg atoms into an ultracold
plasma~\cite{Robinson2000a}. Unlike in a high-temperature plasma
where the kinetic energy of the particles always dominates, an
ultracold plasma can be strongly coupled because the Coulomb
interaction between neighboring particles can exceed their kinetic
energy. Therefore such plasmas are of theoretical interest because
they exhibit qualitatively different features than their
high-temperature counterparts.

Although we have averaged over random distributions of atoms in
this work, it is also possible to construct periodic optical
potentials~\cite{Grynberg1993a,Hemmerich1993a,Kastberg1995a,Verkerk1992a}
and even quasiperiodic optical lattices~\cite{Guidoni1997a}. In
the periodic case it is necessary to, instead of averaging,
consider instead the appropriate dipolar sums for regular
lattices, which are discussed in Refs.~\cite{Cohen1955a} and
\cite{Fujiki1987a}.

\singlespacing
\markright{Bibliography}
\bibliography{thesis}
\bibliographystyle{plain}

\end{document}